\documentclass[
11pt, 
english, 
singlespacing, 
headsepline, 
]{MastersDoctoralThesis} 

\usepackage[utf8]{inputenc} 
\usepackage[T1]{fontenc} 

\usepackage{import}
\usepackage{xcolor}

\usepackage{color}
\usepackage[colorlinks=true,urlcolor=blue,breaklinks,citecolor=blue]{hyperref}

\usepackage[justification=justified]{caption}
\usepackage{float}
\usepackage{graphicx}
\usepackage{amsmath}
\usepackage{amssymb}
\usepackage{braket}
\usepackage{subfigure}
\usepackage{bm}
\usepackage{hyperref} 
\usepackage{doi}
\usepackage{cleveref}
\usepackage{listings}
\usepackage{bbold}

\usepackage[autostyle=true]{csquotes} 

\usepackage{pdfpages}

\usepackage{todonotes}
\usepackage{listings}

\definecolor{codegreen}{rgb}{0,0.6,0}
\definecolor{codegray}{rgb}{0.5,0.5,0.5}
\definecolor{codepurple}{rgb}{0.58,0,0.82}
\definecolor{backcolour}{rgb}{0.921, 0.929, 0.937} 

\lstdefinestyle{mystyle}{
backgroundcolor=\color{backcolour},   
commentstyle=\color{codegreen},
keywordstyle=\color{magenta},
numberstyle=\tiny\color{codegray},
stringstyle=\color{codepurple},
basicstyle=\ttfamily\footnotesize,
breakatwhitespace=true,         
breaklines=true,                 
captionpos=b,                    
keepspaces=true,                 
numbers=left,                    
numbersep=5pt,                  
showspaces=false,                
showstringspaces=false,
showtabs=false,                  
tabsize=2
}

\usepackage{makecell} 

\definecolor{fuchsia}{rgb}{1.0, 0.0, 1.0}

\newcommand{\CSF}{C_\text{SF}}
\newcommand{\CDW}{C_\text{DW}}
\newcommand{\Cs}{C_\text{HI}}
\newcommand{\Cst}{C_\text{HI}}
\newcommand{\chimax}{\chi_\text{max}}
\newcommand{\nmax}{n_\text{max}}
\newcommand{\tr}[1]{\text{tr}\left[#1\right]}

\usepackage{mathtools}

\DeclarePairedDelimiter\floor{\lfloor}{\rfloor}
\newcommand{\code}[1]{\lstinline|#1|}

\newcommand{\relu}{\text{ReLU}}

\usepackage{algorithm}
\usepackage{algpseudocode}

\thesistitle{Investigating Quantum Many-Body Systems with Tensor Networks, Machine Learning and Quantum Computers} 
\supervisor{Prof. Dr. Antonio \textsc{Ac\'in}\\ Prof. Dr. Maciej \textsc{Lewenstein}} 
\examiner{} 
\degree{Doctor of Philosophy} 
\author{Korbinian \textsc{Kottmann}} 
\addresses{} 

\subject{Quantum Physics} 
\university{\href{http://www.university.com}{University Name}} 
\department{\href{http://department.university.com}{Department or School Name}} 
\group{\href{http://researchgroup.university.com}{Research Group Name}} 
\faculty{\href{http://faculty.university.com}{Faculty Name}} 

\AtBeginDocument{
\hypersetup{pdftitle=\ttitle} 
\hypersetup{pdfauthor=\authorname} 
\hypersetup{pdfkeywords=\keywordnames} 
}

\begin{document}

\frontmatter 

\pagestyle{plain} 


\begin{titlepage}
\begin{center}

\vspace{2.5cm} 
\textsc{\Large PhD Thesis}\\[0.5cm] 

\vspace{2.5cm}
\HRule \\[0.4cm] 
{\huge \bfseries \ttitle\par}\vspace{0.4cm} 
\HRule \\[1.5cm] 
 
\begin{minipage}[t]{0.4\textwidth}
\begin{flushleft} \large
\emph{Author:}\\
{\authorname} 
\end{flushleft}
\end{minipage}
\begin{minipage}[t]{0.4\textwidth}
\begin{flushright} \large
\emph{Supervisor:} \\
{\supname} 
\end{flushright}
\end{minipage}\\[3cm]
 
%

 \begin{figure}[H]
\centering
\parbox{2.5cm}{
\includegraphics[width=2.5cm]{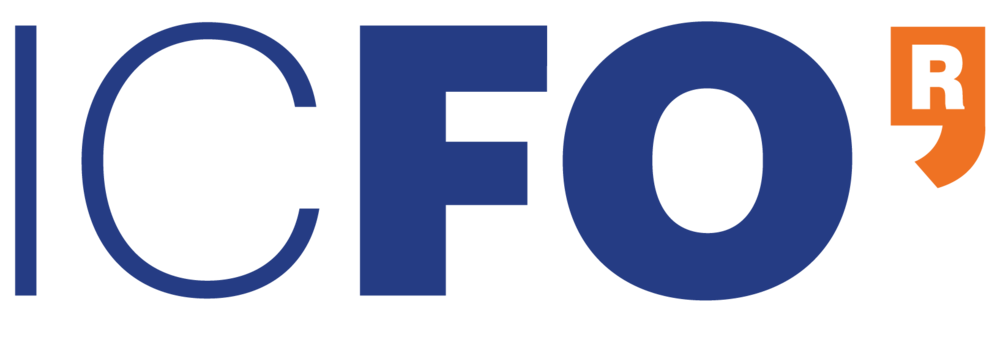} }
\hspace{4cm}
\begin{minipage}{2.5cm}
\includegraphics[width=2.0cm]{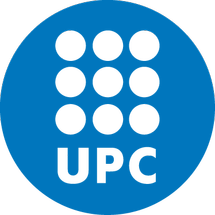}
\end{minipage}
\end{figure}

{\large \today}\\[4cm] 
ICFO - Institute of Photonic Sciences,
Avinguda Carl Friedrich Gauss, 3, 08860 Castelldefels, Barcelona
\\
Universitat Politecnica de Catalunya

\vfill
\end{center}
\end{titlepage}

\newpage


\chapter*{\centering Abstract}
\addchaptertocentry{Abstract}
We perform quantum simulation on classical and quantum computers and set up a machine learning framework in which we can map out phase diagrams of known and unknown quantum many-body systems in an unsupervised fashion.
The classical simulations are done with state-of-the-art tensor network methods in one and two spatial dimensions. For one dimensional systems, we utilize matrix product states (\textsf{MPS}) that have many practical advantages and can be optimized using the efficient density matrix renormalization group (\textsf{DMRG}) algorithm. The data for two dimensional systems is obtained from entangled projected pair states (\textsf{PEPS}) optimized via imaginary time evolution.
Data in form of observables, entanglement spectra, or parts of the state vectors from these simulations, is then fed into a deep learning (\textsf{DL}) pipeline where we perform anomaly detection to map out the phase diagram.
We extend this notion to quantum computers and introduce quantum variational anomaly detection. Here, we first simulate the ground state and then process it in a quantum machine learning (\textsf{QML}) manner. Both simulation and QML routines are performed on the same device, which we demonstrate both in classical simulation and on a physical quantum computer hosted by \textsc{IBM}.

\chapter*{\centering Resumen}
\addchaptertocentry{Resumen}
En esta tesis, realizamos simulaciónes cuánticas en ordenadores clásicos y cuánticos y diseñamos un marco de aprendizaje automático en el que podemos construir diagramas de fase de sistemas cuánticos de muchas partículas de manera no supervisada.
Las simulaciones clásicas se realizan con métodos de red de tensores de última generación en una y dos dimensiones espaciales. Para sistemas unidimensionales, utilizamos estados de productos de matrices (MPS) que tienen muchas ventajas prácticas y pueden optimizarse utilizando el eficiente algoritmo del grupo de renormalización de matrices de densidad (DMRG). Los datos para sistemas bidimensionales se obtienen mediante los denominados estados de pares entrelazados proyectados (PEPS) optimizados a través de la evolución en tiempo imaginario.
Los datos, en forma de observables, espectros de entrelazamiento o partes de los vectores de estado de estas simulaciones, se introducen luego en un algoritmo de aprendizaje profundo (DL) donde realizamos la detección de anomalías para construir el diagrama de fase.
Extendemos esta noción a los ordenadores cuánticos e introducimos la detección de anomalías cuánticas variacionales. Aquí, primero simulamos el estado fundamental y luego lo procesamos utilizando el aprendizaje automático cuántico (QML). Tanto las rutinas de simulación como el QML se realizan en el mismo dispositivo, lo que demostramos tanto en una simulación clásica como en un ordenador cuántico real de IBM.

\chapter*{\centering List of publications}
\addchaptertocentry{List of publications}
Peer-reviewed publications forming part of this thesis:

\begin{itemize}
  \item[] \cite{Kottmann2020} K. Kottmann, P. Huembeli, M. Lewenstein and A. Acín, \textit{Unsupervised Phase Discovery with Deep Anomaly Detection},
Phys. Rev. Letters 125 (17), 170603 (2020),  \\
doi:\href{http://doi.org/10.1103/PhysRevLett.125.170603}{10.1103/PhysRevLett.125.170603}
  \item[] \cite{Kottmann2021supersolid} K. Kottmann, A. Haller, A. Acín, G. E. Astrakharchik and M. Lewenstein, \textit{Supersolid-superfluid phase separation in the extended bose-hubbard model}, Phys. Rev. B 104, 174514
(2021),  
doi:\href{http://doi.org/10.1103/PhysRevB.104.174514}{10.1103/PhysRevB.104.174514}
  \item[] \cite{Kottmann2021} K. Kottmann, P. Corboz, M. Lewenstein and A. Acín, \textit{Unsupervised mapping of phase diagrams of 2D systems from infinite projected entangled-pair states via deep anomaly detection}, SciPost Phys. 11, 25 (2021),   \\
  doi:\href{http://doi.org/10.21468/SciPostPhys.11.2.025}{10.21468/SciPostPhys.11.2.025}
  \item[] \cite{Kottmann2021b} K. Kottmann, F. Metz, J. Fraxanet and N. Baldelli, \textit{Variational quantum anomaly detection: Unsupervised mapping of phase diagrams on a physical quantum computer}, Phys. Rev. Research 3, 043184 (2021),  \\
  doi:\href{http://doi.org/10.1103/PhysRevResearch.3.043184}{10.1103/PhysRevResearch.3.043184}
\end{itemize}

Peer-reviewed publications relevant to this thesis, but not forming part of it:
\begin{itemize}
\item[] \cite{Kaming2021} N. Käming, A. Dawid, K. Kottmann, M. Lewenstein,
K. Sengstock, A. Dauphin and C. Weitenberg, \textit{Unsuper-vised machine learning of topological phase transitions from experimental data}, Machine Learning: Science and Technology 2 (3), 035037 (2021),  
doi:\href{http://doi.org/10.1088/2632-2153/abffe7}{10.1088/2632-2153/abffe7}
\item[] \cite{Szoldra2021} T. Szołdra, P. Sierant, K. Kottmann, M. Lewenstein and J. Zakrzewski, \textit{Detecting ergodic bubbles at the crossover to many-body localization using neural networks}, Phys. Rev. B
104, L140202 (2021),  
doi:\href{http://doi.org/10.1103/PhysRevB.104.L140202}{10.1103/PhysRevB.104.L140202}
\end{itemize}

\chapter*{\centering Prephrase}
\addchaptertocentry{Prephrase}

As \textit{I}, the author, am guiding \textit{you}, the reader, through this thesis, I am going to write in the first person plural form, as \textit{we} make our way through the following chapters.

\hypersetup{linkcolor=black}
\tableofcontents 

\mainmatter 

\pagestyle{thesis} 

\part{Introduction}

\chapter{Motivation}



Developing tools to investigate large-scale interacting quantum systems promises the potential for unprecedented technological advancement in physics, chemistry, material science, medicine, or molecular biology on top of the fundamental understanding of the world. Applications range from optimizing photovoltaic material design \cite{Reddy2012} to drug discovery and design \cite{Shaher2016,Khatami2022}.
The phenomenon of high temperature superconductivity in cuprates \cite{Bednorz1986} and twisted bilayer graphene \cite{Cao2018} is still not understood \cite{Qin2020}.
Further, computational catalysis for large molecules could give access to means to potentially finding a suitable catalyst for Nitrogen fixation (Nitrogenase) \cite{Reiher2017} or carbon capturing \cite{Burg2021}, which are very relevant for the problems in agriculture and climate change that humanity is currently facing. These are few of a variety of applications that make developing methods to study quantum many-body systems highly desirable.

\

In this thesis, we are interested in quantum simulation, that is, studying the properties of quantum many-body systems in a controlled fashion. In particular, we are typically interested in the ground states of Hamiltonians. For example, in computational catalysis, one is interested in the ground state energies of the molecules involved in a catalytic cycle to determine the reaction rates and therefore, how viable a proposed catalyst is. On the other hand, peculiar quantum phases of matter like superfluids, supersolids or superconductors are exhibited at very low temperatures and are therefore described by the ground (and low-excited) states of the system. Further, the discovery of topological phases has extended the possibilities of (quantum) phases of matter that are of fundamental interest in physics and promise potential technological advancements.

There are two branches of quantum simulation: using classical computers or using quantum computers.

One way of utilizing classical computers to perform quantum simulation is to classically compute the wavefunction, that is, the vector in Hilbert space describing the state of the system. This approach suffers from the \textit{curse of dimensionality}, as the Hilbert space grows exponentially with the number of constituents. Take for example a system composed of $N$ local $d$-dimensional degrees of freedom, then the Hilbert space describing a general state of the composite system is of dimension $d^N$. It is known that only a fraction of those states in Hilbert space are physically accessible \cite{Poulin2011}, so the task at hand is to leverage this information to directly target those relevant states. One such approach is given by tensor network algorithms, where suitable Ans\"atze for the known restrictions of physical systems are optimized. The physical principle governing this is typically to target low- to intermediately entangled states, for which tensor network states are precisely designed. Since tensor network states usually require polynomial resources, it is worth noting that, therefore, states for intermediate sized systems of $\mathcal{O}(100)$ constituents but with low entanglement can still be described classically.

For states with high complexity and correlations, the other approach is to leverage a quantum system over which we have full control, a quantum computer, to encode the state of the quantum system we aim to simulate. However, it turns out to be very difficult to coherently control and manipulate interacting quantum systems. There has been tremendous progress in recent years, such that small scale quantum computers are commercially available today. These, however, are still inherently noisy and small, such that they serve more as a toy model and proof of principle for the moment. The hope is that in the short term of the next five to ten years, noise levels can be decreased and system sizes can be increased to be able to achieve a computational advantage for practically relevant problems over competing classical methods, such as, e.g. tensor networks. There have been claims for computational quantum advantage with contemporary hardware using a random circuit sampling approach \cite{Arute2019}, but these have already been caught up by tensor network simulations \cite{Pan2022}. More rigorous is the claim of quantum advantage for experimental Gaussian Boson Sampling \cite{Wu2021,Zhong2021}, though this is not on a universal quantum computer with unknown technological implications.

In the long run, the aim is to build a fault tolerant quantum computer of many qubits that can process very deep circuits to generate states of high complexity and entanglement. In such a device, error rates are low such that the few errors that occur can be corrected with quantum error correction, which requires a large overhead of physical qubits to logical qubits. This would allow for general purpose algorithms like adiabatic quantum computing and quantum phase estimation to investigate the above mentioned relevant systems, but also to provide solutions to optimization problems that are relevant for many industries. Further, quantum computers are relevant as they enable Shor's prime factoring algorithm \cite{Shor1994}, which poses a thread for public-key cryptographic systems\footnote{Most public-key cryptosystems are based on the Rivest–Shamir–Adleman (RSA) algorithm, which relies on the assumption that finding the prime factors of integer numbers is computationally hard, i.e. exponential in the number of integers. Shor's prime factoring algorithm, however, is polynomial in the number of integers and therefore violates this assumption.}.


\

Independently and parallel to these developments, deep learning underwent booming progress in the past decade. Much of its theory was developed already in the second half of the 20th century, but large amounts of data and hardware to rapidly process were not available yet back then. This changed with an ever-growing internet yielding more and more data, and the development of faster and more specialized hardware, i.e. the introduction of graphics processing units (GPUs)\footnote{GPUs are more restricted than general-purpose central processing units (CPUs) as they are specialized in performing very rapid computations of large data in parallel. They were primarily developed for computer games but soon found other applications like deep learning. Other application-specific integrated circuits (ASICs) like the tensor processing unit (TPU) are developed specifically for deep learning by \textsf{Google}.}. This boosted the field of artificial intelligence (AI) to unprecedented successes for tasks like image recognition \cite{Ciregan2012}, natural language processing \cite{attentionisallyouneed}, or playing games \cite{Silver2018}.

\

\paragraph{Vision of this thesis}
We want to use the aforementioned tools to simulate quantum many-body systems and apply deep learning methods to investigate them. The bigger vision we have in mind is an artificial intelligence that performs quantum simulation of various different systems and automatically points out new properties, effects or phases. The main contribution of this thesis to this endeavor is providing methods to map out the phase diagrams from quantum simulation data in an unsupervised fashion requiring no prior knowledge of the system. We demonstrate this on different known physical models as a proof of principle for 1D and 2D tensor network quantum states in classical simulation and quantum states simulated on a quantum computer. The latter is very much in line with recent proposals to use quantum computers to learn from quantum experiments (quantum machine learning with quantum data) \cite{Huang2021}. While these experiments on quantum computers are still in the stage of a proof of concept, we find a phase that has previously been mostly overlooked with classical simulations. This leads us to further physical investigations, discovering previously unknown effects such as a superfluid phase with a hidden broken translational symmetry in the extended Bose Hubbard model.
It is worth noting that in its current form, the employed deep learning methods merely point out regions of interest, but the physical investigation still has to be performed by an expert physicist. Deep learning methods for  gaining physical insights \cite{Iten2020} or interpretability \cite{Dawid2020,Dawid2021,dawid2022modern} might further elevate those efforts but are not subject to this thesis.

\

We start this thesis by introducing tensor network methods in \cref{sec:tensor_networks}, deep learning in \cref{chap:deep-learning} and quantum computing with noisy contemporary hardware via variational quantum algorithms in \cref{chap:qc}. The main results are outlined in \cref{part2}: We start by introducing anomaly detection for physical discovery in \cref{chapter:anomaly}. This method is then applied to the one dimensional Bose Hubbard model in \cref{chapter:bose-hubbard}, where we also perform the physical investigations of the new properties that the machine learning algorithm hinted at. We further demonstrate the viability of deep anomaly detection for two dimensional tensor network data in \cref{chapter:peps}. Finally, we translate this approach to a quantum computer where we perform both the quantum simulation and unsupervised anomaly detection on the same device in \cref{chapter:vqad}, before we conclude in \cref{conclusions}.




\chapter{Introduction to Tensor Network methods for quantum many-body systems}
\label{sec:tensor_networks}

The Hilbert space of all possible quantum states for $N$ particles is exponentially large in $N$. But not all states are physically achievable \cite{Poulin2011}\footnote{In \cite{Poulin2011}, the authors argue that most of the states in Hilbert space can only be produced in an exponential amount of time as they show that the manifold of states that can be realized by a polynomial time-evolution of local Hamiltonians is exponentially small.}. Particularly, it is known that ground states of local and gapped Hamiltonians follow the \textit{area law of entanglement} \cite{Srednicki1993,Plenio2005,Eisert2010} and therefore only occupy an (exponentially) small fraction of the Hilbert space. That is, the entanglement entropy $S(\rho_A)$ of a subsystem $A$ scales with the surface area of the volume that $A$ is occupying in real space, while for a general state it scales with the volume. Matrix product states (MPS) in one spatial dimension and projected entangled pair states (PEPS) in two spatial dimensions are states reproducing the area law of entanglement and are therefore natural Ansätze for ground states of local and gapped Hamiltonians. 

A drastic consequence of the area law is that for one dimensional systems the entanglement entropy for any subsystem is constant, independent of its size. This property was leveraged by White already in 1992 with the invention of the density matrix renormalization group algorithm (DMRG)  \cite{White1992,White1993} to study the ground states of quantum many-body systems. This algorithm is the foundation of modern tensor network methods, as it was realized that one can describe these ground states in terms of matrix product states (MPS) \cite{Fannes1992,Ostlund1995,Dukelsky1998}.
MPS are very powerful as they offer a canonical form that allows for very efficient calculation of observables \cite{Vidal2003,Vidal2004TEBD,Vidal2006,Verstraete2007} and efficient optimization via DMRG with matrix product operators \cite{McCulloch2007}. For a modern review we point to \cite{SCHOLLWOCK2011}, whose line of thought we partly adopt in the following. As we will see later, things get more complicated for tensor network states in higher dimensions due to the lack of a canonical form \cite{Ran2020}. Despite this difficulty, competitive methods in two \cite{Verstraete2004,Orus2013,corboz14_tJ,Corboz2017} and three \cite{Xie2012,Akiyama2019,Williamson2021,Bloch2021} spatial dimensions have been demonstrated.
Furthermore, tensor networks provide state of the art results in quantum chemistry \cite{Baiardi2020} and quantum computation \cite{Pan2022}.

We start this chapter by introducing the general concepts and special properties of matrix product states in \cref{sec:SVD,sec:canonical_form}. We then introduce the DMRG algorithm in \cref{sec:DMRG}, following closely the logic of \cite{SCHOLLWOCK2011}. Sections  \ref{sec:SVD}  to \ref{sec:DMRG} are accompanied by \textsf{Python 3} code along with the mathematical explanations. The notion of 
\textit{infinite} tensor network states, i.e. states in the thermodynamic limit, are introduced in \cref{sec:other_algorithms}, as well as imaginary time evolution. The latter is rather a means for consistency checking of DMRG in 1D, but very relevant for PEPS, which we introduce in \cref{sec:introPEPS}. We conclude with discussing the area law, showing how MPS and PEPS exactly reproduce it and are therefore suitable Ansätze for ground states of gapped and local Hamiltonians in \cref{sec:area_law}.

\section{Singular Value Decomposition}
\label{sec:SVD}

\begin{figure}
\centering
\includegraphics[width=.77\textwidth]{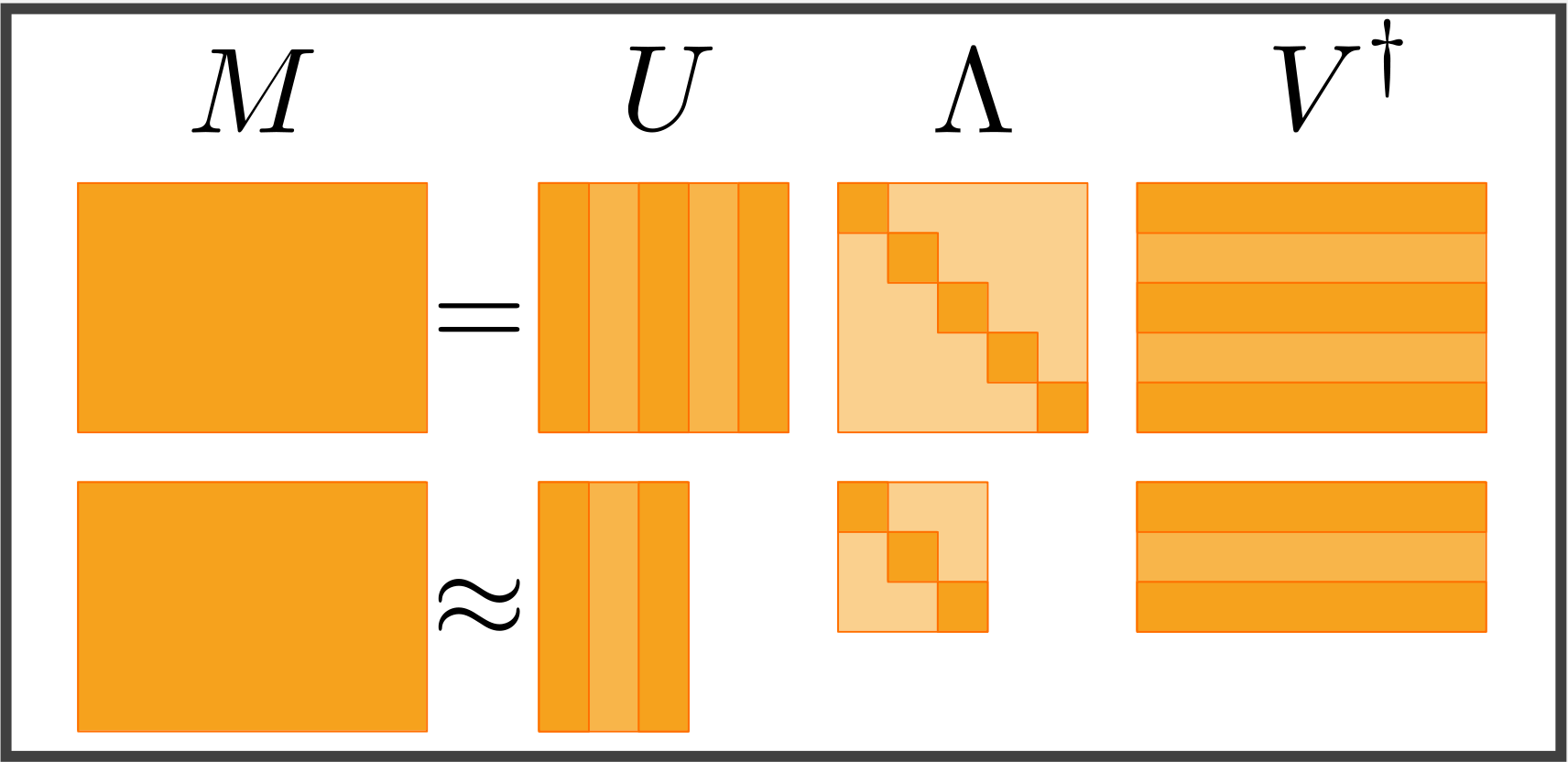}
\caption{Illustration of the general singular value decomposition $M = U \Lambda V^{\dagger}$ of $M$ and its compression by discarding the two smallest singular values and the respective left- and right singular vectors. Note that descending order of the positive singular values is assumed.}
\label{fig:svd}
\end{figure}

Let us start by recalling one of the most important techniques from linear algebra in general, and especially important to tensor network methods, namely the singular value decomposition (SVD) of an arbitrary matrix $M \in \mathbb{C}^{m\times n}$ in terms of
\begin{equation}
\label{eq:SVD}
M = U \Lambda V^\dagger,
\end{equation}
where $\Lambda$ is a diagonal matrix containing the $r = \min(m,n)$ positive \textit{singular values} $\{\lambda_i\}_{i=1}^{r}$ of $M$. The matrix $U\in \mathbb{C}^{m\times r}$ consists of orthonormal \textit{columns} whereas the matrix $V^\dagger \in \mathbb{C}^{r\times n}$ consists of orthonormal \textit{rows} - the left- and right-orthogonal singular vectors of $M$, respectively. In the case of $m=n$, $U$ and $V$ are unitary, however we will almost exclusively be dealing with rectangular matrices in the context of MPS. The case of $n>m$ is illustrated in \cref{fig:svd}.

In practice, SVD can always be achieved by diagonalizing the Hermitian matrix $M M^\dagger \in \mathbb{C}^{m\times m}$ and we can identify the eigenvectors corresponding to the non-zero eigenvalues as $U$, and the square roots of the positive and real eigenvalues as $\Lambda$. We can obtain $V$ in a similar fashion by diagonalizing $M^\dagger M \in \mathbb{C}^{n\times n}$.

One of the many applications of SVD is the compression of matrices. This can be achieved by truncating the singular values below a certain threshold, and discarding the respective singular vectors, as illustrated in \cref{fig:svd}. The resulting approximation $M'$ is optimal in terms of to the Frobenius norm $||M - M'||_F$, where $||A||_F^2 = \tr{A^\dagger A}$ for any $A \in \mathbb{C}^{m \times n}$, which is just the sum of squared singular values (i.e. eigenvalues of $A^\dagger A$). This is in general a very powerful property which serves as the foundation of matrix product state approximations of general quantum many-body states.

\section{Canonical form of an MPS}
\label{sec:canonical_form}
Let us now look at a general quantum state
\begin{equation}
\label{eq:full_state_tensor}
\ket{\Psi} = \sum_{\sigma_1, .. , \sigma_L} c_{\sigma_1, .. , \sigma_L} \ket{\sigma_1, .. , \sigma_L}
\end{equation}
consisting of local, spin-like degrees of freedom $\sigma_i \in \{0, 1, .., d-1\}$. This state is completely characterized by the rank $L$ tensor $c$ with components $c_{\sigma_1, .. , \sigma_L} \in \mathbb{C}^{d^{L}}$. The key trick of MPS is to approximate this exponentially large tensor by a product of smaller tensors. We can formally exactly map the tensor to such a product by consecutively performing SVD
\begin{multline}
c_{\sigma_1, .. , \sigma_L} \stackrel{\text{reshape}}{=} c_{\sigma_1 , (\sigma_2,..,\sigma_L)} \stackrel{\text{SVD}}{=} \sum_{\mu_1} U_{\sigma_1, \mu_1} \Lambda_{\mu_1} V^\dagger_{\mu_1, (\sigma_2,..,\sigma_L)} \\= \sum_{\mu_1} U^{\sigma_1}_{\mu_1} c_{\mu_1}^{\sigma_2,..,\sigma_L},
\end{multline}
where in the first step we reshaped the tensor into a matrix by combining the right-most indices\footnote{Some explicit examples for combining indices: for $d=2$, i.e. spin 1/2, the string of spin variables $\sigma_1\sigma_2\sigma_3$ represents a binary number, i.e. $000 = 0$, $001 = 1$, $010 = 2$, $011 = 3$ and so on. For $d=3$ we could use ternary numbers and so on.}. We can then treat 
\begin{equation}
\Lambda_{\mu_1} V^\dagger_{\mu_1,(\sigma_2,..,\sigma_L)} = c_{\mu_1}^{\sigma_2,..,\sigma_L}
\end{equation}
as a new tensor with \textit{virtual} index $\mu_1$, which is being summed over, and \textit{physical} indices $\sigma_2,..,\sigma_L$, that we from now on write as superscripts to make the distinction. We can repeat this step for $c_{(\mu_1 \sigma_2),(\sigma_3,..,\sigma_L)}$ and all following new tensors all the way through the remaining variables and end up with
\begin{multline}
c_{\sigma_1,..,\sigma_L} = \sum_{\mu_1,..,\mu_{L-1}} U^{\sigma_1}_{\mu_1} U^{\sigma_2}_{\mu_1, \mu_2} \cdots U^{\sigma_{L-1}}_{\mu_{L-2},\mu_{L-1}} U^{\sigma_L}_{\mu_{L-1}}
\end{multline}
and therefore find the original state in its matrix product state form
\begin{equation}
\ket{\Psi} = \sum_{\sigma_1,..,\sigma_L} U^{\sigma_1} U^{\sigma_2} \cdots U^{\sigma_{L-1}} U^{\sigma_L} \ket{\sigma_1,..,\sigma_L}.
\label{eq:MPS_left_canonical}
\end{equation}
We did not write the summation over virtual indices explicitly and left it implicitly as matrix multiplications. This is where the name matrix product state comes from, despite mostly consisting of rank 3 tensors and not matrices.
There are several key features of this representation that we want to point out.
\subsection{Compression} 
First of all, note that by doing this decomposition we actually have not gained anything in terms of dimensionality because the innermost tensor has shape $(d^{L/2},d,d^{L/2})$ (for $L$ even). An efficient compression is achieved by discarding some of the smaller singular values as discussed before. In practice, the user sets a hyper parameter $\chimax \in \mathbb{N}$ called the bond dimension, until which we keep all the singular values. The tensors therefore have constant dimensions $(\chimax,d,\chimax)$\footnote{For site $i$ at the boundaries of the MPS the dimension is $\min(d^i,\chimax)$ and typically one additionally sets a threshold for the smallest singular values to keep, which can further reduce the dimension.}.

\subsection{Norm}
We will verify that $\ket{\Psi}$ is correctly normalized, assuming that the original state in \cref{eq:full_state_tensor} was normalized. Recall that the $U$ matrices in the SVD are left-orthogonal, which implies 
\begin{equation}
\sum_{\sigma_i} (U^{\sigma_i})^\dagger U^{\sigma_i} = \mathbb{1}
\end{equation}
and therefore
\begin{multline}
\braket{\Psi|\Psi} = \sum_{\sigma_1,..,\sigma_L} (U^{\sigma_L})^\dagger .. (U^{\sigma_1})^\dagger U^{\sigma_1} .. U^{\sigma_L} = 1.
\end{multline}
Note that $i=1$ and $i=L$ are special cases due to their reduced dimensionality and should in this notation be regarded as rank 3 tensors with a trivial third dimension, i.e. the last reduction yields $\sum_{\sigma_L,\mu_{L-1}} U^{\sigma_L\dagger}_{1,\mu_{L-1}} U^{\sigma_L}_{\mu_{L-1},1} = \mathbb{1}_{1,1} = 1$, where subscript $(\bullet)_1$  indicates the trivial dimension of size 1.

\subsection{Canonical form}
The process of sweeping through the system as described above has brought the state in its so-called left-canonical form, where we can make use of the left-orthogonality for efficient tensor contractions as shown for the norm. Now note that we can do this process not just for a state in tensor form  \cref{eq:full_state_tensor} but also to a state that is already in MPS form but with different matrices. We can have the same state described by different sets of matrices because we can always insert $\mathbb{1} = MM^{-1}$ for some random invertible matrix $M$ into \cref{eq:MPS_left_canonical}. Sweeping through the system from left to right is one way to fix this \textit{gauge freedom} to left-canonical form. Sweeping from right to left would yield the right-canonical form. But we can actually find an even more convenient form from \cref{eq:MPS_left_canonical} that combines the best of both worlds. We do so by storing the singular values $\Lambda^{[i]}$ after multiplying them onto the next tensor, insert $\Lambda^{[i]}(\Lambda^{[i]})^{-1}$ in \cref{eq:MPS_left_canonical} after the sweep, and identify $(\Lambda^{[i-1]})^{-1} U^{\sigma_i} = \Gamma^{\sigma_i}$ to then obtain the canonical form after Vidal \cite{Vidal2003}
\begin{equation}
\label{eq:vidal-form}
\ket{\Psi} = \sum_{\sigma_1,..,\sigma_L} \Gamma^{\sigma_1} \Lambda^{[1]} \Gamma^{\sigma_2} \Lambda^{[2]} \cdots \Gamma^{\sigma_{L-1}} \Lambda^{[L-1]} \Gamma^{\sigma_L} \ket{\sigma_1,..,\sigma_L}.
\end{equation}
To obtain right- or left-orthogonal matrices is now just a matter of re-grouping $\Gamma$ matrices and the singular values. For example, we can identify $\Lambda^{[i-1]} \Gamma^{\sigma_i} = U^{\sigma_i}$ and $\Gamma^{\sigma_i} \Lambda^{[i+1]}  = V^{\sigma_i\dagger}$ to obtain a mixed canonical form
\begin{multline}
\label{eq:mixed_canonical}
\ket{\Psi} = \sum_{\sigma_1,..,\sigma_L} \left[\Gamma^{\sigma_1} \right] \left[\Lambda^{[1]} \Gamma^{\sigma_2}\right] \cdots \left[\Lambda^{[i-2]} \Gamma^{\sigma_{i-1}} \right] \\
 \Lambda^{[i-1]} \Gamma^{\sigma_i} \Lambda^{[i]}  \cdots \left[\Gamma^{\sigma_{L-1}} \Lambda^{[L-1]}\right] \left[ \Gamma^{\sigma_L} \right]\ket{\sigma_1,..,\sigma_L} \\
 = \sum_{\sigma_1,..,\sigma_L} U^{\sigma_1} U^{\sigma_2} \cdots U^{\sigma_{i-1}} \Lambda^{[i-1]} \Gamma^{\sigma_i} \Lambda^{[i]} V^{\sigma_i+1\dagger} \cdots V^{\sigma_{L-1}\dagger} V^{\sigma_L\dagger}.
\end{multline}

\subsection{Operator expectation values}
\begin{figure}
\centering
\includegraphics[width=.99\textwidth]{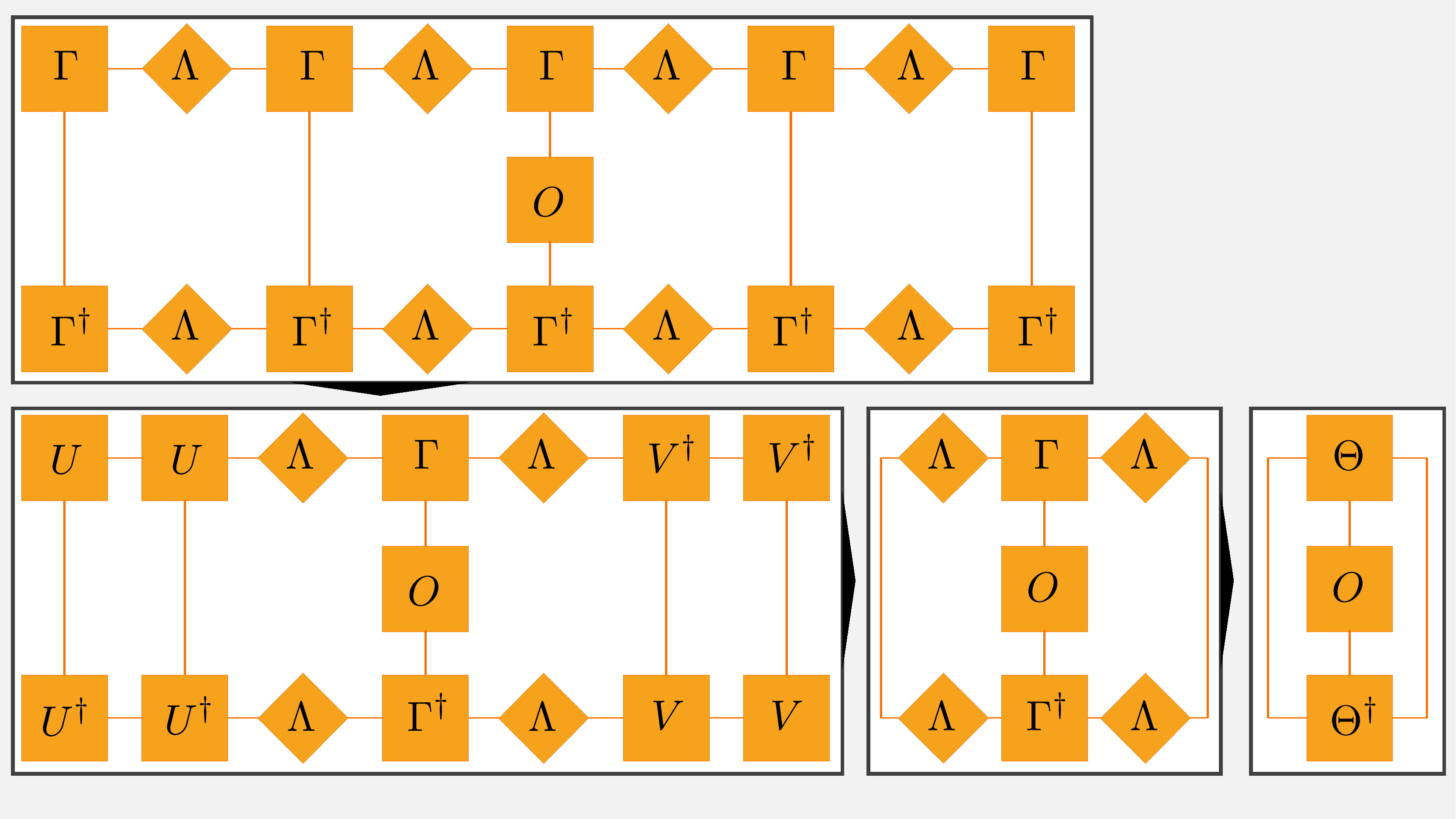}
\caption{Calculating local expectation values $\braket{\Psi|O|\Psi}$ for a $\ket{\Psi}$ in canonical form \cref{eq:vidal-form} reduces to a contraction with the $\Theta$ tensor at the local site. In the first step, we recombine tensors to a mixed canonical form as described in \cref{eq:mixed_canonical}. In the second step, we make use of the left- and right-orthogonality of $U$ and $V^\dagger$ tensors. In the last step we just use the definition of the $\Theta$ tensor, which, knowing this relationship for local observables, plays a special role in MPS.}
\label{fig:mps_exp_value}
\end{figure}

This canonical form is handy for calculating expectation values of local observables, see \cref{fig:mps_exp_value}. For some $d\times d$ dimensional local operator $O^{[i]}$ at site $i$, the expectation value with respect to $\ket{\Psi}$ reduces to 
\begin{equation}
\braket{\Psi|O|\Psi} = \tr{\sum_{\sigma_i\tilde{\sigma}_i}\Theta^{\sigma_i\dagger} O^{\sigma_i\tilde{\sigma}_i} \Theta^{\tilde{\sigma}_i}}
\end{equation}
where we identified $\Theta^{\sigma_i} = \Lambda^{[i-1]} \Gamma^{\sigma_i} \Lambda^{[i]}$. The reasoning for this reduction is graphically illustrated in \cref{fig:mps_exp_value}.

We can use the same logic when calculating correlation functions like $\braket{\Psi|O^{[i]} O^{[j]}|\psi}$: by making use of the canonical form we know that contractions for sites below $i$ and sites beyond $j$ are identities. However, we cannot get around explicitly computing the contractions between site $i$ and $j$. There are many strategies for contracting such a tensor network. The best way to efficiently do so is nicely illustrated in Fig. 17 in \cite{Orus2013} and amounts to moving from left to right to avoid large tensors, i.e. tensors with many legs.

\subsection{Reduced states and entanglement}
The canonical form \cref{eq:vidal-form} is especially handy when we are interested in different marginals (reduced density matrices) of the state. This on the other hand gives us easy access to different entanglement properties, which we will make great use of throughout this thesis.
Let us first start by noting that, by construction, the singular values $\Lambda^{[i]}$ are the Schmidt coefficients of a bipartition $A = \{1,..,i\}$ and $B = \{i+1,..,L\}$ of a matrix product state, i.e.
\begin{align}
&\ket{\Psi} = &\sum_{\mu_i} \Lambda^{[i]}_{\mu_i} \ket{\mu_i}_{A} \ket{\mu_i}_{B} & \\
&\ket{\mu_i}_A =& \sum_{\sigma_1,..,\sigma_i} \left(U^{\sigma_1} \cdots U^{\sigma_i} \right)_{\mu_i} \ket{\sigma_1,..,\sigma_i} & \\
&\ket{\mu_i}_B =& \sum_{\sigma_{i+1},..,\sigma_L} \left(V^{\sigma_{i+1}\dagger} \cdots V^{\sigma_L\dagger} \right)_{\mu_i} \ket{\sigma_{i+1},..,\sigma_L}. &
\end{align}
The states $\ket{\mu_i}_A$ and $\ket{\mu_i}_B$ form an orthonormal set due to the orthonormality of $U$ and $V^\dagger$ matrices. With this we can directly read off the reduced states for subsystems $A$ and $B$ in terms of
\begin{align}
\rho_A = & \sum_{\mu_i} \Lambda^{[i]2}_{\mu_i} |\mu_i\rangle\langle \mu_i|_A & \label{eq:rhoA_general}\\
\rho_B = & \sum_{\mu_i} \Lambda^{[i]2}_{\mu_i} |\mu_i\rangle\langle \mu_i|_B &
\end{align}
With this we can directly obtain the von Neumann entanglement entropy for any bipartition of an MPS in canonical form \cref{eq:vidal-form}
\begin{multline}
\label{eq:S_MPS}
S_{A|B} = -\tr{\rho_A \log(\rho_A)} = - \tr{\rho_B \log(\rho_B)} \\= -\sum_{\mu_i} \Lambda^{[i]2}_{\mu_i} \log\left( \Lambda^{[i]2}_{\mu_i} \right).
\end{multline}
That is why the singular values, i.e. the Schmidt values, and the entanglement energies $\xi$ in $(\Lambda^{[i]}_j)^2 = \exp(-\xi^{[i]}_j)$, are all amiguously referred to as the Entanglement spectrum. It is an interesting quantity in its own right, from which we can learn different properties, as we will see in the main body of this thesis.

\subsection{Python implementation}

We provide a \textsf{Python} implementation of an \lstinline|MPS| class that we later use for our DMRG implementation. For bringing the \lstinline|MPS| into left-canonical form via \lstinline|left_normalize()|, we use QR decomposition instead of SVD, which is more efficient as it does not compute the explicit singular values. A method for bringing the \lstinline|MPS| into right-canonical form is left for brevity and can be found in the full code in \cite{github_dmrg}, which is heavily inspired by the codebase of TeNPy \cite{tenpy}.

\begin{lstlisting}[language=Python]
import numpy as np
from numpy.linalg import qr

def chi_list(L,d,chi_max):
    '''
    Creates a list of the appropriate local bond dimensions.
    This is important for the dimensions near the boundaries. 
    '''
    a = np.ones(L+1,dtype=np.int_)
    for i in range(int(L/2)+1):
        if d**i <= chi_max:
            a[i] = d**i
            a[-i-1] = d**i
        else:
            a[i] = chi_max
            a[-i-1] = chi_max
    return a
    
def create_random_Ms(L,d,chi_max):
    """
    returns a list of random MPS matrices (np_array(ndim=3))
    """
    chi = chi_list(L,d,chi_max)
    return [np.random.rand(chi[i],d,chi[i+1]) for i in range(L)]


class MPS(object):
    """
    Initializes a random, finite dimensional, unnormalized MPS
    
    Parameters
    ------------
    L: Number of Sites
    d: local Hilbert space dimension
    chi: local bond dimension
    
    attributes:
    Ms: list of L ndim=3 tensors M
    Index convention for M: sigma_j, vL, vR
    
    Ss: list of L ndim=1 lists (singular values)    
    """
    def __init__(self,L,d,chi_max):
        self.Ms = create_random_Ms(L,d,chi_max)
        self.Ss = np.random.rand(L,d,chi_max)
        self.L = L
        self.d = d
        self.chi_max = chi_max
        self.chi = chi_list(L,d,chi_max)
    
    def self_norm(self):
        """calculate the norm of the MPS"""
        C = np.tensordot(self.Ms[0].conjugate(), self.Ms[0], ([0,1],[0,1])) 
        for i in range(1,self.L):
            # sum over physical sites 1 .. L
            for j in range(self.d):
                # sum over physical indices s_i
                """
                M1: sl al a'l
                M2: sl al a'l
                """
                temp1 = np.tensordot(C, self.Ms[i], [1,0])
                C2 = np.tensordot(self.Ms[i].conj(), temp1, ([0,1],[0,1]))
            C = C2
        return C[0,0]
        
    def left_normalize(self):
        Ms = self.Ms
        L,d = self.L,self.d
        As = []
        for i in range(L):
            chi1,d,chi2 = Ms[i].shape
            m = Ms[i].reshape(chi1*d,chi2)
            # QR decomposition is like SVD 
            # w/o the explicit singular values
            Q,R = qr(m, mode='reduced')
            A = Q.reshape(chi1,d,min(m.shape)) 
            if i<(L-2): 
                Ms[i+1] = np.tensordot(R,Ms[i+1],1) 
        self.Ms = As
\end{lstlisting}

\section{Matrix Product Operators}
\label{sec:MPO}
\begin{figure}
\centering
\includegraphics[width=.99\textwidth]{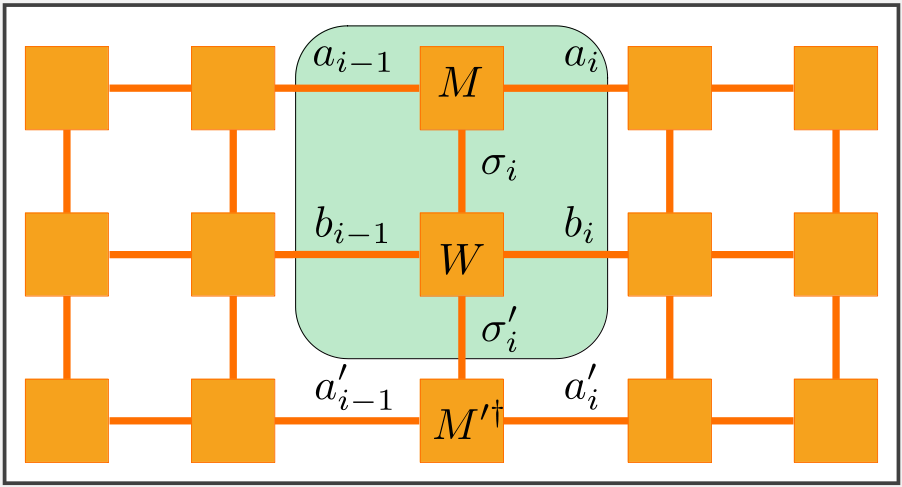}
\caption{Graphical representation of an MPO contracted with two MPS $\braket{\Psi'|O|\Psi}$. Looking solely at the first contraction, we can see how we obtain a new MPS with larger matrices with double stranded bonds as highlighted in the green box. This represents the new matrix after contraction in \cref{eq:Mtilde}.}
\label{fig:MPO}
\end{figure}
Matrix product states are a way to represent multi-partite quantum states. Matrix product operators generalize this concept and provide representations of sums of operators as
\begin{equation}
O = \sum_{\bm{\sigma}\bm{\sigma'}} W^{\sigma_1\sigma'_1} W^{\sigma_2\sigma'_2} \cdots W^{\sigma_L\sigma'_L} \ket{\bm{\sigma'}}\bra{\bm{\sigma}},
\end{equation}
where $\bm{\sigma} = (\sigma_1, \sigma_2, \cdots, \sigma_L)$ is the vector of indices. So just as $M^{\sigma_i}$ can be seen as a vector in $\sigma_i$, where each element is a matrix, $W^{\sigma_i\sigma'_i}$ can be seen as a matrix, where each element is a matrix.

We can apply an MPO to an MPS and obtain a new MPS
\begin{eqnarray}
O\ket{\Psi} = \sum_{\bm{\sigma}\bm{\sigma'}} \left(W^{\sigma_1\sigma'_1} \cdots W^{\sigma_L\sigma'_L}\right) \left(M^{\sigma_1} \cdots M^{\sigma_L} \right) \ket{\bm{\sigma'}} \\
= \sum_{\bm{\sigma}\bm{\sigma'},\bm{a}\bm{b}} \left(W^{\sigma_1\sigma'_1}_{b_0 b_1} M^{\sigma_1}_{a_0 a_1}\right) \left(W^{\sigma_2\sigma'_2}_{b_1 b_2} M^{\sigma_2}_{a_1 a_2}\right) \cdots \ket{\bm{\sigma'}} \\
 = \sum_{\bm{\sigma'}} \tilde{M}^{\sigma'_1} \tilde{M}^{\sigma'_2} \cdots \ket{\bm{\sigma'}}`
\end{eqnarray}
with new, larger matrices
\begin{equation}
\label{eq:Mtilde}
\tilde{M}^{\sigma'_i}_{(b_{i-1}a_{i-1}),(b_i,a_i)} = \sum_{\sigma_i} W^{\sigma_i\sigma'_i}_{b_{i-1} b_i} M^{\sigma_i}_{a_{i-1} a_i}
\end{equation}
with double stranded virtual bonds, as highlighted in \cref{fig:MPO}.

MPOs are efficient representations of Hamiltonians with local interactions as it results in low MPO bond dimension $D_W$ (for virtual MPO bonds $b_i$), which grows with the number of terms and the range of the interactions\footnote{A known exception is exponentially decaying interactions for which there is a compact representation \cite{Frowis2010}.} \cite{McCulloch2007}. Look for example at the simple translationally invariant Hamiltonian $H^{(1)} = \sum_i X_i$, which we can represent by
\begin{equation}
W^{[1]} = (X, \mathbb{1}) ; W^{[i]} = \begin{pmatrix}
\mathbb{1} & 0 \\ X & \mathbb{1}
\end{pmatrix} ; W^{[L]} = (\mathbb{1},X)^T,
\end{equation}
where implicitly the operators in $W^{[i]}$ act on site $i$. Note that by the same logic we can have site-dependent coefficients for each operator term. To see that this indeed represents $H^{(1)}$ we can explicitly contract this MPO for $L=3$ and obtain
\begin{multline}
W^{[1]} W^{[2]} W^{[3]} = (X_1, \mathbb{1}_1) \begin{pmatrix}
\mathbb{1}_2 & 0 \\ X_2 & \mathbb{1}_2
\end{pmatrix} (\mathbb{1}_3,X_3)^T \\
= (X_1, \mathbb{1}_1) (\mathbb{1}_2 \mathbb{1}_3, X_2 \mathbb{1}_3 + \mathbb{1}_2 X_3)^T \\
= X_1 \mathbb{1}_2 \mathbb{1}_3 + \mathbb{1}_1  X_2 \mathbb{1}_3 + \mathbb{1}_1 \mathbb{1}_2 X_3.
\end{multline}

Similarily, for a nearest neighbor Hamiltonian $H^{(2)} = \sum_i X_i Y_{i+1}$ we can construct it via
\begin{equation}
W^{[i](2)} = \begin{pmatrix}
\mathbb{1} & 0 & 0 \\ Y & 0 & 0 \\ 0 & X & 0
\end{pmatrix}
\end{equation}
with $W^{[1]}$ and $W^{[L]}$ being the bottom row and left column, respectively. 

Formally, we can say that each site is described by $W^{[i]}$, including sites $1$ and $L$, but there are two \textit{dummy} vectors $(0,..,0,1)$ and $(1,0,..)^T$ on the left and right boundary, respectively.
Introducing different nearest-neighbor terms amounts to inserting these elements in the left column and bottom row, i.e. for a Heisenberg model of the form 
\begin{equation}
H^{(3)} = \sum_i J_i^x S^x_{i} S^x_{i+1} + J_i^y S^y_{i} S^y_{i+1} + J_i^z S^z_{i} S^z_{i+1} - h_i S^z_{i}
\end{equation}
we obtain
\begin{equation}
W^{[i](3)} = \begin{pmatrix}
\mathbb{1} & 0 & 0 & 0 & 0 \\
S^x & 0 & 0 & 0 & 0 \\
S^y & 0 & 0 & 0 & 0 \\
S^z & 0 & 0 & 0 & 0 \\
-h_i S^z & J_i^x S^x & J_i^y S^y & J_i^z S^z & \mathbb{1}.
\end{pmatrix}
\end{equation}
If on the other hand we are interested in next-to-nearest interactions like in $H^{(4)} = \sum_i X_i Y_{i+2}$ then we need to introduce an identity in the off-diagonal here:
\begin{equation}
W^{[i](4)} = \begin{pmatrix}
\mathbb{1}  & 0 & 0 & 0 \\
Y & 0 & 0 & 0 \\
0 & \mathbb{1} & 0 & 0 \\
0 & 0 & X & \mathbb{1}
\end{pmatrix}.
\end{equation}
Note that just like MPS, the MPO description of an operator $O$ is not unique.
With these four examples we are able to construct all Hamiltonians with local terms, nearest-neighbor interactions and next-nearest-neighbor interactions. For a more formal description of the above recipes in terms of finite automata we refer to \cite{Crosswhite2008}.

\section{Density Matrix Renormalization Group algorithm}
\label{sec:DMRG}
The Density Matrix Renomrmalization Group (DMRG) algorithm was originally proposed by White \cite{White1992,White1993} and found tremendeous success in finding the ground states of local and gapped one dimensional quantum Hamiltonians. Here, we discuss its modern re-inretpretation in terms of matrix product states. We first give an intuitive overview and then follow up with the mathematical details.
The algorithm is based on the variational principle of minimizing the expected energy
\begin{equation}
\min_{\ket{\Psi}} \frac{\braket{\Psi|H|\Psi}}{\braket{\Psi|\Psi}}
\end{equation}
for a trial matrix product state $\ket{\Psi}$ and a matrix product operator $H$. We follow the approach in \cite{SCHOLLWOCK2011} and introduce a Lagrangian multiplier to define the objective function
\begin{equation}
\label{eq:dmrg-obj-function}
\mathbb{O} := \braket{\Psi|H|\Psi} - \lambda \braket{\Psi|\Psi}
\end{equation}
that we want to minimize. Directly minimizing $\mathbb{O}$ is highly non-linear and therefore practically impossible for realistic hyper parameters $\chimax$ and $d$. Instead, one solves the problem \textit{locally} and sweeps through the system until convergence is reached.

We start by fixing all MPS parameters but those at site $i$. We can assume a mixed canonical form that can be obtained by the recipe given in \cref{sec:canonical_form}, such that $M^{\sigma_i}$ is of arbitrary gauge and all remaining sites are left- or right orthogonal, accordingly. The extremum condition of setting the derivative of $\mathbb{O}$ with respect to the elements of $M^{\sigma_i}$ to zero then yields the Eigenproblem
\begin{equation}
\label{eq:Heff}
H_\text{eff} \bm{M} - \lambda \bm{M} = 0
\end{equation}
for the effective Hamiltonian $H_\text{eff}$ (graphically illustrated in \cref{fig:Heff}) for that site and \textit{vector} $\bm{M} = M_{(\sigma'_i,a'_{i-1},a'_i)}$. The size of the vector $\bm{M}$ is $d \chimax^2$. In practice we use $d = \mathcal{O}(10)$ and $\chimax = \mathcal{O}(10-1000)$ and are therefore suitable for nummerical solvers like Lanczos that yield the lowest algebraic Eigenvalue and Eigenvector. The idea of DMRG is then to \textit{sweep} through the system, iteratively optimizing the tensor at each bond locally, while all other tensors are fixed, until some overall convergence criterion is reached. There are a lot of tricks that help speed up this procedure which we elaborate on in the mathematical details below.

\subsection{Mathematical details and Python implementation}

This subsection is dedicated to understanding DMRG in detail and providing a \textsf{Python} implementation. The full code can be found in \cite{github_dmrg}.

\begin{figure}
\centering
\includegraphics[width=.99\textwidth]{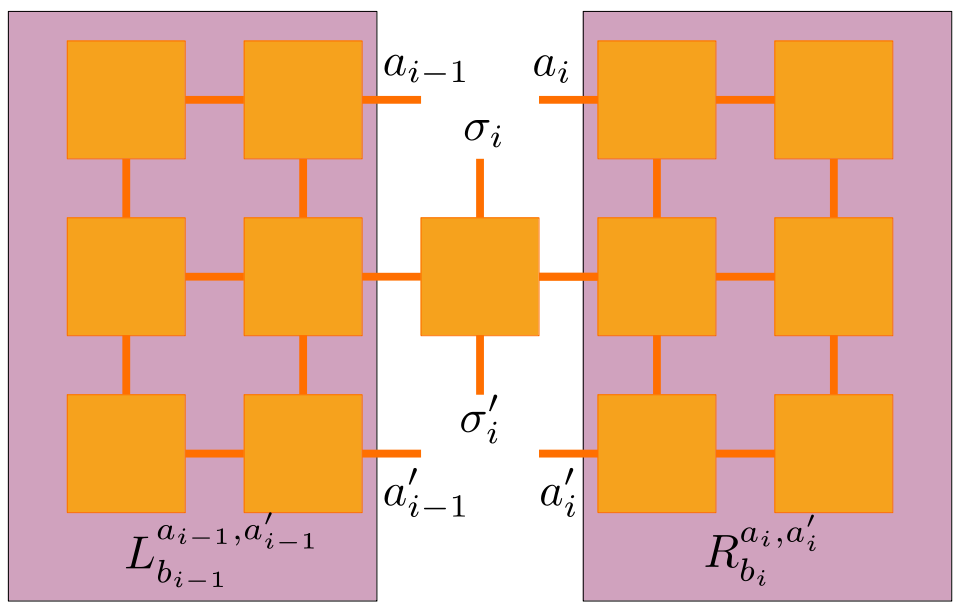}
\caption{Graphical representation of the effective Hamiltonian $(H_\text{eff})_{(\sigma_i, a_{i-1}, a_i),(\sigma'_i, a'_{i-1}, a'_i)}$ at site $i$ consisting of the local MPO element, the left environment $L$ and right environment $R$. The Eigenproblem in \cref{eq:Heff} is finding the Eigenvector $M_{(\sigma'_i, a'_{i-1}, a'_i)}$ corresponding to the smallest Eigenvalue of $H_\text{eff}$.}
\label{fig:Heff}
\end{figure}

Let us start by formally defining the left- and right-environments in an MPS-MPO-MPS contraction
\begin{eqnarray}
L_{b_{i-1}}^{a_{i-1},a'_{i-1}} = \sum_{\substack{a_\ell,b_\ell,a'_\ell \\ \ell \leq i-2}} \prod_{\ell=1}^{i-1} \left( \sum_{\sigma_\ell,\sigma'_\ell} U^{\sigma_\ell*}_{a_{\ell - 1}, a_\ell} W_{b_{\ell-1},b_\ell}^{\sigma_\ell,\sigma'_\ell} U^{\sigma'_\ell}_{a'_{\ell - 1}, a'_\ell} \right) \\
R_{b_{i}}^{a_{i},a'_{i}} =  \sum_{\substack{a_\ell,b_\ell,a'_\ell \\ \ell \geq i+1}} \prod_{\ell=i+1}^{L} \left( \sum_{\sigma_\ell,\sigma'_\ell} (V^\dagger)^{\sigma_\ell*}_{a_{\ell - 1}, a_\ell} W_{b_{\ell-1},b_\ell}^{\sigma_\ell,\sigma'_\ell} (V^\dagger)^{\sigma'_\ell}_{a'_{\ell - 1}, a'_\ell} \right).
\end{eqnarray}
A more comprehensive illustration of this definition is given in \cref{fig:Heff}.
With this definition we can write the Hamiltonian term of $\mathbb{O}$, \cref{eq:dmrg-obj-function}, as
\begin{multline}
\braket{\Psi|H|\Psi} = L_{b_{i-1}}^{a_{i-1},a'_{i-1}} \left( M^{\sigma'_i}_{a'_{i-1},a'_i} W_{b_{i-1},b_i}^{\sigma_i,\sigma'_i} M^{\sigma_i*}_{a_{i-1},a_i} \right) R_{b_{i}}^{a_{i},a'_{i}},
\end{multline}
where the brakets are solely for highlighting the central block at site $i$, and where we use Einstein convention of implicitly summing over all double indices. Now differentiating with respect to $M^{\sigma_i*}_{a_{i-1},a_i}$ simply removes this term and, accordingly, the summation over its incides. We reinterpret the remaining term as 
\begin{equation}
\label{eq:partial-psi-h-psi}
\frac{\partial \braket{\Psi|H|\Psi} }{\partial M^{\sigma_i*}_{a_{i-1},a_i}} = (H_\text{eff})_{(\sigma_i, a_{i-1}, a_i),(\sigma'_i, a'_{i-1}, a'_i)} M_{(\sigma'_i, a'_{i-1}, a'_i)}
\end{equation}
with the definition of $H_\text{eff}$ graphically in \cref{fig:Heff}. The norm term in \cref{eq:dmrg-obj-function} is simply 
\begin{equation}
\braket{\Psi |\Psi} = M^{\sigma_i}_{a_{i-1},a_i}M^{\sigma_i*}_{a_{i-1},a_i},
\end{equation}
again with Einstein convention, due to the mixed canonical form. The derivative simply removes the tensor and summations again and together with \cref{eq:partial-psi-h-psi} we arrive at \cref{eq:Heff}. 

\

We can implement this local site update in terms of the following (incomplete) code. The first definition is the local update routine, which is part of a DMRG object that we specify in more detail later. The second definition is just the effective Hamiltonian from the MPO. Note that this implementation is not very efficient as it explicitly constructs the matrix \lstinline|heffmat|. More efficiently, but more cumbersome, would be to give the \lstinline|H_eff| object instructions on how to multiply vectors with the matrix, which in the \textsc{numpy} / \textsc{scipy} ecosystem can be done using a private \lstinline|_matvec()| method.

\begin{lstlisting}[language=Python]
import scipy.sparse.linalg.eigen.arpack as arp
	def site_update(self,i):
        '''
        Solves the Eigenvalue problem H_eff v = E v
        
        Returns updated matrix M
        '''
        LP = self.LPs[i]
        RP = self.RPs[i]
        W = self.MPO.Ws[i]
        heff = H_eff(LP,RP,W)
        heffmat = heff.Heff_mat
        Mshape_ = self.MPS.Thetas[i].shape
        vguess = self.MPS.Thetas[i].reshape(np.prod(Mshape_))
        e, v = arp.eigsh(heffmat, k=1, which='SA', return_eigenvectors=True, v0=vguess)
        M = v[:,0].reshape(Mshape_)
        return M , e[0]


class H_eff(object):
    '''
    effective Hamiltonian of the contracted right- and left part
    .--a_l-1       a_l--.
    |         s_l       |
    |          |        |
    LP--b_l-1--W--b_l--RP
    |          |        |
    |         s'_l      |
    .--a'_l       a'_l--.
    input::
    	LP[a_l-1,b_l-1,a'_l-1]
    	RP[a_l,b_l,a'_l]
    	W[b_l-1, b_l, s_l, s'_l]
    output::
    	H[(a_l-1, s_l, a_l),(a'_l-1, s'_l, a'_l)]
    '''
    def __init__(self,LP,RP,W):
        self.RP = RP
        self.LP = LP
        self.W = W
        chiL1,chiLw,chiLd = LP.shape 
        chiR1,chiRw,chiRd = RP.shape 
        chiLw, chiRw, d, dd = W.shape
        self.Heff = self.init_Heff()
        self.Heff_mat = self.Heff.reshape(chiL1*chiR1*d, dd*chiL1*chiR1)
    
    def init_Heff(self):
        '''return the ndim=6 tensor that is H_eff[a_l-1,a'_l-1, s_l, s'_l, a_l,a'_l]
        '''
        temp = np.tensordot(self.W,self.RP,axes=([1],[1]))
        temp = np.tensordot(self.LP,temp,axes=([1],[0])) 
        # has indeces:  
        # [a_l-1, a'_l-1, s_l, s'_l, a_l, a'_l]
        # want indices: 
        # [a_l-1, s_l, a_l, a'_l-1, s'_l, a'_l]
        # therefore permute
        return np.transpose(temp,(0,2,4,1,3,5))
\end{lstlisting}

To maintain the mixed canonical form, we normalize the resulting tensor. When sweeping from left to right, we can use the following function, and similarily when going from right to left (for more details see \cite{github_dmrg}). It is performing a SVD and then only keeping the $\chimax$ singular values (or less in case all eigenvalues are smaller than the given threshold \lstinline|eps|).

\begin{lstlisting}[language=Python]
from scipy.linalg import svd
def left_norm(M,eps,chi_max=None):
    chiL, d, chiR = M.shape
    Psi = M.reshape(chiL*d,chiR)
    U,S,Vh = svd(Psi,full_matrices=False)
    nonzeros = S>eps
    if chi_max is None:
        chi_max = min(chiL*d,chiR)
    newchi = min(chi_max,np.sum(nonzeros))
    if newchi == 0:
        newchi = 1
    Ut, St, Vht = U[:,:newchi],S[:newchi],Vh[:newchi,:]
    St = St / np.linalg.norm(St)
    newU = Ut.reshape(chiL,d,newchi)
    return newU,St,Vht
\end{lstlisting}

With this we have everything we need for our full DMRG sweep routine. It is efficient to keep track of the left and right environments for every site and simply build them from the previous step. I.e., after updating site $i$, we can contract $L(i)$\footnote{the left environment at site $i$, i.e. \textit{excluding} site $i$ itself.} with the newly obtained tensor and MPO element of site $i$, to build $L(i+1)$. The same is true going from right to left. Note that in the very beginning, we need to initialize all right environments for the trial state.

\begin{lstlisting}[language=Python]
class DMRG(object):
    '''
    Abstract DMRG engine class
    
    attributes:
    MPS*: the current MPS
    MPO: the model MPO
    RPs*: list of right environments
    LPs*: list of left environments
    (*): get regular updates
    
    Parameters:
    MPS: object, matrix product state
    MPO: object, matrix product operator
    eps: float, epsilon determining the threshold of which singular values to keep, standard set to 10^-10
    chi_max: int, maximal bond dimension of system
    
    
    '''
    def __init__(self, MPS, MPO, chi_max, eps=1e-10):
        MPS.right_normalize()
        self.MPS = MPS 
        self.MPO = MPO
        self.L = self.MPS.L
        self.eps = eps
		self.chi_max = chi_max
        
        ### initialize right- and left parts
        # create empty list of correct size
        self.LPs = [None] * self.L
        self.RPs = [None] * self.L
        # LP can stay empty except for dummy one
        self.LPs[0] = np.ones((1,1,1))
        self.RPs[-1] = np.ones((1,1,1))
        # when starting from left to right (default)
        # we need to initialize all the RPs once
        for i in range(self.L - 1,0,-1):
            # from L-1 to 1
            self.RPs[i-1] = self.next_RP(i)
        
    	def site_update(self,i):
        '''
        Solves the Eigenvalue problem H_eff v = E v
        
        Returns updated matrix M
        '''
        LP = self.LPs[i]
        RP = self.RPs[i]
        W = self.MPO.Ws[i]
        heff = H_eff(LP,RP,W)
        heffmat = heff.Heff_mat
        Mshape_ = self.MPS.Ms[i].shape
        vguess = self.MPS.Ms[i].reshape(np.prod(Mshape_))
        e, v = arp.eigsh(heffmat, k=1, which='SA', return_eigenvectors=True, v0=vguess)
        M = v[:,0].reshape(Mshape_)
        return M , e[0]
    
    def left_to_right(self):
        '''
        Runs through the MPS from left to right
        '''
        eps = self.eps
        chi_max = self.chi_max
        
        for i in range(self.L):
            M = self.MPS.Ms[i]
            # update to new M
            M,e = self.site_update(i)
            ### ..AA[M']BB..
            # left_normalize to A
            A,S,V = left_norm(M,eps,test=self.test)
            # update Ms
            self.MPS.Ms[i] = A
            ### ..AA[A](SVB)B..
            if i < self.L - 1:
                ### ..AA[A](B'=SVB)B..
                SV = np.tensordot(np.diag(S),V,1)
                self.MPS.Ms[i+1] = np.tensordot(SV,self.MPS.Ms[i+1],1)
                self.LPs[i+1] = self.next_LP(i)

    def next_LP(self,i):
        '''
        short version: LP[i] |-> LP[i+1]
        
        extend LP from site i to the next site i+1 
        .------a_j-1----M[j]--a_j
         |               |
        LP[j]--b_j-1----W[j]--b_j
         |               |
        .-----a'_j-1---M*[j]--a'_j
        
        result: np.array[ndim=3] # a_i b_i  a'_i
        '''
        F = self.LPs[i]

        F = np.tensordot(F,self.MPS.Ms[i],axes=[2,0])
        F = np.tensordot(self.MPO.Ws[i],F,axes=([2,0],[2,1]))
        F = np.tensordot(self.MPS.Ms[i].conj(),F,axes=([0,1],[2,1]))
        return F
\end{lstlisting}


\section{Other algorithms}
\label{sec:other_algorithms}
\subsection{infinite DMRG (iDMRG)}
\begin{figure}
\centering
\includegraphics[width=.98\textwidth]{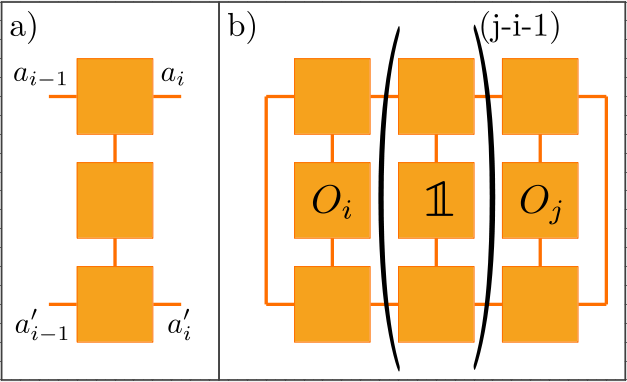}
\caption{a) Transfer matrix $E_0$ \cref{eq:transferm} for some local operator $O$. b) Expectation value $\braket{\Psi|O_i O_j|\Psi}$ for an iMPS with unit cell size $L_\infty=1$. We can calculate correlation functions like this for arbitrary distances exponentiating the transfer matrix by diagonalizing it.}
\label{fig:imps_transferm}
\end{figure}

We can extend DMRG to the thermodynamic limit (iDMRG) in terms of infinite matrix product states (iMPS). Such iMPS are almost identical in structure to their finite counterparts described in \cref{sec:canonical_form}. The main difference is that we assume translational invariance and that the system is described by an infinitely repeating unit cell of size $L_\infty$. So an iMPS is also described by a finite set of rank $3$ tensors $\Gamma^{\sigma_i}$ and singular values $\Lambda^{[i]}$, that are in canonical form like in \cref{eq:vidal-form}. This allows for efficient calculation of local expectation values in terms of the central tensor
\begin{equation}
\braket{\Psi|O|\Psi} = \tr{\sum_{\sigma_i\tilde{\sigma}_i}\Theta^{\sigma_i\dagger} O^{\sigma_i\tilde{\sigma}_i} \Theta^{\tilde{\sigma}_i}},
\end{equation}
exactly the same way as for the finite case as illustrated in \cref{fig:mps_exp_value}.
Here, the index $i$ indicates the position within the unit cell rather than the absolute position of an infinite system. With an iMPS we can calculate the correlation functions of arbitrary distances. For distances within the unit cell $i,j \leq L_\infty$, it is just the explicit contraction between sites $i$ and $j$. Furthermore, we can compute correlation functions for distances beyond the unit cell, and even for arbitrary distances at the same cost. This can be done by definining the transfer \textit{matrix}\footnote{The name transfer \textit{matrix} is of historical origin and may be confusing since the object is clearly not a matrix (rank 2 tensor).} $E_O$ for some operator $O$ (that can be identity) in terms of
\begin{equation}
\label{eq:transferm}
E^{[i]}_O = \sum_{\sigma_i,\sigma'_i} U^{\sigma_i\dagger} O^{\sigma_i,\sigma'_i} U^{\sigma'_i}
\end{equation}
that has \textit{open} virtual indices $a_{i-1}, a_i$ and $a'_{i-1}, a_i$ as illustrated in \cref{fig:imps_transferm} a). Here we assumed left-orthonormal regrouping in terms of $U^{\sigma_i} = \Lambda^{[i-1]} \Gamma^{\sigma_i}$. Correlation functions like $\braket{\Psi|O_i O_j|\Psi}$ can then be evaluated by exponentiating the transfer matrix by the distance between unit cells, as illustrated in \cref{fig:imps_transferm} b), which on the other hand can be done by diagonalizing it.

\textbf{But how do we obtain such ground states in the thermodynamic limit?}

The iDMRG algorithm that we use in this thesis is given by \cite{McCulloch2008}. On a first glance, it may appear similar to the traditional DMRG algorithm by White \cite{White1992,White1993} in that a small system is \textit{grown} by adding sites in the center.
However, there are important conceptual differences. Most notably, we set up a recurrence relation in order to find a fixed point that is translationally invariant by construction. Furthermore, the algorithm proposed by McCulloch \cite{McCulloch2008} has advantages in terms of convergence and stability in comparison to alternative formulations.
For clarity and analogy to the original paper \cite{McCulloch2008}, we assume a two-site unit cell and note that this can be extended to arbitrary unit cell sizes $L_\infty \geq 2$\footnote{In principle $L_\infty=1$ is also possible but it requires some technical adjustments to the algorithm. Further, in practice, we prefer to perform updates on two sites at the same time in order to be able to grow the bond dimension.}. For the construction of the thermodynamic limit result, we set up iterative states $\ket{\Psi_\ell}$ by starting with
\begin{eqnarray}
\ket{\Psi_0} = & U_0 \Lambda_0 V^\dagger_0 & \text{ and} \\
\ket{\Psi_1} = & U_0 U_1 \Lambda_1 V^\dagger_1 V^\dagger_0
\end{eqnarray}
where we neglected all indices and just label the site positions for notational simplicity. These states should be understood as a tool to find the thermodynamic limit state $\ket{\Psi_\infty}$, described below.
Assume we are at the $\ell$-th step in mixed-canonical form
\begin{equation}
\ket{\Psi_\ell} = \cdots U_{\ell-1} U_\ell \Lambda_\ell V^\dagger_\ell V^\dagger_{\ell - 1} \cdots.
\end{equation}
Focusing on the central tensors, we can rewrite 
\begin{eqnarray}
U_\ell \Lambda_\ell V^\dagger_\ell \stackrel{\text{SVD}}{=:} \Lambda^L_\ell V^\dagger_{\ell + 1} V^\dagger_\ell \\
U_\ell \Lambda_\ell V^\dagger_\ell \stackrel{\text{SVD}}{=:} U_\ell U_{\ell +1} \Lambda^R_\ell
\end{eqnarray}
in two different fashions using SVD to obtain the ingredients for a new trial state
\begin{equation}
\label{eq:iDMRG_growing}
\ket{\Psi^\text{trial}_\ell} = \cdots U_{\ell-1} U_\ell U_{\ell + 1} \Lambda^R_\ell \Lambda^{-1}_{\ell-1} \Lambda^L_\ell V^\dagger_{\ell + 1} V^\dagger_\ell V^\dagger_{\ell - 1} \cdots
\end{equation}
after \textit{growing} the state by two sites. 

The central unit cell $U_{\ell + 1} \Lambda^R_\ell \Lambda^{-1}_{\ell-1} \Lambda^L_\ell V^\dagger_{\ell + 1}$ serves as a trial input for the optimization to find $U_{\ell + 1} \Lambda_{\ell + 1} V^\dagger_{\ell + 1}$, which is done analogously to finite DMRG discussed in \cref{sec:DMRG}. This procedure is repeated until the fixed point criteria, $\Lambda_{\ell + 1}$ being sufficiently close to $\Lambda^R_\ell$ and $\Lambda^L_\ell$, are fulfilled. In practice, we often also check for convergence in the energy with some user-specified threshold in difference in energy between iterations. 

The basis of this algorithm is the assumption that the state is defined by an infinitely repeating unit cell $U_\ell \Lambda_\ell V^\dagger_\ell \Lambda^{-1}_{\ell -1}$. What we did in \cref{eq:iDMRG_growing} is just explicitly inserting another such unit cell to \textit{grow} the state and then re-optimizing the new unit cell at the center. This implicitly defines a recurrence relation for which we aim to find the fixed point. The final state is then formally given by
\begin{equation}
\ket{\Psi_\infty} = \cdots \left( U_\ell \Lambda_\ell V^\dagger_\ell \Lambda^{-1}_{\ell -1} \right) \left( U_\ell \Lambda_\ell V^\dagger_\ell \Lambda^{-1}_{\ell -1} \right) \cdots.
\end{equation}

\subsection{Time-Evolving-Block-Decimation (imaginary time evolution)}
\label{sec:TEBD}

\begin{figure}
\centering
\includegraphics[width=.98\textwidth]{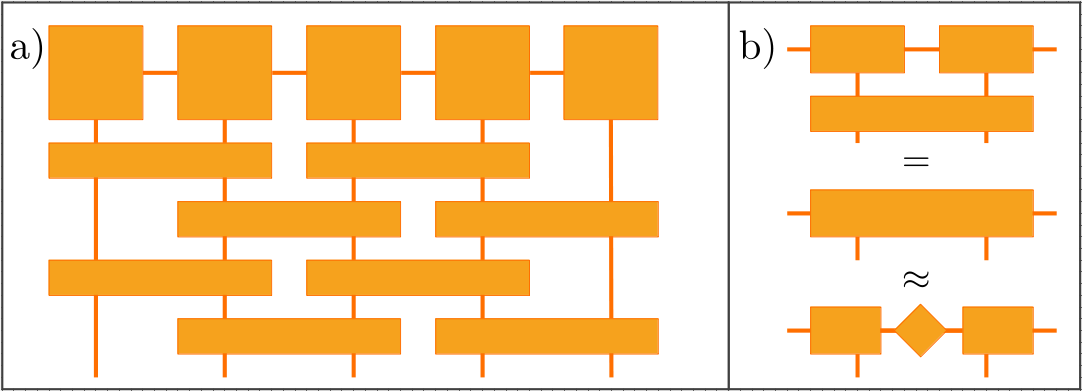}
\caption{a) Schematic of the TEBD algorithm: applying layerwise the even terms $e^{\Delta t H_e}$ and odd terms $e^{\Delta t H_o}$ of the evolution operator in Trotter approximation. b) Retaining canonical form by contraction and SVD at each double site.}
\label{fig:TEBD}
\end{figure}

Time-Evolving-Block-Decimation (TEBD) for MPS was introduced by Vidal in 2004 \cite{Vidal2004TEBD}. Its aim is to apply trotterizations of evolutions $U(t) = \exp\left(-i t H\right)$ of nearest-neighbor Hamiltonians $H = \sum_i H^{(1)}_i + \sum_i H^{(2)}_{i,i+1}$. Being able to perform such evolutions gives access to simulating time evolutions and ground state construction via imaginary time evolution\footnote{Or ground state construction via adiabatic quantum computation, which is beyond the scope of this thesis but briefly summarized here: One starts in the ground state $\ket{\Psi_0(t=0)}$ of some simple Hamiltonian $H(t=0)$ and then slowly interpolates $H(t)$ to the target Hamiltonian $H$. We comment on this again later in \cref{sec:QAOA}.}. The basis of imaginary time evolution stems from statistical mechanics, from which we know that a state in thermal equilibrium is described by 
\begin{equation}
\rho(\beta) = \frac{\exp\left(-\beta H\right)}{Z}
\end{equation}
for an inverse temperature $\beta = 1/(k_B T)$ and \textit{Zustandssumme} $Z = \tr{\exp\left(-\beta H\right)}$ (partition function). Note that formally we can always expand $H = \sum_n E_n |n\rangle \langle n|$ in its energy Eigenbasis. In the limit $\beta \rightarrow \infty \  (T \rightarrow 0)$, only the (pure) contribution corresponding to the ground state survives and we can formally find the ground state via imaginary time evolution
\begin{equation}
\label{eq:imaginary-time-evolution}
\ket{\Psi(T=0)} = \lim_{\beta \rightarrow \infty} \frac{\exp\left(-\beta H\right)\ket{\Psi_\text{trial}}}{||\exp\left(-\beta H\right)\ket{\Psi_\text{trial}}||}
\end{equation}
for some trial state $\ket{\Psi_\text{trial}}$ that needs to have non-zero overlap with the true ground state. Let us briefly derive this. Due to the assumption of non-zero overlap with the ground state we can write $\ket{\Psi_\text{trial}} = c_0 \ket{0} + \sum_{j\neq 0} c_j \ket{j}$, then
\begin{multline*}
\exp(-\beta H) \ket{\Psi_\text{trial}} = c_0 \exp(-\beta E_0) \ket{0} + \sum_{j\neq 0} c_j \exp(-\beta E_j) \ket{j} \\ \propto  \left( \ket{0} + \sum_{j \neq 0} \frac{c_j}{c_0} \exp(-\beta(E_j - E_0)) \ket{j} \right)  \stackrel{\beta \rightarrow \infty}{\rightarrow} \ket{0}
\end{multline*}
as $E_j - E_0 > 0$ $\forall j \neq 0$. Note that there is a factor $c_0 \exp(-\beta E_0)$ that cancels with the normalization factor in \cref{eq:imaginary-time-evolution}.

By identifying $t = -i\beta$, we can see how this corresponds to a time evolution $U(t=-i\beta)$ with respect to an imaginary time and updating the norm as the evolution operator becomes non-unitary.

\textbf{But how do we do it for an MPS?}

It is useful to split the Hamiltonian into an even and odd part $H = H_e + H_o$,
\begin{eqnarray}
H_e := \sum_{i \text{ even}} F_i := \sum_{i \text{ even}} H^{(1)}_i + H^{(2)}_{i,i+1} \\
H_o := \sum_{i \text{ odd}} G_i := \sum_{i \text{ odd}} H^{(1)}_i + H^{(2)}_{i,i+1}
\end{eqnarray}
such that $[F_i, F_j] = [G_i, G_j] = 0$. We can then use the Trotter approximation
\begin{equation}
e^{-i(H_o + H_e)T)} \approx (e^{-i\Delta t H_e} e^{-i\Delta t H_o})^{T/\Delta t}
\end{equation}
in first order\footnote{It is common to use the symmetric second order trotter formula $(e^{-i\Delta t H_e/2} e^{-i\Delta t H_o}e^{-i\Delta t H_e/2})^{T/\Delta t}$.}. Since $e^{-i\Delta t H_{e}} = \prod_{i} e^{-i\Delta F_i}$ and $e^{-i\Delta t H_{o}} = \prod_{i} e^{-i\Delta G_i}$ are just products of disjoint, and therefore commuting, two-body operators, we can individually apply them as illustrated in \cref{fig:TEBD}. 

\

After each application of a local term $\exp(-\Delta t F_i)$ and $\exp(-\Delta t G_i)$ on sites $(i,i+1)$, we perform SVD to a) truncate the state and b) maintain canonical form. The latter allows us to resort to local updates only without having to update all tensors of the state after each update. This also enables infinite TEBD (iTEBD) in an anologous fashion with iMPS as discussed in the previous section. On the other hand, this is significantly different for 2D tensor network states that we discuss in the following \cref{sec:introPEPS}. 

\

All these operations can be done efficiently, but the limiting factor are the two sources of error. The first is from Trotterization and $\varepsilon_\text{trott} \propto (\Delta t)^{2p}T^2$ for $p$-th order Trotterization and $\varepsilon  = 1 - |\braket{\Psi(t)_\text{TEBD}|\Psi_\text{exact}}|$ \cite{Vidal2004TEBD}. This can be reduced by using higher-order approximations and reducing the stepsize. The more problematic error for real time evolution is the truncation error $\varepsilon_\text{MPS}$ from keeping a fixed bond dimension $\chimax$ of internal states. While for ground states it is known that MPS are a good approximation of local and gapped Hamiltonians (more see \cref{sec:area_law}), the same is not true for time evolved states for which we know that entanglement generally grows linearly in $T$ (and therefore leads to exponential computational cost). There are approaches to mitigate these problems and push for longer simulation times \cite{Paeckel2019,Hauschild2018}, but the overall problem remains. One of the possible applications of quantum computers, discussed in \cref{chap:qc}, is to circumvent this problem by simulating the full wave function.

\section{Projected Entangled Pair States (PEPS)}
\label{sec:introPEPS}
A natural extension of matrix product states to higher dimensional systems are projected entangled pair states (PEPS). We here focus for simplicity on 2D rectangular lattices, that already capture the most notable differences to 1D MPS. A good introductory review on PEPS is given in \cite{Orus2013}, that we will partly follow here.

Matrix product states in one spatial dimension are a very special case in that there exists a canonical form\footnote{Tree tensor networks share that property and all tensor networks with \textit{loops} do not possess canonical forms.}. This is not the case for PEPS, which makes handling them much more complicated. All the advantages gained from the canonical form for MPS described in \cref{sec:canonical_form} do not apply anymore and one has to find new strategies to obtain PEPS that describe physically relevant states and extract information from them. The latter is not obvious, and in fact PEPS are somewhat peculiar in that regard as they are known to describe many different families of states (and formal descriptions are known) but are hard to extract information from. This mostly stems from the fact that it is known that contracting two PEPS as illustrated in \cref{fig:peps_contraction0} is computationally inefficient, i.e. for $N$ sites the computational cost is $\mathcal{O}(\exp(N))$ and contraction is \textsf{\# P - hard} \cite{Schuch2007PEPScompute}. This does not mean that all hope is lost for PEPS, but just that we have to resort to approximate methods.

\subsection{Contracting PEPS}

\begin{figure}
\centering
\includegraphics[width=.77\textwidth]{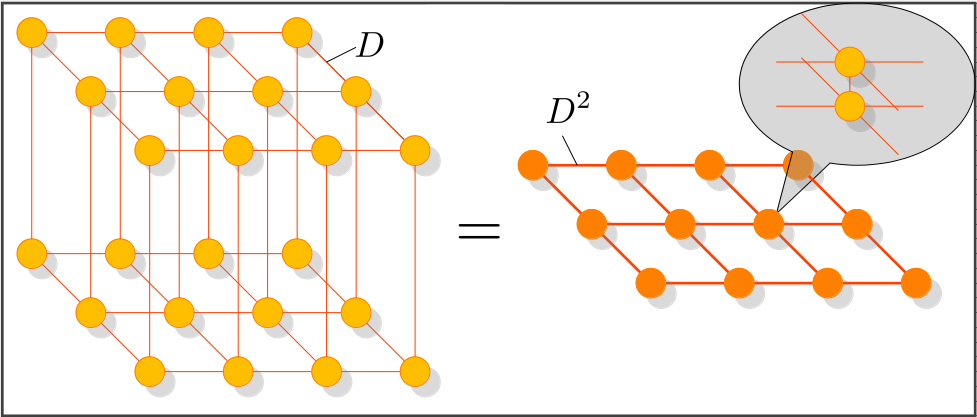}
\caption{Contraction of two 2D square lattice PEPS. We can reduce the problem to the contraction of an MPS at the top, (multiple) MPOs and a second MPS at the bottom. This can be done approximately.}
\label{fig:peps_contraction0}
\end{figure}

\begin{figure}
\centering
\includegraphics[width=.77\textwidth]{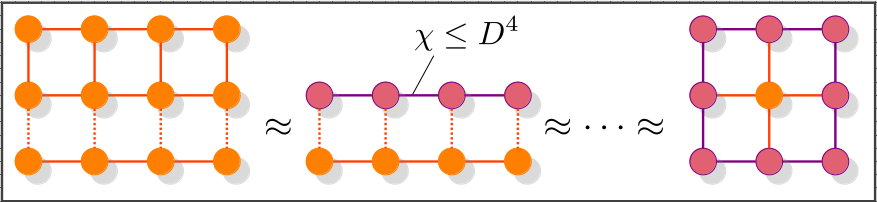}
\caption{Approximately contracting two PEPS with MPS methods described in the previous sections. We can absorb a row of tensors like in an MPO-MPS contraction into a new MPS with bond dimension $D^4$. We can perform this contraction with MPS methods and only keep $\chi$ singular values. We can do this from all sides until we arrive at the contraction of the remaining site (orange) and its \textit{environment} (in purple).}
\label{fig:peps_contraction1}
\end{figure}

Let us start by computing the contraction of two PEPS with bond dimension $D$ as illustrated in \cref{fig:peps_contraction0} \textit{approximately}. As a first step, we combine sites that are contracted and we obtain something that is equivalent to the contraction of two MPSs and (multiple) MPOs with bigger bond dimension $D^2$. The approximate nature comes from the fact that contracting an MPS with bond dimension $D$ and an MPO with bond dimension $D$ yields a state of bond dimension $D^2$ (so $D^4$ for the double connections here) and grows with every further MPO to be applied. One usually defines a second bond dimension $\chi$, sometimes referred to as the boundary dimension, up until which we keep states during the approximate contraction process. Here, we can use the same methods as for MPS contractions. This process is illustrated in \cref{fig:peps_contraction1} and highlights the \textit{environment} of the remaining site.

\begin{figure}
\centering
\includegraphics[width=.77\textwidth]{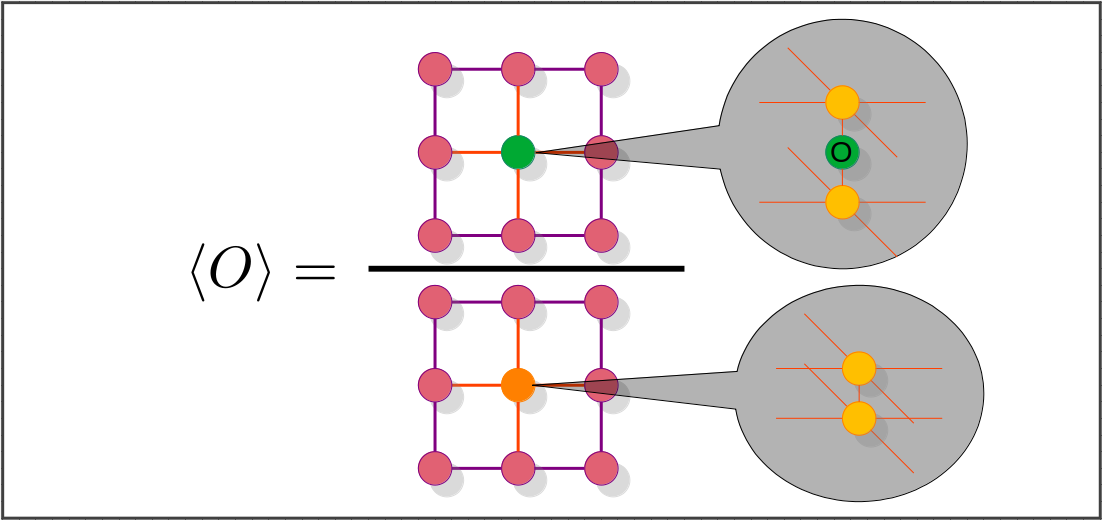}
\caption{The expectation value of an operator can be computed by contracting the sandwiched operator with the site's environment and dividing by its norm, which is the same contraction but without the operator.}
\label{fig:peps_contraction2}
\end{figure}

Now if we want to compute local observables we can repeat the same process but with a local operator \textit{sandwiched} between the two PEPS, as illustrated in \cref{fig:peps_contraction2} and divide by the norm of the PEPS as it is in general not normalized.

\subsection{Simple and full update of a PEPS}

There is a multitude of algorithms to find ground states for PEPS. We here want to focus on imaginary time evolution as described in the second part of \cref{sec:other_algorithms}. The algorithm itself works in the analogous fashion, i.e. by trotterizing the evolution operator and applying 2-site gates locally. 

\begin{figure}
\centering
\includegraphics[width=.77\textwidth]{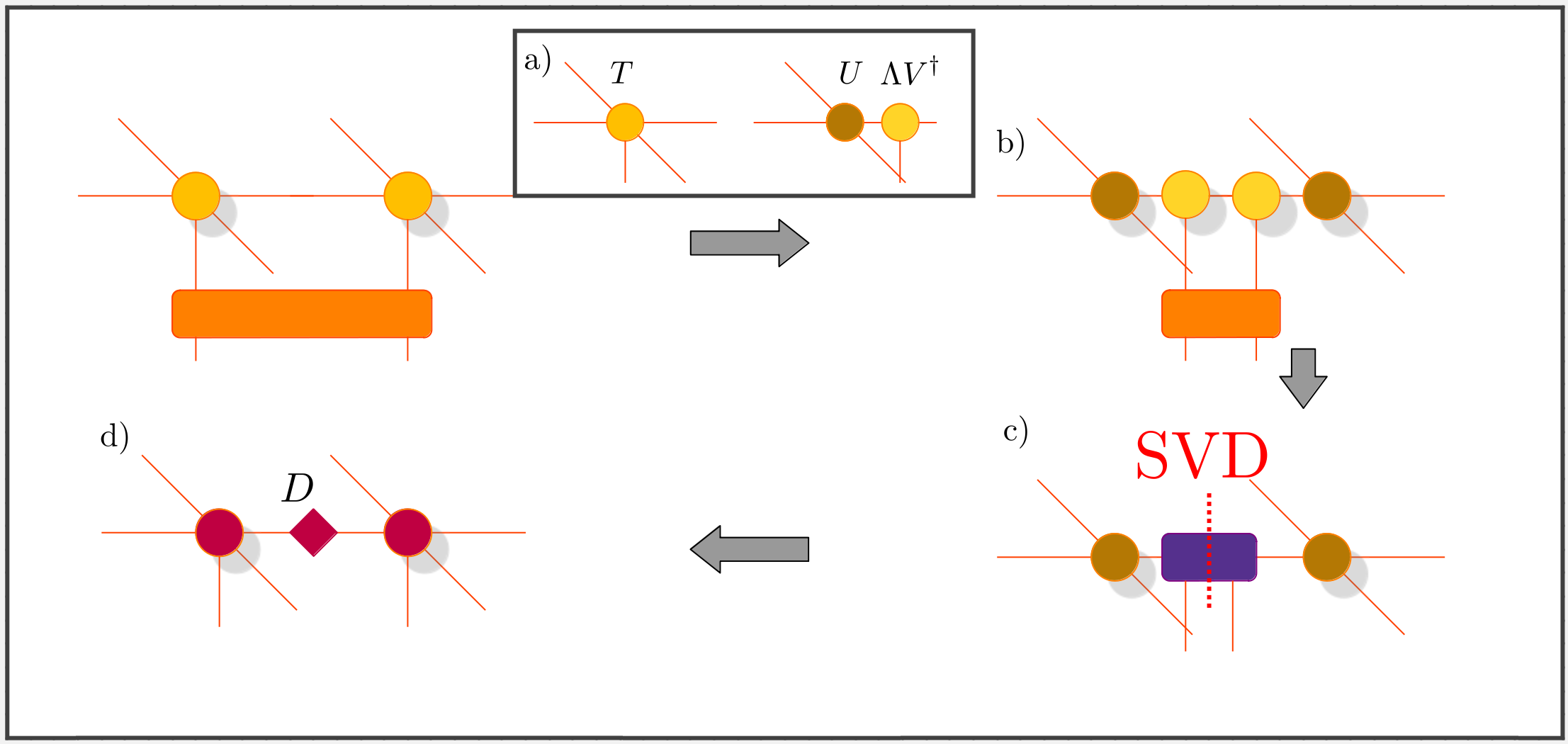}
\caption{Simple Update: Applying a 2-site gate to a PEPS in analogous fashion to MPS. We start by separating the physical and virtual bonds by a first SVD (a), such that we can apply the gate onto the physical parts (b). We then split the new tensor via SVD (c) and only keep the $D$ largest singular values (d). Between c) and d) we multiplied $U$ and $V^\dagger$ to the left and right side, accordingly.}
\label{fig:simple_update}
\end{figure}

We can do this in an analogous fashion to MPS via SVD, as illustrated in \cref{fig:simple_update}. This is referred to as the \textit{simple update} as it only involves the local tensors where the gate is applied. It was introduced in \cite{Jiang2008} though it shall be noted that here we present a slightly optimized version with the trick of separating virtual and physical indices before applying the gate. We can introduce a pseudo-canonical form in analogy to Vidal's form for MPS by explicitly separating singular values $\Lambda$ and local tensors $\Gamma$. To obtain and maintain this form, we multiply the final tensors in \cref{fig:simple_update} by the inverses of the three untouched virtual bonds. This pseudo-canonical form does not fulfill any orthonormality conditions and the singular values do not have a physical interpretation like the entanglement spectrum. However, we will later see that we can still make use of this quantity and infer physical properties via machine learning from them.

The simple update scheme is insufficient to capture the full physical picture as it does not yield an optimal state approximation of the state after application of the gate (and therefore \textit{loses} information due to the fixed and finite bond dimension $D$). This is ultimately due to the lack of a canonical form that allowed for optimal \textit{local} updates in MPS. To capture the full physical picture, one has to perform the \textit{full update} \cite{Jordan2008} that takes into account the full wave function at each update, and is therefore more costly. The idea is to find the best approximation $\ket{\Psi'}$ after applying the gate $g$ onto the state $\ket{\Psi}$ by explicitly minimizing $||\ket{\Psi'} - g\ket{\Psi}||^2 = \braket{\Psi'|\Psi'} - \braket{\Psi'|g|\Psi} - \braket{\Psi|g^\dagger|\Psi'} + \braket{\Psi|g^\dagger g|\Psi}$.
Especially for infinite PEPS (iPEPS), the full update algorithm can be improved by incorporating corner transfer matrix methods and renormalization considerations via the \textit{fast full update algorithm} \cite{Phien2015}.


\section{Area law}
\label{sec:area_law}
The \textit{area law of entanglement} for ground states of gapped and local Hamiltonians states that the entanglement scales with the surface area that  a subsystem is occupying \cite{Srednicki1993,Plenio2005,Eisert2010}. For a $d_\text{euc}$-dimensional hyper lattice in euclidean real space with subsystem $A$ being a hyper cube of volume $L^{d_\text{euc}}$, we have $S(\rho_A) \propto L^{d_\text{euc}-1}$. This is in contrast to general states in Hilbert space that follow a \textit{volume law of entanglement}. This is of practical importance especially for states in 1D and 2D as $S(\rho^{1D}_A) = const$ and $S(\rho^{2D}_A) \propto L$, respectively.

MPS and PEPS have been shown to be suitable Ansätze for ground state calculations of gapped and local Hamiltonians in \cite{Hastings2006, Molnar2015}. One reason for this is the fact that they obey the area law. Let us see why by starting with MPS: Recall from \cref{eq:rhoA_general} in \cref{sec:canonical_form} that $S(\rho_A) = -\sum_i \Lambda^2_i \log(\Lambda^2_i)$ where $\Lambda_i$ are the singular values separating the chain. Since we fix the maximal number of singular values by setting the bond dimension $\chimax$ we obtain $S(\rho_A) \leq \log(\chimax)$\footnote{It is known that the von Neumann entropy is maximal for a uniform distribution, i.e. for $\Lambda^2_i = 1/\chimax$ and hence $\max(S) = -\sum_i \log(1/\chimax)/\chimax = \log(\chimax) = const$.}. 

We can obtain the area law for PEPS by analogous reasoning. A subsystem of $L^2$ is connected through $4L$ links that each can take up to bond dimension $D$ values. Hence, the entanglement entropy for the PEPS is bounded by $S(\rho_A) \leq \log(D^{4L}) = 4L \log(D) \propto L$, so it scales with the circumference $L$ of subsystem $A$.

If we relax the conditions and allow for gapless Hamiltonians the entanglement scaling changes to $S(\rho_A) \propto \log(L)$ in 1D. This can be incorporated by utilizing the multi scale entanglement renormalization (MERA) Ansatz \cite{Vidal2007MERA,Vidal2008MERA}. In practice, we can also use normal MPS and extrapolate in $\chimax$, which is what we do for the critical superfluid and supersolid phases we investigate in the main body of this thesis.

\chapter{Introduction to deep learning}
\label{chap:deep-learning}
Deep learning is a subfield of machine learning in which artificial neural networks (NNs) are used to perform tasks like computer vision, speech recognition or natural language processing by learning from data \cite{Goodfellow2016,nielsenneural}. These tasks are typically part of the field of artificial intelligence (AI), which has seen a great boost with the development of deep learning. This success swapped over to the natural sciences, where on one hand many useful applications of deep learning have been found, and vice versa, methods and theories in natural sciences gave insights to the theory of deep learning \cite{Carleo2019}. 

Deep learning algorithms build on the success of learning algorithms like support vector machines \cite{Cortes1995} by following the same \textit{data-driven} paradigm. For many tasks in AI like computer vision, it is notoriously hard to provide explicit instructions to differentiate images of, say, cats and dogs. In learning algorithms following the data-driven paradigm, one instead leverages many examples of labeled images (training data) to let a computer program \textit{learn} to differentiate them. In the case of deep learning we use general purpose functions in the form of artificial neural networks. Learning then amounts to optimizing the free parameters of the NN by minimizing a \textit{loss function} that quantifies the difference between the outputs and the true values (labels). The overall goal of deep learning is then to be able to \textit{generalize} and correctly make predictions for inputs that have not been part of the training data.

Note that the above actually describes the more specific task of supervised classification. On one hand, there are semi-supervised and unsupervised learning in addition to supervised learning. On the other hand, there is a variety of other deep learning tasks in addition to classification problems. Among others, there is parameter estimation, where the continuous values of a function is learned or generative models for translation or text generation in natural language processing. However, all of these different subfields of deep learning follow a similar approach to supervised classification as stated above, which is why we use it as a representative of deep learning in the following and point out differences when necessary.

A typical deep learning procedure, exemplified here by the task of differentiating cats and dogs, will consist of these four steps:
\begin{enumerate}
\item We assemble a \textit{training set} consisting of example images $x$ of various different cats and dogs, and a correct \textit{label} $y(x)$ for each image.
\item  We build an \textit{artificial neural network}, that is a highly non-linear function with trainable, free parameters $\bm{\phi}$. It maps the images $x$ to $y^\text{out}(x)$, where $y^\text{out}_1(x) = p(\text{dog}|x)$ and $y^\text{out}_2(x) = p(\text{cat}|x)$ aim to approximate the respective probabilities. So in terms of the labels, we have $y(x) = (1,0)$ for an image of a dog and $y(x)=(0,1)$ for a cat.
\item We have to define a loss function that quantifies the difference between the output distribution $y^\text{out}(x)$ and the labels $y(x)$. More generally, it describes our objective in a mathematically differentiable function. \textit{Training} then simply amounts to minimizing this loss function with respect to the free parameters of the neural network.
\item We then test our predictions on a \textit{test set}, consisting of images that have not been used during training. Our goal is that the NN \textit{generalizes} and makes accurate predictions beyond the training data. In the case of supervised classification, additionally, the goal is not to have minimal loss for the test data, but high accuracy of predictions (which is related to the loss but not the same).
\end{enumerate}

\section{Deep learning basics}

In this section we will introduce the basic concepts of deep learning, hinted above, in more detail. We do so by explaining supervised classification as it covers the general concepts present in most deep learning methods.

\subsection{Data sets}

We differentiate between three datasets:
\begin{enumerate}
\item Training data: (labeled) data used during training.
\item Validation data: (labeled) data not used during training, but typically from the same source as the training data. I.e. by splitting a random fraction off the training data.
\item Test data: labeled or unlabeled data not used during training and ideally from a different source.
\end{enumerate}

As an example we take the MNIST handwritten digits dataset as training data \cite{MNIST}. This is a professionally curated set by the US national institute of standards and technology consisting of $10k$ examples for each digit $\{0,..,9\}$, respectively. Now we take a fraction, say, $9/10$ as our training data and $1/10$ for validation. Images of digits written by ourselves could then for example serve as a test data set. 

To differentiate between test and validation set like this is not standard in literature, but will be used for the remainder of this chapter. We will use physical data from numerical simulations in the main body of the text and thus do not have access to a third party source, such that the notion will be ambiguous again in the other chapters.

\subsection{Artificial neurons}
\begin{figure}
\centering
\includegraphics[width=.77\textwidth]{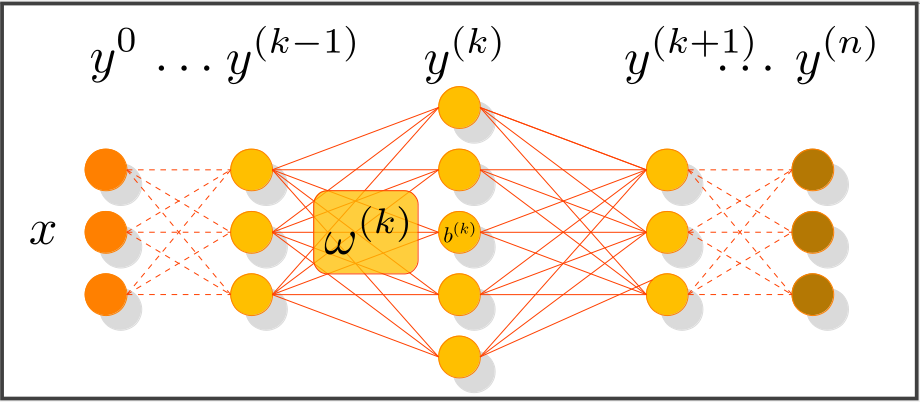}
\caption{Artificial neural network composed of artificial neurons, symbolized by orange circles. The output of each artificial neuron is computed by \cref{eq:perceptron}, the weighted sum of the outputs of the previous layer, subtracted by a local bias and processed through an activation function.}
\label{fig:nn}
\end{figure}

Artificial neural networks (NNs) are composed of different layers with trainable parameters. One core building block of NNs is a fully connected layer made up of artificial neurons. Here, all artificial neurons at layer $k-1$ send a signal to each artificial neuron at layer $k$, as illustrated in \cref{fig:nn}. This is in analogy to how the human brain connects neurons that fire electrical signals. Let us look at the $k$-th layer of a fully connected neural network. Mathematically, the output $y^{(k)}$ of each layer is computed by
\begin{equation}
\label{eq:perceptron}
    y_i^{(k)} = f_\text{act}\left(\sum_j \omega^{(k)}_{ij} y^{(k-1)}_j - b^{(k)}_i\right)
\end{equation}
where $y^{(k-1)}$ is the output from the previous layer and $y^{(0)} = x$ is the input and $y^{(n)} = y^\text{out}(x)$ the output. So an artificial neuron computes a weighted sum subtracted by a neuron-specific bias. These weights $\omega^{(k)}$ and biases $b^{(k)}$ are free parameters and subject to optimization (training) to achieve a certain task. Note that this reduces to a linear transformation of the inputs. In order to gain an advantage in expressibility by connecting multiple consecutive layers, one introduces a non-linear activation function $f_\text{act}(\cdot)$ at each layer, whose functional form depends on the task. In this work, we mainly use the so-called rectified linear unit function $\relu{x} = \left(0 \text{ if } x \leq 0; x \text{ if } x \ge 0\right)$ or $\tanh(x)$. In classification tasks one aims to approximate a probability distribution. Therefore, one typically uses a softmax function $\sigma$ with elements
\begin{equation}
\sigma_i(x) = \frac{e^{x_i}}{\sum_j e^{x_j}}
\end{equation}
on each neuron as the activation function on the final layer. This guarantees all outputs to be $y^{(n)}_i \in [0,1]$ in order to mimic the behavior of a probability distribution.

\subsection{Loss function}
In supervised classification tasks we want the network to approximate a probability distribution for a list of classes. So for example when there are $N_\text{classes}$ classes, we want the network to output $y^\text{label} = (1,0,\cdots)$ if the input image $x$ was of the first class. In order to achieve this task, we define a loss function that captures the success of this endeavor. For classification, where we try to approximate probability distributions, it is natural to use the Kullback–Leibler divergence 
\begin{equation*}
D_\text{KL}(y^\text{label}||y^\text{out}) = \sum_i y^\text{label}_i \log\left(\frac{y^\text{label}_i}{y^\text{out}_i}\right)
\end{equation*}
that quantifies the distance between the true distribution $y^\text{label}$ and the NN output $y^\text{out}$. Later we want to minimize this function summed over all training examples with respect to the NN paramters $\bm{\phi} = (\omega, b)$. Since only $y^\text{out}$ depends on these paramters one typically looks at the cross entropy
\begin{equation}
\mathcal{H}(y^\text{label},y^\text{out}) = -\sum_i y^\text{label}_i \log\left(y^\text{out}_i\right)
\end{equation}
instead. The loss function for our classification task is then
\begin{equation}
\mathcal{L} = \sum_x \mathcal{H}(y^\text{label}(x),y^\text{out}(x))
\end{equation}
for all training examples $x$.
Because $y^\text{label}_i = 1$ only for the correct label and zero otherwise, we note that this amounts to minimum likelihood estimation for the negative log likelihood $-\sum_x\log(p(x))$. This is the loss that is typically minimized in generative models where we try to model the probability distribution from which a dataset is generated in order to generate new samples.


\subsection{Training}
Training then comes down to minimizing this loss function with respect to the trainable parameters $\bm{\phi} =(\omega, b)$ of the network layers. This high-dimensional optimization problem can be tackled with Gradient Descent, where the parameters are iteratively shifted in the direction of the negative gradient, i.e. 
\begin{equation}
    \bm{\phi} \rightarrow \bm{\phi} - \alpha \nabla_{\bm{\phi}} \mathcal{L},
\end{equation}
where $\alpha$ is the so-called learning rate and a hyper-parameter. While this is the foundation of the general optimization strategy, there are several improvements that allow for adapting the learning rate during training. These are typically based on \textit{momentum}, which is a method that takes into account the \textit{speed} at which the optimizer has moved through the loss landscape in previous steps. Another point is to speed up the evaluations and introduce stochasticity by drawing random samples from the training set instead of optimizing over all of them at each evaluation. The latest and most widespread version of momentum based stochastic gradient-descent optimization strategies is ADAM \cite{Kingma2015}, which we will use in the main body of this thesis.

Since $y^\text{out}(x)$ is a chain of linear transformations followed by an activation function, one can calculate $\nabla_{\bm{\phi}} \mathcal{L}$ via the chain rule knowing all the derivatives of the activation functions.

The resulting formulas are nested in the sense that the derivatives depend on the derivatives of the next layer, i.e.
\begin{eqnarray}
\label{eq:BP1}
\mathcal{\delta}^{n} &=& \nabla_{y^\text{out}} \mathcal{L} \odot f'_\text{act}(z^n) \\
\label{eq:BP2}
\mathcal{\delta}^k  &=& ((\omega^{(k+1)})^T \mathcal{\delta}^{k+1}) \odot f'_\text{act}(z^k) \\
\label{eq:BP3}
\frac{\partial \mathcal{L}}{\partial b_{i}^{(k)}} &=&\mathcal{\delta}_i^k \\
\label{eq:BP4}
\frac{\partial \mathcal{L}}{\partial \omega_{ij}^{(k)}} &=& y^{(k-1)}_j \mathcal{\delta}_i^k
\end{eqnarray}
where $\odot$ is element-wise multiplication and $f'$ indicates the derivative. $z^k = \omega^{(k)} y^{k-1} - b$ is the output of the $k$-th layer before the activation. For a pedagogical derivation of the backpropagation formulas (\ref{eq:BP1}) - (\ref{eq:BP4}) we refer to \cite{nielsenneural}. So overall we can compute the derivative for all parameters of the network by passing through the network from the end to the start, which is why this procedure is referred to as the \textit{backpropagation algorithm} \cite{rumelhart1986}. Modern machine learning libraries can automatically compute derivatives with backpropagation by keeping track of the forward pass that needs to be composed of functions whose derivative is known to the program. This makes prototyping new models as easy as putting together \textsc{Lego}\textsuperscript{TM} blocks, which significantly helps to accelerate deep learning research.


\subsection{Convolutional layers}

Another core building block of NNs for Deep Learning are convolutional layers. These layers are particularily well-suited for feature extraction in images. A convolutional layer consists of three steps: \textbf{1)}, a number of filters are \textit{convoluted} with the input, where the filter values are trainable parameters and the number and size of the filters are hyper-parameters provided by the user. \textbf{2)} The second step is a \textit{non-linear activation function}, where we again mostly use the ReLU function. \textbf{3)} In the last step, the dimensionality may be reduced by \textit{pooling} (optional).


\begin{figure}
\centering
\includegraphics[width=.98\textwidth]{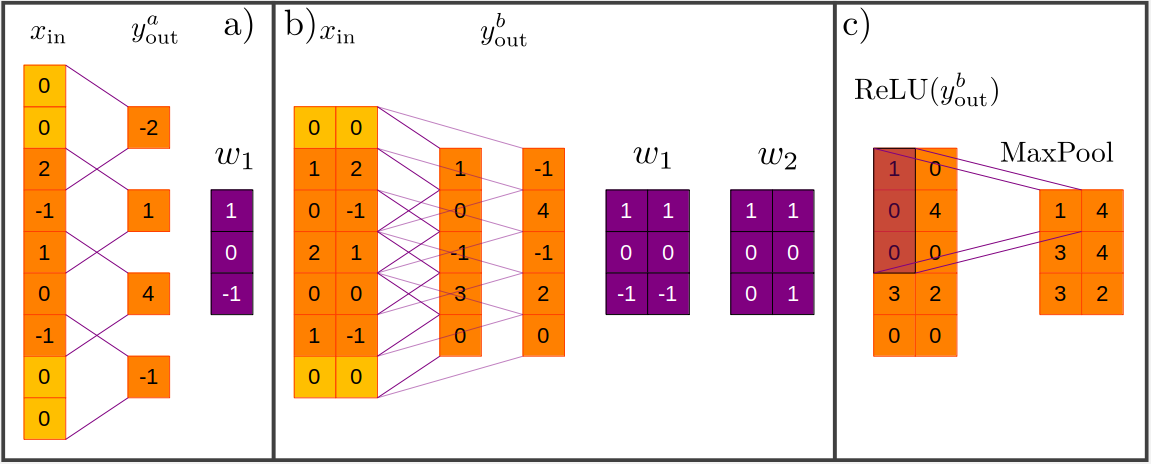}
\caption{One dimensional convolution (a,b) and pooling (c). \\
a) Input $x_\text{in}$ of \lstinline|size = 5| (orange) with a singular channel (horizontal dimension), \lstinline|n_ch = 1|, and zero-padding \lstinline|pad = 2| (yellow). The filter $w_1$ (purple) of \lstinline|dim = 3| (the number of channels of the filter always matches that of the input) is convoluted over the input by summing the elementwise multiplication with parts of the input. These parts are placed on the input with \lstinline|stride = 2|, such that the output has dimension \lstinline|4 = (size - dim + 2 pad)/stride + 1|. \\
b) Input $x_\text{in}$ of \lstinline|size = 5| and \lstinline|n_ch = 2|.  Two filter $w_1$ and $w_2$ (\lstinline|n_filters=2|), each of \lstinline|dim = 3|. Now with \lstinline|stride = 1| and \lstinline|pad = 1|, the output has dimension $(5,2)$, just like the input. Note that each filter generates one output channel. c) Taking the output of b), applying $\text{ReLU}(x) = \{x \text{ if } x\geq 0; \text{else } 0\}$ and then max-pooling with \lstinline|pool_size = 3|, \lstinline|stride = 1| and \lstinline|pad = 0|, resulting in a spatial dimension \lstinline|3 = (size - pool_size + 2 pad)/stride + 1|.}
\label{fig:1dconv}
\end{figure}

Let us explain in more detail what is meant by \textit{convolving} an input with a filter. This is best explained by means of an example, illustrated in \cref{fig:1dconv} for $1$ dimensional convolution. In case a), we have an input $x$ of \lstinline|size = 5| in a single channel (\lstinline|n_ch = 1|), \textit{zero-padding} of \lstinline|pad = 2|, we move the filter by a \lstinline|stride = 2| and we have a filter $w_1$ of \lstinline|dim = 3|. Now the filter is multiplied element-wise with the input and summed over, moving by \lstinline|stride = 2| after each step. This leads to an output of size \lstinline|4 = (size - dim + 2 pad)/stride + 1|. In the second example \cref{fig:1dconv} b) we use \lstinline|n_ch = 2| channels, which is always the depth of the filters as well. Additionally, we use \lstinline|n_filters = 2| such that the output is again with \lstinline|n_ch = n_filters = 2|.

\textit{Pooling} works similarily and is illustrated in \cref{fig:1dconv} c) for the output of b) for MaxPooling. Here we have \lstinline|pool_size = 3|, \lstinline|stride = 1| and \lstinline|pad = 0| such that the spatial dimension is reduced to \lstinline|3 = (size - pool_size + 2 pad)/stride + 1|, analogously to the convolution. So one can reduce the spatial dimensionality both by convolution and pooling, with the difference being whether it is before or after applying the activation function. There are no clear preferences.

The free parameters in a convolutional layer are solely the filters themselves, which are applied along the whole input. This is known as \textit{parameter sharing} and leads to a significant decrease in free parameters in convolutional layers.

\section{Overfitting}

\begin{figure}
\centering
\includegraphics[width=.75\textwidth]{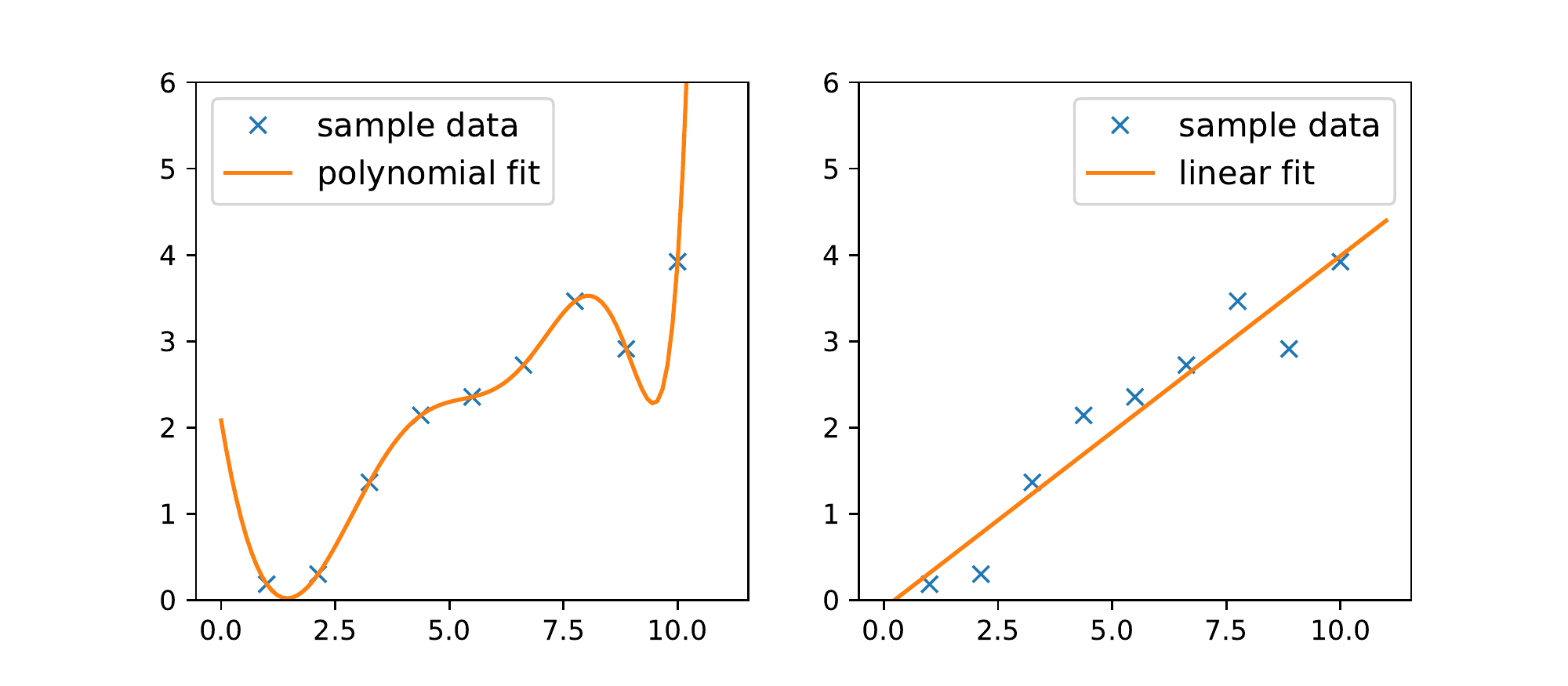}
\caption{Typical overfitting scenario for training data generated by a linear function with added noise. In the left panel, the fit is done using a polynomial of degree equal to the number of training samples. We thereby reach practical zero training loss as the polynomial captures all training samples. At the same time, any point beyond the training region $[1,10]$ will not be captured and the resulting model generalizes poorly. In the right panel, a linear fit is used. The training loss is higher but it will generalize better beyond the training region.}
\label{fig:overfitting}
\end{figure}

One of the fundamental problems in deep learning is \textit{overfitting} that hinders the generalization capabilities of NNs. A typical example for overfitting is when the model has too many degrees of freedom and is able to fit for example the noise of a model. We illustrate this for a simple example in \cref{fig:overfitting}, where samples are drawn from a linear distribution with added normal-distributed noise. We can reach practically zero loss in a least-squares fit by utilizing a polynomial of degree equal to the number of training samples. This comes at the cost of poor generalization capabilities. Any points beyond the training region are not captured well by this model. Alternatively, one is better advised to use a simpler model, e.g. a linear model that generalizes better at the cost of having higher training loss.

In practice one does not know the functional form of the underlying distributions. A way to detect overfitting in deep learning is to keep track of the loss for both the training and validation set during training. A typical overfitting scenario is when the training loss continues to decrease while the validation loss stagnates or increases. There are different strategies to counteract this problem, a simple one being keeping track of the validation accuracy and stopping the training once the turning point is reached (\textit{early stopping}). Another one is to provide more training data in order to be able to train for a longer time without overfitting. But often it is hard to obtain more training data, so in order to be able to train for longer times and potentially achieve better results, we need different strategies. 
The general idea is to reduce the number of weights (dropout) and the \textit{weight} of the weights themselves (regularization). It is not obvious \textit{why} these techniques improve the generalization capabilities and avoid overfitting, but empirical testing quickly yields improvements. Since neural networks are typically overparametrized, the intuition is that obtaining the same training loss with less weights (i.e. a less complex network) should yield a more general solution. This is however quite speculative. Still, regularization and dropout have become standard methods in modern deep learning and we describe them here briefly.

\subsection{L1 and L2 Regularization}
Regularization aims at reducing the \textit{weight} of the weights in a neural network. This can be achieved by adding a positive penalty term that captures the weights in the loss function:
\begin{eqnarray}
\mathcal{L}_\text{L1} = \mathcal{L} + \lambda \sum_{ijn} |\omega^{(n)}_{ij}| \text{ (L1 regularization),} \\
\mathcal{L}_\text{L2} = \mathcal{L} + \lambda \sum_{ijn} (\omega^{(n)}_{ij})^2 \text{ (L2 regularization).}
\end{eqnarray}
Here, $\lambda$ is a regularization hyper parameter that determines how much the regularization term is factored in in the loss function.

\subsection{Dropout}

Dropout is another form of regularization in which a random subset of hidden neurons is deactivated during training. So during one epoch, that is, processing all training examples iteratively in stochastic gradient descent, a specific subset is deactivated. In the next epoch, those neurons are restored and a different subset is deactivated. This effectively mimics training different neural networks and averaging over them, achieving more robust feature extraction \cite{Krizhevsky2017,nielsenneural}. The fraction of deactivated neurons during training is a hyper parameter set by the user. This fraction can be up to a substantial part of all neurons and shows the overparametrized property of neural networks. For example, in the original paper \cite{Hinton2012}, the authors use a dropout rate of up to $50\%$ and achieve much improved results compared to training with no dropout.

\section{Modern deep learning architectures}
The following section is beyond the scope of necessary information for the remainder of this thesis. Yet, it may broaden our horizon and allow us to get a better overview of the field of deep learning. Our overall goal of this chapter is to introduce the transformer model \cite{attentionisallyouneed}, which sets the standard in natural language processing, but is also very important for scientific breakthroughs with deep learning, for example in the protein folding problem  \cite{Jumper2021}. 

We will not go into details of reinforcement learning here, but note that the field has achieved substantial advancement since the introduction of deep learning. Reinforcement learning is used for problems that can be posed as a game, like for example chess. In this case, the board represents the \lstinline|state| and a move is an \lstinline|action|. Previous reinforcement learning algorithms were based on learning large tables of optimal actions for a given state (Q-learning). For many games with a large state space, this becomes intractable. Deep Q-learning solves this issue by replacing the table with a neural network that maps any state to a score for each action. This attempt marks the current state of the art in full information games like chess or go \cite{Silver2018}.

\subsection{Sequence to sequence models}
We so far discussed artificial neural networks consisting of convolutional or fully-connected layers that map an input of fixed size to an output of fixed size. However, in many applications of deep learning we require a certain flexibility in the input and output dimension. Imagine for example a natural language processing task in which we want to translate between English and German with a neural network. This neural network needs to be able to process sentences in the input language of different lengths and output sentences with yet other, different lengths, in the target language. One way to achieve this is through sequence-to-sequence models (seq2seq) consisting of two separate neural networks, the encoder and the decoder. More generally, these models are used for sequential data, where natural language is one example. Others are audio processing, i.e. speech recognition, or video processing.

The encoder and decoder network are typically recurrent neural networks (RNNs). RNNs process sequential data 
\begin{equation}
x = (x^{(0)}, x^{(1)}, \cdots ,  x^{(n-1)})
\end{equation}
in an iterative fashion by going from $x^{(0)}$ to $x^{(n-1)}$. In its most simple form, a RNN is a fully-connected neural network as previously discussed, but it stores a hidden state from the previous time step, such that at time step $t$, the processing explicitly depends on the previous time step $t-1$, i.e. $y(x^{(t)}|x^{(t-1)})$. Therefore, the final output implicitly depends on \textit{all} previous inputs
\begin{equation}
y^\text{out}(x) = y(x^{(n-1)}|x^{(0)}, x^{(1)}, \cdots, x^{(n-2)}),
\end{equation}
which is why RNNs are \textit{autoregressive} models.

More involved models like the long short-term memory (LSTM) \cite{Hochreiter1997} or gated recurrent unit (GRU) \cite{KyungHyun2014} have been developed in order to optimize the usage of the implicit information carried over by the hidden state at each time step. However, it is unrealistic that a hidden state can carry the semantics of a whole sentence. The performance of neural network based natural language processing has been significantly improved with sequence to sequence models with attention \cite{Sutskever2014,Luong2015}.

\begin{figure}
\centering
\includegraphics[width=0.9\textwidth]{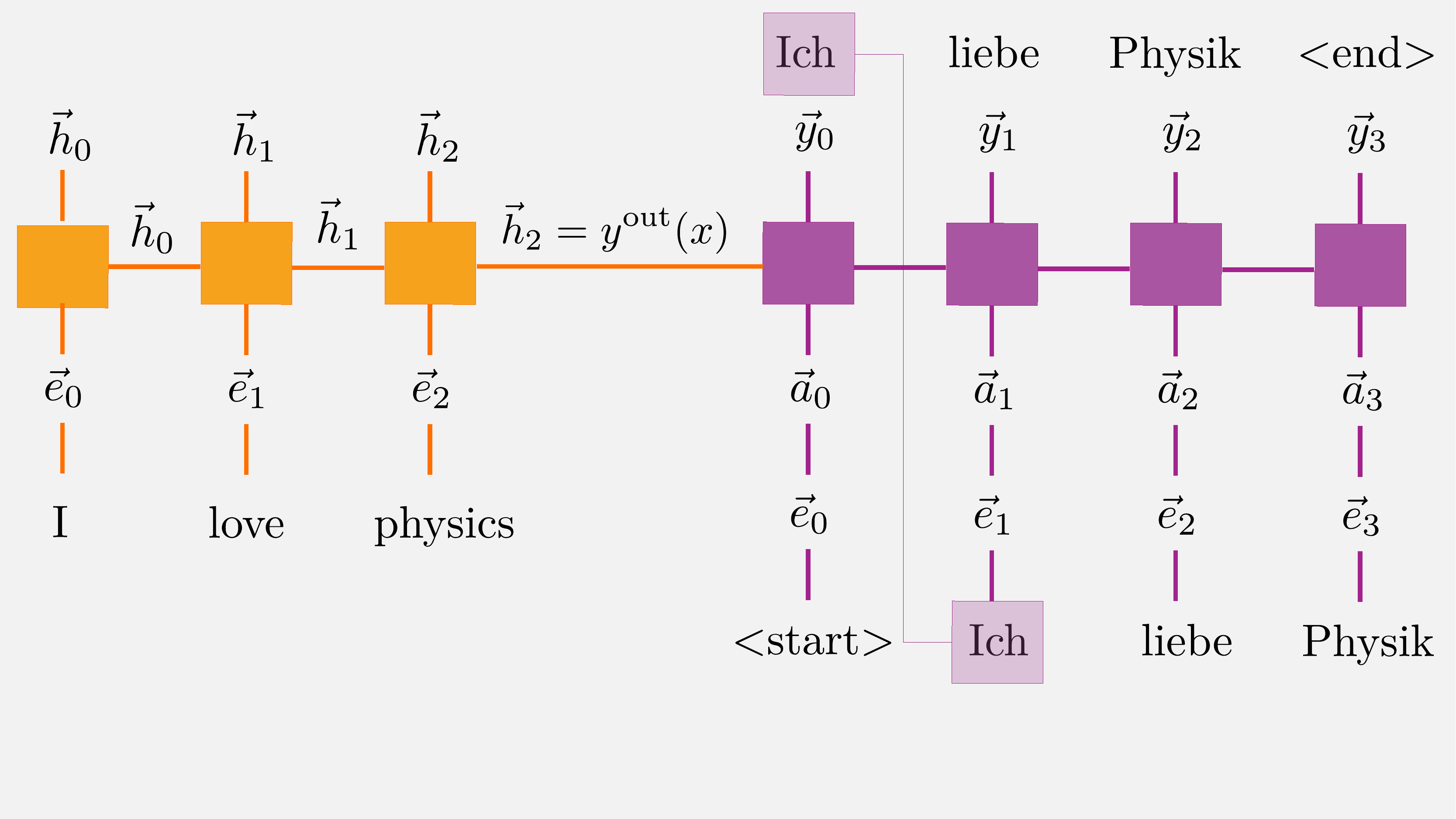}
\caption{A sequence-to-sequence model consisting of an encoder RNN (orange) and a decoder RNN (purple).}
\label{fig:seq2seq}
\end{figure}

Let us go through the example of translating a sentence in \cref{fig:seq2seq}. Our input sequence is \lstinline|x = ["I", "love", "physics"]| and our target output sequence is \lstinline|y(x) = ["Ich", "liebe", "Physik"]|. In the case of natural language, some pre-processing of the input sequence is necessary to obtain numerical vectors that neural networks can process. In a first step, each word in the input sequence is assigned a \textit{token}, which is a one-to-one mapping of all words in the input language dictionary to numbers \lstinline|1, 2, .. , number of words|. As a second step, each token is mapped to an embedding vector of a user-defined dimension. Typically, this is just a large dictionary mapping \lstinline|token| $\mapsto (e_0, e_1, ..)$, where $e_i$ are trainable parameters and $\bm{e} = (e_0, e_1, ..)$ is the embedding vector. So overall, each element $x^{(t)}$ in input $x$ is mapped to an embedding vector $\vec{e}_t$, which is sequentially processed in an RNN, yielding outputs $\vec{h}_t$ at each time step and $y^\text{out}(x)$ at the end of the sequence. For simplicity we assumed that the hidden state is the same as the output at each time step, which is not always the case.

The output of the encoder RNN serves as the initial hidden state for the decoder, which is initialized by a special token \lstinline|<start>| which indicates the start of a sentence. This token is then mapped for the output language to an embedding, which is processed through the decoder RNN. The output of each time step in the decoder is normalized with a \lstinline|softmax| function to mimic a probability vector of the size of the output language dictionary, yielding the most likely first word given the \lstinline|<start>| token and the encoder output. The most probable output words are then fed back as inputs until a special token \lstinline|<end>| is predicted. During training, one typically uses \textit{teacher forcing} by feeding back the correct prediction, independent of the potentially wrong prediction of the network during training.

So on one hand RNNs allow for variable size input sequences, and on the other hand seq2seq models allow for input and output sequences to be of different sizes.

Now it is entirely not clear why the above procedure should work. And given a model in the form described until now it is unlikely to achieve great results. The game changer comes with the attention mechanism, which is a way to relate the input and output items to allow for refined semantics between words. For each prediction $y_i$ in the decoder model, we compute an attention vector $\vec{a}_i$, which is a weighted sum 
\begin{equation}
\vec{a}_i = \sum_t \alpha_t \vec{h}_t
\end{equation}
of the hidden states of the encoder. The weights $\alpha_t$ are the so-called attention scores, where various different ways of computing them have been proposed. It is typically a geometric measure that relates $\vec{e}_i$ with all $\vec{h}_t$. The simplest form is simply a dot product $\alpha_t = \vec{e}_i \cdot \vec{h}_t$ given matching dimensions. Additionally, a trainable matrix defining an alternate norm can be inserted and activation functions applied. Typically, the attention scores are normalized by applying a \lstinline|softmax| function to have a convex combination of hidden states. The attention vector itself can serve as input to the decoder RNN, as depicted in \cref{fig:seq2seq}, or alternatively, it can be concatenated with the embedding vector.

\subsection{Transformer model}
The architecture described in the previous section can achieve impressive results. However, its ability to process arbitrary length sequences due to its sequential nature, which we advertised as one of the big advantages of these models, is at the same time its biggest drawback. This is because modern CPUs and GPUs have stagnated in terms of processing frequencies and the biggest speed-ups can be achieved through parallelizing computation. Due to the sequential nature of the processing in seq2seq models, this hinders training of very large datasets. 
The current state of the art in natural language processing is provided by transformer models which circumvent this problem and allow for parallel processing. This comes at the cost of not being able to input arbitrary sized sequences. However, the trick is to pad each input sequence with special tokens $0$ to achieve the same, potentially large, input size. Because the processing in a transformer is relatively light as it uses a lot of parameter sharing, we can allow for very large (fixed) input lengths and therefore restore this flexibility \textit{in practice}\footnote{For very large inputs this remains a problem.}.

The transformer model relies almost entirely on the previously described attention mechanism, hence the snappy title "Attention Is All You Need" of the original article \cite{attentionisallyouneed}. It is still a seq2seq model in the sense that it consists of an encoder and a decoder. The encoder performs \textit{self-attention}, that is, it relates all embeddings of the input sequence with each other and outputs the attention vectors. We can process all embeddings of the input sequence at once, which amounts to simple matrix multiplications of 
\begin{equation}
M_e = (\vec{e}_0,\vec{e}_1,\cdots)
\end{equation}
with itself to generate the scores. After applying a softmax row-wise for normalizing the weights, we can perform another matrix multiplication to achieve the weighted sums of the attention mechanism. Typically, inputs take on different roles as \textit{query} $Q$, \textit{value} $V$ and \textit{key} $K$ in this process, which includes a trainable mapping, e.g. $Q = Q(M_e)$. The resulting output is then simply
\begin{equation}
\text{Attention}(M_e) = \text{softmax}\left(Q K^T\right)V,
\end{equation}
where $Q$, $K$, and $V$ are all results of the input embeddings $M_e$.
Optionally the weights are rescaled before softmax is applied \cite{attentionisallyouneed}.

At the decoder, the inputs are processed in the same fashion with self-attention. However, to avoid future-inference at time $t$ from time $\tau > t$, \textit{masking} is utilized, which is a process that sets all future tokens to be ignored (typically with a special token $0$).
The decoder hidden states are then \textit{attended} by the encoder outputs in the same fashion as described before between encoder and decoder.
Architectural details may vary from model to model. The original paper additionally uses feed-forward neural networks at the end of the encoder and decoder \cite{attentionisallyouneed}. For translation or prediction tasks, we want to output probability vectors, so we need to apply a softmax function to the last layer. Another detail we have left out so far is the addition of a \textit{positional encoding} to the input embeddings. That is, adding a fixed (but potentially trainable) vector depending on the position in the sequence to the embeddings to restore (or rather mimic) a sense of sequential ordering.

All the processing in the encoder and decoder are parallelizable and fairly simple being almost exclusively matrix multiplications. This enables processing of very large datasets, which on the other hand allow for surprisingly realistic text generation or translation with large transformer models \cite{BERT,GPT}.


\chapter{Introduction to variational quantum algorithms}
\label{chap:qc}
Quantum computing has been a continuously developing field ever since the breakthrough discovery of Shor's algorithm in 1994 \cite{Shor1994}.

Most quantum algorithms, like the ones discussed in the textbook by Nielsen and Chuang  \cite{nielsen_chuang_2010}, require a fault tolerant quantum computer \cite{Aharonov1999}. That is, a quantum computer whose elementary operations are executed with such low error rates, that the few errors that still occur can be corrected with quantum error correction. Beyond Shor's algorithm, such a quantum computer would allow general purpose quantum simulation algorithms like quantum phase estimation \cite{Kitaev1995, nielsen_chuang_2010} or adiabatic quantum computation \cite{Farhi2000} to simulate \textbf{practically relevant} systems with the potential to enable unprecedented technological advancements \cite{Reiher2017, Burg2021, Delgado2022}. Fault tolerant quantum computers, however, are for the moment out of reach, though there has been substantial progress in experimental quantum error correction \cite{Semeghini2021,Satzinger2021}.

To speed up these efforts and make practical quantum computers available, big industry players started investing in their own quantum computing programs, acquired full research labs from universities, as well as quantum startups forming and taking part in the academic research environment. This commercialization, in part fueled by what some argue to be a \textit{quantum hype} \cite{Ezratty2022}, led to the availability of contemporary quantum computers \textit{in the cloud}, that is, quantum computers in labs of the providing company that are operated remotely by researchers and customers around the world. Some, like a big part of \textsc{IBM Quantum}'s devices, are free to use for researchers.

All these machines are inherently noisy and are therefore severely restricted in the gate count of operations that can be executed before the system decoheres or errors accumulate. Additionally, these machines are restricted in their overall size and connectivity, which is why the term \textit{noisy intermediate scale quantum} (NISQ) devices was coined \cite{Preskill2018}. The aforementioned fault tolerant quantum algorithms are therefore not suitable for these machines. Instead, one typically performs \textit{variational quantum algorithms} \cite{Bendetti2019,bharti2021noisy} on contemporary, noisy hardware. 

This approach shares similarities with the data-driven approach in training machine learning models, discussed in the previous chapter, in that a cost function is minimized to achieve a certain goal \cite{McClean2015, Stokes2019, Cerezo2020_variational}. The typical approach is having a variational quantum circuit with trainable parameters $\theta$ prepare a variational quantum state $\ket{\Psi(\theta)}$, and then a cost function which is the expectation value of some observable with respect to said state. One of the most natural algorithms in that setting is the variational quantum eigensolver (VQE)  \cite{Peruzzo2014}, where the cost function 
\begin{equation}
\mathcal{L} = \braket{\Psi(\theta) | H | \Psi(\theta)}
\end{equation}
is the expected energy and the goal is to approximate the ground state energy of the system.


In this chapter, we introduce variational quantum circuits by exemplary going through the steps of VQE, which covers the essence of the majority of variational algorithms and will be important for the main body of this thesis. Like in deep learning, discussed in the previous chapter, it is not clear how to design such variational circuits a priori. One attempt to salvage this problem is an adaptive approach called ADAPT-VQE \cite{2019adaptvqe}, that we discuss next. We further introduce the quantum approximate optimization algorithm (QAOA) \cite{Farhi2014} as a special case of VQE. We conclude by discussing the serious practical and fundamental restrictions of VQE in terms of scalability due to noise and Barren plateaus, the phenomenon of an exponentially vanishing gradient of the loss function \cite{McClean2018,Cerezo2020}.


\section{Variational quantum eigensolver}

The goal of the variational quantum eigensolver algorithm is to optimize a parametrized circuit $V(\theta)$ such that it minimizes the expected energy
\begin{equation}
E(\theta) = \braket{0|V(\theta)^\dagger H V(\theta)|0}
\end{equation}
with respect to the variational state $V(\theta)\ket{0}$. Note that since $V(\theta)$ is a unitary operation there is no need to explicitly normalize this expectation value.

This formulation is very similar to DMRG that we discussed in \cref{sec:tensor_networks}. This time, instead of using a classical approximation of the ground state as an Ansatz, we use a variational circuit that approximately prepares the ground state wavefunction on a quantum computer. For DMRG we could make use of the canonical form that allows for very efficient exact and local optimizations such that we iteratively sweep through the system to reach the global minimum. Here, we resort to gradient descent, which is typically used in machine learning as discussed in the previous chapter. However, we do not have direct access to the gradient since our wave function is encoded in physical qubits on a quantum computer. Yet, the gradient can be computed as an expectation value on the same device via the parameter shift rule \cite{Mitarai2018,Schuld2018}.

Let us briefly derive this. For simplicity, let $V(\theta)$ be a single qubit Pauli rotation $R_x(\theta) = \exp(-i\theta \sigma_x/2)$. The partial derivative then becomes
\begin{equation}
\label{eq:dEdtheta_direct}
\frac{\partial E}{\partial \theta} = \frac{\partial}{\partial \theta} \braket{0|R_x^\dagger(\theta) H R_x(\theta)|0} = \frac{i}{2}\braket{0|R_x^\dagger(\theta)[\sigma_x,H]R_x(\theta)|0}.
\end{equation}
We can then use the general identity for Pauli operators and Hermitian $H$ \cite{Mitarai2018},
\begin{equation}
\label{eq:mitarai}
[\sigma_x,H] = -i\left( R_x^\dagger\left(\frac{\pi}{2}\right) H R_x\left(\frac{\pi}{2}\right) - R_x^\dagger\left(-\frac{\pi}{2}\right) H R_x\left(-\frac{\pi}{2}\right)\right),
\end{equation}
to derive the parameter-shift rule for variational quantum circuits
\begin{equation}
\frac{\partial E(\theta)}{\partial \theta} = \frac{1}{2} \left(E\left(\theta + \frac{\pi}{2}\right) - E\left(\theta - \frac{\pi}{2}\right) \right).
\end{equation}
Since \cref{eq:mitarai} is valid for any pauli operator and Hermitian operator $H$, this can readily be extended to general variational circuits consisting of Pauli rotations. More generally, the parameter shift rule is valid for any parametrized unitary that is generated by a self-inverse operator (like Pauli matrices). Therefore, we obtain a rule for many possibilities of parametrized gates. Note that in the \textsf{IBM} machines that we employ later, the set of native gates only contains Pauli rotations as parametrized gates.

The set of native gates is the universal (and potentially overcomplete) set of operations that the physical device can execute. They are universal in the sense that any unitary operation can be decomposed in terms of them \cite{Barenco1995}. One common example would be
\begin{equation}
\label{eq:native_gates}
\mathcal{N} = \left\{R_z, X, \sqrt{X}, \text{CX}, \mathbb{1} \right\}
\end{equation}
with $R_z(\theta) = e^{-i \theta \sigma_z}$, $X = \sigma_x$, $\text{CX} = \mathbb{1} \otimes |0\rangle\langle 0| + X \otimes |1\rangle\langle 1|$ (CNOT gate) and
\begin{equation}
\sqrt{X} = \frac{1}{2} \begin{pmatrix}
1+i & 1-i \\
1-i & 1+i
\end{pmatrix},
\end{equation}
the \textit{square root of} $X$ in the sense that $(\sqrt{X})^2 = X$. 

With $R_z$ and $\sqrt{X}$ we can build arbitrary single qubit rotation gates. For this, first note that
\begin{equation}
Y = \sqrt{X}^\dagger Z \sqrt{X}
\end{equation}
and that $\exp(-i \theta \vec{n} \cdot \vec{\sigma}) = \cos(\theta) \mathbb{1} + i \sin(\theta) \vec{n} \cdot \vec{\sigma}$ for any unit vector $\vec{n}$ and $\vec{\sigma} = (\sigma_x, \sigma_y, \sigma_z)$. Then it is straight forward to see that
\begin{equation}
R_y(\theta) = \sqrt{X}^\dagger R_z(\theta) \sqrt{X}.
\end{equation}
Another example for 2-qubit gates is decomposing a SWAP gate in terms of three CX gates
\begin{multline*}
\text{SWAP}_{01} = \begin{pmatrix}
1 &0 &0 &0 \\
0 &0 &1 &0 \\
0 &1 &0 &0 \\
0 &0 &0 &1 \\
\end{pmatrix}
 \\= \begin{pmatrix}
1 &0 &0 &0 \\
0 &1 &0 &0 \\
0 &0 &0 &1 \\
0 &0 &1 &0 \\
\end{pmatrix} \begin{pmatrix}
1 &0 &0 &0 \\
0 &0 &0 &1 \\
0 &0 &1 &0 \\
0 &1 &0 &0 \\
\end{pmatrix}\begin{pmatrix}
1 &0 &0 &0 \\
0 &1 &0 &0 \\
0 &0 &0 &1 \\
0 &0 &1 &0 \\
\end{pmatrix}  \\= \text{CX}_{01} \text{CX}_{10} \text{CX}_{01}.
\end{multline*}
So for a quantum computer with the native gate set $\mathcal{N}$ in \cref{eq:native_gates}, any parametrized circuit will ultimately be decomposed into rotations and fixed gates such that we can estimate the gradient of $E(\theta)$ in $2n_\text{params}$ evaluations of $E$.

Computational overheads for contemporary quantum computers are still rather large, so this is often still too costly in practice. Alternatively, we can use stochastic methods to approximate the gradient like simultaneous perturbation stochastic approximation (SPSA) \cite{Spall1992}, which is less expensive in function evaluations.

\section{ADAPT-VQE}

The composition of $V(\theta)$ restricts the search for the ground state to the manifold of states that can be reached with $V(\theta)$. For DMRG it is known that MPS span the correct manifold of target states for ground states of gapped and local Hamiltonians. Similarily, in variational circuits there are problem inspired Ansatze, especially for quantum chemical Hamiltonians (see e.g. \cite{bharti2021noisy} for a review).

Due to noise restrictions in physical quantum computers, it is often desired to use a minimal Ansatz.  The Adaptive Derivative-Assembled Pseudo-Trotter Ansatz Variational Quantum Eigensolver (ADAPT-VQE) algorithm is a way to adaptively construct such an approximation to a minimal Ansatz \cite{2019adaptvqe}. Its original formulation was derived for quantum chemical methods, but we can formulate it for a general operator pool of which we want to construct our Ansatz. A unitary operator $U$ is generally generated by a Hermitian operator $\mathcal{H}$ in the way that $U = \exp(-i\theta \mathcal{H})$. Let us define a finite operator pool $\mathcal{P} = \{U_i\}$ consisting of parametrized unitary operators of this form. Imagine we already have an Ansatz 
\begin{equation}
V^{\ell-1}(\bm{\theta}) = U^{\ell-1}(\theta_{\ell-1}).. U^2(\theta_2) U^1(\theta_1)
\end{equation}
with optimized parameters $\theta_i$. We use superscripts in $U$ to indicate the ordering in $V$ and subscripts to indicate the enumeration in the operator pool (which here is left unspecified).
We now want to ask how much adding the $i$-th operator of the pool at the $\ell$-th position, $U^\ell_i(\theta_\ell) = \exp(-i\theta_\ell \mathcal{H}^\ell)$, would affect the expectation value $\braket{0|V^\dagger H V|0}$. So we append $U^\ell_i(\theta_\ell)$ to the left of $V$ and compute the derivative around $\theta_\ell=0$, for which $V$ would stay unaffected. This calculation is analogous to that of the parameter-shift rule and yields
\begin{equation}
\label{eq:adaptive-derivative}
\frac{\partial \braket{0|V^{\ell \dagger} H V^\ell|0}}{\partial \theta_\ell}\bigg\rvert_{\theta_\ell = 0} = i \braket{0|V^{\ell-1\dagger} [\mathcal{H}^\ell, H] V^{\ell-1}|0}.
\end{equation}
So it is the expectation value of the commutator between the Hamiltonian and the generator of $U$ with respect to the current trial state $V^{\ell-1}\ket{0}$\footnote{Note that since we evaluate at $\theta_\ell = 0$ we obtain $V^{\ell-1}$. It is the task of the following step to determine the nonzero value of $\theta_\ell$.}.
We then compute \cref{eq:adaptive-derivative} for all $U\in \mathcal{P}$ and pick the one with largest magnitude in order to grow our Ansatz to $V^\ell$. All parameters $\theta_1,..,\theta_\ell$ are then optimized again. This process is repeated until either a maximum number of gates is achieved or the absolute value of all derivatives are below a user-specified threshold.


\section{QAOA}
\label{sec:QAOA}
We want to briefly comment on the quantum approximate optimization algorithm (QAOA) \cite{Farhi2014}, which is a special case of VQE. Here, the problem is of classical nature and encoded in the minimization of a spin Hamiltonian $H_P$. A typical example is the maxcut problem for graphs \cite{Fahri2014}. These kind of problems can be solved with adiabatic quantum computing. That is, one prepares the system in the ground state of a simple Hamiltonian $H_x = - \sum_i \sigma_x$, $\ket{+}^{\otimes N}$ and \textit{adiabatically} evolves the state according to a time dependent Hamiltonian $H(t)$ that interpolates between $H_x$ and $H_P$. According to the adiabatic theorem, the evolved state remains in the ground state of the system, as long as the gap of the Hamiltonian does not close and the evolution is \textit{slow}. In practice, this is harder to do since the slow evolution requires long coherence times and applying many time evolution gates, since evolution on a digital quantum computer amounts to applying trotterized time evolution operators, just like in \cref{sec:TEBD}. QAOA aims to approximate this procedure by evolving according to $H_P$ and $H_x$ in an alternating fashion
\begin{equation}
V(\bm{t}) = e^{-it_1 H_P} e^{-it_2H_x} e^{-it_3 H_P} e^{-it_4H_x} ...,
\end{equation}
where the the times $\bm{t} = (t_1,t_2,..)$ are variational parameters that are optimized with respect to the objective function is the expectation value $\braket{+|V^\dagger H_P V|+}$. In that sense, QAOA is a special case of VQE with a problem inspired Ansatz circuit.

\section{Restrictions of VQE}


Variational algorithms like VQE suffer from serious limitations in scalability. On one hand, noisy hardware accumulates errors and does not allow for coherent application of many consecutive gates. This can in principle be counteracted with improved error rates, error mitigation and error correction. 

A more fundamental problem is the occurrence of barren plateaus that make training deep circuits on quantum computers with many qubits practically impossible \cite{McClean2018}. This is because it has been shown that for generic variational circuits and Hamiltonians, the expectation value of the gradient is zero and its variance vanishes exponentially for any circuit of depth $\mathcal{O}(\text{poly}(N))$, where $N$ is the number of qubits \cite{Cerezo2020}. In the case of generic Ans\"atze, that is problem-independent circuits with no heuristic for initialization, only for shallow circuits with depth $\mathcal{O}(\log(N))$ and local cost functions is there a chance to be able to optimize the circuit. Further, it has been shown that gradient-free optimization does not salvage this problem \cite{Arrasmith2021}. And while other approaches promise trainability \cite{Pesah2021,Volkoff2021}, the problem for large scale VQE simulations persists.

Barren plateaus are not a problem for structured problems where good heuristics are known. Examples for this are QAOA as discussed above and ADAPT-VQE for quantum chemistry with the operator pool consisting of unitary coupled cluster operators. In both cases, physical insights give heuristics for structuring and initializing the Ansatz, which avoids barren plateaus and therefore allows trainability. However, poor scaling in molecule sizes pose serious limitations for VQE calculations of molecules of relevant sizes \cite{Kuhn2019}.

This ultimately means that variational algorithms are likely not scalable to system sizes that are beyond what is reachable with tensor network methods and serve more as a proof of principle \cite{Liu2022}. While the latest estimates for practically relevant quantum computations like computational catalysis require $\mathcal{O}(1000)$ logical qubits \cite{Burg2021}, variational algorithms are still one of the only practical means to explore the capabilities of contemporary noisy hardware.


\part{Results}
\label{part2}

\chapter{Deep Anomaly Detection for unsupervised phase discovery}
\label{chapter:anomaly}
In this chapter, we are going to introduce the concept of \textit{deep anomaly detection}, which will be the foundation of the works described in the following chapters. This method has been successfully employed in our works in \cite{Kottmann2020,Kottmann2021,Kottmann2021b,Kaming2021,Szoldra2021} and was already adapted by others in \cite{Zhang2021,Acvevedo2021,Munoz2021}. Anomaly detection with and without deep neural networks is an active field of research in computer science. Here, we follow an approach based on \textit{autoencoders} \cite{borghesi2019anomaly} and adopt it to the discovery of (quantum) phase transitions. It shall be noted that autoencoders have been used in a similar fashion to discover physical concepts from data \cite{Iten2020}.

\begin{figure}
\centering{
\includegraphics[width=.6\textwidth]{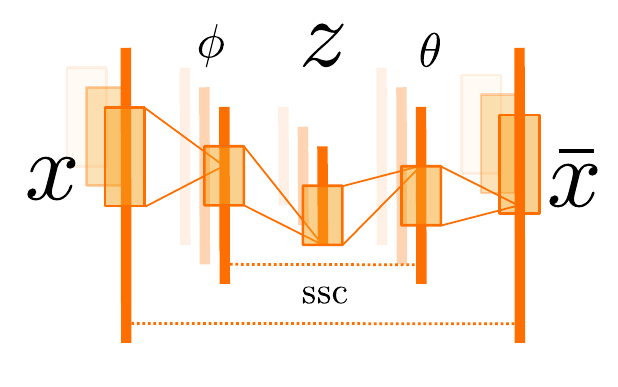}
}
\caption{An input $x$ is mapped through an encoder network, parametrized by $\phi$ to the latent space variable $z$, which is again mapped to the output $\overline{x}$ through a decoder network, parametrized by $\theta$. Autoencoders always have the same input and output dimension, but the internal architecture can vary. In this specific example both encoder and decoder consist of $2$ convolutional layers.}
\label{fig:AE_main}
\end{figure}

\paragraph{Autoencoders}

\textit{Autoencoders} are a special neural network architecture where the input dimension matches that of the output dimension, see \cref{fig:AE_main}. It consists of an encoder network, parametrized by $\phi$, that maps the input $x$ to a latent space variable $z$, and a decoder network, parametrized by $\theta$, that maps the latent space to the output $\overline{x}$. In the context of classical machine learning, autoencoders have been introduced for various tasks. For example, an autoencoder can be trained by pairs of colored images and their black and white counterpart to artificially color images. Further, pairs of images and noisy versions of those images can be used to train an autoencoder for de-noising. Furthermore, we can find an abstract representation of data by simply training an autoencoder to reproduce the input. When the latent space dimension is smaller than the input dimension, this then accounts to data-specific compression. This is very different from general purpose compression as it is specific to the data it has been trained on. Colloquially, an example would be an autoencoder that has been trained to compress images of dogs that then fails to compress images of cats. This is the key idea of how we use autoencoders for anomaly detection.

\begin{figure}
\centering{
\includegraphics[width=.6\textwidth]{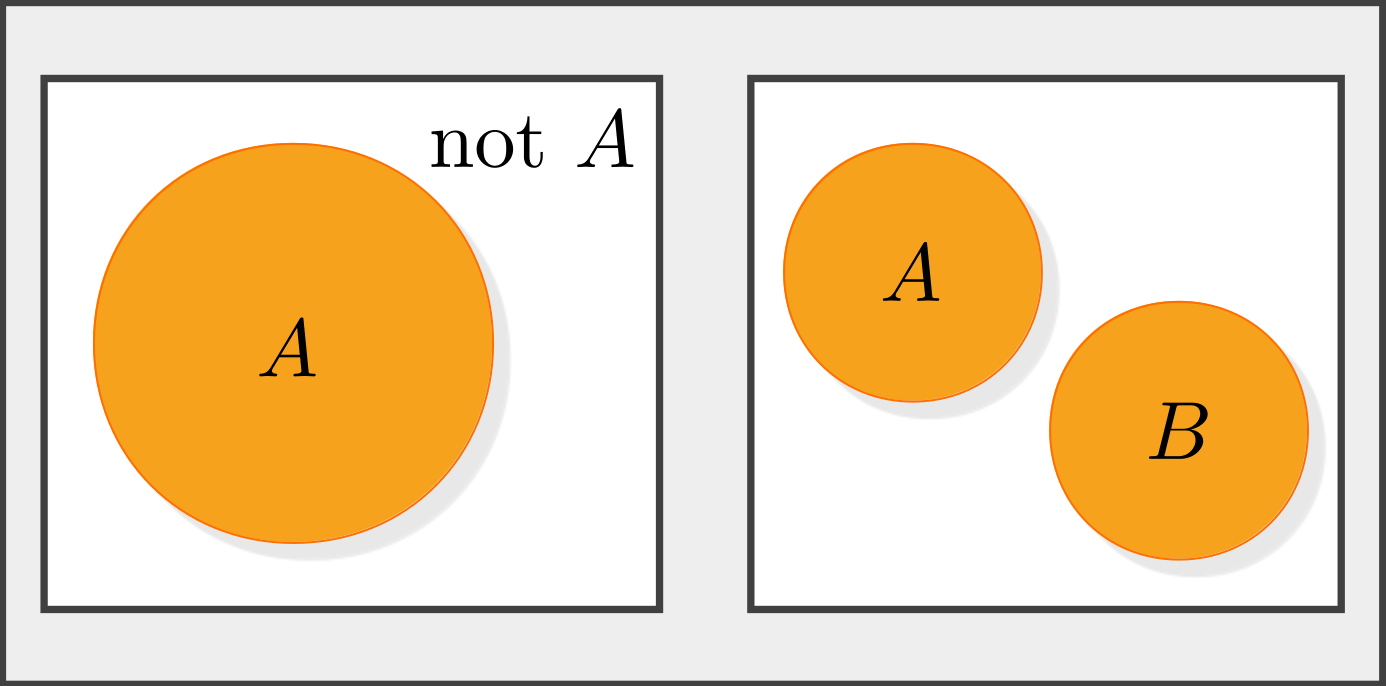}
}
\caption{Conceptual difference between anomaly detection - differentiating a class A and its complement, and supervised classification - differentiating two classes A and B (or multiple classes).}
\label{fig:A-notA}
\end{figure}

\paragraph{Anomaly Detection} Anomaly detection is used in scenarios where we want to differentiate a class of data from its complement. A typical example in classical machine learning would be to differentiate \textsc{valid} and \textsc{fraudulent} credit card transactions. A financial institution that is interested in finding credit card fraud likely has access to an abundance of valid transaction data, wheras no or very little of fraudulent transactions. Therefore, a supervised classification approach, where a neural network is trained to classify transactions as valid or fraudulent, would most likely fail due to this strong imbalance in training data. Instead, one would want to differentiate \textsc{valid} transactions and \textsc{not valid} transactions with anomaly detection. This conceptual difference between supervised learning and anomaly detection is graphically illustrated in \cref{fig:A-notA}. We do this by training an autoencoder on the abundance of available data, that we call \textit{normal} data $\mathcal{N}$, to reproduce itself. I.e. the corresponding loss function $\mathcal{L}$ for this task is some metric between the input and output $d(x,\overline{x})$\footnote{We denote the output of the autoencoder as $\overline{x}$, which is to be interpreted as a function of $x$.}. Throughout this thesis we will mainly use the $\ell^2$ norm for this purpose, i.e.
\begin{equation}
\mathcal{L} = ||x - \overline{x}||^2
\end{equation} 
with $||x||^2 = \sum_{i=1}^{\text{dim}(x)} x_i^2$.
By minimizing this loss, we find an abstract representation in the latent space of the autoencoder that is specific to the \textit{normal} data, from which we can reproduce the original data. For data that does not share the characteristics of the \textit{normal} data, this process will fail and we can detect \textit{anomalous} data with an increase in $\mathcal{L}$ - this is often referred to as the \textit{anomaly syndrome}.

\paragraph{Mapping out phase diagrams} We can use anomaly detection to map out all phase transitions in the phase diagram of a many-body system in an unsupervised fashion, i.e. without any prior knowledge of the system itself. In this thesis we focus on quantum phase transitions, yet it shall be noted that this approach works with thermal phase transitions as well. We consider a scenario where we are given unlabeled data of a phase diagram in some physical parameter space $\lambda_i \in \mathcal{D} \subset \mathbb{R}^n$ for a Hamiltonian $H(\{\lambda_i\})$\footnote{We do not even assume the knowledge of the Hamiltonian $H$. However, in practice we use it to generate the data.}.

The data $x$ at $\vec{\lambda}$ in the phase diagram corresponds to the ground states of the system. This can be in the form of the whole wavefunction, observables like correlation functions calculated from the ground state, reduced states, or entanglement spectra. We can then extract the phase boundaries by performing \cref{alg:anomaly_detection-phase_diagram}.

Here, $\text{AnomalyDetection}(\mathcal{T})$ corresponds to training an autoencoder with the data from the training region $\mathcal{T}$ and outputting the loss phase diagram, i.e. the loss $\mathcal{L}$ for all $x(\vec{\lambda})$ in the phase diagram $\mathcal{D}$. Instructions like \textit{extract phase boundary} are ambiguous as various different methods can be used to achieve this. One very simple example would be setting a fixed threshold as a hyper parameter set by the user and classify the data according to the loss being above or below that threshold. Similarily, \textit{extract regions of high loss} can for example be achieved by picking the maxima of the loss phase diagram and drawing circles of constant size around it. This is to say that there are general subroutines that execute these instructions, however, in the following thesis we will mainly do these steps by hand (or rather eye) due to the small number of phases and the $1D$ and $2D$ nature of most phase diagrams we look at.

\

As is common in machine learning, the success of \cref{alg:anomaly_detection-phase_diagram} will strongly depend on the quality of the data and how well suited the autoencoder architecture is. Finding a suitable architecture is typically a matter of trial and error, though we will find that for most of our applications very simple architectures like purely dense, purely convolutional and mixed dense-convolutional autoencoders perform similarily.

It shall also be noted that this is by far not the only possibility to map out phase diagrams in an unsupervised fashion. A very similar approach is presented in \cite{Greplova2020}, where the authors train a neural network to predict physical parameters from the input. These predictions fail at and around phase transitions, which can then be used as an anomaly syndrome\footnote{The authors do not use different terminology, but the relation to anomaly detection discussed here is evident.} to find transition points.

\begin{algorithm}
\caption{Unsupervised mapping of phase diagram}\label{alg:anomaly_detection-phase_diagram}
\begin{algorithmic}
\State $k = 0$

\State Draw random training region $\mathcal{T}_0 \subset \mathcal{D}$ from phase diagram $\mathcal{D}$
\State \textsc{result}$_0$ $=$ $\text{Anomaly Detection}(\mathcal{T}_0)$

\State Extract phase boundary from \textsc{result}$_0$ and append to \textsc{list of boundaries}

\While{$\textsc{list of boundaries} \neq \emptyset$}
\State $k = k + 1$
\State Extract regions of high loss $\left\{A^{k}_i\right\}$ from previous \textsc{result}$_{k-1}$

\If{$\nexists A^{k}_i$ such that $A^{k}_i \cap \bigcup_{l=0}^{k-1} T_l = \emptyset$}
\State \textbf{break}
\EndIf
\State Sort $\left\{A^{k}_i\right\}$ by maximal loss value in descending order
\State $\mathcal{T}_k = A^k_i$ for first $i$ that satisfies $A^{k}_i \cap \bigcup_{l=0}^{k-1} T_l = \emptyset $
\State \textsc{result}$_k$ $=$ $\text{Anomaly Detection}(\mathcal{T}_k)$
\State Extract phase boundary from \textsc{result}$_0$ and append to \textsc{list of boundaries}
\EndWhile
\ \\
\Return{\textsc{list of boundaries}}

\end{algorithmic}
\end{algorithm}

\chapter{Discovering new features in the extended Bose Hubbard model}
\label{chapter:bose-hubbard}
In this chapter we will apply the aforementioned anomaly detection algorithm \ref{alg:anomaly_detection-phase_diagram}, discussed in \cref{chapter:anomaly}, to map out the phase diagram of the extended Bose Hubbard model. We will discover unexpected new features that we then thoroughly analyze. The content of this chapter is a unification of our publications entitled \textit{Unsupervised phase discovery with deep anomaly detection} \cite{Kottmann2020} and \textit{Supersolid-Superfluid phase separation in the extended Bose-Hubbard model} \cite{Kottmann2021supersolid}.

We elaborate on the importance of Bose Hubbard models in \cref{sec:motivation0} and summarize its physical properties in \cref{sec:bose-hubbard-physics}. Anomaly detection is used to map out the phase diagram for various different quantities representing the ground states in \cref{sec:machine-learning-setup,sec:discovering-new-features}. We find new features in accordance with \cite{Batrouni} and expand the results in the following aspects. In \cref{sec:phase-separation}, we perform a numerical analysis of the mechanical stability in terms of second derivatives of the energy and the Gibbs potential as a function of the density $n$, that leads to phase separation. In \cref{sec:oscillations}, we investigate the spatial oscillations of the entanglement spectra for the homogeneous SF phase in the regime of parameters corresponding to the phase separation/phase coexistence. Here, all single-particle observables seem to be spatially homogeneous, while entanglement Rényi entropies and entanglement spectra exhibit oscillations. The spatial period of these oscillations, as well as the period of the Schmidt gap closing, is of the order of 10-20 lattice constants, i.e. has nothing to do with the periodicity of the CDW or SS, which is 2 lattice constants. In \cref{sec:luttinger}, we use Luttinger liquid theory to explain the presence of oscillations in the entanglement spectrum thus ruling out the possibility that these oscillations appear due to topological effects. Both in SF and SS phases, the excitation spectrum is governed by gapless linear phonons which make the Luttinger liquid description applicable and therefore allow predictions of the long-range behavior of the correlation functions.

\section{Motivation}
\label{sec:motivation0}

\paragraph{Physical motivation} Bosonic Hubbard models remain in the focus of interest in condensed matter and ultracold quantum matter physics since the seminal paper of Fisher {\it et al.}~\cite{Fisher}. In recent years, considerable attention was devoted to extended/non-standard Hubbard models (for a review cf.~\cite{Gajda}). They are of fundamental interst as Extended Bose Hubbard models provide perhaps the simplest models that include beyond on-site interactions. They exhibit a plethora of quantum phases in 1D: Mott insulator (MI), Haldane insulator (HI), superfluid (SF), supersolid (SS), and charge density wave (CDW). Furthermore, quantum simulators of these models and their variants are experimentally feasible in various platforms: ultracold atoms or molecules in optical lattices \cite{lewenstein2012ultracold}, systems of trapped ions, or Rydberg atoms.

\paragraph{Machine learning motivation} Compared to previous unsupervised attempts in \cite{nieuwenburg2017learning, liu2017self,wang2016discovering,wetzel2017unsupervised,chng2017unsupervised}, this method needs only one or few training iterations and has better generalization properties from employing deep neural networks \cite{kawaguchi2017generalization, valle2018deep}. This allows for efficient fully automatized phase discovery in the spirit of self-driving laboratories \cite{Hase2019}, where artificial intelligence augments experimentation platforms to enable fully autonomous experimentation. Intuitively, the method explores the phase diagram until an abrupt change, an anomaly, is detected, singling out the presence of a phase transition. The intuition is similar to the approach introduced in~\cite{Zanardi2006}, where the authors proposed to detect quantum phase transitions by looking at the overlap between neighbouring ground states in the phase diagram. Here, the machine is used to detect these anomalies. Moreover, as we explain next, it does it from scalable data.

In principle, there are many possible choices as input data for training our method, including the full state vector. To improve scalability and reach large system sizes, we propose to use quantities that arise naturally in the state description and do not require complete state information. For instance, we obtain ground states with tensor networks, from which we use the tensors themselves or the entanglement spectrum (ES) as input data. These quantities arise naturally from the state description without further processing and contain crucial information about the phase, like ES for example \cite{Deng2011,Shinjo2019,Tsai2019}. We stress, however,  that the choice of preferred quantities to be used for ML may in general vary and depend on the simulation method. In fact, we show that our method also works well with physical data accessible in experiments such as low-order correlation functions.

\section{Bose Hubbard model}
\label{sec:bose-hubbard-physics}
This work deals with the physics of the extended Bose Hubbard model in 1D and focuses on three of the most challenging and discussed phenomena of contemporary physics: supersolidity, phase separation, and entanglement. The extended Hubbard model in 1D has been studied extensively \cite{Rossini2012,Kuehner1997,Kuehner1999,Mishra2009,Urba2006,
Ejima2014,Cazalilla2011,Batrouni2006,Deng2011,Berg2008} and with a focus on supersolidty for incommensurate fillings~\cite{Kawaki2017,Kuehner1997,Kuehner1999,Mishra2009}. The authors in \cite{Deng2011} claimed to have found supersolidity for filling $n=1$  without in-depth discussion. We later elaborate how this was a misconception and there is only a phase-separated region consisting of a supersolid and superfluid part at this filling. The complete phase diagram of the model was described by Batrouni {\it et al.} (see~\cite{Batrouni} and references therein; our work expands the results of Ref.~\cite{Rossini2012}). These authors studied the phase diagram of the one-dimensional bosonic Hubbard model with contact ($U$) and nearest-neighbor ($V$) interactions focusing on the gapped HI phase which is characterized by an exotic nonlocal order parameter. They used the Stochastic Green Function quantum Monte Carlo as well as the Density Matrix Renormalization Group (DMRG) algorithm to map out the phase diagram. Their main conclusions concern the existence of the HI at filling $n=1$, while the SS phase exists for a very wide range of parameters (including commensurate fillings) and displays power-law decay in the one-body Green function. In addition, they found that at fixed integer density, the system exhibits phase separation in the $(U,V)$ plane.\\

We apply the the density matrix renormalization group (DMRG) algorithm in terms of Tensor Networks, i.e. Matrix Product States (MPS), described in \cref{sec:tensor_networks}, to study the ground-state properties of the extended Bose-Hubbard model,
\begin{multline}
\label{eq:H}
H = -t \sum_i \left( b^\dagger_i b_{i+1} + b^\dagger_{i+1} b_i \right) \\
+ \frac{U}{2} \sum_i n_i(n_i -1) + V \sum_i n_i n_{i+1},
\end{multline}
with nearest neighbour interaction on a one dimensional chain with $L$ sites. Here, $n_i = b^\dagger_i b_i$ is the number operator for Bosons defined by $[b_i,b_j^\dagger] = \delta_{ij}$. The model is characterized by three energy scales: the nearest-neighbor tunneling amplitude $t$, on-site interactions of strength $U$, and nearest-neighbor interactions tuned by $V$.
We set the energy scales in units of the tunneling coefficient by setting $t=1$ and continue with dimensionless quantities. Typically, we are interested in varying the on-site interaction $U$ and nearest-neighbour interaction $V$. We explicitly enforce filling $n :=\sum_i \braket{n_i}/L_\infty = 1$ by employing $U(1)$ symmetric tensors \cite{Silvi2017}, which we implement using the open source library TeNPy \cite{tenpy}. We perform simulations both in a microcanonical ensemble (fixed number $N$ of particles) and a canonical ensemble (fixed chemical potential $\mu$ and fluctuating number of particles). 

\begin{figure}[tb]
\centering
	\includegraphics[width=.77\textwidth]{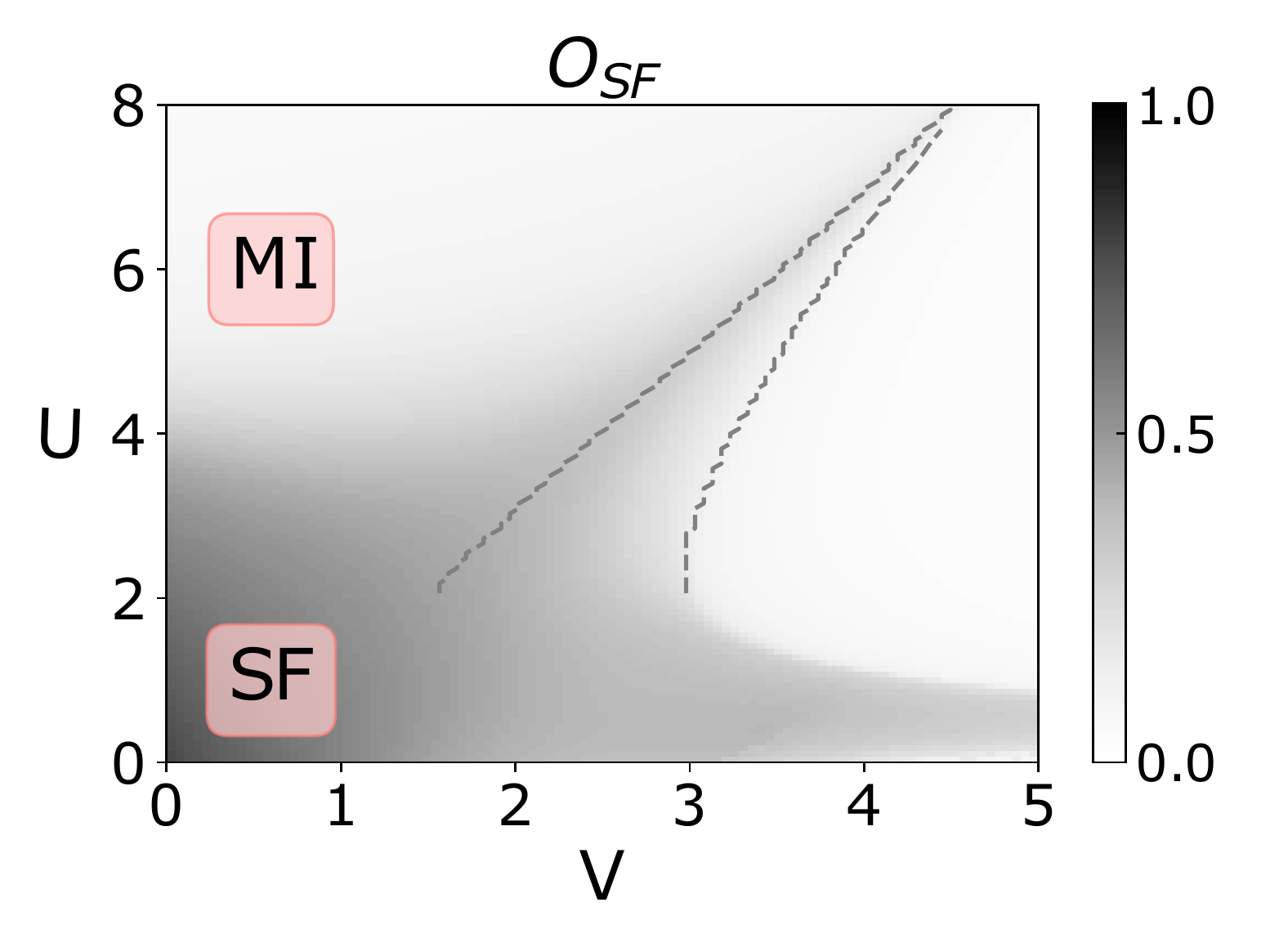}
	\includegraphics[width=.77\textwidth]{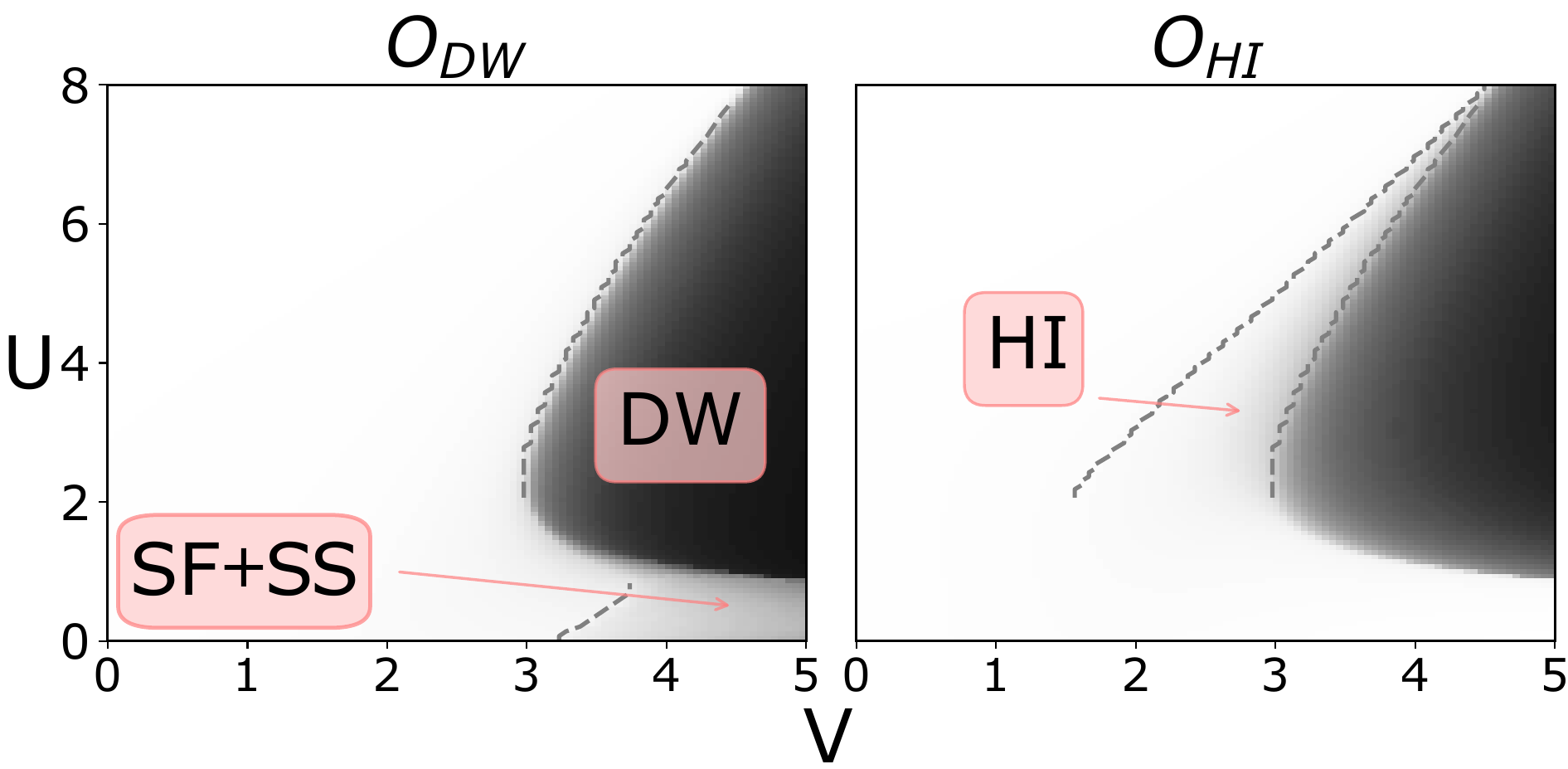}
\caption{\label{fig:phase_diagram} Extended BH phase diagram with five distinct phases obtained by the correlators \cref{prleq:CSF,prleq:CDW,prleq:Cs}. MI: Mott Insulator, SF: Super Fluid, SS: Super Solid, DW: Density Wave, HI: Haldane Insulator. The dashed lines indicate the transition points observed from diverging correlation lengths between MI-HI-DW and non-zero $\mathcal{S} := \max_{k\neq0} \left|\sum_j \braket{n_j} e^{-ikj}/L\right|^2$ between SF and SF+SS.}
\end{figure}

One way to physically classify these phases is to look at the correlators
\begin{align}
\CSF(i,j) &= \braket{b^\dagger_i b_j} \label{prleq:CSF} \\
\CDW(i,j) &= \braket{\delta n_i (-1)^{|i-j|} \delta n_j} \label{prleq:CDW} \\
\Cs(i,j) &= \braket{\delta n_i \exp\left( -i \pi \sum_{i \leq l \leq j-1} \delta n_l\right) \delta n_j}\label{prleq:Cs}
\end{align}
with $\delta n_i = n_i - \bar{n}$. $\CSF$ discriminates the Mott-insulating (MI) phase and the superfluid (SF) phase, where it decays exponentially and with a power-law, respectively. The correlators for density-wave (DW) and Haldane-insulating (HI) phases decay to a constant value in the respective phases. More details about the characterization of the system can be found in \cite{Rossini2012}. The non-local string term in \cref{prleq:Cs} is characteristic of topological order, where the translational symmetry remains protected with a transition in the Luttinger liquid universality class from MI and gets broken with a transition in the Ising universality class to DW \cite{Berg2008}. We visualize the phase diagram by computing $O_{\bullet} = \sum_{i,j} C_{\bullet}(i,j)/L_\infty^2$ in \cref{fig:phase_diagram} in the thermodynamic limit for a repeating unit cell of $L_\infty=64$ sites with a maximum bond dimension $\chimax = 100$ and assuming a maximum occupation number $\nmax=3$, which results in a local dimension $d=\nmax+1=4$. We use data from these states for the following machine learning analysis using deep anomaly detection.

\section{Machine Learning setup}
\label{sec:machine-learning-setup}

We perform deep anomaly detection to map out phase diagrams as described in \cref{chapter:anomaly}. I.e., for each ground state $\ket{\psi}$ we take corresponding data $x$, such as its entanglement spectrum (ES), central tensors or low order correlation functions. That data has characteristic features that the autoencoder (AE) learns to encode into the latent variable $z$ at the bottleneck \cite{Iten2020}, from which it is ideally able to reconstruct the original input. The loss $\mathcal{L}$ directly indicates the success of this endeavour, which we improve by employing symmetric shortcut connections (SSC, see \cref{fig:AE_main}), inspired from \cite{Dong2016} to typical losses $<5\%$. Now, the intuition is that, when confronted with data from unknown phases, the AE is unable to encode and decode $x$. This leads to a higher loss, from which we deduce that the states do not belong to the same phase as the ones used to train the AE.

Deep learning architectures are known to generalize well \cite{kawaguchi2017generalization, valle2018deep}, such that it suffices to train in a small region of the parameter space. Compared to known supervised deep learning methods this anomaly detection scheme does not rely on labeled data. We choose training data from one or several regions of the phase diagram, and ask how the loss of a test data point from any region of the phase diagram compares to the loss of these training points. This can be performed with no a priori knowledge and in a completely unsupervised manner. The computationally most expensive step is the training and with our method it has to be performed only once to map the whole phase diagram, as opposed to multiple trainings like in \cite{nieuwenburg2017learning, liu2017self}. Furthermore, it does not require a full description of the physical states in contrast to \cite{Zanardi2006}, where full contraction is necessary. Thus, for higher dimensional systems, \cite{Zanardi2006} is infeasible as contraction is known to be generally inefficient for 2d tensor network states (commonly referred to as PEPS, see \cite{Orus2013}). We will expand on this notion of using anomaly detection with PEPS data further in the next chapter, \ref{chapter:peps}.

The specific architecture in use consists of two 1d-convolutional encoding and decoding layers with SSCs (\cref{fig:AE_main}), implemented in TensorFlow \cite{tensorflow2015-whitepaper}.
To ensure the reproducibility of our results, we made the source code available under an open source license~\cite{githubrep}.

\paragraph{Data} 
Recall from \cref{sec:tensor_networks} that an MPS in canonical form \cite{Vidal2003} is written in terms of 
\begin{multline}
\label{prleq:ansatz_mps}
\ket{\Psi} = \sum_{\bm{\sigma}} \Gamma^{\sigma_1} \Lambda^{[1]} \cdots \Lambda^{[i-1]} \Gamma^{\sigma_i} \Lambda^{[i]} \cdots  \\
\Lambda^{[L-1]} \Gamma^{\sigma_{L}} \ket{\sigma_1 \ldots \sigma_i \ldots \sigma_{L}}.
\end{multline}
At site $i$, $\{\Gamma^{\sigma_i}\}$ is a set of $d$ matrices and $\Lambda^{[i]}$  the diagonal singular value matrix of a bipartition of the chain between site $i$ and $i+1$, i.e. the Schmidt values (see \cite{SCHOLLWOCK2011}). As input data $x$ corresponding to the ground state $\ket{\Psi}$, we find best results by using the entanglement spectra $\exp(-\Lambda^{[i]}/2)$, which are known to contain relevant information about quantum phase transitions \cite{Deng2011}. We can also use the local tensors $\Theta^{\sigma_i} = \Lambda^{[i-1]} \Gamma^{\sigma_i} \Lambda^{[i]}$, from which all local observables can be computed and implicitly contain the entanglement spectra. We will use the tensor from the middle of the chain and refer to it as the central tensor $\Theta$. Further, we also use low-order correlation functions like $\CSF$ in \cref{prleq:CSF} that are straight-forward to access in an experiment.

\section{Discovering new features using deep anomaly detection}
\label{sec:discovering-new-features}

\begin{figure}[tb]
\begin{center}
\includegraphics[width=.77\textwidth]{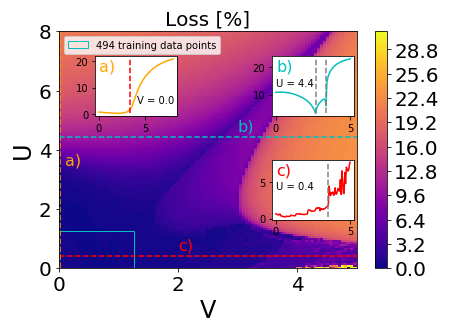}
\end{center}
\caption{2D loss map of the AE after training near the parameter space origin (blue square frame). The insets a), b) and c) show the loss along the dashed lines. Vertical green dashed line in inset a) indicates critical $U_c = 3.33$ \cite{Kuehner1999}. Vertical grey dashed lines in inset b) and c) are the transitions from \cref{fig:phase_diagram}. The phase boundaries are determined by a rise in loss (inset a) and c)). The anomalous regions are already well-separated by decreasing losses because of the critical behaviour at the phase boundaries (inset b)), which share similarities with the critical SF phase. Higher loss indicates that this region is more different from the training region in the blue square, lower loss indicates similarity.}
\label{fig:2D_pbc_ES_SF}
\end{figure}

We start by using the ES as our data. Assuming no a priori knowledge, we start by training with data points at the origin of the parameter space $(U,V) \in [0,1.3]^2$, which in our case accounts to training in SF. By testing with data points from the whole phase diagram we can clearly see the boundaries to all other phases from SF in \cref{fig:2D_pbc_ES_SF}. The BKT transition between SF and MI is matched by an abrupt rise in loss (\cref{fig:2D_pbc_ES_SF}, inset a)). In this particular case, we can already determine the different phases inside the anomalous region due to their different loss levels and the appearance of two valleys at the phase boundaries between MI, HI and SF (\cref{fig:2D_pbc_ES_SF}, inset b)). Physically, we can explain these valleys by the criticality of these Luttinger and Ising type transitions, which lead to a slowly decaying ES at the boundary, just like in the critical SF phase. 

\begin{figure}[H]
\begin{center}
\includegraphics[width=.77\textwidth]{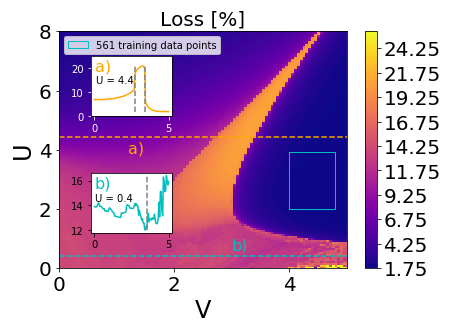}
\end{center}
\caption{2D loss map of the AE after training with ES from blue square frame. The insets a), b) show the loss along the dashed lines. The training in the region of high loss from fig. 3 in the main text confirms the boundaries. The MI and HI phases are well separated because the loss in HI is much higher in comparison (inset a)). This isn't necessarily the case as indicated in inset b) for SF and SF+SS.}
\label{fig:2D_pbc_ES_DW}
\end{figure}

It is not necessarily always the case that one can differentiate the different phases inside the high-loss anomalous region. Thus, we propose picking homogeneous and high contrast anomalous regions after the initial training as a systematic approach. In \cref{fig:2D_pbc_ES_DW} we do this for $(U,V) \in [4,4.8]\times[2,4]$, which is the region where the loss after the initial training was highest and accounts to the DW phase. We can confirm the previously determined boundaries to the anomalous region, which is very sharp due to the Ising type transition. Further, we can again separate MI and HI due to different loss levels but without a valley in between (\cref{fig:2D_pbc_ES_DW}, inset a)).

\begin{figure}
\begin{center}
\includegraphics[width=.77\textwidth]{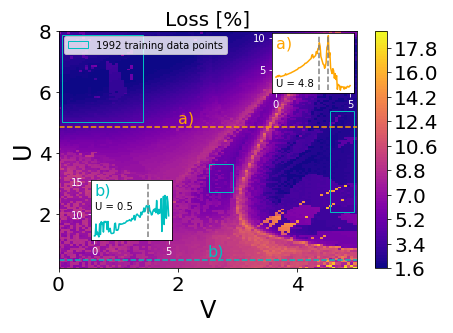}
\end{center}
\caption{Instead of the ES, we use the central tensor $\Theta$ as input data for the AE and use 2D convolutional layers. The same AE can encode both MI, HI and DW data.}
\label{fig:2D_pbc_tensor_MI-HI-DW}
\end{figure}

We also test our method on other input data. Instead of ES, we can also use a tensor $\Theta^{[i]\sigma_i}_{v_{i-1},v_i} = \sum_{a,b} \Lambda^{[i-1]}_{v_{i-1},a} \Gamma^{\sigma_i}_{a,b} \Lambda^{[i]}_{b,v_i}$ or the correlator $\braket{b_i^{\dag}b_i}$. The tensor $\Theta^{[i]\sigma_i}_{v_{i-1},v_i}$ comes from the chain, from which one can compute all single-site expectation values \cite{SCHOLLWOCK2011}. This quantity has three indices, which is why we interpret it as a colored image, because the two virtual indices $v_{i-1}$ and $v_i$ can be interpreted as the two dimensions of the image and the physical index $\sigma_i$ can be interpreted as the color channel. Instead of 1d convolution, we now use 2d convolution with otherwise identical architecture. Even though translational invariance is broken in DW, we find it suffices to use only one tensor $\Theta$ from the center of the unit cell. This is because, despite the broken translational symmetry, entanglement is still distributed uniformly in the unit cell, which is implicitly encoded in $\Theta$. Furthermore, we find that the network is capable of encoding more than one phase in the training dataset, seen in \cref{fig:2D_pbc_tensor_MI-HI-DW}. We still find the boundaries between MI, HI and DW due to the criticality of the transitions (see \cref{fig:2D_pbc_tensor_MI-HI-DW}, inset a)), similar to the valleys in \cref{fig:2D_pbc_ES_SF}. 

\begin{figure}
\begin{center}
\includegraphics[width=.77\textwidth]{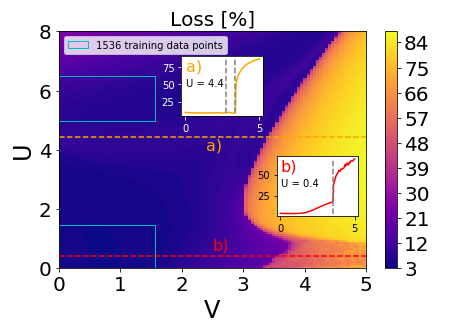}
\end{center}
\caption{2D loss map of the AE after training in the two blue square frames in the SF and the MI phase. The insets a) and b) show the loss along the dashed lines. Instead of the ES, we use the physically accessible correlator $\CSF$ as input data. The HI is not recognized as this correlator does not contain information about the topological order of this phase.}
\label{fig:2D_pbc_cor_MI-SF}
\end{figure}

Furthermore, we use experimentally accessible correlators. In \cref{fig:2D_pbc_cor_MI-SF}, instead of unprocessed data from simulation, we calculate $\{\CSF(i,j)\}_{i,j=1}^{64}$ and train in MI and SF simultaneously. We interpret rows as color channels for 1d convolution. Because $\CSF$ does not contain any information about the topological order in HI, the method does not recognize this region as we would expect (\cref{fig:2D_pbc_cor_MI-SF}, inset a)). Overall, the boundaries match perfectly with a sharp increase onto a plateau at the transition points. This opens the possibility to use physical observables from experiment with the caveat of requiring physical knowledge a priori.

\section{Phase Separation}
\label{sec:phase-separation}
\begin{figure}
\centering
\includegraphics[width=.77\textwidth]{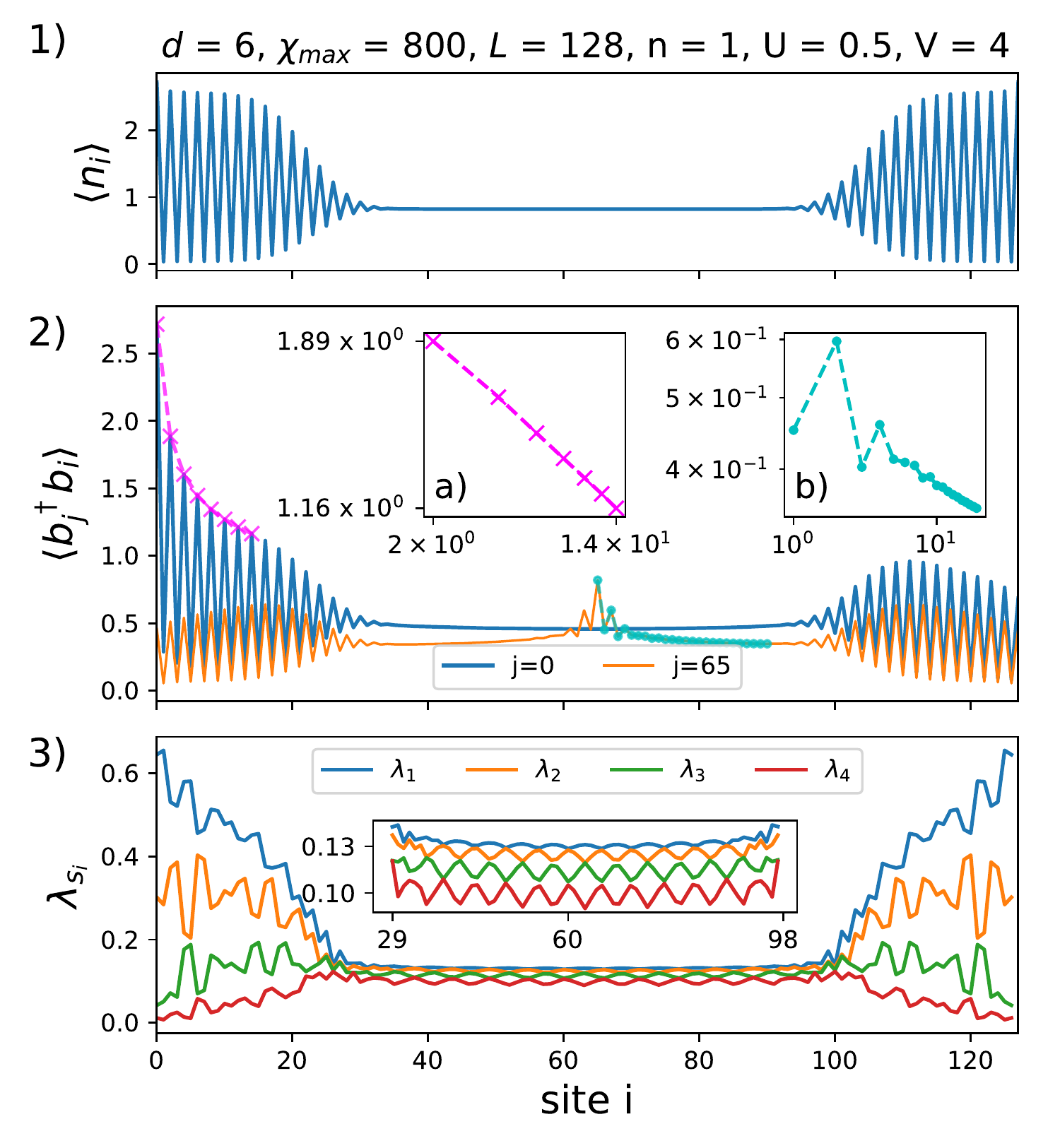}
\caption{
Main characteristics of the phase-separated ground state. 
Panel 1) Density profile. The system is separated into two phases described by a flat density typical to a fluid (SF phase) and a periodic structure typical to a solid (SS phase). 
The solid patterns of alternating occupation are pinned at the edges due to the use of open boundary conditions, leaving the superfluid uniform density in the middle. 
The volume occupied by each of the phases depends on the filling $n$, which here is $n=1$. 
Panel 2) Off-diagonal single-particle correlation function $\braket{b_j^\dagger b_i}$ in SF and SS phases.
Slow power-law decay [seen as straight lines in a log-log plot in insets 2(a) and 2(b)] allows the system to be coherent at distances larger than the lattice spacing which is a one-dimensional analog of Bose-Einstein condensation, and implies that both phases are superfluid.
3) The entanglement spectrum shows different periodicities in the two different phases.
The first four largest elements of the entanglement spectrum $\{\lambda_i\}$ are plotted in descending order $\lambda_1 \geq \lambda_2 \geq \lambda_3 \geq \lambda_4$.}
\label{fig:SS_SF_separation}
\end{figure}

By close inspection of \cref{fig:2D_pbc_ES_SF,fig:2D_pbc_cor_MI-SF}, we see a region with noticeable contrast for small $U$ and large $V$ ($(U,V) \sim (0.5,4)$), indicating the presence of a separate phase. This is interesting because, initially, we did not expect to find a fifth phase in the diagram. Upon further physical investigation, we find a phase-separated state between SF and supersolid (SF + SS), see \cref{fig:SS_SF_separation} 1.
The superfluid region shows a power-law decay of $\CSF$ and a uniform density, whereas the supersolid region features staggered local densities with a simultaneous presence of coherence as indicated by the power-law decay of $\CSF$ (see \cref{fig:SS_SF_separation} 2)).

\begin{figure}
\centering
\includegraphics[width=.77\textwidth]{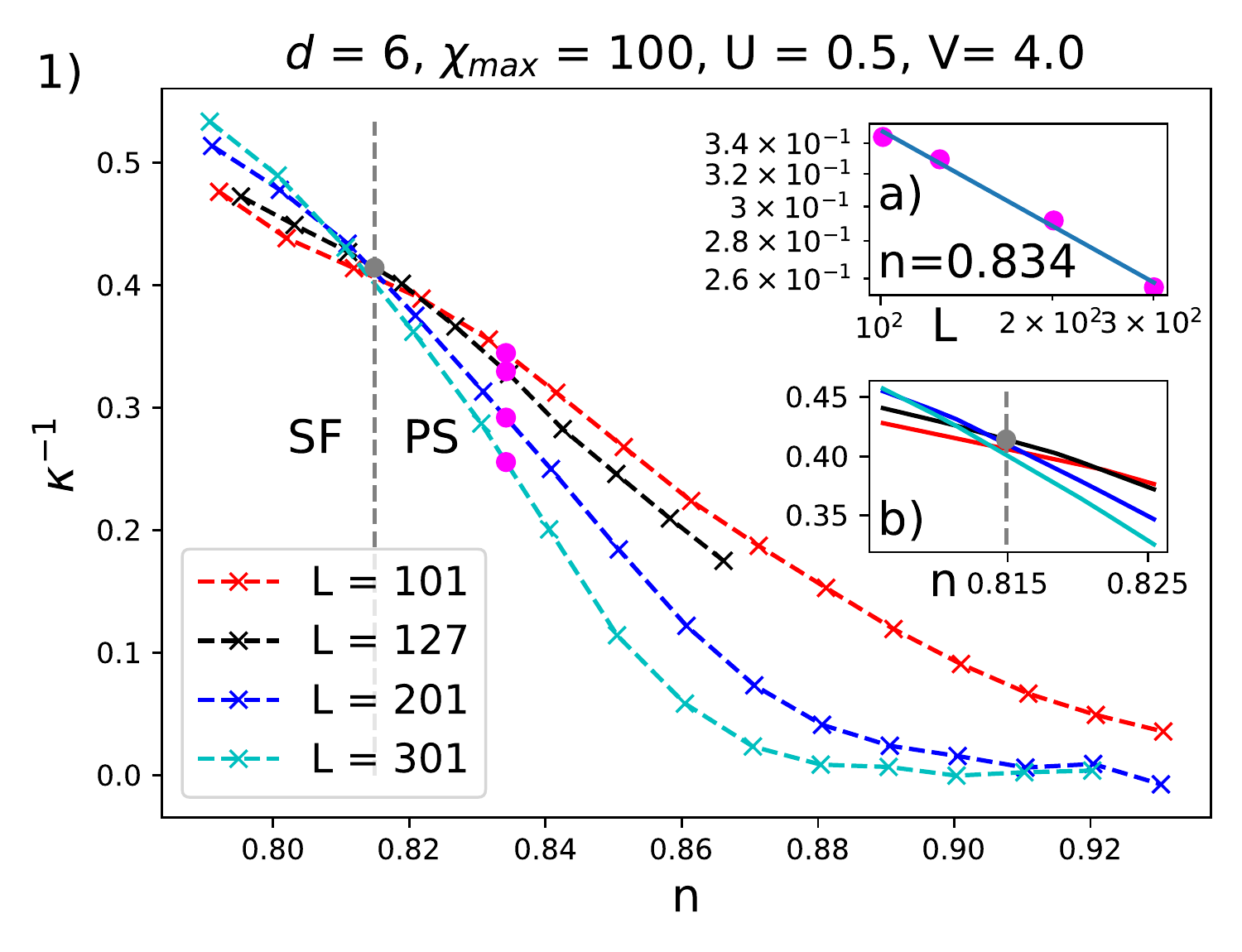}
\includegraphics[width=.77\textwidth]{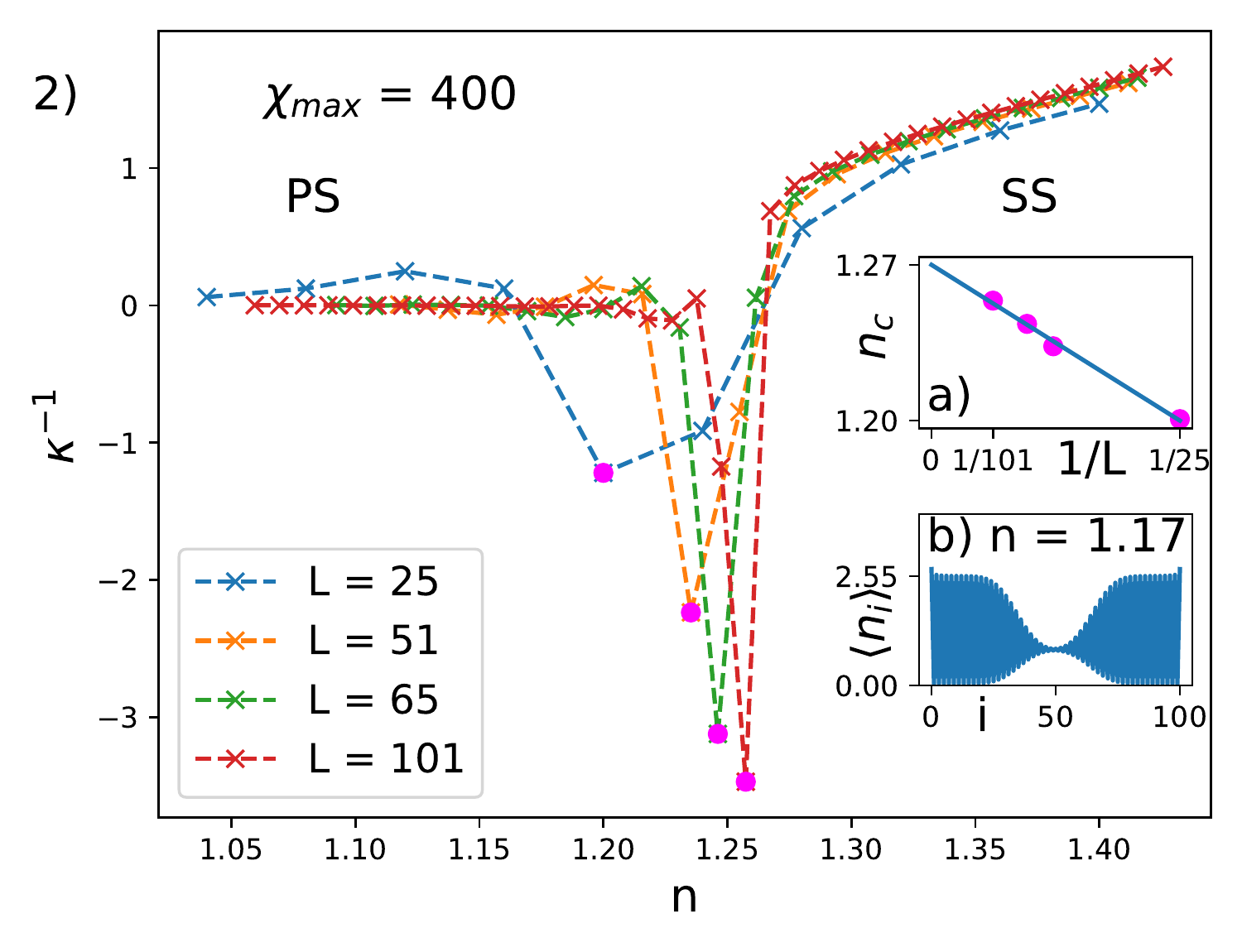}
\caption{
Finite-size study of the inverse compressibility $\kappa^{-1}$, Eq.~(\ref{eq:kappam1}), as a function of the filling $n$ for $(U,V) = (0.5,4)$ as in fig.~\ref{fig:SS_SF_separation}.
Vanishing thermodynamic value of $\kappa^{-1}$ \cref{eq:kappam1} signals instability towards phase separation. 1) SF-PS transition: The transition point is estimated at critical filling $n^\text{SF-PS}_c = 0.815$ defined as the position of the intersection of lines corresponding to different system sizes.
In the phase-separated region, $n>n_c$ region, the value of the inverse compressibility $\kappa^{-1}$ is lowered as the system size is increased (lines correspond to $L=101; 127; 201; 301$, from top to bottom) and vanishes in the thermodynamic limit.
Inset~(1a): example power-law decay of the inverse compressibility $\kappa^{-1}$ as a function of system size $L$ in the phase-separated regime, $n>n_c$ 
Inset~(1b): Zoom-in on the intersection. 2) PS-SS transition: Dependence of the inverse compressibility on filling $n$ is scaled with the system size $L$ leading to the estimated value for the critical density equal to $n^\text{PS-SS}_c = 1.27$ (Inset~(2a)). Inset~(2b): Example density in PS state close to the transition to SS. Note that in the solid part the average density $n\approx 2.55/2$ matches the critical filling $n^\text{PS-SS}_c$.}
\label{d2E-S_f}
\end{figure}

A phase transition happens if the equation of state, describing the dependence of the chemical potential on the density, has two minima. The single-minimum scenario converts to the two-minima one when an inflection point, $d\mu/dn = 0$ appears. Indeed, it is known that the divergence in compressibility leads to a phase separation \cite{Grilli1991,Emery1993,Misawa2014} and this criteria has been used to locate its position numerically. Thus, occurrence of the phase separation can be understood as a mechanical instability of the system, signaled by a vanishing inverse compressibility \cite{Emery1990,Ammon1995,Coulthard,Moreno2010}
\begin{equation}
\label{eq:kappam1}
\kappa^{-1} = n^2 \frac{\partial^2 \mathcal{E}}{\partial n^2} \approx n^2 \frac{\mathcal{E}(n +\Delta n) + \mathcal{E}(n - \Delta n) - 2 \mathcal{E}(n)}{\Delta n^2}
\end{equation}
where $\mathcal{E} = E_0/L$ is the ground state energy density and $n = \sum_i \braket{n_i}/L$ the average particle density. For these calculations, we fix $L$ and vary $n=N/L$ in an equidistant manner $N\in\mathbb N$, such that $\Delta n = (N_1 - N_0)/L$ for different fillings. The system becomes mechanically unstable and phase separation occures when the compressibility becomes infinite (or $\kappa^{-1}=0$) \cite{Moreno2010}. We show that this is exactly the case and report the finite-size scaling of the SF-PS transition in fig.~\ref{d2E-S_f}. We estimate the transition point at the crossover for different finite system sizes as $n_{c}^\text{SF-PS} \approx 0.815$ (see~\cref{d2E-S_f} inset 1b) for a detailed view). For larger fillings, $n>n_c$, the inverse compressibility $\kappa^{-1}$ tends towards zero in the thermodynamic limit (see~\cref{d2E-S_f} inset 1a)), signaling spinodal decomposition leading to the phase-separated ground states for intermittent fillings.
We estimate the critical filling as $n_{c}^\text{PS-SS} \approx 1.27$ from extrapolating the points for which the second derivative changes abruptly (see~\cref{d2E-S_f} a)).
This filling coincides with the average density of the SS part in the PS configuration $n\approx 2.55/2$ in the vicinity of PS transition, as shown in~\cref{d2E-S_f} inset 2b). 
To rule out artifacts from the restricted local Hilbert space dimension, we achieve consistent results for maximal local occupation number $d=4,6,9$ and found no significant differences between $d=6$ and $d=9$ (see \cref{appendix:supersolid}). As a compromise between performance and accuracy, we fixed $d=6$ for all presented calculations.

The surface energy between SS and SF phases is minimized in configurations with only two domains. Open boundary conditions, employed in DMRG calculations, pin the solid region to the edges while the superfluid one is observed in the center [see fig.~\ref{fig:SS_SF_separation}].
The solid region appears at random positions within the unit cell in iDMRG calculations, where unit cells are repeated periodically, as one would expect in a phase-separated ground state.

\begin{figure}
\centering
\includegraphics[width=.77\textwidth]{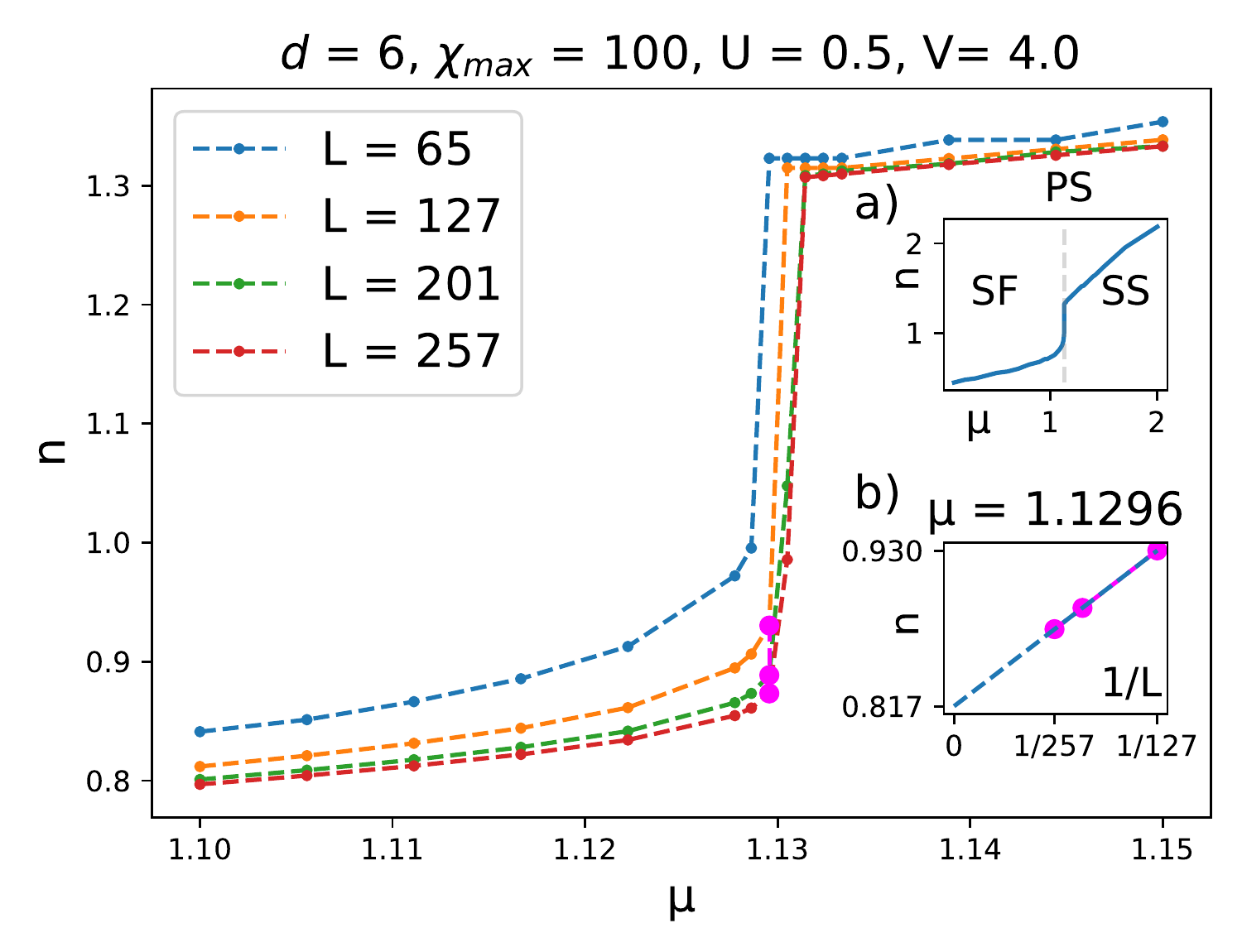}
\caption{Filling $n$ versus chemical potential $\mu$ for finite chains of length $L$ with OBC and no explicit $U(1)$ symmetry. The apparent discontinuity at roughly $\mu \approx 1.13$ signals spinodal decomposition for fillings between $n=0.817$ and $1.31$. 
Insets (a) Wider range in $\mu$ showing the extent of the SS phase, 
(b) Finite-size extrapolation for the filling values in the main plot at $\mu = 1.1296$ to estimate the lower critical filling $n_c=0.817$.
}
\label{fig:f_mu_n-max-5}
\end{figure}

\begin{figure}
\centering
\includegraphics[width=.77\textwidth]{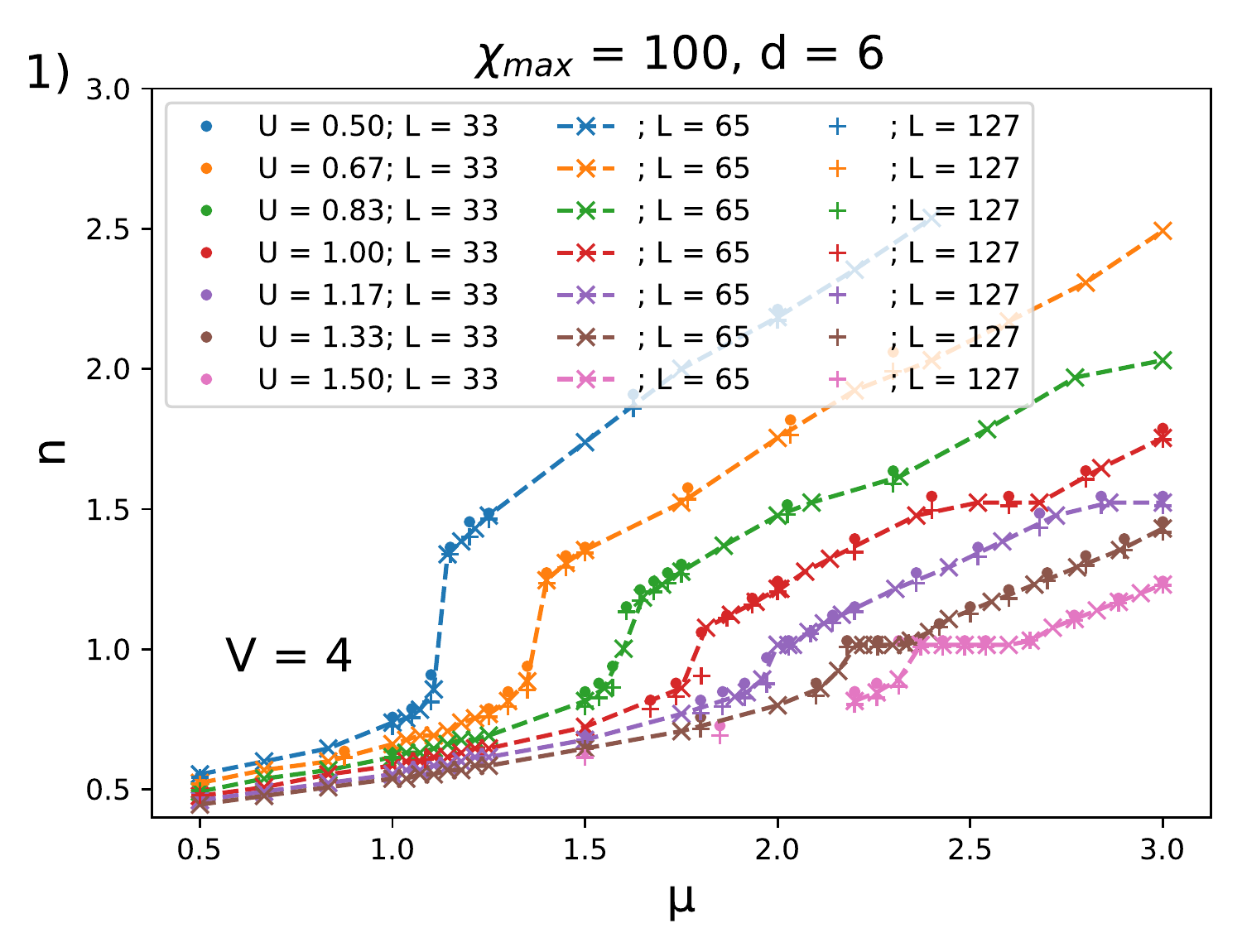}
\includegraphics[width=.77\textwidth]{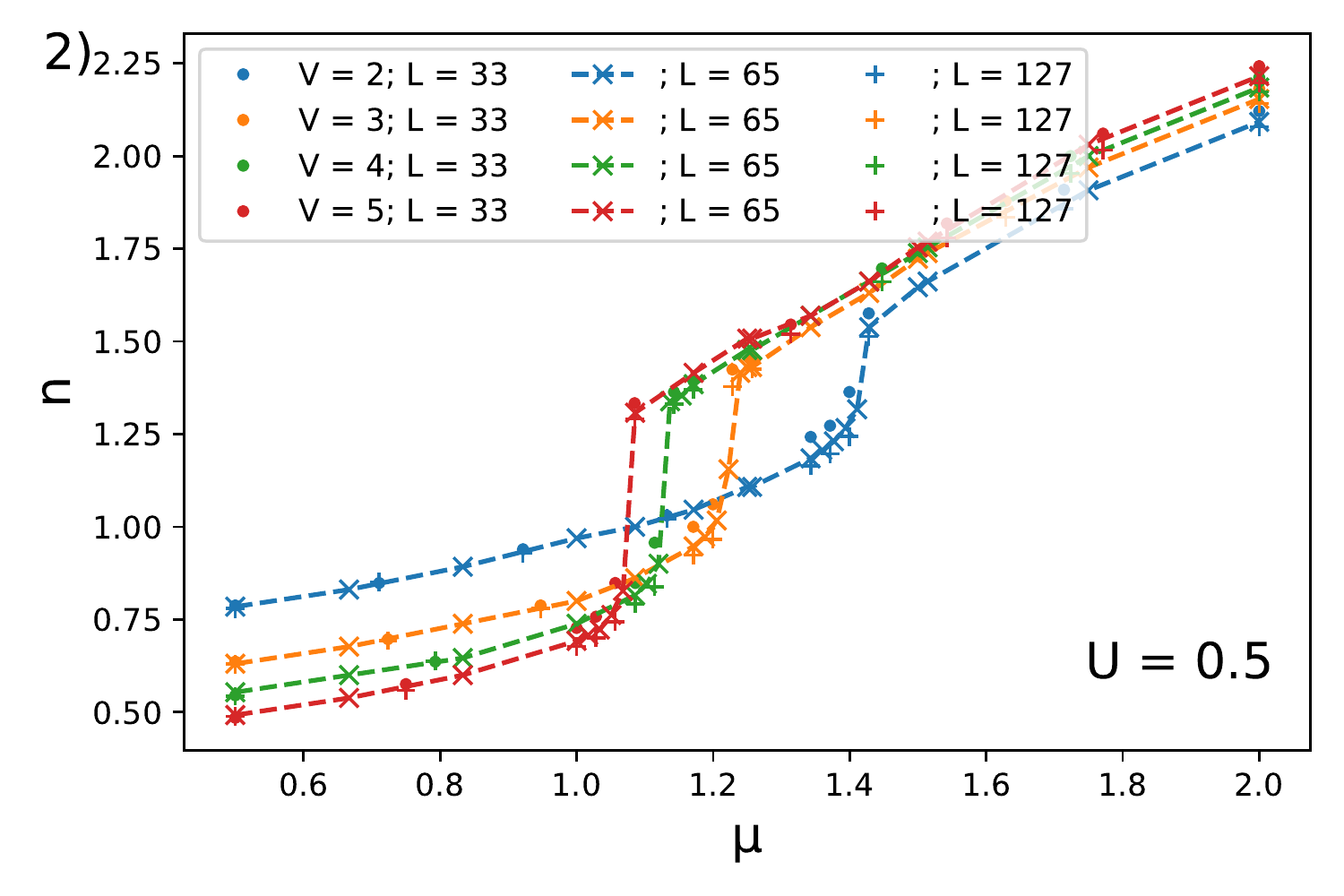}
\caption{Dependence of filling $n$ on the chemical potential $\mu$ for different system sizes $L$ and different values of $U$
The phase separation is seen as a discontinuity in $n(\mu)$. 1) Fixing $V=4$ we alter $U$ and see that beyond $U_c \approx 1$ the systems form a CDW, seen by the plateaus at filling $n=1$ (lines are ordered in increasing order of $U$ from the top line to the bottom one).
2) For smaller $V$, PS occurs at higher fillings such that it is not present in the $n=1$ phase diagram for small $V$ anymore (lines are ordered in increasing order of $U$ from the top lines to the bottom ones).
}
\label{fig:n-mu-U}
\end{figure}

An alternative way to narrow down the appearance of phase separation is via altering the chemical potential $\mu := \partial \mathcal{E} / \partial n$ (note that $\kappa = n^{-2} \partial n / \partial \mu$). In fig.~\ref{fig:f_mu_n-max-5} we show the filling $n(\mu)$ obtained with open boundary conditions for finite chains as we vary the chemical potential $\mu$. Notably, we observe a discontinuity at $\mu_c \approx 1.13$, exactly at the point where the compressibility $\kappa$ becomes infinite. We extrapolate the critical fillings to be between $n_{c} \in [0.82,1.31]$. This is in agreement with the densities we obtained in the previous calculation.
We show in fig.~\ref{fig:n-mu-U} how the dependence $n(\mu)$ changes if we alter $(U,V)$. In fig.~\ref{fig:n-mu-U}(1) discontinuities in $n(\mu)$ are clearly visible, signaling formation of a PS state below a critical $U_c(V=4) \approx 1$. For larger values of $U$, the system forms a CDW phase at commensurate fillings, signaled by the formation of plateaus with constant $n(\mu)$ (i.e. $n=1$ here).
From \cref{fig:n-mu-U} (2) we observe that the phase separation occurs for larger average densities $n>1$ if the nearest neighbor interaction is weak. 
Therefore, the reported effects go beyond the usual commensurate effects between lattice geometry and average density.

\section{Spatial oscillations in SF phase}
\label{sec:oscillations}
\begin{figure}
\centering
\includegraphics[width=.77\textwidth]{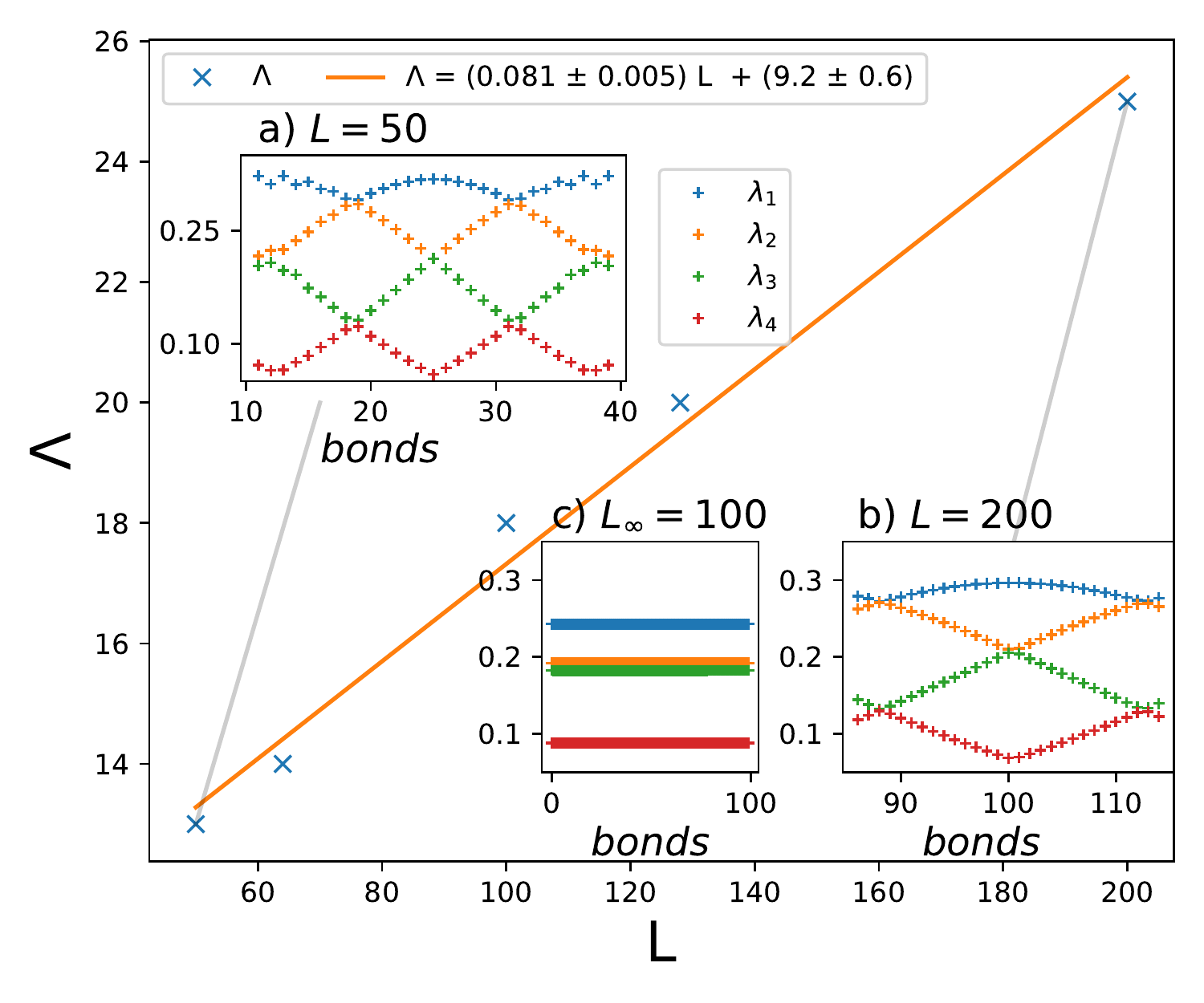}
\caption{
Spatial period $\Lambda$ taken from the Entanglement spectrum (ES) at $(U,V) = (0.5,3)$ with OBC and $\chimax = 100$, $d = 6$ as a function of the system size $L$. This state was reported as SS in Ref.~\cite{Deng2011}, but it is actually an SF. Further, the spatial oscillations vanish in the thermodynamic limit as the spatial period grows linearly in system size. However, we show that for incommensurate fillings like in \cref{fig:OBC_f-0-77,fig:iDMRG_f-0-65_Csl-Ssl-ES} this kind of oscillations do survive in the thermodynamic limit.
a,b) ES $\lambda_{s_i}$ for bonds at the center of a system of length $L=50$ and $L=200$, respectively. c) ES for iDMRG simulation in the thermodynamic limit with a unit cell size $L_\infty = 100$ showing no spatial oscillations.
}
\label{fig:reproduce_santos}
\end{figure}

\begin{figure}
\centering
\includegraphics[width=.77\textwidth]{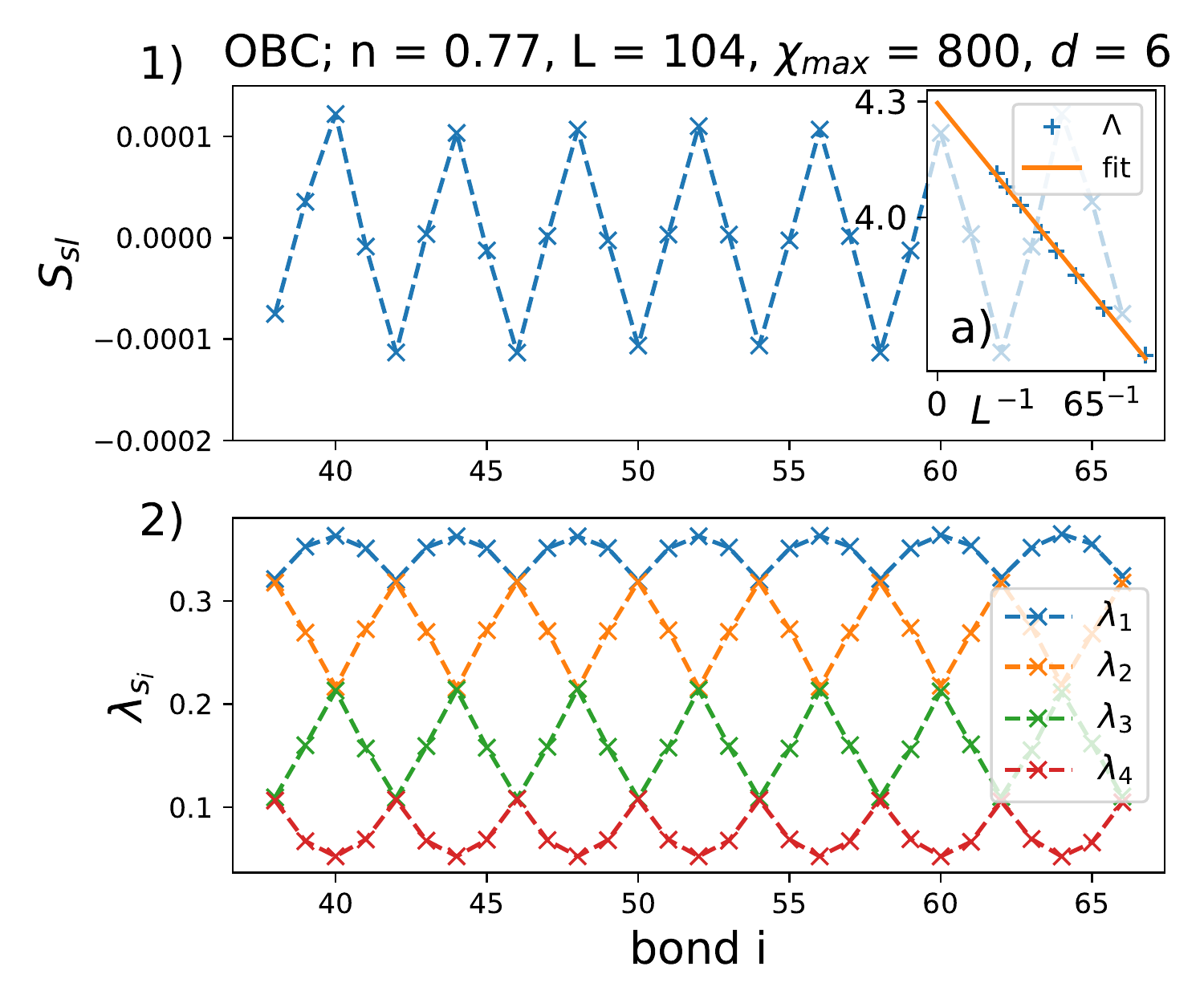}
\caption{Entanglement properties of the superfluid bulk for filling $n=0.77$ and $(U,V) = (0.5,4)$ close to phase-separation. 1) Oscillatory sub-leading part $S_\text{sl}$ of the Rényi-2 entropy $S_2$ in \cref{Renyi_scaling}. 2) The four largest squared Schmidt coefficients $\lambda_{s_i}$ shown spatially along the bonds. The inset 1a) shows spatial frequencies from Fourier analysis of $\lambda_{s_i}$. The extrapolation yields a spatial period $\Lambda = 4.3$ in the thermodynamic limit.}
\label{fig:OBC_f-0-77}
\end{figure}

\begin{figure}
\centering
\includegraphics[width=.77\textwidth]{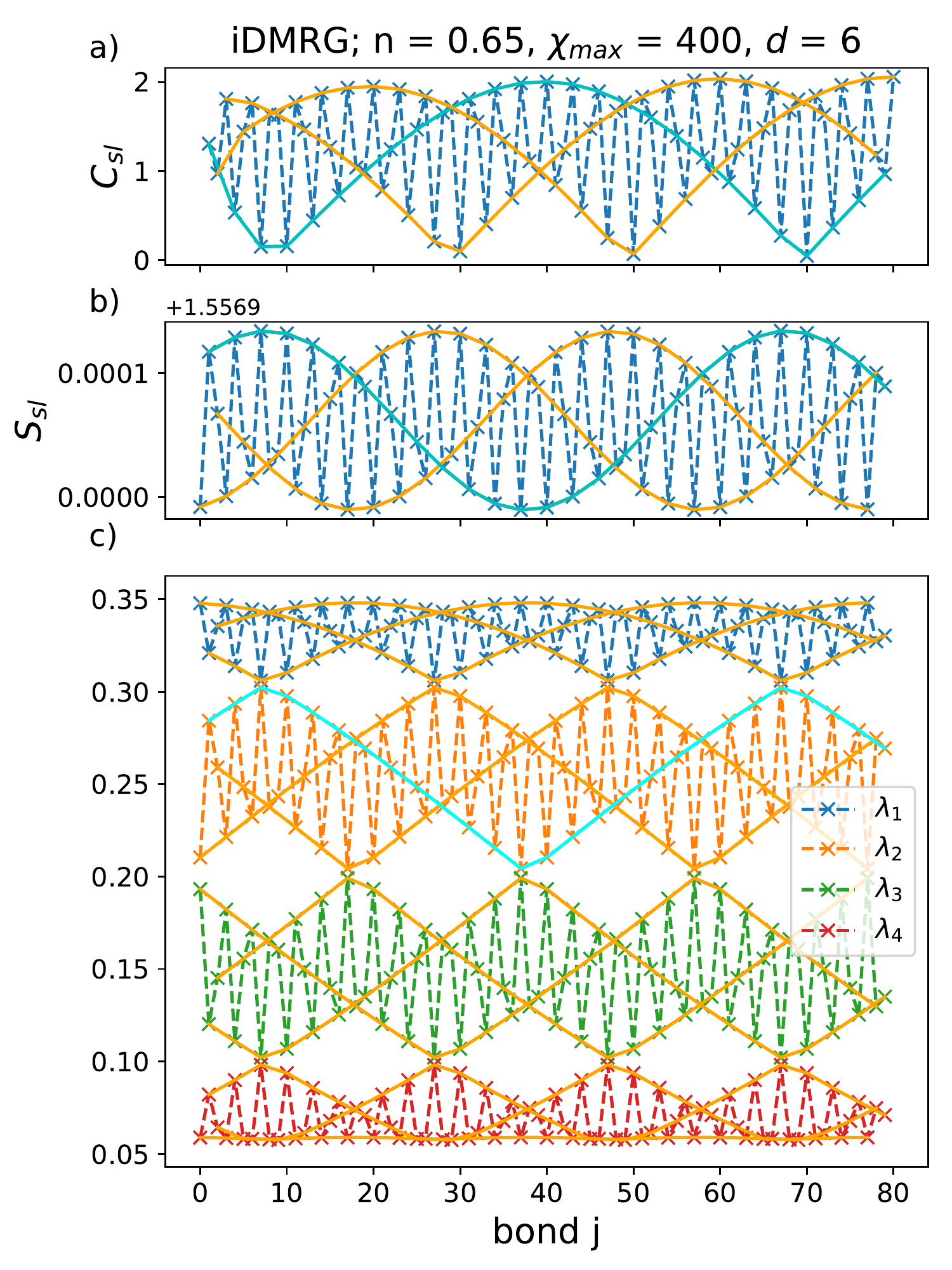}
\caption{
a) Sub-leading part $C_\text{sl}$ containing the oscillatory part of the string-order correlator $\Cst$ \cref{prleq:Cs}. b) Oscillatory part of the Rényi-2 entropy $S_2$. c) The spatial dependence of the four largest squared Schmidt coefficients $\lambda_{s_i}^2$ on the bonds of the MPS. For all a),b) and c): The orange lines are guides to the eye with every third data point plotted to highlight the envelope. We highlight one line in cyan color in all three subplots as a guide to the eye, making it apparent that the period is the same in all three quantities.}
\label{fig:iDMRG_f-0-65_Csl-Ssl-ES}
\end{figure}

\begin{figure}
\centering
\includegraphics[width=.77\textwidth]{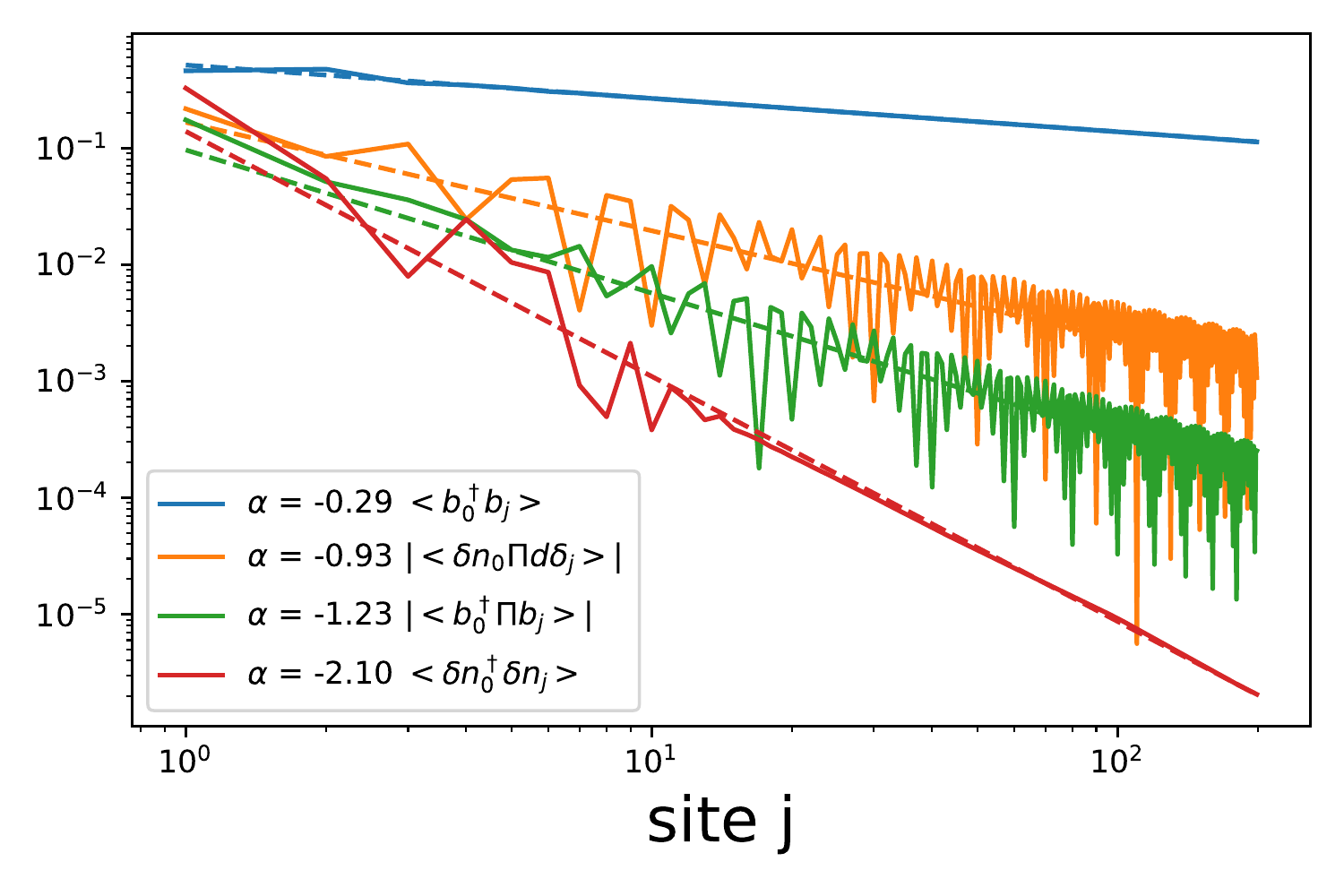}
\caption{Comparison of different common correlators for the same state as in \cref{fig:iDMRG_f-0-65_Csl-Ssl-ES} in the SF phase with oscillating ES. 
We denote the non-local string term as $\Pi = \exp\left(-i \pi \sum_{0 \leq l < j} \delta n_l\right)$ with $\delta n_l =  n_l - n$. 
We see that only the correlators with this string term show the oscillations matching the ES. The critical exponent $\alpha$ is obtained from linear fitting the respective correlator in a double logarithmic scale (dotted lines in corresponding colors).}
\label{fig:iDMRG_correlators_f-065}
\end{figure}

For the homogeneous superfluid at fillings below the phase-separated phase, enhanced spatial oscillations appear in the entanglement spectrum (ES) and other observables (see \cref{fig:reproduce_santos,fig:OBC_f-0-77,fig:iDMRG_f-0-65_Csl-Ssl-ES,fig:iDMRG_correlators_f-065}).
These signatures are present in SF states for weak on-site interactions $U$ over a broad range of $V$.
Such oscillatory patterns were reported earlier in Ref.~\cite{Deng2011} at $(U,V) = (0.5,3)$ and $n=1$.
However, for the superfluid at integer fillings, we observe the absence of oscillations in the thermodynamic limit such that they cannot be linked to a bulk feature of the given phase and must be related to finite-size effects, instead.
We demonstrate this in \cref{fig:reproduce_santos}, where we show the spatial period of the oscillatory patterns in the entanglement spectrum (examples thereof are visible in insets a) and b)) as a function of the system size, for which we observe a linear increase, i.e. a vanishing frequency for $L\rightarrow\infty$. 
This is in agreement with iDMRG simulations (thereby directly approximating the ground state in the thermodynamic limit), for which we do not find oscillations at all. It shall also be noted that the system is in a superfluid phase for these parameters, and not a supersolid phase as claimed in \cite{Deng2011}.
The commensurate scenario at $n=1$ is in strong contrast to incommensurate fillings, for which oscillations in the entanglement spectrum are a robust feature of the bulk.

In the inset of \cref{fig:OBC_f-0-77}~a), we display a finite-size extrapolation of the spatial period, which is extracted from the leading frequency in the Fourier transform of the oscillatory part of the entanglement spectrum (see \cref{fig:OBC_f-0-77}~2))).
Notably, $\Lambda(1/L\rightarrow0)\approx4.3$ assumes a finite value in the thermodynamic limit.

The oscillations of the entanglement spectrum cannot be detected by standard local observables and two-body correlations, but, interestingly, they appear prominently in non-local observables like the string-order correlator, for which the long-range power-law decay is modulated by oscillations of the same frequency (c.f. \cref{fig:iDMRG_f-0-65_Csl-Ssl-ES}~a)).
In order to extract the oscillatory part of $C_\text{sl}$ we fit it with a power-law decay $\Cst(i,j) = c/|i-j|^\alpha C_\text{sl}$ and divide the correlator by the envelope $c/|i-j|^\alpha$ to show the remaining oscillatory part.
In contrast, common correlators without the non-local string term do not show this oscillatory behavior as depicted in  \cref{fig:iDMRG_correlators_f-065}. All of the correlators here decay algebraically, in particular the ones involving the non-local string term, indicating lack of long range trivial and topological order, but power-law correlations. The correlators with string term exhibit oscillations in the tails reminding us of the oscillations of the ES.

A complementary way to resolve these spatial oscillations is given by the Rényi entropy 
  \begin{equation}
    S_\alpha(\ell)= -\ln({\rm Tr}\rho^\alpha(\ell))/(1-\alpha),
    \label{eq:renyi_entropy}
  \end{equation}
accessible in experiments for the special case $\alpha=2$.
$S_2(\ell)$ depends on the purity $\rho^2(\ell)$ for a lattice block of size $\ell$, which can be detected in the framework of trapped ions through quantum state tomography~\cite{linke2018} or through direct measurement of the quantum purity~\cite{islam2015}.

The asymptotic decay of the Rényi entropies $S_\alpha(\ell)$ for critical systems is well-known~\cite{calabrese2004,calabrese2009} and given by 
\begin{eqnarray}
S_\alpha(\ell) &=& \frac{\alpha+1}\alpha\frac c{6b} \ln\left( d[\ell|L] \right) + S_{\rm sl} + \gamma, \label{Renyi_scaling}\\
d[\ell|L] &=& \left|L/\pi\sin\left( \pi \ell/L \right)\right|.
\end{eqnarray}
The leading contribution to $S_\alpha(\ell)$ is proportional to $\ln\left( d[\ell|L] \right)$ and describes a universal scaling law with prefactor factor $c$ called the central charge of the conformal field theory (in a fermionic system, it is equal to the number of Fermi points).
An additional factor $b$ distinguishes the case of periodic ($b=1$) and open ($b=2$) boundary conditions and $\gamma$ constitutes a non-universal constant.
Subleading terms are denoted by $S_{\rm sl}$ and, in general, oscillate in space.
The subtle oscillations of the entanglement spectrum are obviously carried over to the subleading terms of the Rényi entropies, which we present in \cref{fig:iDMRG_f-0-65_Csl-Ssl-ES,fig:OBC_f-0-77}.

We note that we observe the same properties for the homogeneous SS \textit{above} the filling for phase separation. The main difference is that for this SS, there is a spatial solid pattern in the density. Taking this into account, the remaining properties are the same as is shown in \cref{appendix:supersolid}.

Overall, the spatial oscillations in the entanglement spectrum, that can be uncovered by looking at the string order correlators or Rényi entropies, are a clear manifestation of a broken translational symmetry. This point is further strengthened by simulations with iDMRG: Upon choosing a suitable unit cell size, iDMRG can converge, as is the case in \cref{fig:iDMRG_f-0-65_Csl-Ssl-ES}, and yields results in agreement with the bulk of finite size simulations. For unit cell sizes incommensurate with the spatial period, iDMRG has trouble converging. We show further details about this in \cref{appendix:supersolid}. This feature distinguishes the superfluid phase under investigation from the superfluid phase at filling $n=1$. In the following, we convince ourselves that this phase is still well described by Luttinger liquid theory and shows the same critical finite-size scaling behavior.

\section{Luttinger liquid description}
\label{sec:luttinger}
Long-range properties of gapless one-dimensional systems are well captured by the Luttinger liquid theory and are governed by the Luttinger parameter $K$. 
This theory is based on using an effective low-energy Hamiltonian and can be used to calculate the small-momentum and long-range behavior of the correlation functions. 
The Luttinger theory, being an effective one, takes the Luttinger parameter as an input and an independent calculation is needed to relate the value of $K$ to the microscopic parameters of the lattice model.
In the following, we use two independent ways to calculate $K$, which is useful for the characterization of the system properties. Furthermore, it serves as a stringent test for the internal consistency of the numerics.

\begin{figure}
\centering
\includegraphics[width=.77\textwidth]{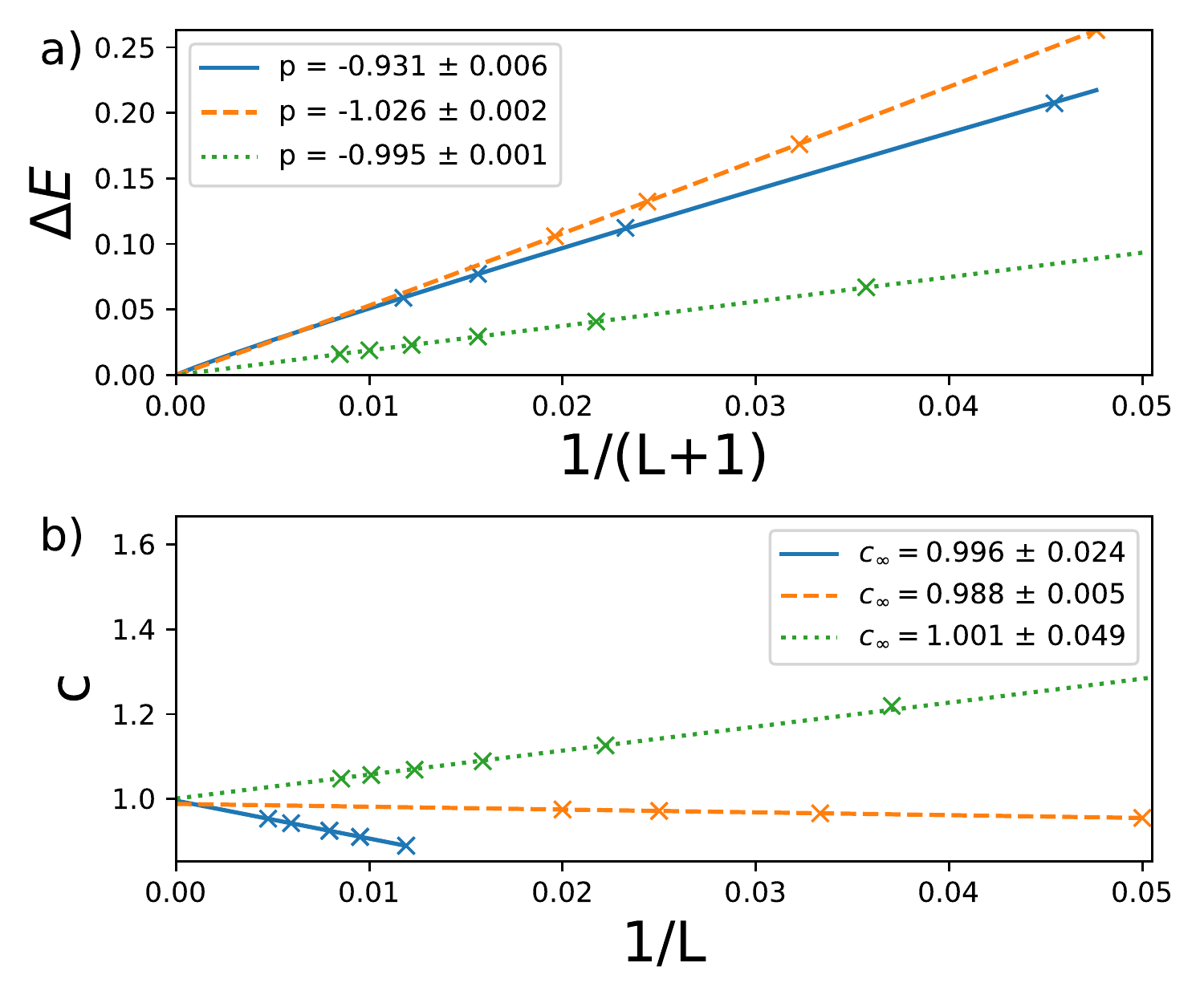}
\caption{Critical scaling for a superfluid and supersolid states with and without spatial oscillations.
Solid blue lines: SF at $(U,V) = (0.5,4)$ with $n=0.62$ and spatial oscillations. Dashed orange lines: SF at $(U,V) = (0.5,0.5)$ at integer filling without spatial oscillations. Dotted green lines: SS at $(U,V) = (0.5,4)$ with $n=1.333$ and spatial oscillations. Despite very different spatial features, all states seem to be well described within the same field theoretic description. a) Finite-size scaling of the level splitting $\Delta E = E_1 - E_0$ vanishing in the thermodynamic limit and yielding the critical exponent $p$ from $\Delta E \propto 1/(L+1)^p$, roughly matching the expected $p_\text{LL}=-1$ for a Luttinger liquid. b) Central charge $c$ \cref{fig:central_charge_entropy} extrapolated to the thermodynamic limit $c_\infty$ from $c = c_\infty + \text{const}./L$ matching the central charge $c_\text{LL} = 1$ for a spinless Luttinger liquid.}
\label{fig:deltaE_c}
\end{figure}

We check various other quantities and compare them with known parameters for an SF without spatial oscillations.
Within the Luttinger liquid description, the lowest-lying excitation spectrum is considered to be linear in momentum $k$, i.e. $E(k) = \hbar k v_s$, where $v_s$ is the speed of sound.
Furthermore, the speed of sound is related to the compressibility through $mv_s^2 = n \partial \mu/\partial n = (n\kappa)^{-1}$\cite{LLvolumeIX}.
In a finite-size (open boundary) system of size $L$, the minimum allowed value of the momentum is inversely proportional to the length of the wire, i.e. $k_{min} = \pi/(L+1)$. 
As a result, the excitation spectrum has a level splitting
\begin{equation}
\Delta E = \frac{\pi\hbar v_s}{L+1}
\label{eq:Delta E}
\end{equation}
which vanishes in the thermodynamic limit, $L\to\infty$.
We confirm this antiproportional scaling $\Delta E \propto (L+1)^p$ in fig.~\ref{fig:deltaE_c}(a) by a fit of the critical exponent $p = -0.931 \pm 0.006$, which is consistent with the expected value of $p_\text{LL} = -1$.

We further extract the central charge $c$ from the von Neumann entropy (Rényi entropy in the limit $\alpha \rightarrow 1$)
\begin{equation}
\label{fig:central_charge_entropy}
  S_1 = \frac{c}{6} \log(d[\ell|L])
\end{equation}
for which we obtain an extrapolated value of $c(L\rightarrow\infty) \approx 0.99$ throughout the superfluid phases, which is in perfect agreement with the predicted result $c_\text{LL} = 1$ for a spinless Luttinger liquid (see~\cref{fig:deltaE_c} panel~b)).

\begin{figure}
\centering
\includegraphics[width=.77\textwidth]{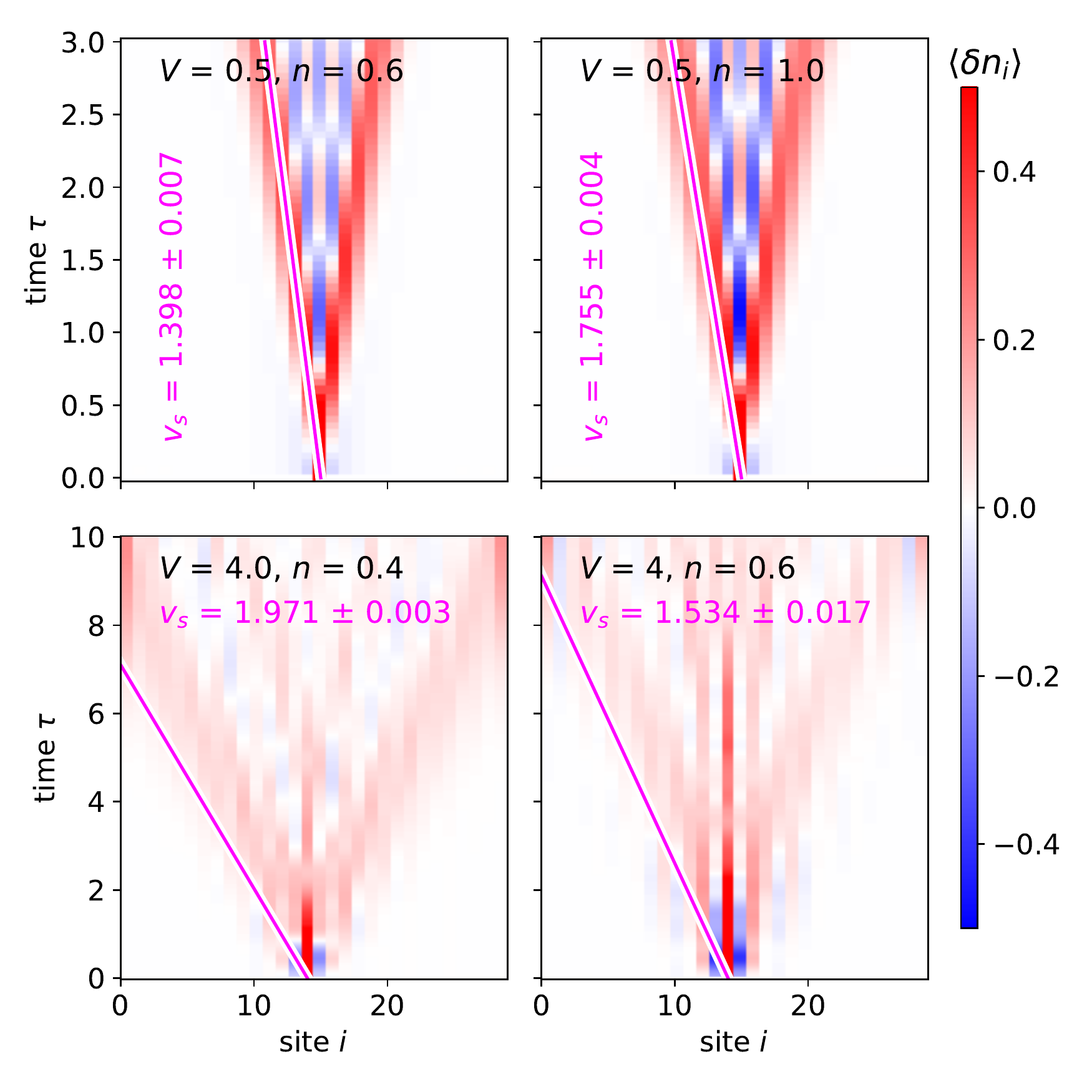}
\caption{Dynamical analysis of a local perturbation. 
We create a particle at a central site of the ground-state wave function and track the time evolution of the space-resolved density.
The propagating defect $\braket{\delta n_i} = \braket{\Psi(\tau)|n_i|\Psi(\tau)} - \braket{\Psi_0|n_i|\Psi_0}$ 
shows a typical ``light-cone'' structure, for which the red and blue coloring corresponds to positive and negative excess with respect to the average ground-state density. 
The boundary of the cone propagates with the speed of sound (highlighted in magenta), obtained by fits of the level splitting according to~\cref{eq:Delta E}. 
In all panels, we fixed $U=0.5$ and parameters $L=30$, $\chimax = 400$, and $d=6$.
}
\label{fig:lightcones}
\end{figure}

To check the validity of \cref{eq:Delta E}, we compute the speed of sound $v_s$ at four distinct points in parameter space $(V,n) \in \{(4,0.6), (4,0.4), (0.5,1), (0.5,0.6)\}$ with fixed $U=0.5$ and compare them with dynamical simulations. For this, we disturb the ground state at the middle of the chain $\ket{\delta \Psi_0} = b^\dagger_{L/2} \ket{\Psi_0}$ and compute its time evolution $\ket{\Psi(t)} = \exp\left(-i\tau H \right) \ket{\delta \Psi_0}$ via trotterization (TEBD) \cite{White2004TEBD,Vidal2004TEBD}. We then compute the density distribution at each time step and substract the ground state density, $\braket{\delta n_i} = \braket{\Psi(\tau)|n_i|\Psi(\tau)} - \braket{\Psi_0|n_i|\Psi_0}$. The resulting lightcones are displayed in \cref{fig:lightcones} and match well the overlayed fitted speed of sound from \cref{eq:Delta E} (in magenta). Further we compare these values of the speed of sound obtained via $v_s = 1/\left( \hbar \pi n^2 \kappa K \right)$, incorporating the Luttinger parameter $K$ from~\cref{eq:KfromS} and the compressibility $\kappa$ from~\cref{eq:kappam1}. We extrapolate results to the thermodynamic limit and find a reasonable agreement within $18\%, 14\%, 1\%$ and $9\%$, again for $(V,n) \in \{(4,0.6), (4,0.4), (0.5,1), (0.5,0.6)\}$ and fixed $U=0.5$, respectively. This is a stringent test of the internal consistency of the method, as thermodynamic relation between the compressibility obtained from equation of state and the speed of sound is tested.

In the following, we rely on Luttinger liquid theory to describe the asymptotic behavior of correlation functions.
To do so, we employ an abelian bosonization analysis~\cite{Giamarchi1992},
\begin{align}
b^\dag_x\rightarrow\psi^\dag(x)\sim\sum_{m=-\infty}^{\infty}{\rm e}^{2\pi im(nx + \phi) + i\theta(x)}
\label{eq:bosonization}
\end{align}
in which $[\phi(x),\partial_{x'}\theta(x')]=i\delta(x-x')$ satisfy canonical commutation relations.
Here the $\sim$ symbol denotes equality up to a prefactor, which depends on the momentum cutoff employed to derive the low-energy description~\cite{Cazalilla2004}.
The low-energy effective Hamiltonian of the extended Bose Hubbard model in 1D results to
\begin{align}
H = \frac1{2\pi}\int{\rm d}x\left(uK(\partial_x\phi)^2+\frac uK(\partial_x\theta)^2\right) + \mathcal O_{\rm sg}
\end{align}
in which $\mathcal O_{\rm sg}$ denotes additional sine-Gordon type operators as a result of the density-density interactions which are responsible for the opening of energy gaps (e.g. in the MI and HI phase).
For the characterization of the superfluid phase, these operators are irrelevant and can be disregarded.

The local density is given by
\begin{align}
n_x\rightarrow\rho(x)=\left(n+\partial_x\phi(x)\right)\sum_l{\rm e}^{2\pi il(n x+\phi(x))}
\end{align}
in which $n$ denotes the average density.
Note that the slowly oscillating contributions correspond to $l=0$, which allows one to identify $\delta\rho(x)=\rho(x)-n\approx\partial_x\phi$ as the field encoding the local density fluctuations.
This allows to approximate the argument of the $\Pi$ operator to $\sum_{x<l<x'}\delta n_l\rightarrow \phi(x')-\phi(x)$.

Correlation functions of the rescaled fields $\phi'=\phi{\sqrt K}$ and $\theta'=\theta/(\sqrt K)$ are readily obtained by a generating functional of the corresponding quantum mechanical partition function~\cite{GogolinNersesyanTsvelik} and result in the asymptotic expressions
\begin{align}
\braket{\phi(x) \phi(x')} = -\frac1{2K}\log\left(|x-x'|\right),\\
\braket{\theta(x) \theta(x')} = -\frac K{2}\log\left(|x-x'|\right).
\end{align}
By using the identity
\begin{align}
\braket{\rm e^{\rm i\sum_k b_kf(x_k)}} = \rm e^{-\frac12\sum_{k,k'}b_kb_{k'}\braket{f(x_k)f(x_{k'})}},\quad
f\in\{\phi,\theta\}
\end{align}
we arrive at the following asymptotic forms of the correlation functions, keeping only the dominant contributions
\begin{align}
\label{eq:bdb_fit}
|\braket{\psi^\dagger(x) \psi({x'})}|
&\approx
|\braket{\rm e^{i[\theta(x)-\theta(x')]}}|
\propto|x-x'|^{-K/2},\\
\label{eq:1_fit}
|\braket{\psi^\dagger(x) \Pi \psi(x')}|
&\approx
|\braket{\rm e^{i\theta(x)}\rm e^{i[\phi(x')-\phi(x)]}\rm e^{-i\theta(x')}}|
\nonumber\\
&\propto|x-x'|^{-1/2(K+1/K)},\\
\label{eq:2_fit}
|\braket{\delta\rho(x)\Pi\delta\rho({x'})}|
&\approx
|\braket{\partial_x\phi\rm e^{i[\phi(x')-\phi(x)]}\partial_{x'}\phi}|
\nonumber\\
&\propto|x-x'|^{-1/(2K)-2},\\
\label{eq:3_fit}
|\braket{\rho(x)\rho(x')}_{\rm conn.}|
&\approx
|\braket{\partial_x\phi\partial_{x'}\phi(x')}|
\propto|x-x'|^{-2}.
\end{align}

The Luttinger liquid predictions for the long-range asymptotic of the correlation functions are verified in fig.~\ref{fig:iDMRG_correlators_f-065} and a very good agreement is found. 
Note that the oscillations observed in \cref{fig:OBC_f-0-77} and \cref{fig:iDMRG_f-0-65_Csl-Ssl-ES} are consistent with the field theoretic description if sub-leading corrections are not neglected.

\begin{figure}
\centering
\includegraphics[width=.77\textwidth]{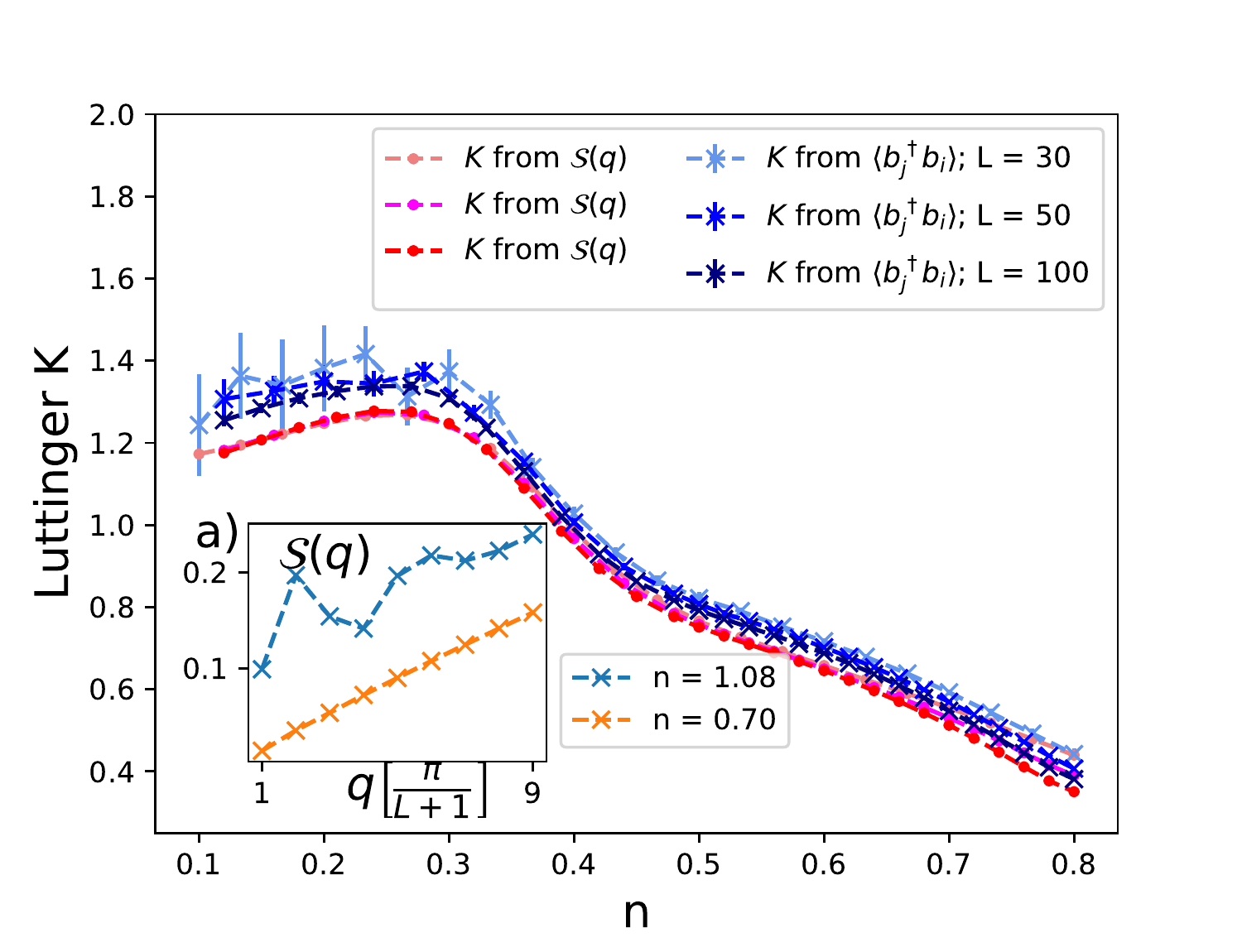}
\caption{
Luttinger parameter $K$ dependence on the filling $n$, calculated with OBC with $\chimax = 400$ and $d = 6$ at $U=0.5$ and $V=4$.
Two independent estimations are used, from the long-range asymptotic of the off-diagonal single-particle correlation function $\braket{b^\dagger_i b_j}$ \cref{eq:bdb_fit} and from the small momenta of the structure factor $\mathcal{S}(q)$ via \cref{eq:KfromS}.  
a) Structure factor $\mathcal{S}(q)$ for small momenta at different fillings. With the onset of phase separation, $\mathcal{S}(q)$ deviates from linear dependence at its origin and \cref{eq:KfromS} becomes invalid.}
\label{fig:f-K_master}
\end{figure}

Thus, the Luttinger liquid is capable of capturing correctly the long-range properties. 
At the same time, a microscopic simulation is needed to connect the parameters of the microscopic Hamiltonian to the effective parameters of the Luttinger liquid model. 
In particular, it is of great practical value to find such a relation for the Luttinger parameter $K$.
We extracted the Luttinger parameter $K$ from correlation functions in \cref{eq:bdb_fit,eq:1_fit,eq:2_fit}. 
However, we expect that oscillatory subleading terms are more important in \cref{eq:1_fit,eq:2_fit}, causing large error bars for fits of the leading order only, and we resort to a detailed comparison between the value of $K$ obtained from \cref{eq:bdb_fit,eq:KfromS} only in \cref{fig:f-K_master}.

Alternatively, the Luttinger parameter $K$ can be extracted from the slope of the linear part of the structure factor \cite{Astrakharchik2016,praga}
\begin{equation}
\mathcal{S}(q) = \sum_{ij} e^{-iq(i-j)} (\braket{n_i n_j} - \braket{n_i}\braket{n_j})/(L+1).
\end{equation}
In the framework of the Tomonaga-Luttinger description, we can compute the Luttinger parameter via
\begin{equation}
\label{eq:KfromS}
\frac{1}{2 \pi K} = \lim_{q\rightarrow 0} \frac{\mathcal{S}(q)}{q},
\end{equation}
where $q$ and $S(q)$ depend on the system size $L$ and the boundary conditions, see \cite{Ejima2011,arik2016}. We obtain $S(q)/q$ by performing a fit of the lowest momenta where $S(q) \propto q$ is linear. 
If the lowest-lying excitation spectrum is exhausted by linear phonons, the Luttinger liquid description is applicable and the Luttinger parameter defined according to Eq.~(\ref{eq:KfromS}) is independent of the actual size of the system $L$ if it is large enough.
We see that both estimations of $K$ match well in the SF phase, as seen in \cref{fig:f-K_master}. 

The knowledge of the Luttinger parameter $K$ allows one to apply the effective description as provided by the Luttinger liquid to static and dynamic long-range properties. In particular, low-momentum behavior of the momentum distribution can be obtained as a Fourier transform of off-diagonal single-particle correlation function~(\ref{eq:1_fit}) resulting in a divergent $n(k) \propto
|k|^{1-K/2}$ behavior for $K<2$. 
That is, for all cases shown in Fig.~\ref{fig:f-K_master}, the occupation of zero-moment state diverges in the thermodynamic limit which is a reminiscence of Bose-Einstein condensation in one dimension. 
Another special value of the Luttinger parameter is $K=1/2$, below which an SF state might be sustained a unit filling as opposed to a Mott insulator which is realized for any finite height of the optical lattice\cite{zwerger2003mott,RevModPhysBloch}.
In the considered system, small values of $K$ correspond to a large filling fraction $n$, and further increasing $n$ leads to phase transition.

In conclusion, we do not find signatures that suggest an alternate field-theoretic description for the SF with spatial oscillations linked to a ``symmetry enriched quantum criticality''~\cite{verresen2019gapless,thorngren2020intrinsically}.
The most striking evidence for the absence of (quasi) zero energy edge modes is provided by the finite-size scaling of the energy level splitting $\propto(L+1)^{-1}$. Due to the bulk-boundary correspondence and as outlined in \cite{thorngren2020intrinsically}, an intrinsically gapless topological phase would host edge excitations that provide strong corrections to the lowest splitting, i.e. \cref{eq:Delta E}, which we do not observe throughout the superfluid phase.
Furthermore, we do not see any spontaneous boundary occupation, nor did we observe non-vanishing edge-to-edge correlations in the thermodynamic limit.
Instead, we demonstrated the applicability of the standard Luttinger liquid description by numerical estimates of the excitation spectrum, the central charge, and the Luttinger liquid parameter.

\section{Conclusions}
\label{sec:conclusions}
In this chapter, we investigated the extended Bose Hubbard model in one dimension with tensor network and machine learning methods. We first showed how to map out the phase diagram with no a priori knowledge with entanglement spectra, central tensors and observables as data using anomaly detection. We found an unexpected new phase in the filling $n = 1$ phase diagram and confirmed the findings of \cite{Batrouni} that it is a phase-separated phase and at the same time clarified misconceptions of \cite{Deng2011} regarding the same region. Further, we rigorously demonstrated the \textit{hidden} broken translational symmetry of the homogeneous superfluid and supersolid phases at incommensurate fillings and show that it is a true physical property in the thermodynamic limit. It is \textit{hidden} in the sense that it is only visible in the entanglement structure or non-local string-order correlation functions. The latter is often related to topological properties of the ground state, for which we find no evidence, i.e. we have not observed any edge states, or bulk-edge correspondence in these regions, nor could we see variations from expected scaling in this universality class. Furthermore, we confirmed that the model agrees in the superfluid phase with the predictions made within the standard framework of Luttinger-Tomonaga theory. We did so by showing a match of the predicted speed of sound with dynamical simulations (using TEBD) and providing a relation between the Luttinger parameter and the microscopic parameters of the model.

In view of recent progress with experiments on dipolar atoms, Rydberg atoms, and trapped ions, as well as novel methods of detection of entanglement entropies and spectrum, our results open an interesting playground to test CFT and Luttinger liquid properties in experiments.
Our simple bosonization approach is fully applicable in the superfluid phase only. A more powerful field theory predicting the phase transition between superfluid and supersolid would be of general interest.
The outlook for future studies includes investigations of the same model in 2D, and extension to true long-range interactions, with a particular focus on dipolar ones, where the experiments are on the way.

\chapter{Unsupervised mapping of phase diagrams of 2D systems from iPEPS via deep anomaly detection}
\label{chapter:peps}
The following chapter is a taken over from our publication \textit{Unsupervised mapping of phase diagrams of 2D systems from infinite projected entangled-pair states via deep anomaly detection} \cite{Kottmann2021} with slight modifications to put it into context of this thesis. We are again applying \cref{alg:anomaly_detection-phase_diagram}, but this time look at data from two-dimensional systems. This necessitates substantially more complicated physical simulation algorithms and changes the data format. I.e. the singular values that arose naturally in previous simulations have a clear physical interpretation as the entanglement spectrum, whereas in 2D this is not the case. Here we show that singular values taken from 2D tensor networks still contain sufficient information to map out the phase diagram. Curiously, we find that training with one single example is sufficient, which raises the question of the necessity of neural networks in the first place.

\section{Introduction}
With the introduction of a new data-driven computation paradigm, machine learning (ML) techniques have been very successful in performing recognition tasks and have had a big impact on industry and society. ML has been successfully applied to a variety of physical problems, and vice versa, physics has inspired new directions to explore in understanding or improving ML techniques \cite{Carleo2019}. Among the most prominent and successful applications of ML in physics is the classification of phases in many body physics \cite{wang2016discovering,VanNieuwenburg2016,carrasquilla2017machine,schindler2017probing,liu2017self,wetzel2017unsupervised,chng2017unsupervised,koch-janusz2017mutual, huembeli2018identifying, huembeli2019automated, deng2016exact,zhang2017machine, broecker2017machine, Tsai2019,Shinjo2019, theveniaut2019neural,Dong20182,Kawaki2017,Funai2018}. 
Of particular interest are unsupervised methods that require no or little prior information for labeling ~\cite{VanNieuwenburg2016, liu2017self,wang2016discovering,wetzel2017unsupervised,chng2017unsupervised, nussinov2016inference}. 
In particular, phase diagrams have been mapped out in a completely unsupervised fashion with no prior physical knowledge from 1D tensor network data \cite{Kottmann2020} and experimental data \cite{Kaming2021} via anomaly detection as described in \cref{alg:anomaly_detection-phase_diagram}. 

In this work, we extend the application of anomaly detection for phase characterization to 2D quantum many body systems with projected entangled-pair states (PEPS). PEPS have been introduced as an efficient ansatz for ground states for 2D Hamiltonians \cite{Verstraete2004,Nishio2004,Verstraete08} and extended to simulate the thermodynamic limit with infinite PEPS (iPEPS) \cite{Jordan2008}. Various methods to optimize the iPEPS tensors exist, including (fast-) full update imaginary time evolution algorithms~\cite{Jordan2008,Phien2015} and energy minimization approaches~\cite{corboz16b,vanderstraeten16,liao2019}. Computationally cheaper alternatives were introduced with the simple update~\cite{Vidal2003,Jiang2008} and cluster update~\cite{wang11b} algorithms that perform optimizations locally at the cost of numerical accuracy. Progress in the systematic study of continuous phase transitions has recently been achieved based on a finite correlation length scaling analysis~\cite{corboz18,rader18,hasik21}. 

The intention of the scheme presented in this work is not to improve numerical accuracy of determining phase boundaries, but to obtain a qualitative phase diagram with \textit{low computational cost} and \textit{no physical a priori knowledge}. The former, low computational cost, is achieved by employing the simple update optimization algorithm with contractions omitted throughout the whole process. Physical knowledge in this scheme is redundant as we resort to quantities obtained directly from the iPEPS wave function from simulations as inputs for the machine learning method; in this case singular values between bonds or reduced density matrices. In other words, we do not need to choose and compute suitable observables that contain sufficient information about the phase boundaries for the machine learning processing. In 1D systems, the singular values between bonds have a clear physical interpretation, as they characterize the entanglement properties between the subsystems at each end of the bond. In 2D, however, there is no such interpretation and it is non-trivial to show that singular values at bonds are still sufficient to determine phase boundaries. In the considered approach, phase boundaries are characterized without the need to know the order parameter or symmetry groups of the phases. In fact, in a scenario where we are given data from iPEPS to analyze, in principle we do not even need to know the Hamiltonian.

In contrast to supervised methods, where at least a rough idea of the regions of different phases (and the number of separate phases) is needed for labeling a training set, we do not need to know anything about the phase diagram by using unsupervised anomaly detection. This is because in this scheme a region of the diagram is chosen to represent normal data and is then tested for the whole diagram. Initially, this normal region is chosen randomly and may contain states from one or multiple phases. When states from different phases than it has been trained on are tested, they are marked as anomalies. Between those states and the training region, there is a transition from normal to anomalous data that corresponds to the phase transition. In the next training iteration, the normal region is put where anomalies have been found in the previous round and the process repeated until no previously unseen anomalous region is found. This process only needs $\mathcal{O}(N_\text{phases})$ iterations where $N_\text{phases}$ is the number of phases in the diagram. This is in contrast to \textit{learning by confusion} schemes where the phase diagram is scanned by iteratively shifting the labeling and retraining \cite{VanNieuwenburg2016,liu2017self}. 

In spirit, the approach presented here is similar to the method described in \cite{Zanardi2007,zhou08}. There, phase transitions are determined by looking at the overlap (fidelity) between neighboring ground states in the phase diagram, with a drop in the fidelity at quantum phase transitions as the overlap between states from different phases is small (zero) for finite (infinite) system sizes. The big advantage of the approach presented here is that we avoid computationally expensive contractions of the tensor network, which, in contrast, is needed to compute overlaps.

As an example, we examine the 2D frustrated bilayer Heisenberg model, a challenging problem which suffers from the negative sign problem \cite{Stapmanns2018}. The model contains two 1st order and one 2nd order phase transition and is therefore a good benchmark for the success of this method.
This manuscript is organized as follows: The general approach of applying anomaly detection with neural networks to map out phase diagrams is described in \cref{pepssec:anomaly}. Details on iPEPS are described in \cref{pepssec:numerical_methods} and a brief overview of the 2D frustrated bilayer Heisenberg model is given in \cref{pepssec:Hamiltonian}. The results are then presented in \cref{pepssec:numerical_results}, followed by our conclusions in \cref{pepssec:conclusions}.


\section{Anomaly Detection}
\label{pepssec:anomaly}

We follow the approach in described in the previous  \cref{chapter:anomaly,chapter:bose-hubbard}, where it was shown that phase diagrams can be mapped out from different data types in an unsupervised fashion via \textit{anomaly detection}. The scheme works in the following way: We employ a special neural network architecture, called an autoencoder, to efficiently decode and encode data of the type it has been trained on (data specific compression). For the training\footnote{\textit{Training} in machine learning refers to data-specific optimization. This is described in more detail below around \cref{eq:L}.}, we define a training dataset containing \textit{normal} data. The autoencoder is trained to efficiently reproduce data with the same or similar characteristics. Anomalies are detected by deviations of a loss function between input and output of the autoencoder, compared to the region it has been trained on and amount to separate phases in the diagram.
This training has to be performed only $\mathcal{O}(N_\text{phases})$ times where $N_\text{phases}$ is the number of phases present in the phase diagram. Moreover, this procedure does not necessitate any prior physical knowledge about the system as one starts with an arbitrary parameter range, typically at the origin of the parameter space. From there, abrupt changes in the reproduction loss are saved as possible phase boundaries and the next training iteration is done in the region with the highest loss after the previous training. Note that in principle one does not even need to know the underlying Hamiltonian, it suffices to be provided with data and the corresponding physical parameters.

We employ autoencoder neural network architectures implemented with \textsf{TensorFlow} \cite{tensorflow2015-whitepaper}. An autoencoder is composed of an encoder and a decoder. The encoder takes the input $x$ and maps it to a latent space variable $z$. This latent space variable is then mapped by the decoder to the output $y(x)$. Both encoder and decoder are composed of multiple consecutive layers, parametrized by free parameters $\theta$. We tried different architectures comprising different combinations of fully-connected and convolutional layers. We find no specific model dependence, with different architectures performing similarly such that simple \textit{vanilla} autoencoders composed solely of fully-connected layers suffice and are used throughout this paper. For details about the implementation see \cite{notebooks}.
The goal of the autoencoder is to reproduce $x$, i.e. matching the output of the network with its input $y(x) = x$, which is achieved by minimizing the reproduction loss
\begin{equation}
\label{eq:L}
    L(\theta) = \sum_i ||x_i-y_i(x_i)||^2
\end{equation}
with respect to the free parameters $\theta$ ($y(x)$ implicitly depends on $\theta$). The sum reaches over the training examples defined for the training iteration. Here we have chosen the loss to be the mean squared error as it is simple and effective for our task, but in principle there is a variety of possible and valid loss functions depending on the task and data at hand. The optimization task of minimizing $L$ is achieved by gradient descent $\theta \mapsto \theta - \alpha \nabla_\theta L(\theta)$, i.e. computing the gradient of $L$ and changing the free parameters in the opposite direction for some stepsize $\alpha$ (hyper parameter given by the user). For neural networks, there is an efficient implementation called backpropagation \cite{kelley1960gradient}. We employ \textsf{ADAM}, a modern optimization scheme with adaptive stepsizes based on gradient descent with backpropagation for faster optimization \cite{Kingma2015}.

\section{Infinite projected entangled-pair states}
\label{pepssec:numerical_methods}
An iPEPS~\cite{Jordan2008} is a tensor network ansatz to represent 2D ground states directly in the thermodynamic limit and can be seen as a generalization of 1D infinite matrix product states (iMPS) to 2D. The ansatz consists of a unit cell of rank-5 tensors repeated periodically on a lattice. Here we use a unit cell with two tensors arranged in a checkerboard pattern on a square lattice, with one tensor per dimer in the bilayer model introduced in Sec.~\ref{pepssec:Hamiltonian}. Each tensor has one physical index representing the local Hilbert space of a dimer and four auxiliary indices, which connect to the four nearest-neighbor tensors. The accuracy of the variational ansatz is systematically controlled by the bond dimension $D$ of the auxiliary indices. To improve the efficiency of the calculation we use tensors which exploit the U(1) symmetry of the model~\cite{singh2010,bauer2011}. 

In this paper the optimization of the tensors is done based on an imaginary time evolution, which involves a truncation of a bond index at each time step (see Refs.~\cite{Jordan2008,corboz2010,Phien2015} for details). While an optimal truncation requires a full contraction of the 2D tensor network (called the full update~\cite{Jordan2008}), which is computationally expensive, there exist also local, approximate schemes avoiding the full contraction. In the simple-update approach~\cite{Vidal2003,Jiang2008}, which we use in this work, the truncation is performed by a local singular value decomposition of two neighboring tensors. A more accurate, but still local scheme is provided by the cluster update introduced in Ref.~\cite{wang11b}, in which the truncation is done based on a cluster of tensors, where the accuracy is controlled by the cluster size.

We start the iPEPS optimizations from random initial states, thereby avoiding the need for any knowledge of the system a priori. Depending on the initial state, the iPEPS may converge to a local minimum. To improve the convergence behavior, the state is initially evolved for a few steps at large bond dimension and large imaginary time step, then projected to $D=1$, and then evolved at the target $D$. Optimization runs are discarded and repeated when convergence is not reached after a certain number of steps. Finally, the state is further evolved at the target dimension at a smaller time step until convergence is reached. We found that this scheme improves the efficiency and quality of the results, compared to an evolution only at the target $D$, especially close to the first order phase transition line of the model.

As input data for the anomaly detection, we use the singular values of the four auxiliary bonds obtained from the simple-update approach. In 1D iMPSs in canonical form, they correspond to the Schmidt coefficients, characterizing the entanglement between the two sides of the system connected by the bond. In 2D, however, there exists no canonical form and the singular values do not correspond to the Schmidt coefficients because of the loops in the tensor network ansatz. Still, we will show that the singular values contain information that can be used for and interpreted by the machine learning algorithm to characterize the underlying ground state.

For comparison we also consider the 2-site reduced density matrix as input data, which is computed by contracting the 2D tensor network using the corner transfer matrix method~\cite{nishino1996,orus2009-1,corboz14_tJ}.

\section{Model}
\label{pepssec:Hamiltonian}
We test the anomaly detection scheme in combination with iPEPS for the $S=1/2$ 2D frustrated bilayer Heisenberg model -  a challenging problem, where a large part of the phase diagram is inaccessible to Quantum Monte Carlo due to the negative sign problem~\cite{Stapmanns2018}. 
The model can be represented as a two-dimensional lattice of coupled dimers, formed by the two $S=1/2$'s of the adjacent layers. The Hamiltonian  reads
\begin{equation}
    H = \sum_i J_\perp \vec{S}_{i,1} \cdot  \vec{S}_{i,2}  + \sum_{\substack{i,m=1,2 \\j = i + \hat{x},i + \hat{y}}} \left[ J_{\|} \vec{S}_{i,m} \cdot \vec{S}_{j,m} + J_{x} \vec{S}_{i,m} \cdot \vec{S}_{j,\bar{m}} \right]
\end{equation}
where $J_{\|}$ is the nearest-neighbor intralayer coupling and $J_{\perp}$ ($J_{x}$) the nearest (next-nearest) neighbor interlayer coupling,  $i$ is the index of a dimer, $j$ runs over the nearest-neighbor dimers, $m$ denotes the two layers, and $\bar{m}$ the layer opposite to $m$.

In the limit of strong $J_\perp$  the ground state is a dimer singlet (DS) state with vanishing local magnetic moment. For $J_{x}=0$ the model is unfrustrated with an ordered bilayer antiferromagnetic (BAF) ground state for $J_\perp/J_\| < 2.5220(2)$~\cite{wang06}, separated from the DS phase by a continuous transition. The limit $J_{x}=J_{||}$ corresponds to the fully-frustrated Heisenberg bilayer model with a dimer-triplet antiferromagnetic (DTAF) ground state for $J_{\perp}/ J_\| < 2.3279(1)$~\cite{muller00}, in which spins on a dimer are parallel (in contrast to the BAF phase where the spins on a dimer are antiparallel). The ground state phase diagram of the full model was mapped out with iPEPS in Ref.~\cite{Stapmanns2018} and is shown in \cref{fig:random-init} 1a) by the white-dashed lines. It hosts a quantum critical endpoint at which the line of continuous transitions between the BAF and DS phase terminates on the first order line separating the DTAF phase from the DS and BAF phases.

\section{Numerical Results}
\label{pepssec:numerical_results}

\begin{figure}
    \centering
    \includegraphics[width=.98\textwidth]{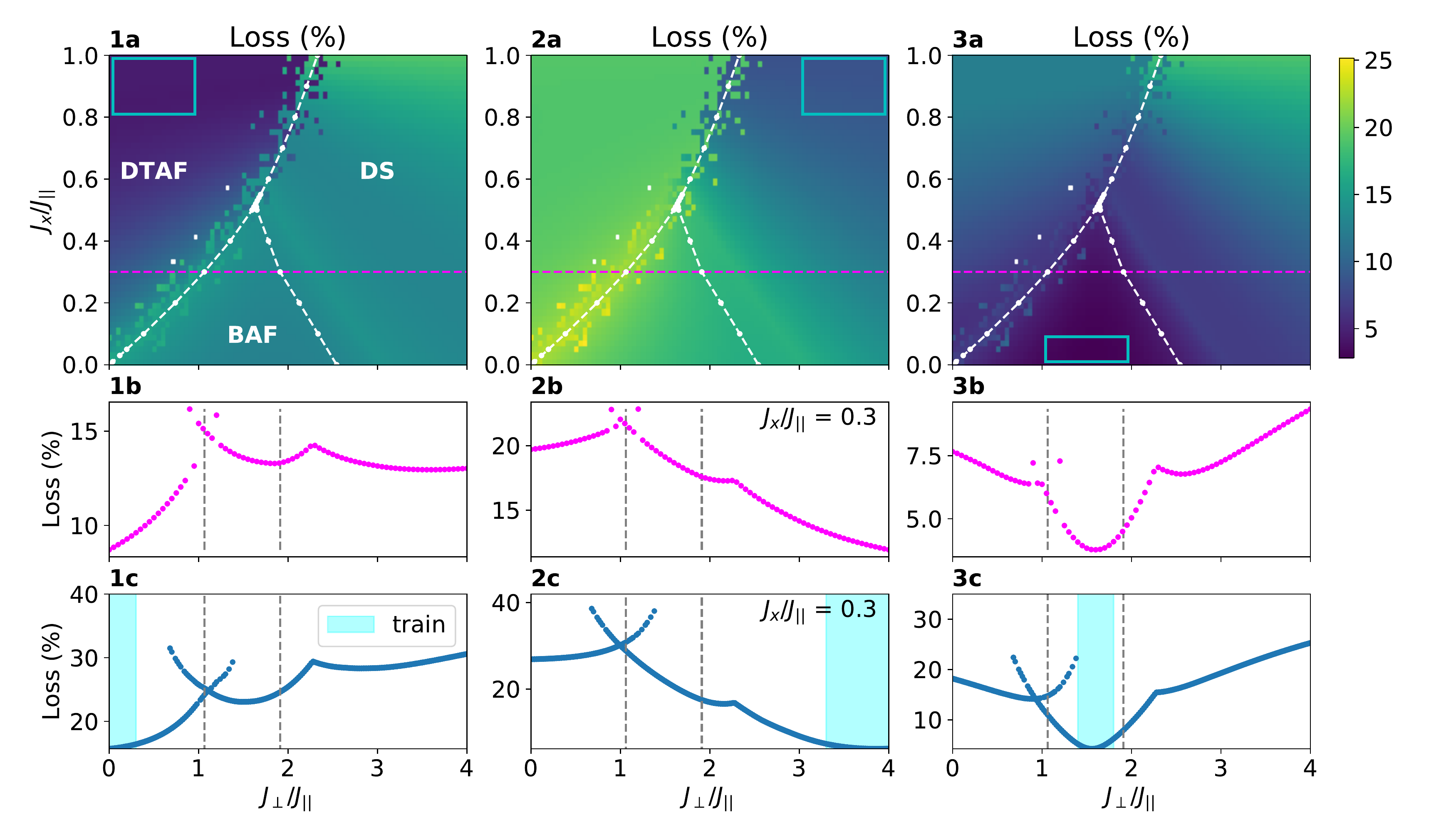}
    \caption{Three training iterations to map out all three phases of the phase diagram without contraction and prior knowledge of the system, i.e. using simple update algorithm starting from random initial iPEPS. The cyan rectangles show the training regions. Overlayed in white are theoretical predictions, extrapolated to the infinite $D$ limit with full update optimization from \cite{Stapmanns2018}. Deviations of the second order BAF-DS transition are expected due to finite $D=6$ and simple update optimization. 1a) Starting the training at the top-left (in DTAF phase) yields the first-order transition line. Even the second order transition line is already pronounced inside the region of higher loss beyond the first order line. 2a) Second iteration in the region of highest loss (DS phase) from the previous picture showing the part of the first order line adjacent to the DS phase and the second order line. 3a) Confirming and completing the picture by training in the BAF phase. The second row (1b-3b) is a single cut as indicated by the magenta line in the phase diagram above. The third row (1c-3c) shows the loss after training and evaluating extra single cut data with five independent simulations per data point. 
    Around the first order transition two branches are obtained due to the characteristic hysteresis behavior.}
    \label{fig:random-init}
\end{figure}

We now use the anomaly detection scheme described in \cref{pepssec:anomaly} with data from iPEPS, described in \cref{pepssec:numerical_methods} to map out the phase diagram of the model without prior physical knowledge in \cref{fig:random-init}. Three training iterations suffice to map out the boundaries of all three phases of the system. 
We start in the top-left corner of the phase diagram corresponding to the DTAF phase in \cref{fig:random-init} 1a) and obtain the first order transition line. The noise around the line is due to hysteresis effects in the vicinity of the first order phase transition: Depending on the random initial tensors, the converged states end up in one of the two adjacent phases. We will later see how a sharp transition line can be obtained by measuring the energy of the states. Note how the second order transition line is already indicated within the anomalous region of high loss. The second training is performed in the region of highest loss from the previous iteration in the top-right corner corresponding to DS in \cref{fig:random-init} 2a). The second order transition line to the BAF phase is again signaled by a bump in the loss diagram. To complete and confirm both lines, training is performed in the BAF phase in \cref{fig:random-init} 3a). In \cref{fig:random-init} (1c-3c) we present data for the single cut at $J_x/J_{||} = 0.3$ with five independent simulations for each value of $J_\perp/J_{||}$,  illustrating the characteristic hysteresis behavior around the first order DTAF-BAF transition.

All the results in \cref{fig:random-init} are overlaid with the previous iPEPS simulation results from \cite{Stapmanns2018}. Note that those results are much more precise since the iPEPS were optimized with the more expensive full update algorithm and the data has been extrapolated to the infinite $D$ limit, whereas here we only consider simple-update data at finite bond dimension $D=6$. Thus a quantitative deviation can be expected, especially for the location of the second order BAF-DS transition line. However, the main goal here is to  get a cheap and fast overview of the phase diagram, which serves as a useful starting point for more accurate numerical investigations (e.g. based on a finite correlation length scaling analysis~\cite{corboz18,rader18}). We note that the ML approach  could also be combined with the more accurate cluster update scheme~\cite{wang11b}, which is still a local approach, or with the full-update~\cite{Jordan2008,Phien2015} or energy minimization schemes~\cite{corboz16b,vanderstraeten16,liao2019,crone20}, which require full contractions~\footnote{In the latter two cases the singular values can be extracted using the approach described in Ref.~\cite{phien15a}.}.

\begin{figure}
    \centering
    \includegraphics[width=.98\textwidth]{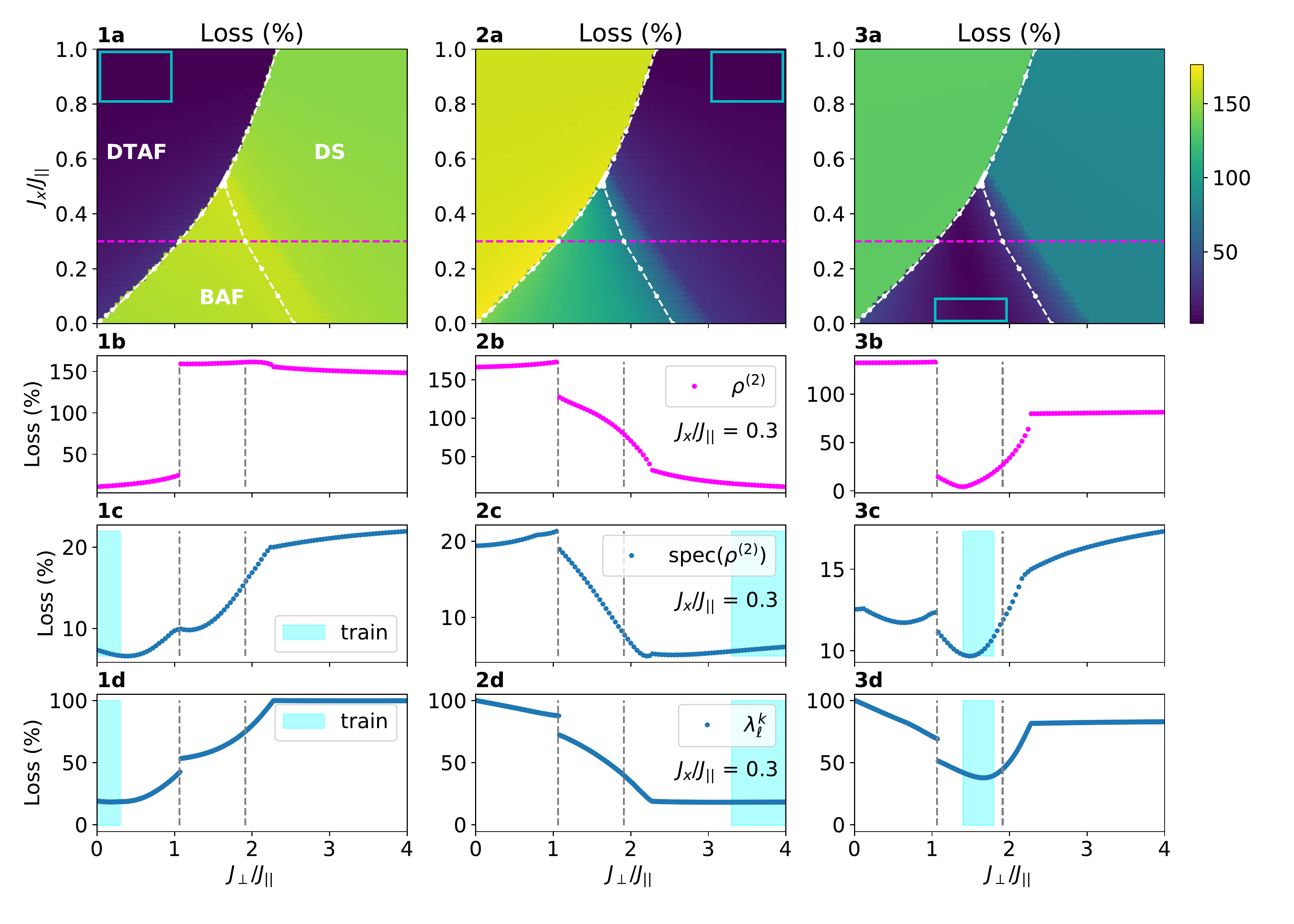}
    \caption{Three training iterations to map out all three phases of the phase diagram. Here, the reduced 2-site density matrix is used as input data. In comparison to \cref{fig:random-init}, the first order transition line is much sharper as the ground states were post-selected from energy considerations. The second row (b) shows the line at $J_\perp/J_{||} = 0.3$ as indicated by the dotted magenta line in row (a). In row (c), the eigenvalues of the reduced 2-site density matrix in log-scale are used. The training is done just for the single cut at $J_\perp/J_{||} = 0.3$.
    Row (d) uses again the singular values like in \cref{fig:random-init}.}
    \label{fig:post-selected}
\end{figure}

To get an even clearer picture of the predicted phase boundaries, we compute the energy of the states to post-select the best ground states. The ground-state optimization is initialized from three different initial iPEPS (a representative point in each phase) and only the lowest in energy is kept. The boundaries in the resulting pictures are now much more pronounced at the cost of invoking contractions to calculate the energy, yet still not taking any physical knowledge of the order into account. For the sake of showcasing the method with different inputs, we here use the 2-site reduced density matrix $\rho^{(2)}$ but also confirm again the viability with singular values.

We proceed in an analogous fashion and perform three trainings to map out the phase diagram in \cref{fig:post-selected}. The phase boundaries appear even sharper compared to \cref{fig:random-init}, especially for the first order transition line with a corresponding discontinuity (jump) in loss at the transition. In \cref{fig:post-selected} c) we confirm that the results are qualitatively the same when using the singular values as an input instead. 

We note that, while in practice the singular values are found to be gauge-invariant (see also Ref.~\cite{phien15a}), the reduced density matrix is not necessarily unique. If the state breaks, e.g., SU(2) spin symmetry, then different random initial states will lead to different reduced density matrices (since the local magnetic moments can be aligned along different directions). In Fig.~\cref{fig:post-selected}, this is not an issue, because each anomaly detection is based on a single initial state (which fixes the direction of the magnetic moments), and by using U(1) symmetric tensors, the magnetic moments are automatically parallel to the z-axis. Alternatively, one can also consider the eigenvalue spectrum of the reduced density matrix as input data, which is gauge-invariant, as shown in \cref{fig:post-selected} (1c-3c).

\begin{figure}
    \centering
    \includegraphics[width=\textwidth]{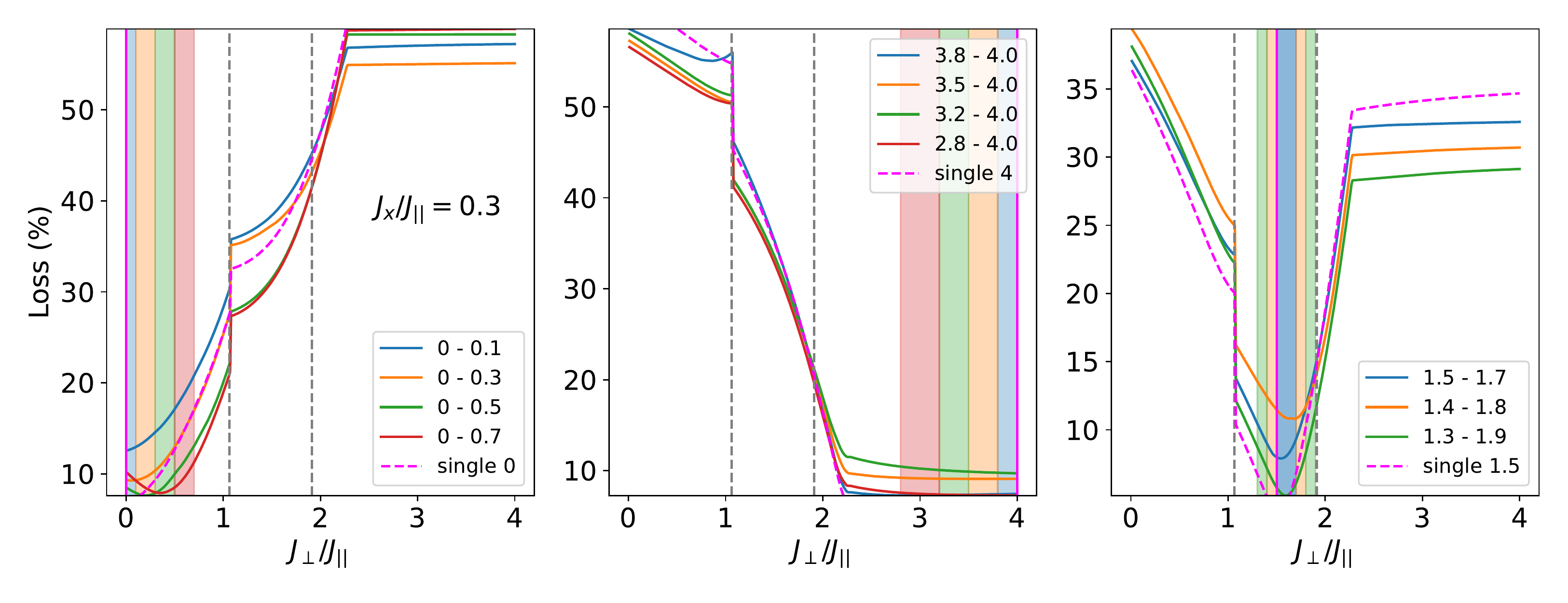}
    \caption{Varying the training region to show that the outcome of the algorithm is not sensitive to the number of training examples $N_\text{ex.}$ and the extent of the training region. The number of epochs $N_\text{epochs}$ is chosen such that $N_\text{epochs} \cdot N_\text{ex.} = \text{const.}$, i.e. the neural network \textit{sees} the same amount of examples throughout all training iterations. We see that we can even map out the regions with just one single training example (dotted magenta curve).}
    \label{fig:062}
\end{figure}

It is a common mantra in machine learning that more data always yields better results. In the present study, where the machine learning task is to find the phase boundary from singular values, we find this to actually not be the case. Also, the result of the algorithm is not sensitive to the extent of the training region, i.e. how far in parameter space the examples during training reach. It seems that one example of the phase already captures the characteristic features and that the data within the phase is so homogeneous that adding more examples does not improve the result. To show this, we take a single cut at $J_x/J_{||} = 0.3$ and vary the extent of the training regions in \cref{fig:062} for singular value data with $D=10$. In all cases the predicted transitions are the same and the result is insensitive to the chosen training region. We can put this to the extreme and only use one single training example $N_\text{ex.}=1$ and still obtain the same results. In all the cases of training regions in \cref{fig:062}, the number of epochs $N_\text{epochs}$, that is the number of times the neural network processes the full training set, is chosen in such a way that $N_\text{epochs} \cdot N_\text{ex.}$ is held constant, such that during training the same number of examples are processed for a fair comparison.

\begin{figure}
    \centering
    \includegraphics[width=.75\textwidth]{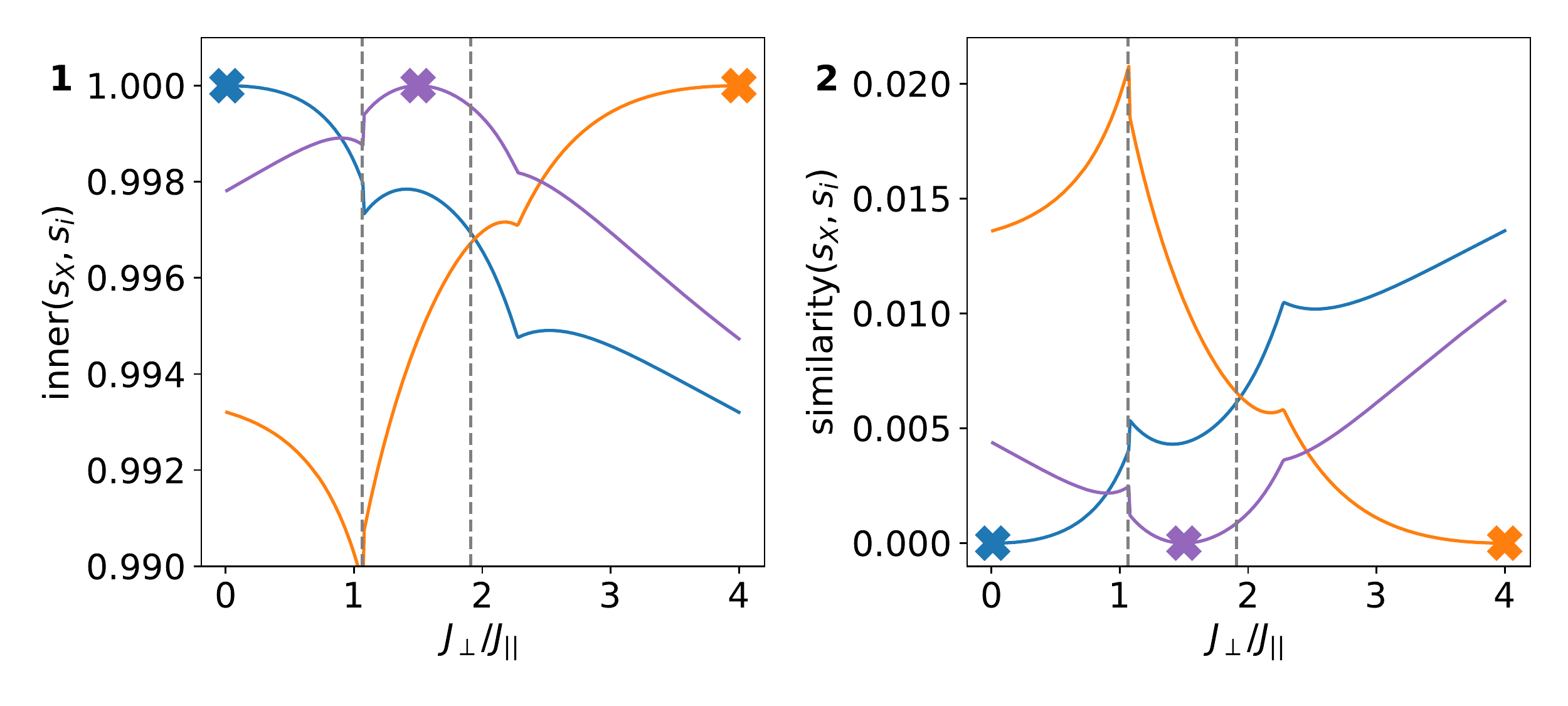}
    \caption{Detecting the phase boundaries with purely geometric data driven methods at $J_x/J_{||} = 0.3$. 1) Inner product between the singular values $S_i$ along the parameter space and a fixed point (indicated by the X). 2) Geometric similarity measure equivalent of the loss for the autoencoder \cref{eq:L}.}
    \label{fig:referee_test}
\end{figure}

This raises the question of the necessity of the neural network machinery for the anomaly detection. In \cref{fig:referee_test} we show simple, purely geometric and data-driven approaches that indicate the phase boundaries in the spirit of anomaly detection without using neural networks. In the first case, we compute the inner product between normalized singular value vectors $s_i$ for different physical parameters. Here, the inner product is just the \textit{standard} inner vector product 
\begin{equation}
    \text{inner}(s_i,s_j) = \sum_k s_i^k s_j^k.
\end{equation}
The normalization is done such that $\text{inner}(s_i,s_i) = 1$. Using inner products, there is a clear interpretation of the overlap values and we can see that the contrast in \cref{fig:referee_test} 1) is of order $\sim 0.01$ and therefore arguably small. 
We get better results when using a geometric similarity measure
\begin{equation}
    \text{similarity}(s_i,s_j) = \sum_k |s_i^k - s_j^k|^2
\end{equation}
between a fixed normalized singular value vector along the physical parameter space in \cref{fig:referee_test} 2). Note that this is equivalent to the loss in \cref{eq:L}, used for the autoencoder. These results are now very similar to the ones obtained with the autoencoder in \cref{fig:062}. An interesting open question to answer in future work is whether such a data-driven geometric analysis in the spirit of machine learning but without neural networks suffice in general or if this is specific to the model and data at hand.


\section{Conclusions}
\label{pepssec:conclusions}
In this work, we showed how to combine anomaly detection, a method for unsupervised ML, with iPEPS, a tensor-network ansatz for variational optimization, to map the phase diagram of 2D systems. By employing ML, we circumvented the necessity for defining and calculating suitable observables to identify the phases. Furthermore, no prior physical knowledge was required to run the unsupervised anomaly detection (i.e. no labels needed).
We saw that a successful training can be achieved with an arbitrarily small amount of examples, therefore making the amount of data generated a matter of aesthetics by the user. Based on this, we saw that for the present model and data, purely geometric and data-driven analyses sufficed and raised the question whether such approaches are feasible in general for finding phase transitions from data.

It shall also be mentioned that the dimension of the data being used was small, $D \times 4$ in the case of the singular values, such that in this case there was no necessity for dedicated machine learning hardware like graphical processing units (GPUs) and all trainings were performed in less than $10$ seconds on a commercial laptop with an \textsf{Intel} i7-4712HQ CPU, see \cite{notebooks}. 
Here we used the resource economic simple update algorithm to obtain the iPEPS ground states, but we note that the ML approach could also be combined with  more accurate (but computationally more expensive) optimization approaches. 
In summary, we provided a very fast and efficient approach to qualitatively map out the phase diagram of 2D systems with no prior physical knowledge of the underlying system, offering a powerful way to obtain quick insights into the physics of new models. 

\chapter{Variational Quantum Anomaly Detection}
\label{chapter:vqad}
This chapter is taken from our publication \textit{Variational Quantum Anomaly Detection: Unsupervised mapping of phase diagrams on a physical quantum computer} \cite{Kottmann2021b}. We extend the application of anomaly detection to discover phases on a quantum computer, i.e. performing a quantum machine learning (QML) routine on the same device that is simulating the quantum system. The overall concept is very similar to \cref{alg:anomaly_detection-phase_diagram}, described in \cref{chapter:anomaly}, but with some technical subtleties. Most notably, the input and output to the QML routine are quantum states, so comparing them is not straight-forward. We circumvent this problem by finding a different cost function as our anomaly syndrome, as we discuss later in \ref{vqadsec:proposal}. We demonstrate the success of the method for a system with a topologically non-trivial phase in simulation, and for the transverse-field Ising model on a (noisy) physical quantum computer in \cref{vqadsec:results}.

\section{Introduction}

With the rise of deep learning in the 2010s, the term \textit{quantum machine learning} was mostly used to refer to leveraging quantum computers for linear algebra tasks such as matrix inversion in sub-polynomial time via the Harrow-Hassidim-Lloyd algorithm \cite{Harrow2008, Biamonte2017}. One famous use-case was the quantum recommendation system algorithm with an exponential quantum speed-up at the time \cite{Kerenidis2016}, which inspired classical analogs of the algorithm with the same, sub-polynomial, complexity (termed as \textit{quantum-inspired} machine learning algorithms) \cite{Tang2018}. Today, quantum machine learning refers to using quantum circuits as neural networks \cite{Perez-Salinas2020}, or kernel functions \cite{Schuld2021} to perform classical machine learning tasks like supervised learning \cite{Farhi2018,Rebentrost2013}. There are cases, where quantum models have provable advantages over classical models \cite{Liu2020}, but it has been argued that these instances are special cases and no quantum speed up is to be expected for quantum machine learning with classical data \cite{Kubler2021}.

On the other hand, applying classical machine learning to quantum physics has been a great success story \cite{Carleo2019}, most prominently for the classification and mapping of phase diagrams \cite{Carrasquilla2016,VanNieuwenburg2016,Huembeli2018}. These methods rely on classical data and are therefore restricted by the available classical simulation methods. With physical devices surpassing system sizes that are classically tractable \cite{Arute2019}, there is need for methods to investigate physical quantum states with quantum computers.

In this paper, we propose a quantum machine learning algorithm for \textit{quantum data}. The data are ground states of quantum many-body systems that are prepared by a quantum simulation subroutine and serve as the input for \textit{Variational Quantum Anomaly Detection} (VQAD). Our quantum anomaly detection scheme belongs to the category of variational quantum algorithms where the circuit \textit{learns} characteristic features of the input state \footnote{The term \textit{learning} is commonly used in (quantum) machine learning and data-driven problem solving to refer to data-specific optimization.}. This can in principle be leveraged for obtaining physical insights of the system from training \cite{Iten2018} and is in contrast to previous proposals that are based on kernel methods (one-class support vector machines) \cite{Liu2017,Liang2019}. In the present study, we use it to map out an unknown phase diagram of a system without requiring knowledge about the order parameter or the number and location of the different phases.

In anomaly detection, the task is to differentiate \textit{normal} data from \textit{anomalous} data, opposed to supervised learning tasks, where a fixed set of classes with labels for training are differentiated. On the other hand, the task of anomaly detection requires an \textit{anomaly syndrome}, i.e., an observable that is trained to be of a certain value (typically $0$) when \textit{normal} data is input, and be significantly larger for \textit{anomalous} data it is tested on. In classical machine learning, anomaly detection has already been used to extract phase diagrams in an unsupervised fashion from simulated and experimental data \cite{Kottmann2020,Kaming2021, Kottmann2021}. VQAD allows us to perform anomaly detection directly \textit{on} a quantum computer, and, with programmable devices readily available, we demonstrate it experimentally on a real device.

\section{Proposal}
\label{vqadsec:proposal}

\begin{figure}
\centering
    \includegraphics[width=.8\textwidth]{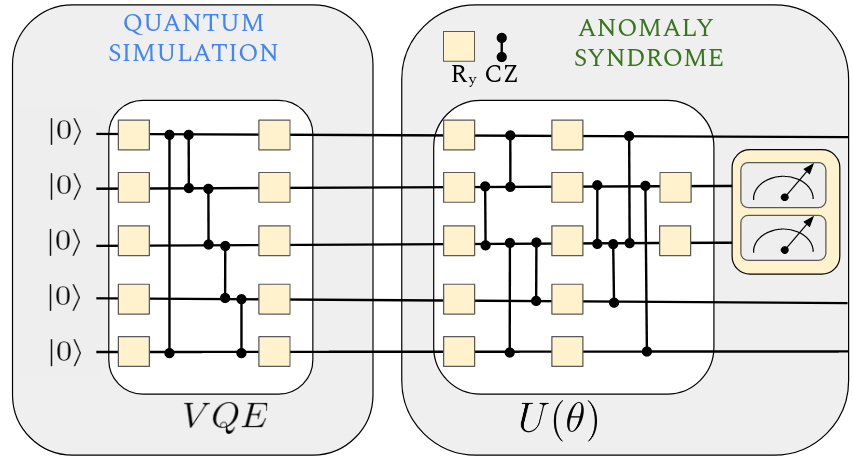}
    \caption{Overview of our proposal. First, the quantum states are prepared via VQE. Then, they are processed through the anomaly syndrome, consisting of a parameterized unitary $U(\theta)$ and a measurement of a subset of qubits, referred to as trash qubits. $R_y$ indicates a parameterized y-axis rotation and $CZ$ a (fixed) controlled-z gate.}
    \label{fig:overview}
\end{figure}

The task of detecting anomalies in ground states of quantum many-body Hamiltonians can be loosely divided into two sub tasks:~Preparing the ground state for specific Hamiltonian parameters, and computing an anomaly syndrome indicating whether the state corresponds to a \textit{normal} example or an anomaly. An overview of our proposed algorithm is shown in \cref{fig:overview}. 
The problem of state preparation on quantum computers is one of ongoing research, and in principle, one can use any state preparation subroutine for preparing the ground state. Here, we choose the Variational Quantum Eigensolver (VQE) as it has the lowest hardware requirements while achieving reliable results on current devices \cite{Peruzzo2014,Kandala2017}. The VQE algorithm iteratively minimizes the expectation value of a Hamiltonian with the ansatz circuit to find the ground state by optimizing the parameters of the circuit via a quantum-classical feedback loop. We choose a minimal ansatz as depicted in \cref{fig:overview} that is sufficient for simulating the Ising Hamiltonian discussed in Sec.~\ref{sec:exp}. A shallow ansatz allows us to run both, the quantum simulation, and the quantum anomaly detection on real noisy devices. For more complex systems, the problem of finding a suitable hardware efficient ansatz can be addressed for example by the adaptive VQE algorithm \cite{2019adaptvqe}. In this work we employed the VQE implementation provided by the Qiskit library \cite{Qiskit} and optimized it using simultaneous perturbation stochastic approximation (SPSA) \cite{Spall1998}. For all technical details we refer to App.~\ref{app}.

\begin{figure}
    \centering
    \includegraphics[width=.5\columnwidth]{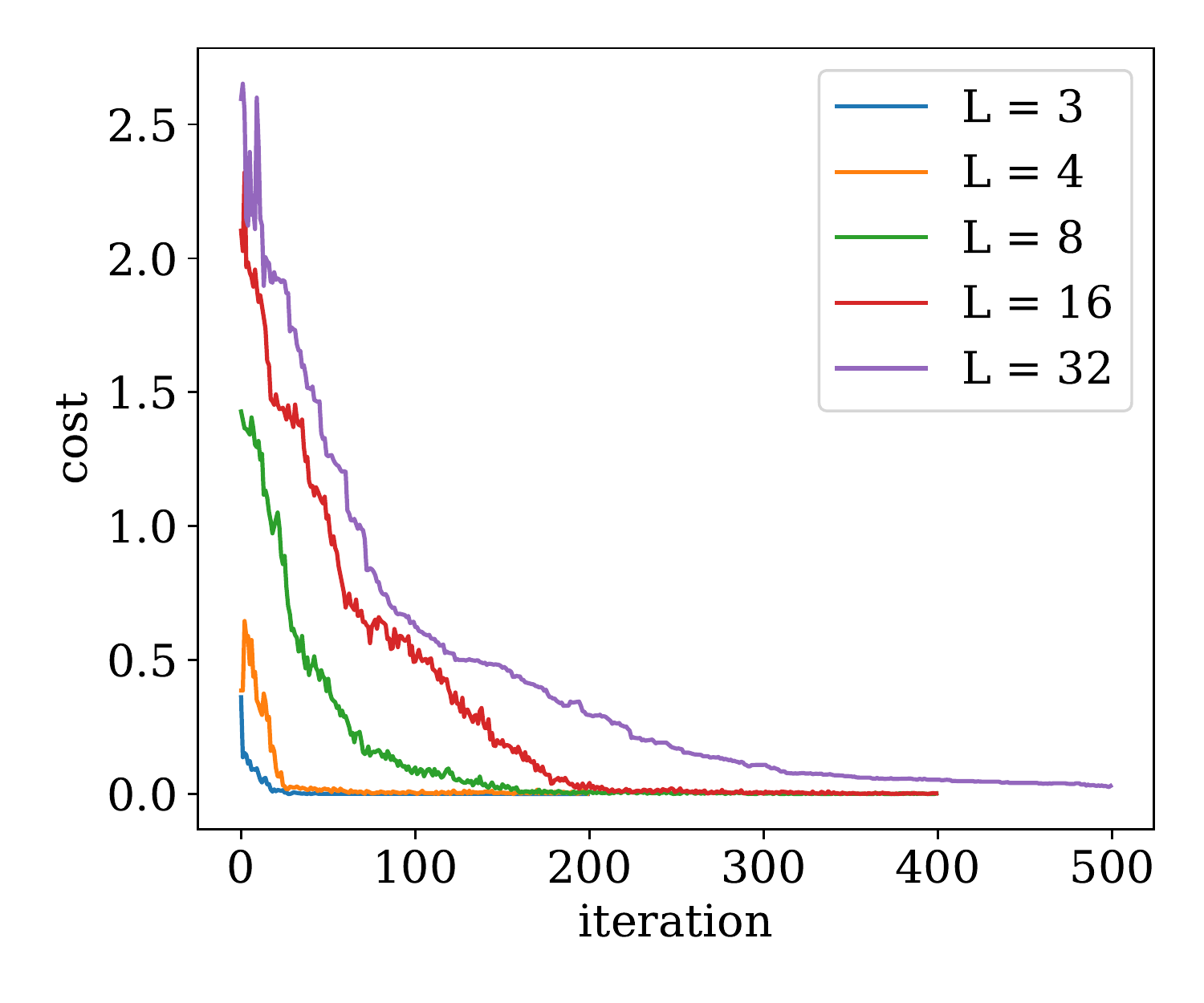}
    \caption{Scaling of the training cost of the anomaly syndrome ansatz. Successful training of the proposed anomaly syndrome ansatz for $L \in \{3,4,8,16,32\}$ corresponding to $n_t \in \{1,2,3,4,5\}$ trash qubits (and therefore $n_t$ circuit layers). The result for $L=32$ was obtained through MPS simulations with a maximal bond dimension of BD = $100$. We used $1000$ shots per evaluation and achieve a perfect cost value of $0.00$ for all system sizes, however the run for $L=32$ shown here finished at $0.03$.
    }
    \label{fig:ansatz_scaling}
\end{figure}

Once the ground state is prepared on the quantum device a subsequent circuit serves as the anomaly syndrome. Our circuit ansatz is inspired by the recently proposed quantum auto-encoder, which similar to its classical counterpart can be used for compression of classical and quantum data \cite{Romero2017,Prieto2021}. It is composed of several layers each consisting of parameterized single qubit y-rotations and controlled-z gates. After the final layer a predefined number $n_t$ of \textit{trash} qubits is measured in the computational basis. The objective is to decouple the trash qubits from the rest of the system, effectively compressing the original ground state into a smaller number of qubits. The circuit parameters are then optimized to faithfully compress states that are considered \textit{normal}. However, when the optimized circuit is tested on anomalous states not seen during training, it is expected that the circuit fails to decouple the trash qubits from the rest of the system. To quantify the degree of decoupling we use the Hamming distance $d_{H}$ of the trash qubit measurement outcomes to the $|0\rangle^{\otimes n_{t}}$ state, i.e., the number of 1s in a bit-string of measurement outcomes \cite{Prieto2021}. The cost function $C$ can then be defined as the Hamming distance averaged over several circuit evaluations $C = 1/N\sum_i^N d_{Hi}$, where $N$ is the number of performed measurements or shots. The cost function can also be rewritten in terms of expectation values of local Pauli-z operators $Z_j$
\begin{equation}
	C = \frac 1 N \sum_{i=1}^N d_{Hi} = \frac{1}{2} \sum_{j=1}^{n_{t}}\left(1-\left\langle Z_{j}\right\rangle\right) .
\end{equation}
The VQAD circuit achieves perfect compression if the trash qubits are fully disentangled from the remaining qubits and mapped into the pure $|0\rangle^{\otimes n_{t}}$ state resulting in a cost equal to zero.

The specific circuit ansatz for the anomaly syndrome is shown in \cref{fig:overview} for the case of $n_t=2$ trash qubits. Each layer of the circuit starts with parameterized single-qubit y-rotations applied to every qubit followed by a sequence of entangling controlled-z gates. The currently available NISQ devices are inherently noisy and the computations are subject to gate errors. To minimize the number of two-qubit gates we apply the controlled-z gates only between trash qubits and non-trash-qubits as well as between trash qubits themselves instead of an all-to-all entangling map \cite{Prieto2021}. This entangling map is physically motivated as the goal of the circuit is to disentangle the trash qubits from the rest, with the trash qubits resulting in the $|0\rangle^{\otimes n_{t}}$ state. In a single layer each non-trash qubit will be coupled to exactly one trash qubit. This entangling scheme is repeated in the subsequent layers until every non-trash qubit has been coupled to each trash qubit exactly once, i.e. the number of layers of the circuit is equal to $n_t$. After the final layer, additional single-qubit y-rotations act on the trash qubits before they are measured.

Barren Plateaus are the fundamental obstacle prohibiting training of variational circuits with increasing numbers of qubits \cite{McClean2018}. It was previously shown that using local cost functions and circuits featuring a number of layers scaling at most logarithmically in the system size can prevent the occurrence of Barren Plateaus \cite{Cerezo2020}. Additionally for realistic devices, gate errors lead to decoherence, making quantum simulation on real devices a challenging task even for small systems and low depths \cite{CerveraLierta2018}. The former calls for a minimal number of layers while the latter calls for a minimal number of gates overall. Therefore, we seek a minimal solution for our variational circuit that we want to implement on a readily available NISQ-era quantum computer. On the other hand, it is desirable to have an ansatz as general as possible to be able to capture a wide range of problems (see \textit{circuit complexity} \cite{Bernstein1997,Brandao2019}).

For the anomaly syndrome in this paper, we propose an ansatz that aims at compromising between being general enough to compress the ground states of the investigated systems while still being trainable. One way to make our circuit scalable for larger systems is to choose the number of trash qubits $n_t = \floor*{\log_2 L}$, where $L$ is the total number of qubits. Together with the fact that our cost function is composed of only local operators, the training is expected to not suffer from Barren Plateaus. We empirically confirm successful trainability, i.e., achieving a cost of $0.0$ for ground states of the systems discussed later in the manuscript, for $L \in \{3,4,8,16, 32\}$, corresponding to $n_t \in \{1,2,3,4,5\}$, respectively. In \cref{fig:ansatz_scaling}, a ground state of the Ising model \cref{IsingHam} at $g_x, g_z, J = (0.3,0,1)$ is taken as a realistic example and we can confirm successful trainability in all cases.

Note that in principle the trash qubits can be placed anywhere in the circuit, however, when performing computations on a real quantum device it proved advantageous to explicitly take the qubit connectivity structure of the device into account in order to reduce the number of required SWAP operations. Specifically here, we placed the trash qubits in the middle of the IBMQ devices.



\begin{figure*}
	\centering
	\includegraphics[width=\textwidth]{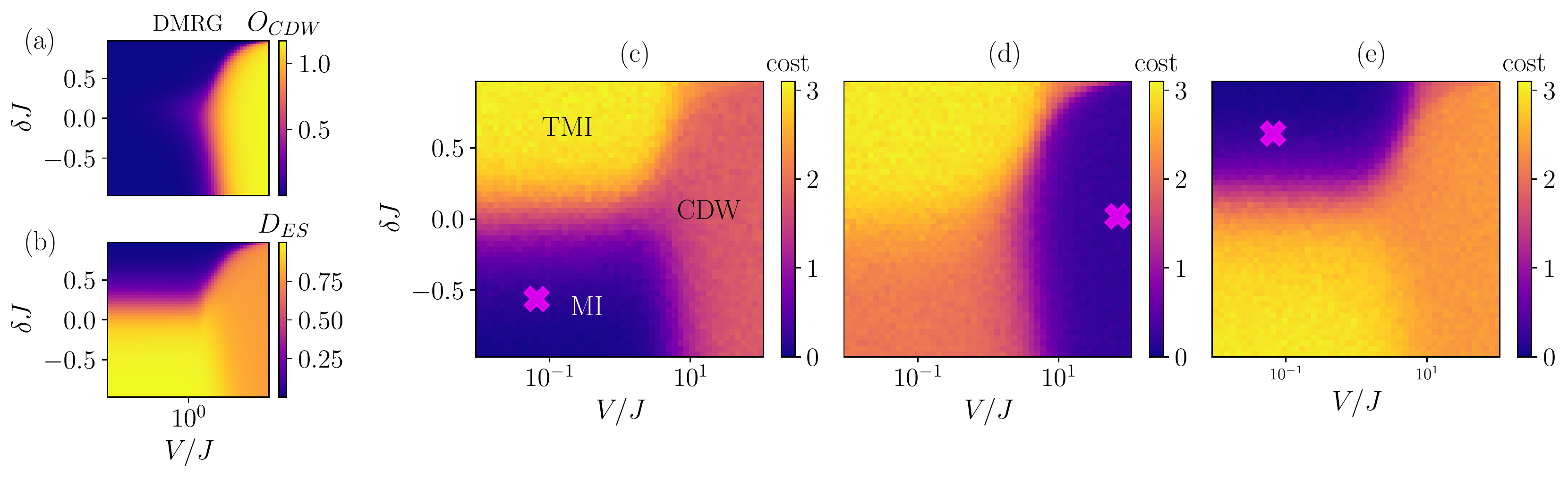}
	\caption{(a)-(b) Phase diagram of the DEBHM from Eq.~\eqref{eq:ham_bh} using (a) the order parameter $O_{CDW}$ defined in Eq.~\eqref{eq:OCDW}, and (b) the degeneracy of the entanglement spectrum, $D_{ES}$, defined in Eq.~\eqref{D_ES}. The results were obtained from DMRG simulations for a system of length $L=12$ at half filling $\bar{n} = 0.5$. We fix the maximum bond dimension $BD = 50$ and the maximum number of bosons per site to $n_0 = 1$. (c)-(e) Cost/anomaly syndrome of a VQAD trained on a single ground state (indicated by a cross) of the $L=12$ DEBHM using $n_t=6$ trash qubits in the (c) MI phase, (d) CDW phase, and (e) TMI phase. The cost at each data point is the Hamming distance averaged over 1000 measurement shots using an ideal quantum device simulator.}
	\label{fig:bh}
\end{figure*}

The training and inference procedure is identical to the classical anomaly detection schemes for mapping out phase diagrams \cite{Kottmann2020}. In the first step, one randomly chooses a training region in the phase diagram that represents \textit{normal} data, which is an arbitrary definition. Note that no prior knowledge about the phase diagram is therefore required. The circuit representing the anomaly syndrome is then trained on ground states of the training region, and tested on the whole phase diagram. States in the same phase as the training data are \textit{normal} and can be disentangled, leading to a low cost. \textit{Anomalous} states can be inferred through an increase in the cost function signaling that the corresponding ground state cannot be disentangled by the optimized circuit. From the resultant cost profile, we can deduce the phase boundary between the phase the circuit has been trained on and any other phases in the diagram. 
This procedure is then repeated by training in the anomalous region from the previous iteration until all phase boundaries are found. An example is provided in \cref{fig:bh}.

Anomaly detection is a semi-supervised learning task. The setting is typically that one is provided with one class of data that is well known, \textit{normal} data, and aims at finding outliers of that distribution, \textit{anomalous} data. An archetypical example is credit card fraud where a big database of normal transactions is provided and one aims at finding fraudulent ones. We consider anomaly detection semi-supervised as labeled data (x, “normal”) is provided for training while (x, “anomalous”) is to be inferred. Here, however, we arbitrarily define (x, “normal”) and iteratively find the different classes (phases of matter). The definition of (x, “normal”) is arbitrary and does not necessitate prior knowledge. Furthermore, it is merely a means to an end to find the different classes. In that sense, the way anomaly detection is used to map out the phase diagram can be regarded as an unsupervised learning method.

Note that in previous works, where the same task has been tackled with classical machine learning techniques, it has been shown that a single ground state was sufficient to successfully train the model \cite{Kottmann2021}. This feature stems from the fact that ground states within the same phase share similar properties and there is very little variance when changing the physical parameters inside one phase. We observe this feature also in the training of the VQAD.

\section{Results}
\label{vqadsec:results}

\subsection{Simulations with ideal quantum data}
\label{sec:sim} 
In order to test the performance of VQAD, we first study the one-dimensional extended Bose Hubbard model with dimerized hoppings (DEBHM) \cite{sugimoto19},
 \begin{eqnarray}\label{eq:ham_bh}
 H&=&-\sum_{i=1}^{L-1}(J+\delta J(-1)^i) (b_i^\dagger b_{i+1} + \text{h.c.})+ \nonumber \\
 &&+\frac{U}{2}\sum_i^L n_i(n_i-1)+V\sum_{i}^{L-1} n_i n_{i+1},
\end{eqnarray} 
where $b^\dagger_i(b_i)$ is the bosonic operator representing the creation (annihilation) of a particle at site $i$ of a lattice of length $L$. The tunneling amplitudes $J-\delta J$ $(J+\delta J)$ indicate hopping processes on odd (even) links connecting nearest-neighbor sites, while $V$ represents the nearest-neighbor (NN) repulsion. Here, we take the hardcore boson limit, i.e. the on-site repulsion $U/J \rightarrow \infty$, such that the local Hilbert space is two-dimensional and each site can only accommodate $0$ or $1$ bosons. This model can be effectively mapped into a spin-$1/2$ system \cite{tao13}.

Previous studies of the DEBHM model at half filling ($\bar{n} = 0.5$) have demonstrated the existence of three distinct phases \cite{sugimoto19}. For small and intermediate values of $V/J$ and $\delta J>0$, we find a topological Mott insulator (TMI) displaying features analogous to a symmetry protected topological phase appearing in the dimerized spin-1/2 bond-alternating Heisenberg model \cite{tao13}. On the other hand, for negative values of $\delta J$ we expect a trivial Mott insulator (MI), while in the regime where the nearest-neighbor repulsion dominates, a charge density wave (CDW) appears.

In \cref{fig:bh}(a)-(b), we study the phase diagram of the model in Eq.~\eqref{eq:ham_bh} in terms of the parameters $\delta J$ and $V/J$, using the density matrix renormalization group algorithm (DMRG) \cite{schollwock11,white92, tenpy}. In order to differentiate between the Mott insulating phases and the CDW, one can compute the CDW order parameter
\begin{eqnarray}\label{eq:OCDW}
O_{CDW} = \sum_{i=1}^{L/2}(-1)^{i} \delta n_i,
\end{eqnarray} 
which detects staggered patterns in the density. In \cref{fig:bh}(a) we report a vanishing value of $O_{CDW}$ everywhere but in the region with large values of $V/J$, which corresponds to the CDW \footnote{In the definition of $O_{CDW}$, we consider only half of the sites of the system because the DMRG algorithm outputs a symmetric state, which is a superposition of the two degenerate ground states.}. To characterize the TMI we study the entanglement spectrum (ES), which is expected to be doubly degenerate in a topologically non-trivial phase \cite{pollmann10} due to the existence of edge states. The entanglement spectrum $\{\lambda_i\}$ is defined in terms of the positive real-valued Schmidt coefficients $\{\alpha_i\}$ of a bipartite decomposition of the system by $\alpha_i^2 = \exp(-\lambda_i)$. We determine its degeneracy using
\begin{equation}\label{D_ES}
    D_{ES} = \sum_{i} (-1)^i e^{-\lambda_i}.
\end{equation}
In \cref{fig:bh}(b), we show that the quantity $D_{ES}$ vanishes only for small NN interaction strengths $V$ and positive values of $\delta J$, which correponds to the TMI. The trivial MI and CDW phases do not show a degeneracy and hence do not host topological edge states.
%

In the following, we test the capabilities of the VQAD with ideal states obtained from DMRG simulations. The anomaly syndrome is trained using a single representative ground state within one of the phases such that the cost measured at the trash qubits is minimised and the states of this phase can be efficiently compressed by the circuit. Afterwards, the trained circuit processes all states from the full phase diagram, ideally with similarly low cost in the same phase and significantly higher cost in other phases.

In \cref{fig:bh}(c)-(e) we show the resultant cost diagram for three circuits, each optimized at a different point in the phase diagram. Indeed, ground states outside of the training phase give rise to a large cost and hence are correctly classified by the VQAD as anomalous. Surprisingly, a single ground state example (indicated by the cross) was sufficient to successfully train the VQAD and infer all three phases. Similar results were recently reported for the case of classical anomaly detection using neural network auto-encoders \cite{Kottmann2021}.

\begin{figure}
    \centering
    \includegraphics[width=.77\textwidth]{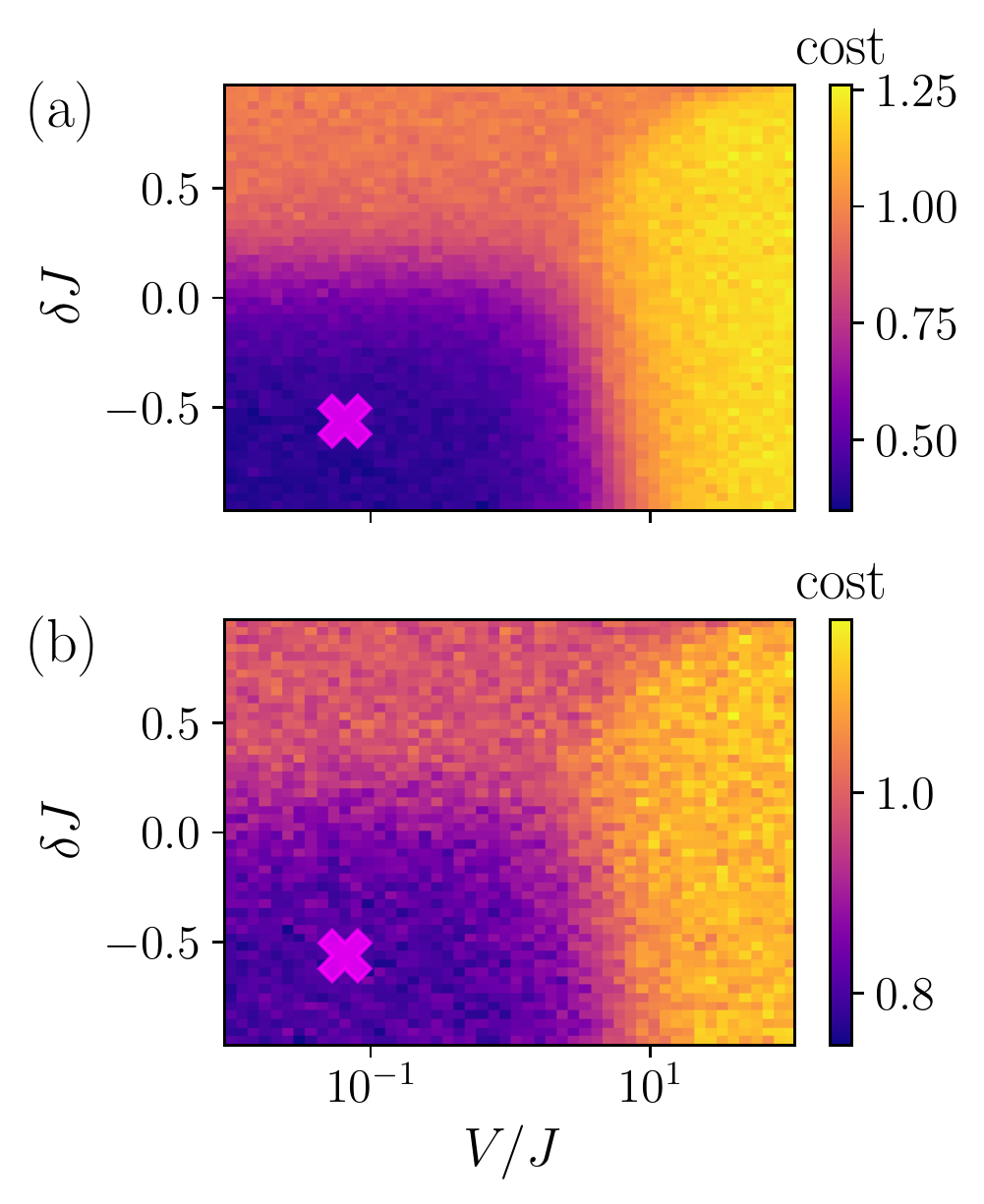}
    \caption{Cost of a VQAD trained on a single ground state in the MI phase (marked by the cross) of the DEBHM with $L=12$ sites and $n_t=2$ trash qubits. The gates of the VQAD circuit are subject to depolarizing noise with $p_{\text{err}}=0.001$ (single-qubit gates) and (a) $p_{\text{err}}=0.01$, (b) $p_{\text{err}}=0.07$ (two-qubit gates). The chosen values are motivated by the error probabilities of real devices.}
    \label{fig:bh_noisy}
\end{figure}

To demonstrate the robustness of the VQAD against noise present in currently available NISQ devices we apply a depolarizing noise channel after each gate with error probabilities $p_{\text{err}}=0.001$ (single-qubit gates) and $p_{\text{err}}=0.01,0.07$ (two-qubit gates) and show two exemplary cost profiles of the trained anomaly detector in \cref{fig:bh_noisy}. Since the noise becomes more prominent with larger circuit depths, we used the two-layer VQAD circuit ansatz with only two trash qubits in this case. While it is not possible to reach a cost of zero in the training phase, the optimization still converges and all three phases can be successfully inferred. Hence, this suggests that even if the VQAD is not able to fully disentangle the trash qubits, the phase diagram can still be recovered from the resultant cost profile.

\subsection{Experiments on a real quantum computer}
\label{sec:exp} 
We have seen that with ideal quantum data, VQAD can map out non-trivial phase diagrams including topologically non-trivial phases with and without noise in the anomaly syndrome. Next, we discuss its performance in real-noise simulations, that is with noise profiles and qubit connectivities from a real quantum device. Furthermore, we perform the quantum simulation subroutine, i.e., the ground state preparation via VQE, on the same circuit.
For this task, we consider the paradigmatic transverse longitudinal field Ising (TLFI) model \cite{fogedby78}
\begin{equation}
	H=J\sum_{i=1}^{L} Z_{i} Z_{i+1}-g_{x} \sum_{i=1}^{L} X_{i}-g_{z} \sum_{i=1}^{L} Z_{i},
	\label{IsingHam}
\end{equation}
where $X_{i},Z_{i}$ are the Pauli matrices on site $i$, $J$ is the coupling strength, and $g_x, g_z$ are the transverse and longitudinal fields, respectively. For $g_z=0$ the model is exactly solvable and shows a quantum phase transition from a ferromagnetic (antiferromagnetic) phase for $g_x/J<1$ and $J$ negative (positive) to a paramagnetic one for $g_x/J>1$ \cite{sachdevqpt}. In the following we set $J=1$ and vary the longitudinal and transverse fields. In this regime the model is not exactly solvable and the phase diagram has been extensively studied numerically \cite{sen00,ovchinnikov03}. The antiferromagnet-paramagnet quantum phase transition is best characterized by the order parameter which in this case is the staggered magnetization
\begin{equation}
\hat{S}=\sum_{i=1}^L (-1)^i \frac{Z_i}{L}.
\end{equation}

We simulate the ground states of the Hamiltonian in \cref{IsingHam} using VQE for $L=5$. On a noisy device, long-range entangling gates are performed by consecutive local two-qubit gates (SWAP operation), increasing the actual circuit depth. A large number of consecutive gates leads to decoherence due to gate errors and destroys the results. With the circuit presented in \cref{fig:overview} for the VQE subroutine, we found a trade-off between expressibility and noise tolerance with a circular entanglement distribution and only one layer. Additionally, we performed measurement error mitigation \cite{Bravyi2021}, which can further improve the results of the cost function as seen in \cref{fig:mitigation} in App.~\ref{app}.

\begin{figure}
    \centering
    \includegraphics[width=.77\textwidth]{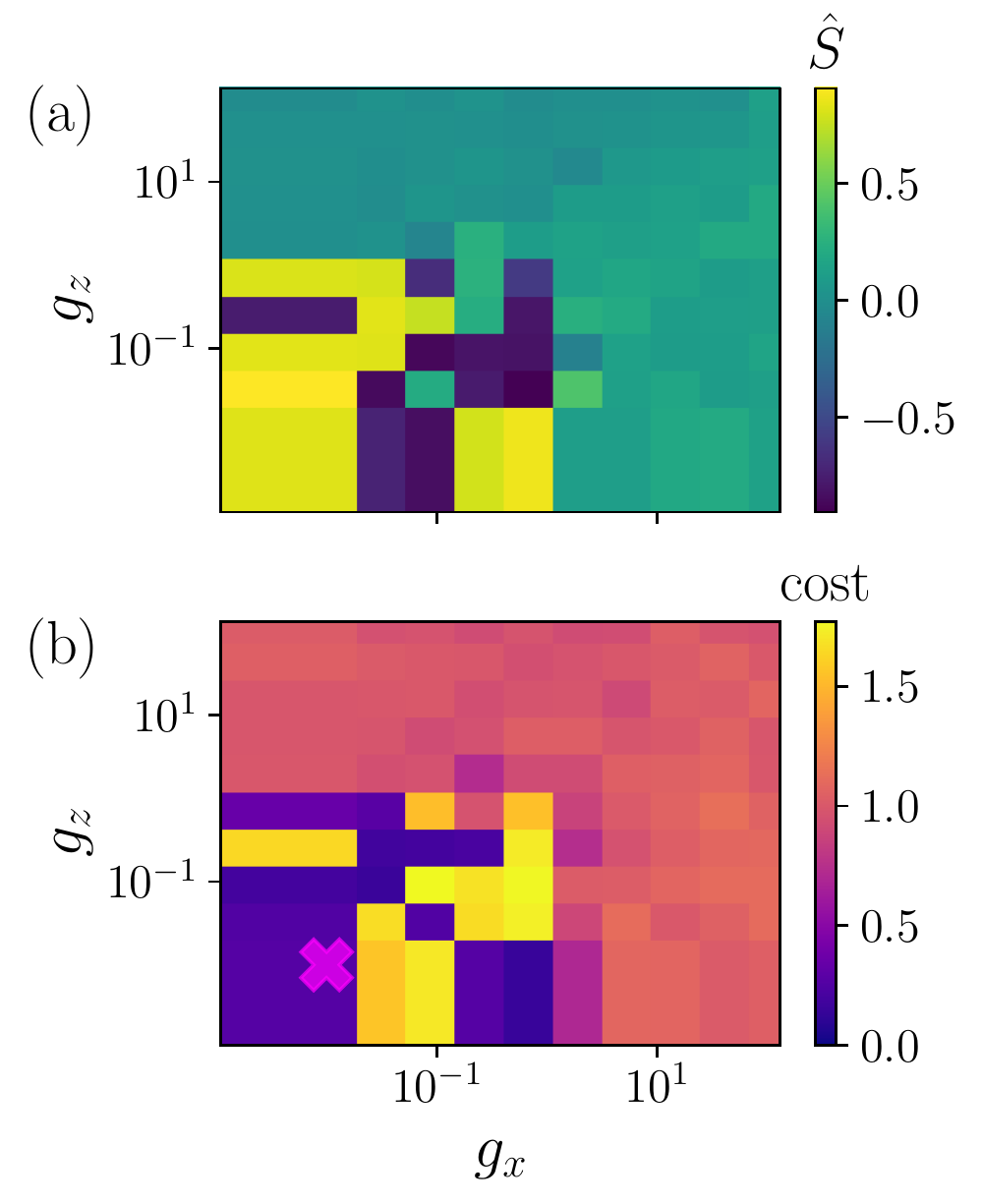}
    \caption{Real-noise simulations of the staggered magnetization $\hat{S}$ (a) and the anomaly syndrome (b) for the TLFI model. We trained the anomaly syndrome in the ordered phase on a state with positive $\hat{S}$, indicated by the purple cross. Inside the ordered phase, there is a perfect correlation between low cost states for positive $\hat{S}$, and very high cost where VQE converged to a negative $\hat{S}$. The paramagnetic phase is detected by a plateau in the anomaly syndrome.}
    \label{fig:antiferro2D}
\end{figure}

For small values of $g_x$ and $g_z$, in the ferromagnetic ordered phase, the ground states $\psi\simeq\ket{10101}$ ($\langle\hat{S}\rangle=1$) and $\psi\simeq\ket{01010}$ ($\langle\hat{S}\rangle=-1$) have a similar energy, which is why the optimization can get stuck in local minima. Hence, in the ordered phase, VQE can converge to both a state with positive or negative staggered magnetization, or an equal superposition of the two as can be seen in \cref{fig:antiferro2D}(a). The VQAD simulation results in \cref{fig:antiferro2D}(b) show a perfect correlation between positive $\braket{\hat{S}}$ and low cost, and vice versa, negative $\braket{\hat{S}}$ and high cost - which, intuitively,  can be expected \footnote{In a very hand-wavy way, we can understand this as we train the circuit $U$ to perform $U \ket{10101} = \ket{\Psi} \otimes \ket{00}_\text{trash}$ such that $U \ket{01010} = \ket{\Psi} \otimes \ket{11}_\text{trash}$ if we input a state with opposite ordering.}. The disordered phase is detected from the plateau of high cost ($\sim 1$).

\begin{figure}
    \centering
    \includegraphics[width=\columnwidth]{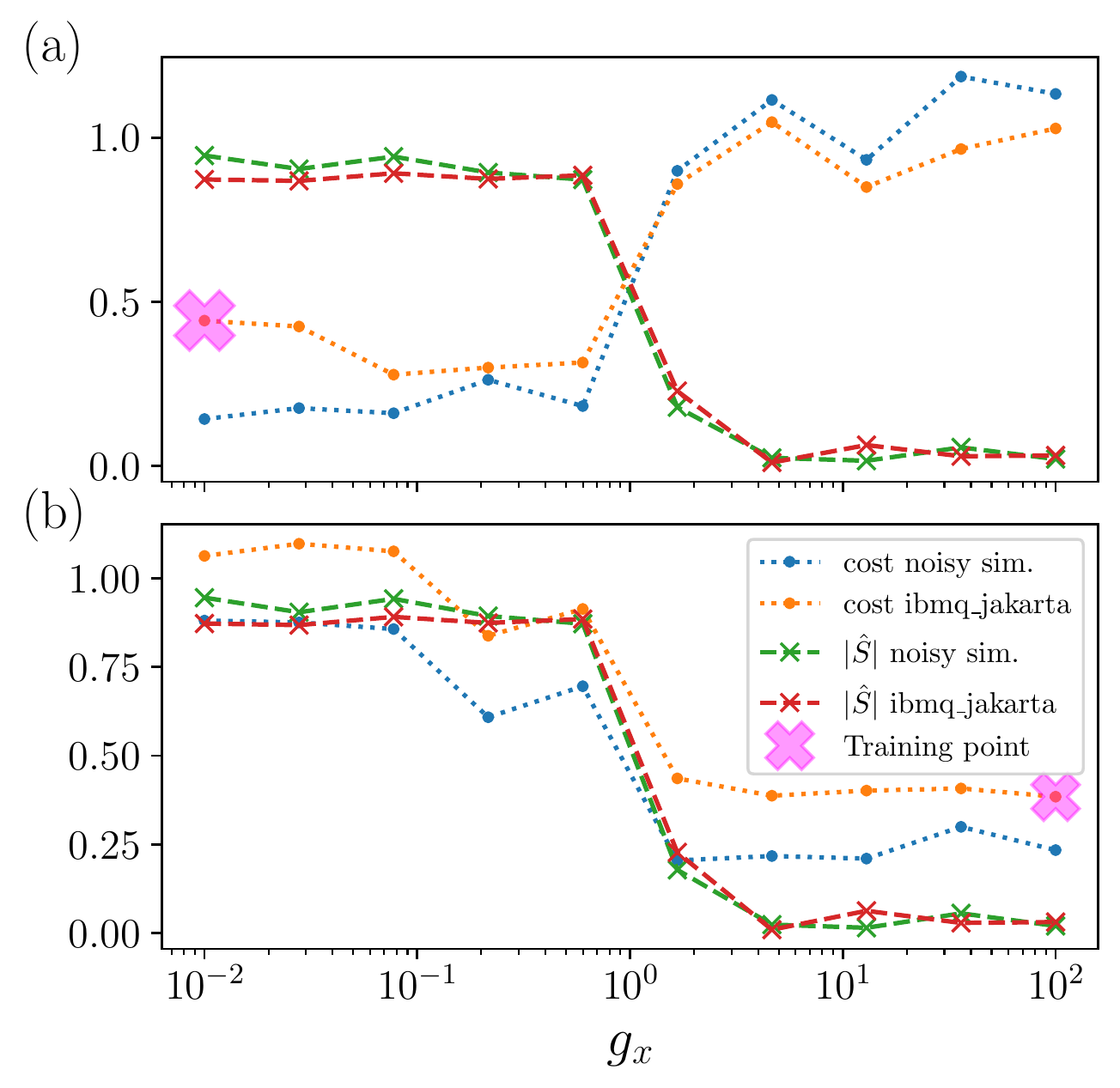}
    \caption{Real device VQAD experiments:
    We show the order parameter $\hat{S}$ compared to the VQAD results both for execution on \code{ibmq_jakarta} and noisy simulators with the same noise profile. We trained on a single ground state in the ordered (a) and paramagnetic (b) phase. 
    For sampling $\hat{S}$, we use the same parameters for the VQE circuit in simulation and experiment. All values for $\hat{S}$ in the paramagnetic phase are negative, hence, for better visualization we plot its absolute value $|\hat{S}|$.
    For training the anomaly syndrome, the optimized parameters from the simulation are taken as an initial guess.
    }
    \label{fig:ibmq_main}
\end{figure}

We see that VQAD also performs well under realistic conditions, so we next test the algorithm on a physical device. For this task, we use the $L=5$ qubits on \code{ibmq_jakarta} \cite{Bravyi2021}. To avoid jumps in the staggered magnetization in the ordered phase and improve convergence of the VQE optimization, we reuse already optimized parameters at neighboring points in the phase diagram as a good initial guess. Due to a large computation time overhead per execution on the real device, we additionally prepared pre-optimized parameters for both subroutines from a realistic noisy simulation, and use these as initial guesses for the optimization on the device. We found that for computing the staggered magnetization it is actually not necessary to re-run the VQE optimization on the physical device, and we can achieve faithful results by directly using the optimized parameters from the simulation as seen in \cref{fig:ibmq_main}. The resulting cost values for the optimized circuit, plotted in \cref{fig:ibmq_main}, clearly distinguish the two phases, with the cost from the experiment showing solely an almost constant offset compared to the noisy simulation.

\section{Outlook}

We showed that our proposed algorithm is capable of mapping out complex phase diagrams, including topologically non-trivial phases. We further demonstrated that the algorithm also works in realistic scenarios for both real-noise simulations and on a real quantum computer. Hence, we provide a tool to experimentally explore phase diagrams in future quantum devices, which will be especially useful when physical devices surpass the limit of what can be classically computed.

Currently, the main bottleneck of VQAD is the presence of noise in real devices. We were able to improve our anomaly detection scheme by employing measurement error mitigation and adopting the circuits according to the physical device. These results are promising, and with current efforts on enhancing device performances, error mitigation and circuit optimization strategies in the community, we are hopeful to see even further improvements soon.

In this work we focused on using VQAD to extract the phase diagram of quantum many-body systems. A possible future extension would be to apply it to the problem of entanglement witnessing and certification in many-body scenarios without tomography. Furthermore, the use of an autoencoder-like architecture has the advantage over kernel-based schemes in that there exists tools of interpreting the feature space in classical autoencoders to gain physical insights \cite{Iten2018}, which can be a possible future extension for the quantum case discussed here.

\section{Technical details}
\label{app} 

\begin{figure}
    \centering
    \includegraphics[width=.75\columnwidth]{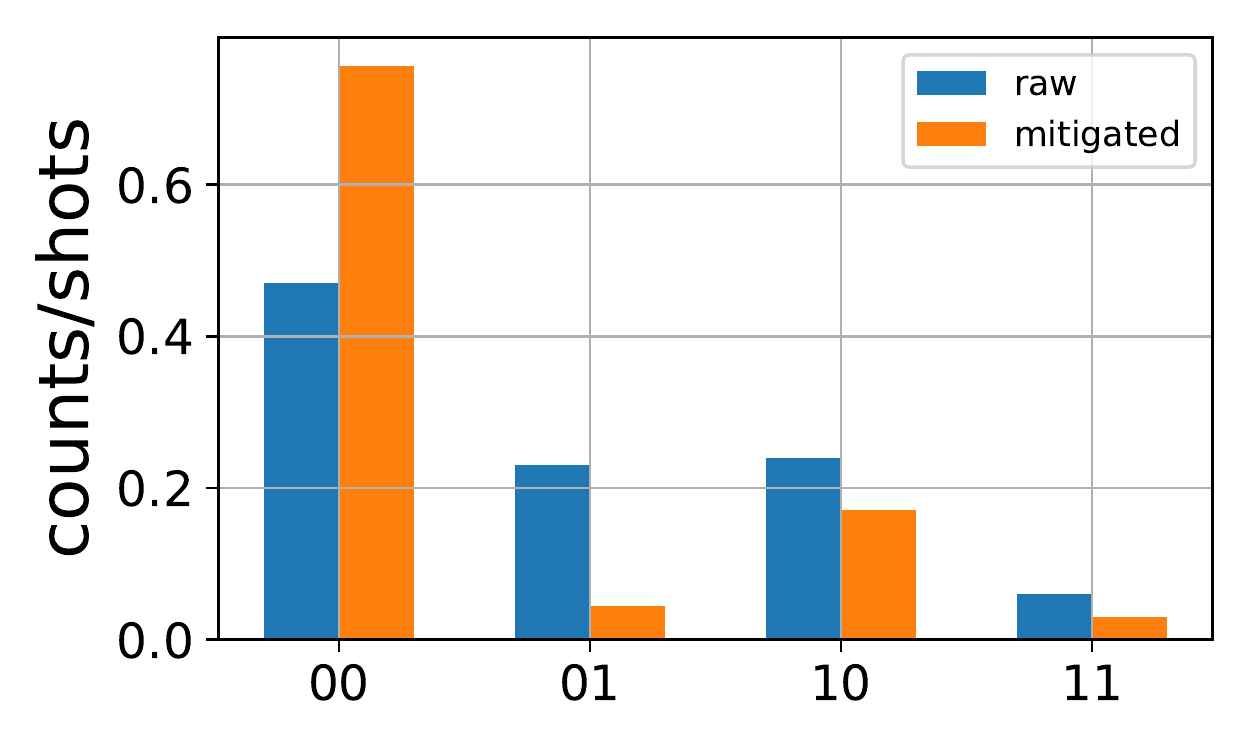}
    \caption{Comparison of the trash qubit measurement outcomes with and without measurement error mitigation. The anomaly syndrome circuit has been trained with and without error mitigation on a ground state of the TLFI model in the ordered phase in real-noise simulations. Ideally, all of the $1000$ shots would result in the $00$ bit string. By mitigating the measurement errors we improve the results towards this desired outcome.}
    \label{fig:mitigation}
\end{figure}
The code to run the simulations and experiments discussed in the main text can be found in our repository on \textsf{GitHub} \cite{github}. The optimization of the circuit parameters was performed using simultaneous perturbation stochastic approximation (SPSA) \cite{Spall1998,Kandala2017}.
To obtain the results presented in \cref{fig:bh} of Sec.~\ref{sec:sim}, a VQAD circuit ansatz composed of 6 layers (6 trash qubits) was employed resulting in $6L+6$ parameters. For the noisy simulations and real-device execution discussed in Sec.~\ref{sec:exp}, we used the ansatz in \cref{fig:overview}, counting $2L$ and $2L+2$ parameters for the quantum simulation and anomaly syndrome, respectively. In classical real-noise simulations, we used $500$ VQE optimization iterations for the initial ground state optimization, and $200$ iterations for all subsequent optimizations where the previously optimized parameters were taken as initial guesses. For the anomaly detection circuit, we found converged results with less than $100$ optimization iterations. As an example, calculating the expectation value of the magnetization takes roughly $2-10$ seconds on a commercial laptop (here: i7-4712HQ), while the real-device execution takes about $30$ seconds. Furthermore, we used measurement error mitigation \cite{Bravyi2021} provided by the Qiskit library to improve the results of the VQAD simulations in the presence of noise as illustrated in \cref{fig:mitigation}.

\part{Conclusion}
\label{conclusions}
\chapter{Discussion}

We have provided a tool to study phase diagrams of quantum many-body systems in an unsupervised fashion requiring no prior physical knowledge about the phases of the Hamiltonian. We demonstrate this on various different platforms and models.

Foremost, we have employed matrix product states to simulate the extended Bose Hubbard model. We were able to reproduce the full phase diagram at integer filling, including the mostly overlooked phase-separating phase between a supersolid and a superfluid part of the system. This led us to further physical investigations of this region and we found previously unknown features of this model with a homogeneous superfluid and supersolid with hidden broken translational symmetry. Even though this hidden broken symmetry can be unraveled with the same string observables that are typically employed for symmetry protected topological phases, we provide convincing evidence that this effect is most likely not of topological nature. Still, it would be very interesting to observe these phases of matter experimentally to confirm our findings. Continued analytical, numerical and experimental investigations may shed light on the nature of this new effect.

Further, we demonstrate the viability of this anomaly detection approach with data from projected entangled pair states of a two dimensional system. We show that even though the singular values between nodes do not have a clear physical interpretation like the entanglement spectrum for matrix product states, they can still be used to infer information about the quantum phase of matter using machine learning. A notable observation is that a training set of just one single data point suffices for this training. This on the other hand suggests that deep learning might be superfluous when training with singular values from classically simulated quantum states, which are harder to obtain experimentally. A natural question to ask is for which phases and which data types purely geometric approaches struggle or generally fail and deep learning becomes a necessity. Answering this question in general is most likely difficult, but testing with a variety of examples may already give a lot of insights.

Finally, we extend this method to the quantum machine learning (QML) realm. Here, quantum data in form of simulated ground states serve as input to our  QML routine, where a parametrized quantum circuit is optimized in a quantum-classical feedback loop to disentangle the \textit{trash qubits} of the system from the rest of the state. This data-specific disentangling scheme can then be used in the same fashion as an anomaly syndrome when states of different structures, i.e. from different quantum phases of matter, are input. We demonstrated this both in noisy simulation, showing scalability up to 32 qubits and on a real quantum device with 5 physical qubits. The most obvious continuation of this is scaling it up to larger system sizes, i.e. employing tensor network methods for simulation. The viability of variational quantum algorithms is heavily debated due to the onset of barren plateaus for larger system sizes. The complexity (depth) of an Ansatz to partially disentangle the input state is unclear, so it would be interesting to test this for larger system sizes and see whether there is a trade off between trainability and complexity for disentangling the state.

\

Our initial vision was to have an artificial intelligence go through a catalog of known and unknown quantum Hamiltonians and point out interesting effects. We here provided a method to map out phase diagrams in an unsupervised fashion. There are several extensions of this work that may one day let this initial dream come to fruition. While we provide a general procedure with \cref{alg:anomaly_detection-phase_diagram}, in practice we perform all these steps by hand. Encapsulating this procedure in a truly automated fashion would be a very interesting and useful engineering task. Moreover, this method points out phases which can then be compared with a known catalog of phases to spot novelties that may have previously been overlooked. Setting up such a catalog would be a matter of collecting information from publications of the last $\sim 30$ years, for which natural language processing tools and web scrapers could be useful.
After all, a fully automatized artificial intelligence agent that performs these tasks with no human interference seems out of sight since the two most crucial parts, quantum simulation and interpretation, still necessitate expert knowledge. Still, setting up such a framework can vastly reduce the required work per person and enable accelerated scientific discovery.

\cleardoublepage

\phantomsection

\addcontentsline{toc}{chapter}{Bibliography}

\bibliographystyle{SciPost_bibstyle}
\bibliography{lit.bib} 


\appendix 

\chapter{Additional information on the Superfluid-Supersolid phase separation phase and its surroundings}
\label{appendix:supersolid}

\section{Phase Separation checks}
\label{appendix:PS-size-scaling}
\begin{figure}
\centering
\includegraphics[width=.77\textwidth]{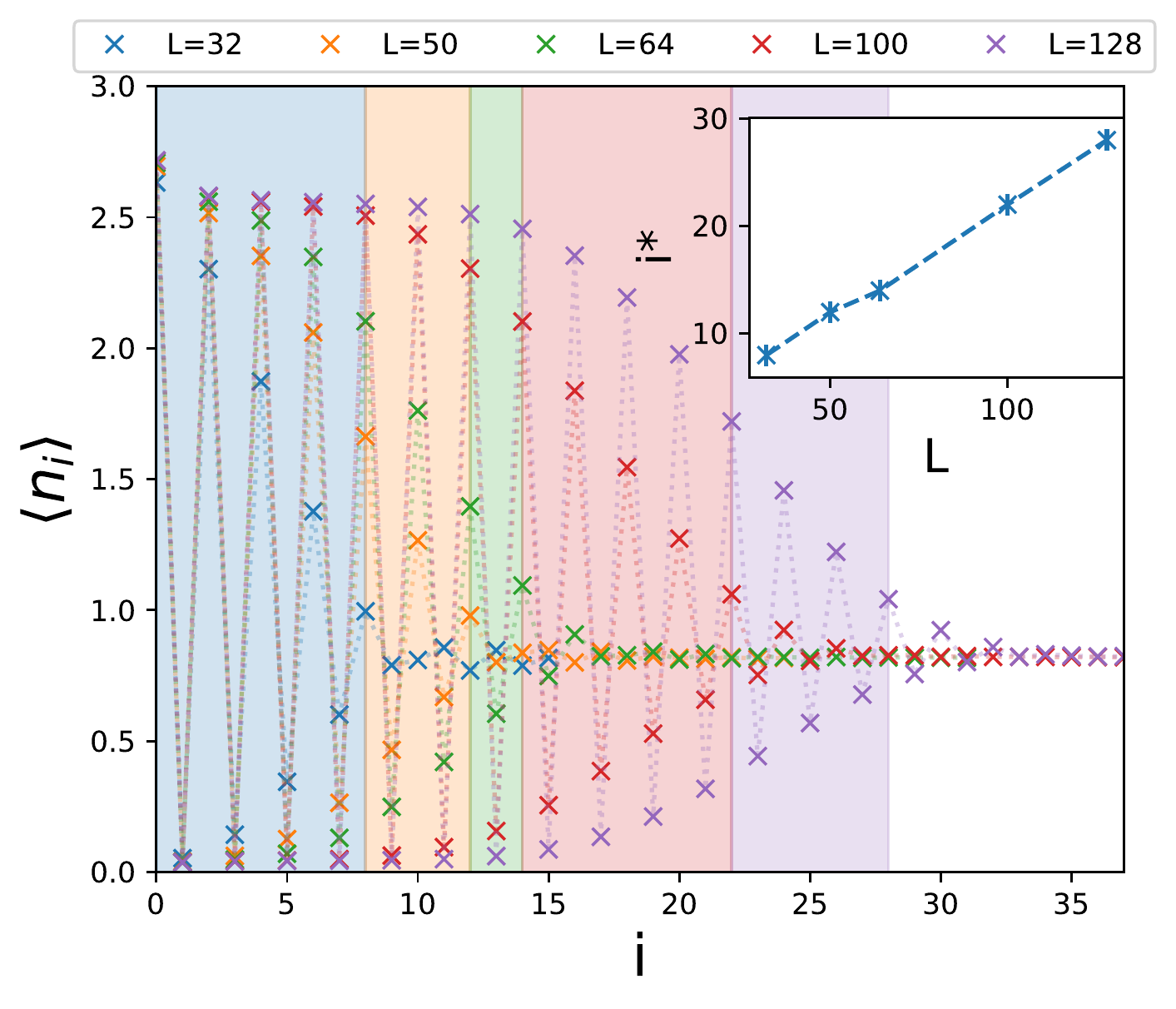}
\caption{
Linear dependence of the extent of the supersolid part in the phase-separated phase in system size. The physical parameters $(U,V,n) = (0.5,4,1)$ are fixed and the bond dimension increases with the system size with $\chimax \in [400,1000]$. The extent of the SS part $i^*$ is indicated by the colored region and estimated by calculating the maximum of the second derivative of $\braket{n_i}$ with $i$ even. The inset shows $i^*$ vs system size $L$ showing a linear dependence as expected.
}
\label{fig:appendix-PS-size-scaling}
\end{figure}

One important check for the phase separation phase is to look at how the size of the respective phases scales with $L$. In \cref{fig:appendix-PS-size-scaling} we show the extent of the supersolid phase on the boundary for different system sizes and find a linear dependence, as expected. I.e., this rules out the supersolid part being a boundary effect.

\section{Local Hilbert space dimension}
\label{appendix:d-scaling}
\begin{figure}
\centering
\includegraphics[width=.77\textwidth]{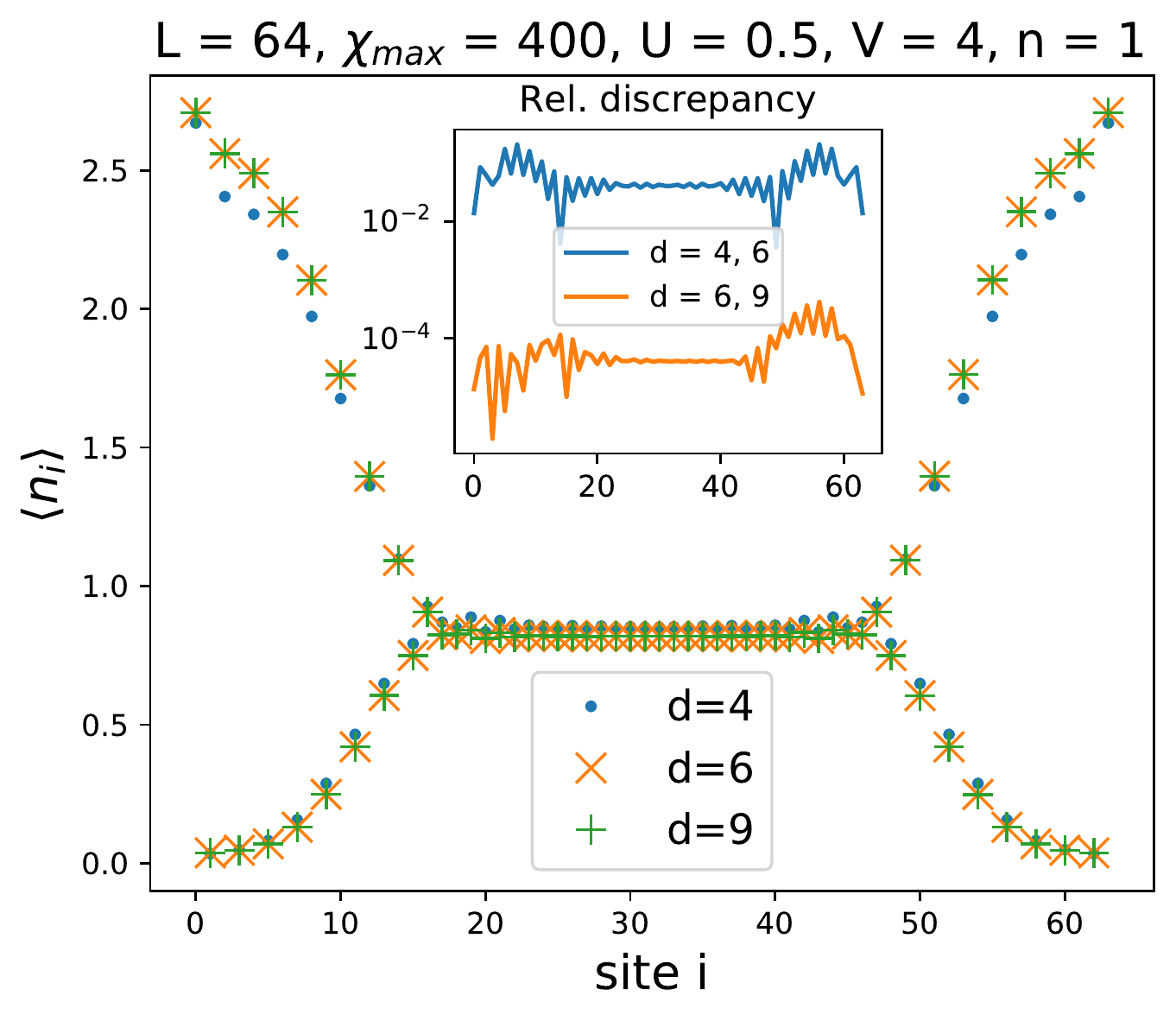}
\caption{
Comparison of the density distribution for different local Hilbert space truncations. We find no qualitative difference between $d=6$ and $d=9$. Inset: Quantitative analysis by comparing the relative discrepancy $|(\braket{n_i}^{d_1} - \braket{n_i}^{d_2})/\braket{n_i}^{d_2}|$. The absolute value of the peak deviation for $d=6,9$ is approx. $0.00025$, corresponding to a peak discrepancy of $0.04$\%.
}
\label{fig:appendix-d-scaling}
\end{figure}

\begin{figure}
\centering
\includegraphics[width=.77\textwidth]{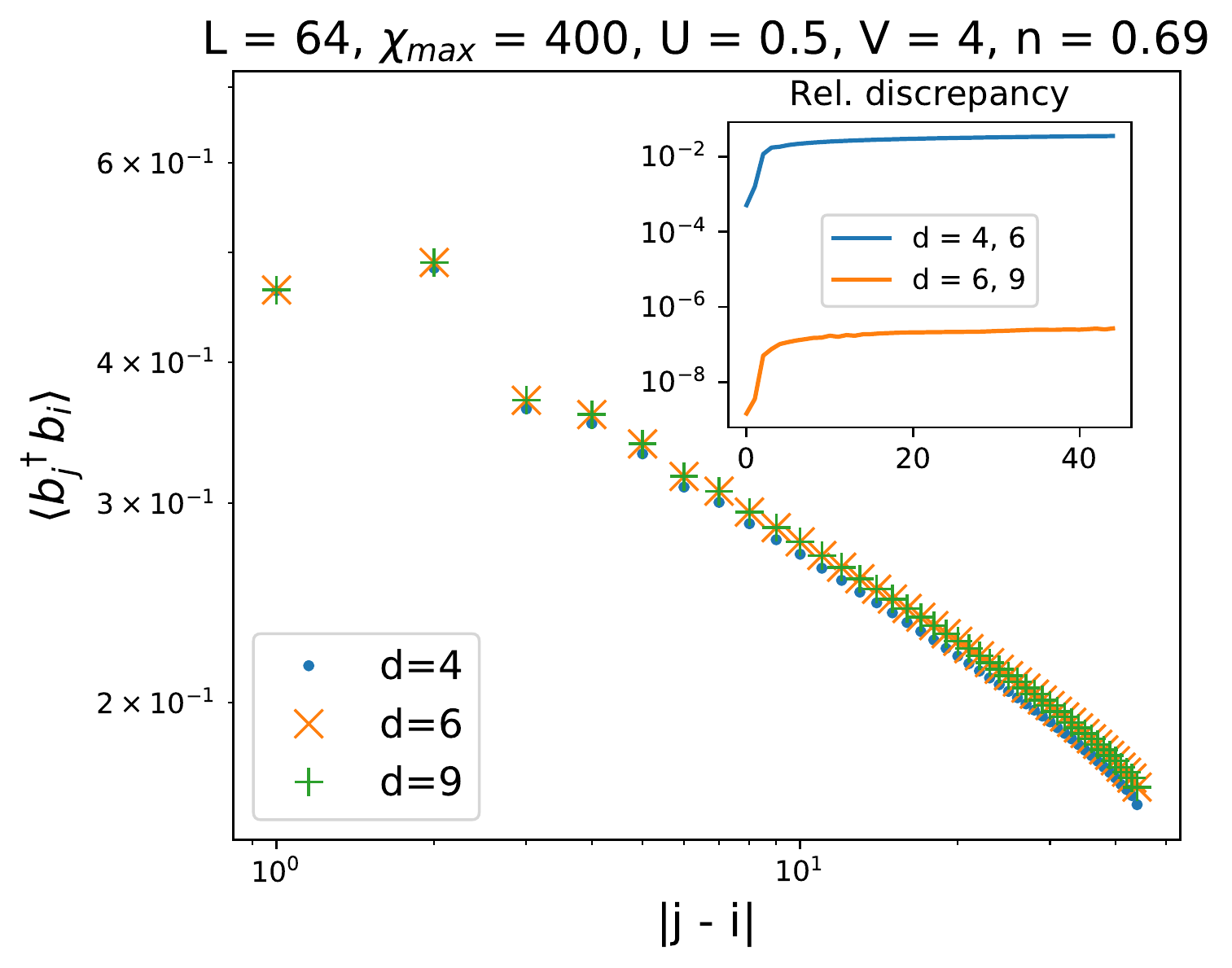}
\caption{
Comparison of $\braket{b_i^\dagger b_j}$ in the bulk for different local Hilbert space truncations. The physical parameters are set for the superfluid phase. Again, we find no qualitative difference between $d=6$ and $d=9$. Inset: Quantitative analysis by comparing the relative discrepancy.
}
\label{fig:appendix-d-scaling2}
\end{figure}

We give evidence to our claim in the main text, that we find no sgnificant difference between simulations with local Hilbert space dimension $d=6$ and $d=9$. As a first relevant example, we compare the density distribution for a fixed filling $n=1$ in the phase separated phase in \cref{fig:appendix-d-scaling}. Qualitatively, the cases $d=6$ and $d=9$ yield no visible deviation. We confirm this also quantitatively in the inset, where we see relative deviations on the order of $10^{-4}$. As a second example, we compare the correlation function $(\CSF)_{ij} = \braket{b_j^\dagger b_i}$ for a superfluid ground state with again the same local Hilbert space dimensions $d$ in \cref{fig:appendix-d-scaling2}. We find again no qualitative differences between $d=6$ and $d=9$, as well es quantitative discrepancies on the order of $10^{-8}$. Overall, we conclude that $d=6$ is sufficient for the effects that we study in the main text.

\section{Comparison SF and SS}
\begin{figure}
\centering
\includegraphics[width=.77\textwidth]{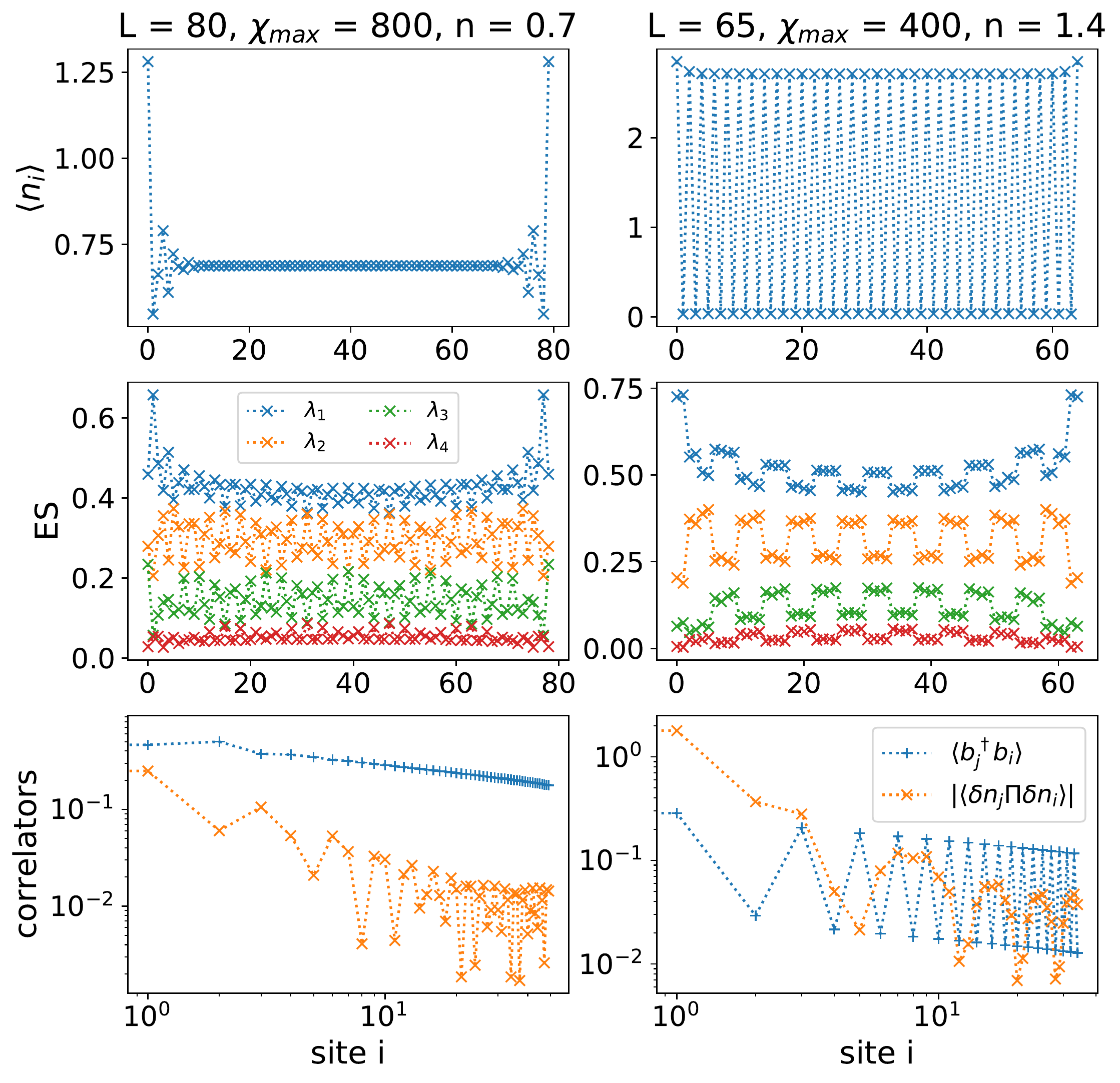}
\caption{
Comparison of the SF (left panel) and SS (right panel) phase at incommensurate fillings with oscillating patterns. a) The main difference of both phases is the flat vs. the solid pattern in the spatial density. b) The entanglement spectrum (ES) shows spatial oscillations. In the case of the SS, the oscillating nodes are pairs of sites. c) Both phases are superfluid with an algebraic decay of $\CSF = \braket{b_j^\dagger b_i}$. Further, the hidden spatial oscillations can be unveiled by measuring $\Cst = \braket{\delta n_j \Pi \delta n_i}$ with $\Pi = \exp\left(-i \pi \sum_{0 \leq l < j} \delta n_l\right)$ and $\delta n_l =  n_l - n$. Note that the frequency and shape of these oscillations changes with the filling $n$, what we display are just two examples.
}
\label{fig:appendix-SF-SS}
\end{figure}

We give more details on the characterization of the (homogeneous) superfluid and supersolid phases at incommensurate fillings. 
The main property that distinguishes the superfluid for fillings below the phase separation phase and supersolid for fillings above the phase separation phase, is the solid pattern in density as can be seen in \cref{fig:appendix-SF-SS}. However, they share unexpected properties that manifest themselves in the entanglement spectrum or string correlators as is discussed in the main text for the SF phase. The main property is that the entanglement spectrum shows spatial oscillations. In the case of the SS, the oscillating nodes are pairs of two sites. Further, both phases are superfluid with an algebraic deca in $\CSF = \braket{b_j^\dagger b_i}$. In both cases, the hidden pattern can be unveiled by looking at the string order correlator $\Cst = \braket{\delta n_j \Pi \delta n_i}$ with $\Pi = \exp\left(-i \pi \sum_{0 \leq l < j} \delta n_l\right)$ and $\delta n_l =  n_l - n$.

\section{Translational invariance of the superfluid phase}
\begin{figure*}
\centering
\includegraphics[width=.95\textwidth]{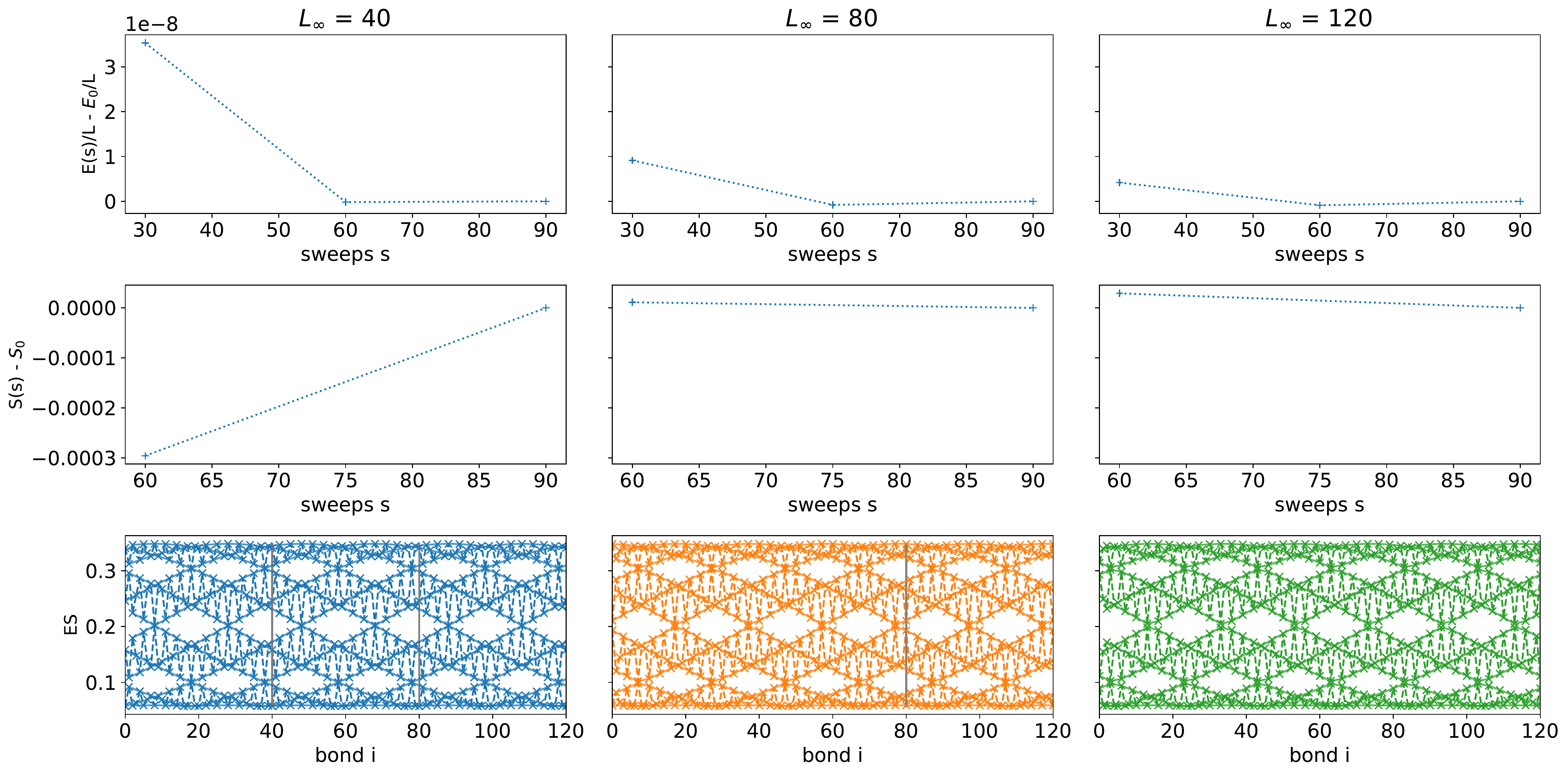}
\caption{
iDMRG convergence for the filling $n=13/20=0.65$. We check the convergence of the energy per site $E/L_\infty$ and mean entanglement entropy $S$ with respect to the number of sweeps through the unit cell for sizes $L_\infty = 40, 80, 120$. The bottom row shows the emerging entanglement spectrum that matches perfectly in all three cases. The used bond dimension is $\chimax = 400$ and the convergence criteria are to either have the relative change in energy be smaller than $10^{-10}$ or reaching a maximum number of $1000$ sweeps. The vertical grey line indicates where the unit cell is repeated.
}
\label{fig:appendix-iDMRG-0.65}
\end{figure*}

\begin{figure*}
\centering
\includegraphics[width=.95\textwidth]{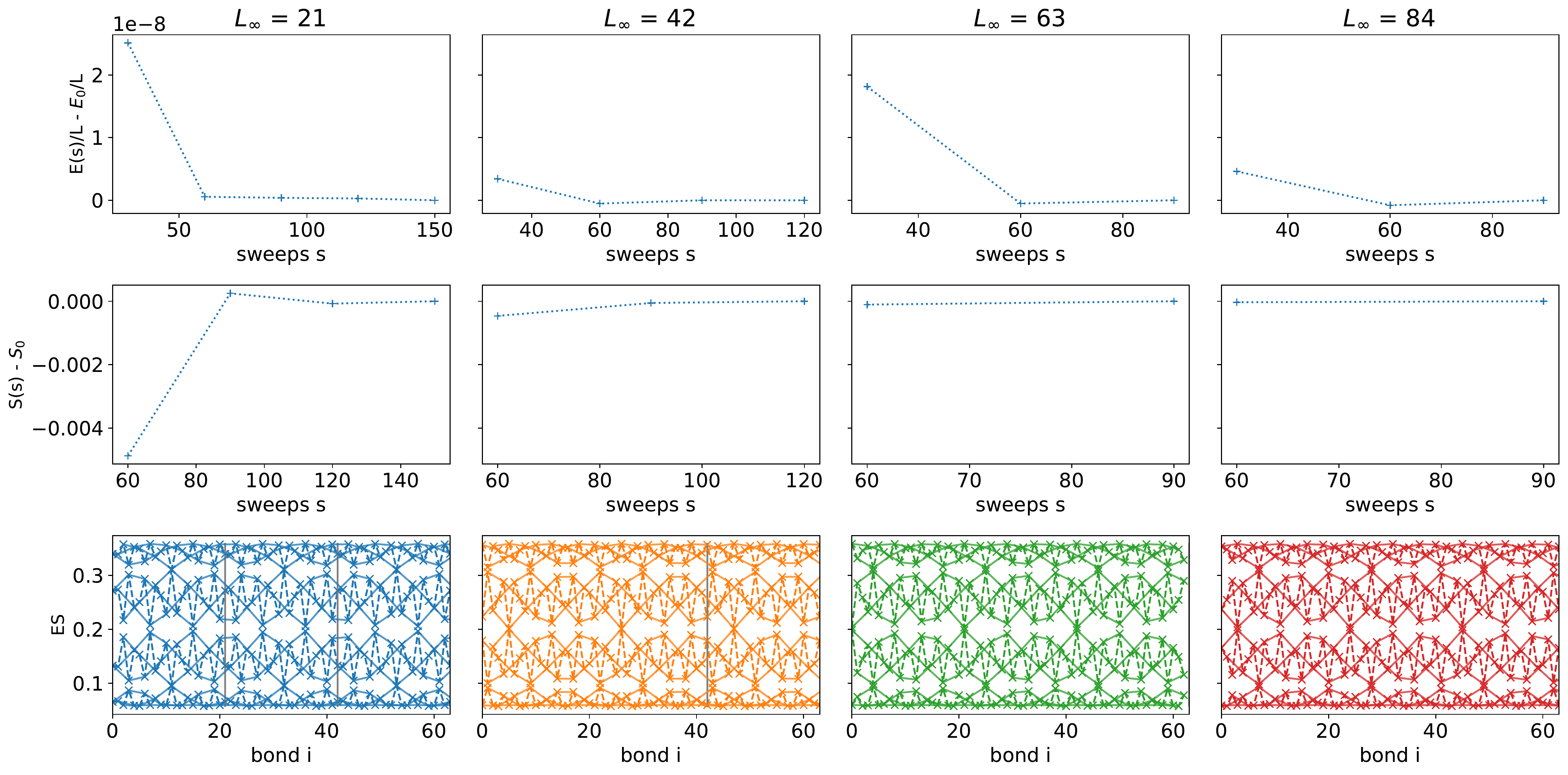}
\caption{
iDMRG convergence for the filling $n=13/21\approx 0.619$. We check the convergence of the energy per site $E/L_\infty$ and mean entanglement entropy $S$ with respect to the number of sweeps through the unit cell for sizes $L_\infty = 21, 42, 63, 84$. The bottom row shows the emerging entanglement spectra that match well in all four cases. The used bond dimension is $\chimax = 400$ and the convergence criteria are to either have the relative change in energy be smaller than $10^{-10}$ or reaching a maximum number of $1000$ sweeps.
}
\label{fig:appendix-iDMRG-0.62}
\end{figure*}

\begin{figure*}
\centering
\includegraphics[width=.77\textwidth]{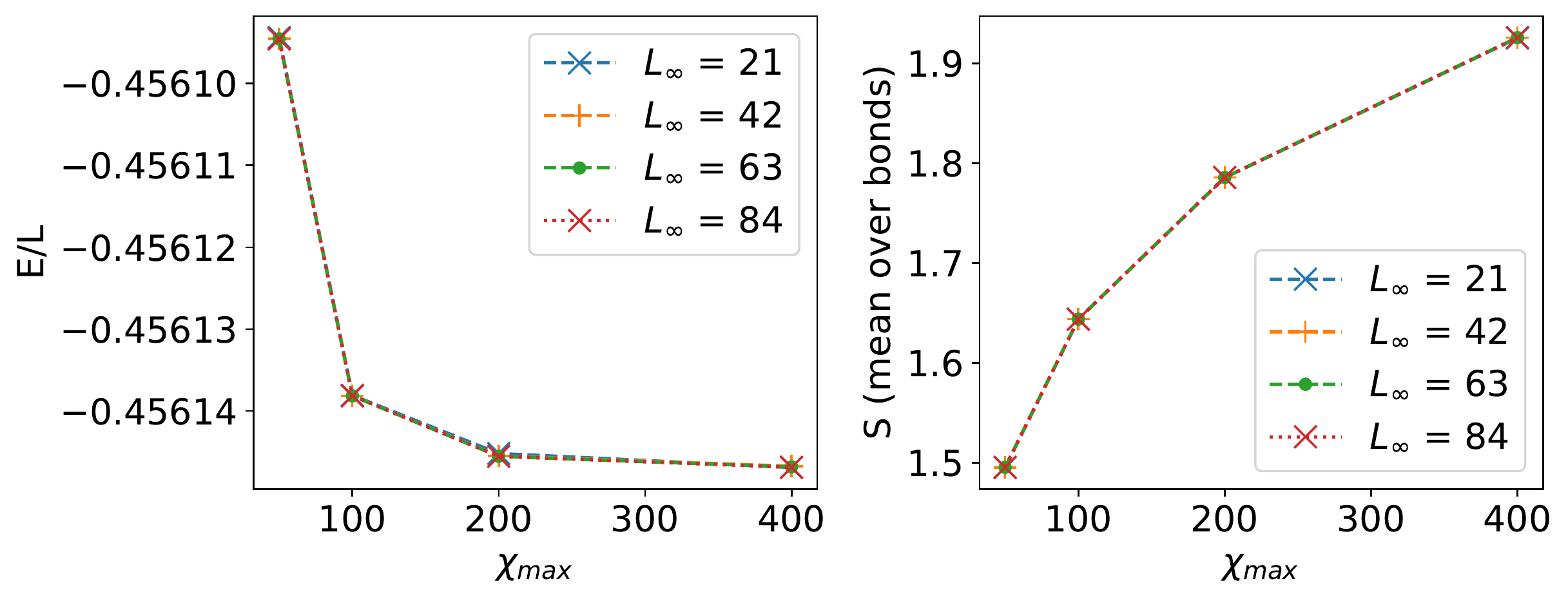}
\caption{
Energy per site and mean entanglement entropy $S$ for different bond dimensions $\chimax$. In terms of energy, we are reaching saturation for the bond dimension $\chimax = 400$ that we are using. In terms of the entanglement entropy, we see a logarithmic grows as is expected for critical phases.
}
\label{fig:appendix-iDMRG-0.62_bonds}
\end{figure*}

\begin{figure*}
\centering
\includegraphics[width=.95\textwidth]{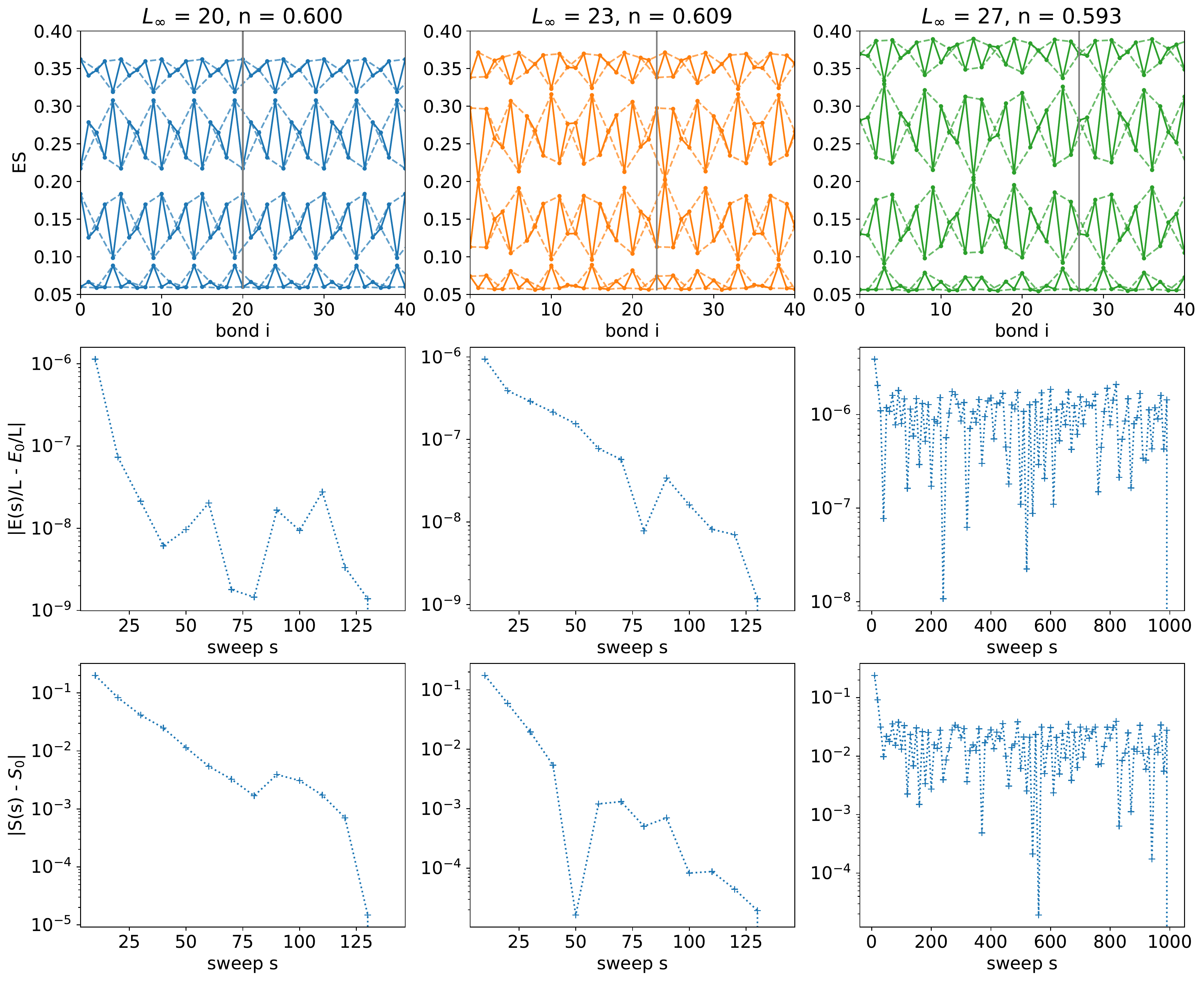}
\caption{
iDMRG convergence for the filling $n=3/5 = 0.6$. The convergence criteria of relative change in energy be smaller than $10^{-10}$ is not met for $L_\infty = 27$ with the optimization terminating at the maximum number of sweeps $1000$. For $L_\infty=20, 23$ the criteria is met, however the optimization is not monotonic. From the entanglement spectra, it seems there is some strain and the unit cell sizes cannot accommodate the desired spatial period the system is trying to establish.
}
\label{fig:appendix-iDMRG-0.6}
\end{figure*}

\begin{figure*}
\centering
\includegraphics[width=.99\textwidth]{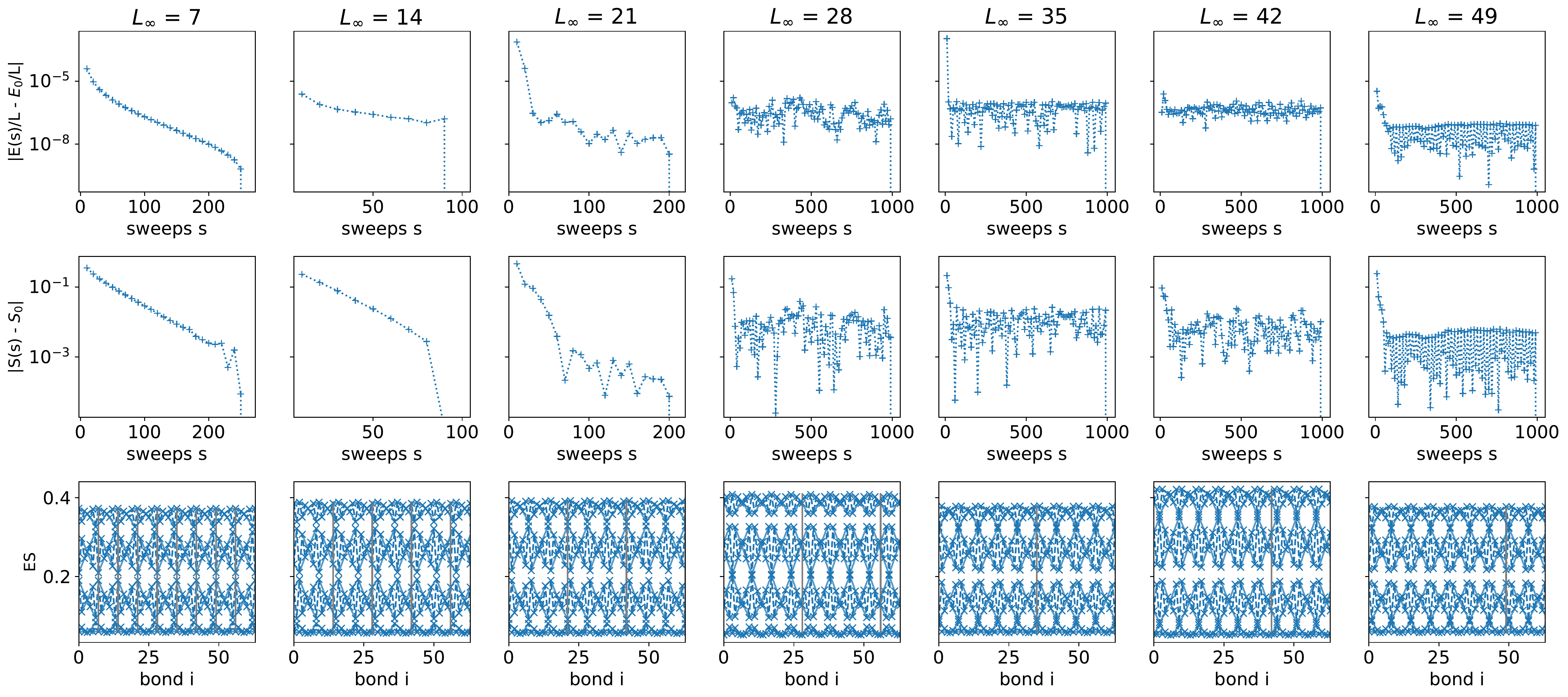}
\caption{
iDMRG convergence for the filling $n=4/7\approx0.571$ for system sizes that are multiples of $7$. Again, we find that strain is causing convergence problems. For $L_\infty = 7, 14, 21$ it meets the convergence criteria but fails for $L_\infty = 28$ and beyond. The convergence for $L_\infty = 7, 14, 21$ might be misleading and might only arise because of a lack of space for strain to emerge.
}
\label{fig:appendix-iDMRG-0.571}
\end{figure*}

\begin{figure*}
\centering
\includegraphics[width=.7\textwidth]{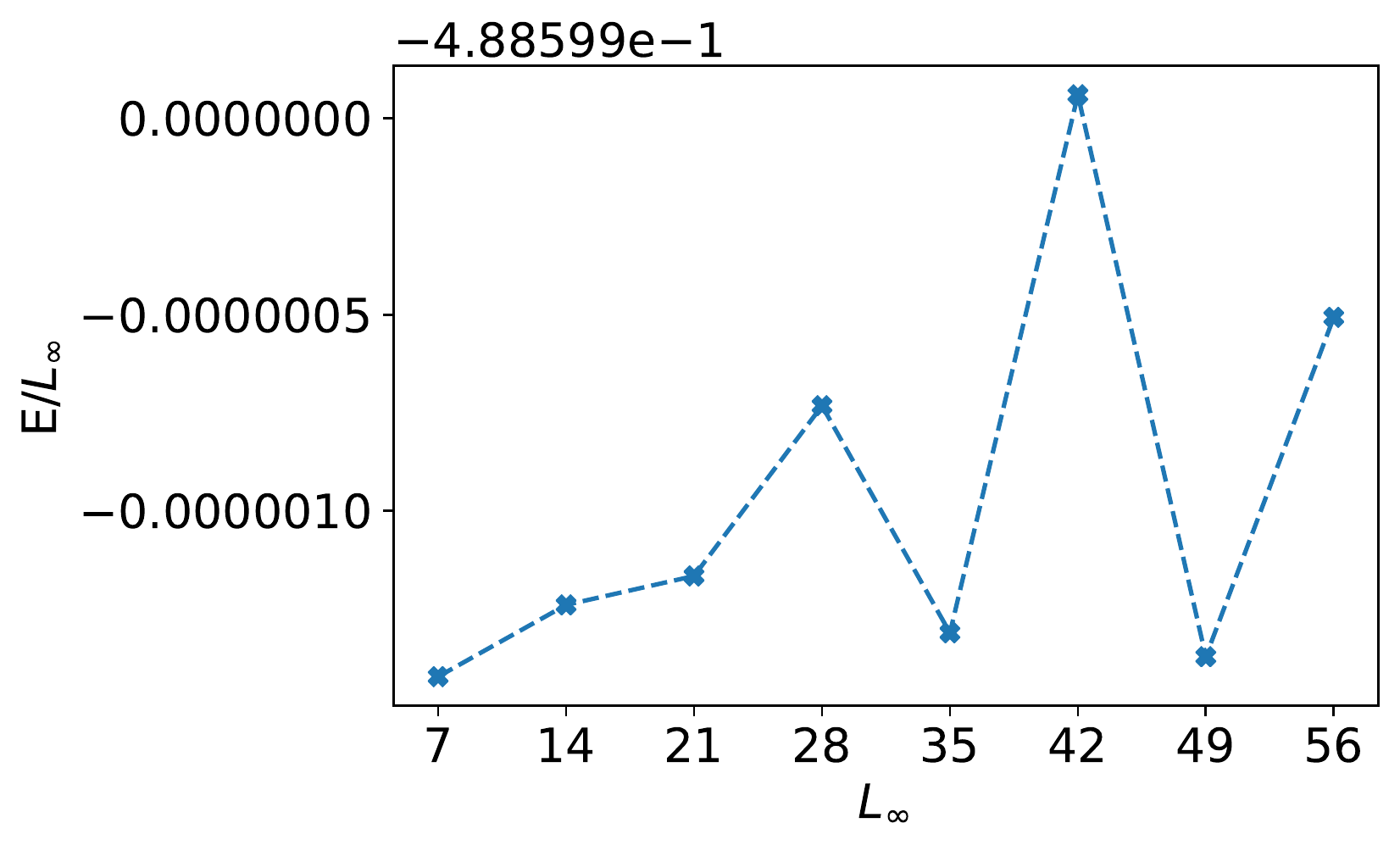}
\caption{
Energy per site for $n=4/7$ with the unit cell being a multiple of $7$. Corresponding convergences are displayed in \cref{fig:appendix-iDMRG-0.571}. Despite strain and convergence problems, the finale ground state energies are still well within range of each other.
}
\label{fig:appendix-iDMRG-0.571-energy}
\end{figure*}

\begin{figure*}
\centering
\includegraphics[width=.95\textwidth]{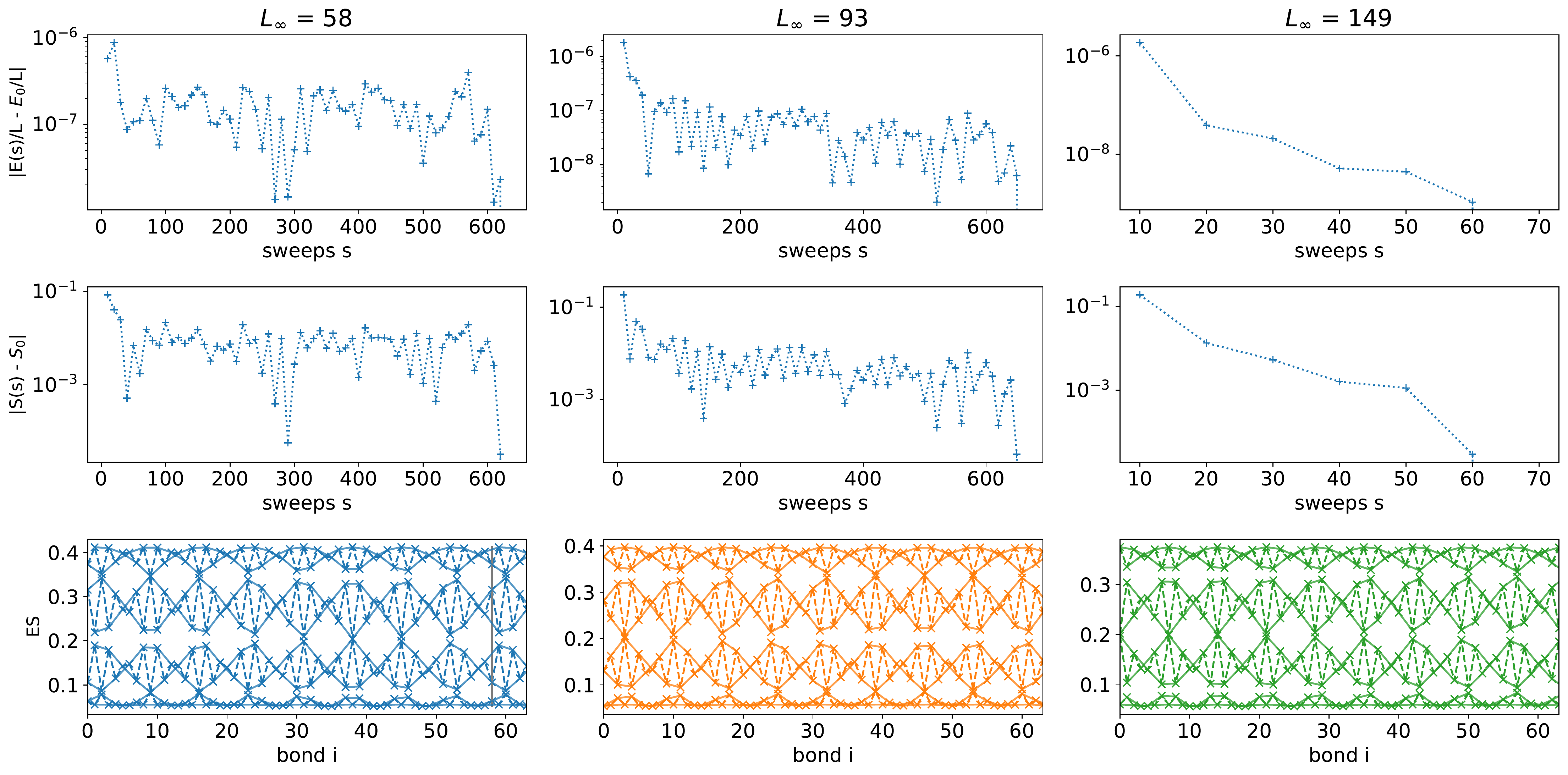}
\caption{
iDMRG convergence for $n=\approx4/7\approx0.571$ with system sizes that are not multiples of $7$. We look at $n=33/58\approx0.569$, $n=53/93\approx0.570$ and $n=85/149\approx0.570$. In all cases the convergence criteria is eventually met. However, the irregular entanglement spectra patterns indicate that also here the unit cell sizes cannot accommodate well the desired spatial pattern.
}
\label{fig:appendix-iDMRG-0.571-incommensurate}
\end{figure*}

\begin{figure*}
\centering
\includegraphics[width=.95\textwidth]{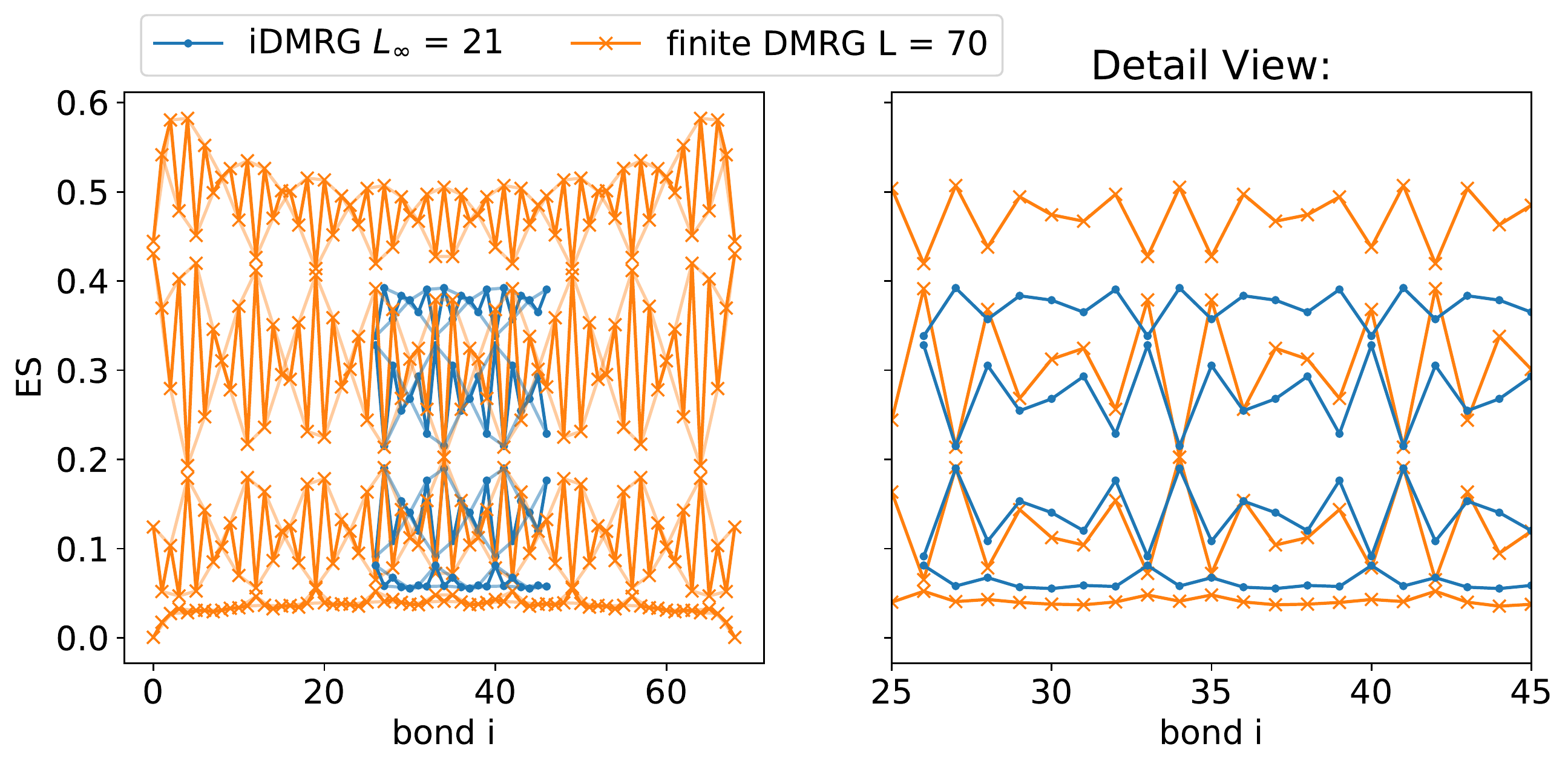}
\caption{
Comparison of finite and infinite DMRG simulations for $n=4/7$ with strain. Even though the unit cell size is not optimal for the targeted filling, the spatial pattern of the infinite system is still matching well with the bulk of the finite system.
}
\label{fig:appendix-iDMRG-0.571-comparison}
\end{figure*}

The emergence of the spatial pattern in the superfluid and supersolid phase close to the phase separation is a manifestation of a broken translational symmetry. To further strengthen this point, we perform some checks with iDMRG. We expect that iDMRG is running into problems when the chosen unit cell size is incommensurate with the spatial period of the entanglement spectrum of the system. This is indeed what we find. The spatial period depends on the targeted filling. For some fillings, and thus for some spatial periods, it has proven harder or easier to find a suitable unit cell size.
We start with the case $n=13/20=0.65$ that is shown in the main text. We tried with $L_\infty = 40, 80, 120$ and reached good convergence in all cases as is displayed in \cref{fig:appendix-iDMRG-0.65}. There is no sign of strain as the emerging pattern is regular and repeats perfectly. In this case, the bond dimension is chosen to be. Similarily, we get good results for $n = 13/21 \approx 0.619$ as shown in \cref{fig:appendix-iDMRG-0.62} for $L_\infty = 21, 42, 63, 84$. We also check the dependence on the bond dimension in \cref{fig:appendix-iDMRG-0.62_bonds} and find sufficient convergence for $\chimax = 400$, that we use throughout this analysis.
We can see how strain is manifested in the ES pattern and how this leads to convergence problems in the example of $n=3/5=0.6$ in \cref{fig:appendix-iDMRG-0.6}. For $L_\infty = 27$ it does not converge at all, the energy is oscillating until the maximum number of sweeps (1000) is reached. For $L_\infty=20$ and $23$, the convergence criteria is met, but we can still see some non-monotonicities - however on a much smaller scale. It seems there is a certain tolerance to strain when the missmatch is not too large.
For filling $n=4/7=0.579$ it proved very difficult to find a suitable unit cell size. We check for unit cells that are multiples of $7$ in \cref{fig:appendix-iDMRG-0.571}. While the convergence criteria is not met for $L_\infty = 28$ and beyond, we still find comparable ground state energies for uneven numbers of sites, as seen in \cref{fig:appendix-iDMRG-0.571-energy}. We also tried for system sizes that are not multiples of $7$ in \cref{fig:appendix-iDMRG-0.571-incommensurate}. While the convergence criteria is eventually met here, the irregular pattern in the entanglement spectrum indicates that also here the unit cell sizes are not optimal to accommodate the desired spatial pattern. We conjecture that with increasing unit cell sizes, a suitable size can be approached, but is in practice difficult or infeasible to simulate due to large system sizes. Curiously, even for these incommensurate sizes, the match between infinite and finite DMRG is still well, as can be seen in \cref{fig:appendix-iDMRG-0.571-comparison}.

To summarize: When simulating the system with iDMRG, one encounters problems with convergence whenever the unit cell size is incommensurate with the spatial period of the system, and converges well when the unit cell size is commensurate. Overall, this behavior is what one would expect of a system with broken translational symmetry strengthens the point that the effect is indeed physical.

\newpage


\newpage 

\

\newpage

\chapter*{\centering Acknowledgements}
\addchaptertocentry{Acknoweldgements}
I would like to thank my supervisors Maciej and Toni for their generous support and the academic freedom they grant, all the members of my groups for the kind, collaborative and supportive environment, and my academic collaborators Patrick Huembeli, Friederike Metz, Joana Fraxanet, Niccolo Baldelli, Niklas Käming, Anna Dawid, Alexandre Dauphin, Christof Weitenberg, Andreas Haller, Gregory Astrakharchik, Philippe Corboz, Pedro Cruz, Luke Mortimer, Tomasz Szoldra, Piotr Sierant, and Jakub Zakrzewski for their patience and insights during meetings and discussions.

\vspace{10pt}

A special thanks goes out to Johannes Hauschild, whose relentless efforts to develop, maintain and support TeNPy has substantially enabled me to perform my research.

\vspace{10pt}

I want to thank Adri, Teo and Erik for being awesome students and Paolo Stornati for being a great co-supervisor. I also want to thank Daniel Gonzalez, Joseph Bowles, Patrick Huembeli, Emanuele Tirrito, Alexandre Dauphin, Gregory Astrakharchik, Pietro Massignan, Anna Dawid, and Paolo Stornati for their patience and helping me whenever I had a question.

\vspace{10pt}

I want to thank the people at Xanadu for hosting me in Toronto and making the last months of my PhD so special. Particularly, I had a great time working directly with Josh, Antal, Romain, Edward, Albert, Maurice and Luis and have learned a great deal in such a short time from them.

\vspace{10pt}

I want to acknowledge the support of my friends and colleagues at ICFO, Paolo Abiuso, who has been the constant rock through this PhD for me from day one - Matteo Scandi, with whom I share a home and a cat, and whose excellent cooking has sparked joy even on the grimmest of days - Joana Fraxanet, with whom I can speak as openly and freely about anything at any time - Jacopo Surace, who challenges my fundamental values in the most casual way day in day out - "the old generation" consisting of Ivan Supic, Flavio Baccario, Alexia Salavrakos, Boris Bourdoncle, Joseph Bowles, Matteo Lostaglio, Mohammad Mehboudi, Alejandro Pozas-Kerstjens, Felix Huber, Patrick Huembeli and Jan Kolodynski that made my arrival in Barcelona exceptionally smooth - and Ivan Milenkovic, my best mate in Barcelona.

\vspace{10pt}

I want to thank my friends Fabi, Markus, Ines, Maria, Henning, Freddy, Leo and Engel for their friendship, open doors, open ears and moral support despite the long distance, and helping me keep my sanity during the last four years, especially during the pandemic.

\vspace{10pt}

I want to thank Morghan for her strength, kindness and righteousness, which I strive to match one day.

\vspace{10pt}

I truly appreciate all of you.

\vspace{10pt}

\begin{otherlanguage}{german}

Ich hab das große Glück und Privileg in einer stabilen und liebevollen Familie in einer der wohlhabensten und sichersten Gegenden der Welt geboren worden zu sein. Ich wähne mich extrem glücklich ein Teil einer so starken Großfamilie wie den Mayers zu sein und möchte ganz besonders meiner Tante Diemut für ihre niemals müde Unterstützung danken. Ich bin ebenfalls dankbar für meine Gotta Marianne, sowie den Blaschkos Franz, Brigitte, Felix und Jan, die ebenso zu diesem einzigartigen heimatlichen Umfeld gehören.

Ich kann mir kein geduldigeren, liebenswürdigeren und großzügigeren großen Bruder als Jakob vorstellen, mein Vorbild menschlich wie akademisch. Wann immer schwierige Entscheidungen im Leben anstehen, weiß ich mit wem ich mich beraten kann.

Zuletzt möchte ich Mama und Papa danken, von denen ich weiß, dass sie immer für mich da sind falls ich sie brauch, die mir immer vertraut und an mich geglaubt haben, und die mir stets ein Gefühl von maximaler Sicherheit vermittelt haben, was mir schlussendlich maßgeblich durch einen Karrierepfad geholfen hat der voll von Unsicherheiten ist.

\end{otherlanguage}

\end{document}